\documentclass[final,1p]{elsarticle}
\usepackage{afterpage,epstopdf}
\usepackage{longtable}\def\chem#1#2{$\rm{}^{#1}\kern-0.8pt#2$}

\begin{document}

\def\reac#1#2#3#4#5#6{$\rm\,{}^{#1}\kern-0.8pt{#2}\,({#3}\,,{#4})\,{}^{#5}\kern-0.8pt{#6}\,$}
\def\gsimeq{\,\,\raise0.14em\hbox{$>$}\kern-0.76em\lower0.28em\hbox{$\sim$}\,\,}
\def\lsimeq{\,\,\raise0.14em\hbox{$<$}\kern-0.76em\lower0.28em\hbox{$\sim$}\,\,}
\def\DJ{D\hspace{-7pt}-\hspace{4pt}}
\def\PRpp#1{[{#1} pages]}

\begin{frontmatter}
\title{NACRE II: an update of the NACRE compilation of charged-particle-induced thermonuclear reaction rates for nuclei with mass number $A < 16$}
\author{Y. Xu$^{a,}$\fnref{fn1}}
\author{K. Takahashi$^{a,b}$}
\author{S. Goriely$^{a}$}
\author{M. Arnould$^{a,}$\corref{MA}}
\address{$^a$Institut d'Astronomie et d'Astrophysique, Universit\'e Libre de Bruxelles, Belgium \\
  $^b$GSI Helmholtzzentrum f\"ur Schwerionenforschung, Darmstadt, Germany }
\author{M. Ohta$^{c,d}$}
\author{H. Utsunomiya$^{d}$}
\address{$^c$Hirao School of Management, Konan University, Kobe, Japan\\
          $^d$Department of Physics, Konan University, Kobe, Japan}
\cortext[MA]{Corresponding author. E-mail address: marnould@ulb.ac.be (M. Arnould)}
\fntext[fn1]{Current address: Cyclotron Institute, Texas A\&M University, College Station, TX 77843, USA }

\begin{abstract}
 An update of the NACRE compilation [Angulo et al., Nucl. Phys. A  656 (1999) 3] is presented.  This new compilation, referred to as NACRE II, reports thermonuclear reaction rates for 34 charged-particle induced, two-body exoergic reactions on nuclides with mass number $A<16$, of which fifteen are particle-transfer reactions and the rest radiative capture reactions.
  When compared with NACRE, NACRE II features in particular (1) the addition to the experimental data collected in NACRE of those reported later, preferentially in the major journals of the field by early 2013, and (2) the adoption of potential models as the primary tool for extrapolation to very low energies of astrophysical $S$-factors, with a systematic evaluation of uncertainties.

 As in NACRE, the rates are presented in tabular form for temperatures in the $10^6 \lsimeq T \leq 10^{10}$ K range.  Along with the 'adopted' rates, their low and high limits are provided. The new rates are available in electronic form as part of the 
Brussels Library (BRUSLIB) of nuclear data.  The NACRE II rates also supersede the previous NACRE rates in the Nuclear Network Generator (NETGEN) for astrophysics. [http://www.astro.ulb.ac.be/databases.html.]
\end{abstract}
\end{frontmatter}
\clearpage
\section{Introduction}
\label{SectIntro}
The series of publications of William A. Fowler and his collaborators, starting with their pioneering work in 1967 \cite{FCZ67} and extending up to 1988 \cite{CF88} (often referred to as CF88), demonstrated in the most vivid way the pivotal role played by compilations of charged-particle induced thermonuclear reaction rates at sub-Coulomb energies in the fields of stellar structure and evolution models as well as of nucleosynthesis investigations.

The so-called NACRE (Nuclear Astrophysics Compilation of REactions) database \cite{NACRE} marked the beginning of a second generation of such astrophysics-oriented compilations.   Since its publication in 1999, the compilation has indeed been used in great many astrophysical model calculations.  Its aim was to supersede the previous compilations not only by using newly available experimental data as inputs, but also by introducing fundamentally new aspects to their format.  Among others, NACRE featured (1) explicit references to the sources of the experimental data (and to some theoretical works) considered; (2) a documentation on the procedure of evaluation of those data; (3) the assessment of uncertainties in the reaction rates, with the lower and higher limits being presented along with the 'adopted' values, and (4) a tabular presentation of the rates for temperatures in the $10^{6} \lsimeq $T$ \leq 10^{10}$ K range.  (See \cite{NACRE} for more details).

Slightly more than half of the CF88 rates were re-compiled in NACRE on the basis of a careful evaluation of experimental data that became available by mid-June 1998.  Comprised in NACRE is  an ensemble of 86 charged-particle induced reactions on stable targets up to Si involved in Big Bang nucleosynthesis and in the non-explosive H- and He-burning modes, complemented with a limited number of reactions of special astrophysical significance on the unstable \chem{3}{H}, \chem{7}{Be}, \chem{13}{N}, \chem{22}{Na} and \chem{26}{Al} nuclides.\footnote{The NACRE homepage http://pntpm.ulb.ac.be/Nacre is also accessible through the BRUSLIB web-site http://www-astro.ulb.ac.be.}

Since NACRE, many cross sections of astrophysical interest have been measured or re-measured, and additional efforts have been put forth toward better predictions of the required reaction rates. In particular: thermonuclear reaction rates of relevance to the Big Bang nucleosynthesis have been re-evaluated with the use of the R-matrix method \cite{Desc04}; the current status of experimental and theoretical studies of astrophysical $S$-factors of solar fusion reactions have been thoroughly surveyed \cite{RMP}; an extended re-evaluation of reaction rates of charged-particle induced reactions on nuclei with mass number in the $14 \leq A \leq 40$ range together with the associated uncertainties has been prepared with the help of a Monte-Carlo simulation [6 - 9]. 

In parallel to those developments, it was thought in 2004 that the time was ripe for an update and an extension of NACRE.  This project, referred to as NACRE II, was launched through a formal collaboration between the Konan University (Kobe, Japan) and the Universit\'e Libre de Bruxelles (Brussels, Belgium).  Several preliminary accounts of NACRE II have appeared sporadically [10 - 16].\footnote{The Konan-ULB Collaboration formally ended at the end of March, 2009. An unfortunate incident, however, incapacitated the compilation work from being completed at that time. The NACRE-II project was resumed in the present form in September, 2009.} 
 In the meantime, the initial work programme has been adapted as much as possible in order to minimize the overlap with the other compilation works. The consequent compilation has a scope and/or pursues a course different from those of \cite{Desc04,RMP}, and complements the most recent compilation [6 - 9] by considering reactions on 'target' nuclides with mass numbers $A < 16$. More specifically: the current version of NACRE II surveys 34 two-body exoergic reactions (15 particle transfer and 19 radiative capture reactions), and adopts potential models to phenomenologically describe and extrapolate resonant and non-resonant reaction cross sections at low energies of interest.      

 Section 2 briefly reviews the theoretical models underlying the present compilation, and the procedure followed for evaluating the reaction rates.  Section 3 is composed of subsections, each of which is designated for the results for a specific reaction.  Section 4 presents a short summary. Appendices supplement the main text with the tables of the adopted values of the model parameters, and of the reverse two-body reaction rates. 

%
\section{The Method}
\subsection{The quantities in quest}
\label{sectQuantity}
The thermonuclear reaction rates of a two-body reaction $A(a,b)B$, which are in quest for astrophysical modellings, are canonically expressed by the Maxwellian-averaged rate $<\sigma v>$ times the Avogadro number $N_{\rm A}$ (e.g.\,\cite{Clayton,RRbook}),
\begin{equation}
 N_{\rm A}<\sigma v> = N_{\rm A} \frac{(8/\pi)^{1/2}}{\mu^{1/2} (k_{\rm B}T)^{3/2}} \int_{0}^{\infty} E\, \sigma(E)\, {\rm exp}[- E/(k_{\rm B}T)]\, {\rm d}E, 
\label{eqrate}
\end{equation}
\noindent
where $\sigma(E)$ is the reaction cross section at the centre-of-mass incident energy $E = \mu v^{2}/2$ with $v$ being the relative velocity and $\mu = m_A m_a / (m_A + m_a)$ the reduced mass with $m_A$ and $m_a$ standing for the masses of target ($A$) and projectile ($a$) nuclei, while $k_{B}$ is the Boltzmann constant and $T$ is the temperature.

In dealing with charged-particle induced reactions at very low energies below the Coulomb barrier, it is more convenient, and indeed customary, to introduce the astrophysical $S$-factor, $S(E)$, which is classically defined as
\begin{eqnarray}
           S(E) = E\, \sigma(E)\, {\rm exp}[2 \pi \eta],
\label{eqSfac}
\end{eqnarray}
\noindent
where $\eta=Z_{A}Z_{a}e^{2}/(\hbar v)$ is the Sommerfeld parameter, with $Z_{A} Z_{a}e^{2}$ being the product of the nuclear charges of $A$ and $a$. This definition allows the $S$-factor to  exhibit a much weaker energy dependence at low energies than that of the cross section itself, easing the comparison between predictions and experimental data, as well as the extrapolation at lower energies. 
  
In practice, the rate is commonly expressed in $\rm{cm^{3} mol^{-1} s^{-1}}$, such that 
\begin{eqnarray}
N_{A}<\sigma v> = 3.73\times10^{10} \hat{\mu}^{-1/2}T_{9}^{-3/2} \int_{0}^{\infty}  E \sigma(E) {\rm exp}[-11.605 E/ T_{9}]\, {\rm d}E, 
\label{eqrate1}
\end{eqnarray}
\noindent
when $E$ and $\sigma(E)$ are in units of MeV and barn, $\hat{\mu}$ denotes the reduced mass in atomic mass unit (amu = $M_{\rm atm}(^{12}$C)/12 = 931.494 MeV/c$^{2}$), and $T_{9}$ is the temperature in units of $10^{9}$ K. Correspondingly, $\eta$ in Eq.\,(\ref{eqSfac}) equals to $0.1575 Z_{A}Z_{a} (\hat{\mu}/E)^{1/2}$.       

A quick reference to the energy around which the integrand of Eq.\,(\ref{eqrate}) becomes maximum can be made when the $S$-factor is constant or nearly so, leading to the so-called Gamow peak energy 
\begin{eqnarray}
E_{0} = (\hat{\mu}/2)^{1/3} {\big (} \pi e^{2}Z_{A}Z_{a} k_{B}T/\hbar{\big )}^{2/3},
\label{eqgamow}
\end{eqnarray}
\noindent
the values of which are plotted in Fig.\,1 for proton- and $\alpha$-induced reactions in the temperature range of $10^{6} \sim 10^{9}$ K.
 The concept of the Gamow peak looses its significance at higher temperatures as its width increases approximately as $0.7 \sqrt{e_{0} T_{9}}$ MeV, where $e_{0}$ is $E_{0}$ in MeV.
  The heights of the corresponding Coulomb barriers are added for comparison. It is clear that the reaction cross sections at energies far below the Coulomb barrier are generally in quest. They are next to impossible to measure in the laboratory, and thus have to be obtained by extrapolation from the values measured at higher energies, preferably with the help of some theoretical considerations.
\begin{figure} [t]
\centering{
\includegraphics[height=0.55\textheight,width=0.95\textwidth]{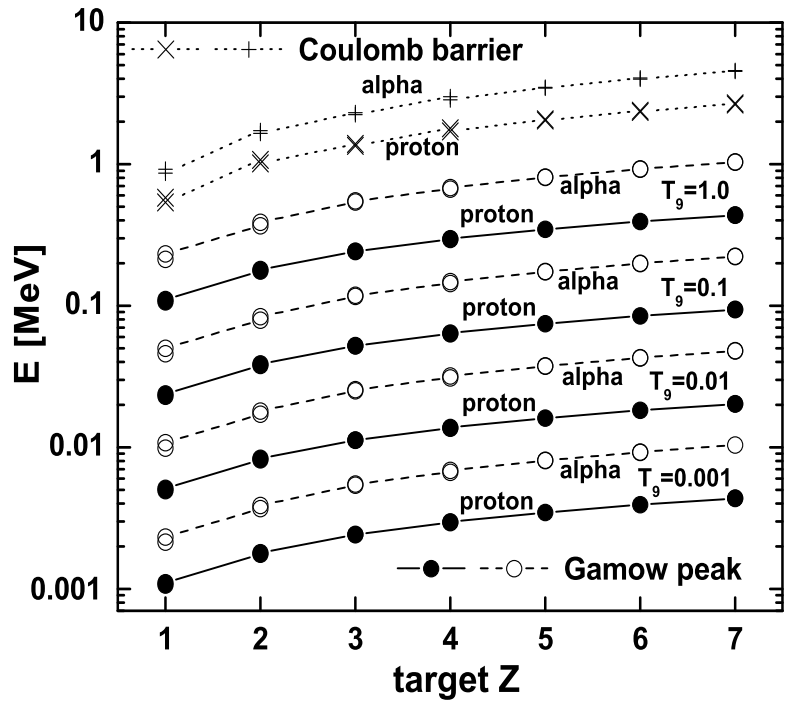}
\vspace{-0.5truecm}
\caption{Gamow peak energies for proton- and $\alpha$-induced reactions at temperatures of $10^{6}-10^{9}$ K in comparison with the Coulomb barrier heights against the atomic numbers $Z$ of the targets. Except $^{3}$H, all the isotopes considered here are stable. Just for illustration, the Coulomb barriers are evaluated at the radii of $1.1 \times(A_{A}^{1/3}+A_{a}^{1/3})$ fm, where $A_{A}$ and $A_{a}$ are the target and projectile mass numbers.
}}
\label{figgamow}
\end{figure}
%
\subsection{Reaction mechanisms}
\label{sectMechanism}
Two extreme mechanisms are considered for low-energy nuclear reactions: the "compound nucleus process" and the "direct reaction process" (with the so-called "pre-equilibrium process" lying in between). In the former, the projectile merges with the target to excite many degrees of freedom  that have time to statistically equilibrate. This first stage, the formation of the compound nucleus, is followed by its decays by particle or photon emissions. The radiative captures of thermal neutrons are typical examples. The observed resonances are very narrow, reflecting the long times needed for the formation (and decays) of the compound nucleus. 

As opposed to the compound nucleus process, the direct reaction proceeds via the excitation of only a few degrees of freedom on a much shorter time scale reflecting the time taken by the projectile to traverse the target. This process has been traditionally associated with energies that are high enough for the mean-free path of the incident particle to be comparable with the size of the nucleus. In these conditions, the particle ejection occurs preferentially at forward angles.  It has become clear, however, that the direct reaction process is also important, and often dominant, in charged-particle induced reactions at the very low energies of astrophysical interest (e.g.\,\cite{oberhummer}). Given the difficulty to penetrate the Coulomb barrier, the reaction may occur before the projectile could tunnel through deep inside the target nucleus. The formation of a compound state is suppressed accordingly, which is reinforced in very light nuclei by the paucity of quasi-bound states. 
%
\subsection{The model of our choice: the potential model}
\label{sectChoice}
There are a few different approaches for describing the nuclear reactions of astrophysical interest at sub-Coulomb barrier energies. What may be considered as two extremes among oft-used ones are the R-matrix method on the purely phenomenological side and the so-called "microscopic cluster models" such as the resonating group method or the generator coordinate method on the side closer to first principles. Both approaches have been subject to much scrutiny in the last decades.  In between the two extremes lies the potential model. All of these and some other models clearly have advantages and disadvantages (see [19 - 21] 
 for concise summaries). Furthermore, some {\it ab initio} many-body approaches 
 have also been applied in certain cases, the references for which shall be noted in the respective subsections in Sect.\,3. 

In the present work, we adopt the potential model as the major tool for supplementing the experimental data. Doing so, we hope that most of the non-resonant as well as resonant contributions to cross sections at very low energies could be effectively described by a direct reaction model. In practice, we adopt the zero-range Distorted-Wave Born Approximation (DWBA) for particle transfer reactions, and an extended "Direct Capture" model for radiative capture reactions. In what follows, we reserve the term "Potential Model (PM)" exclusively for radiative capture reactions, whereas the term "DWBA" is used for transfer reactions.
\subsubsection{The radial wave functions}
\label{sectRadial}
Various wave functions requested in the potential models that follow are obtained by solving two-body Schr\"{o}dinger equations, the radial parts of which may be expressed in the relative coordinate $r$ as
\begin{eqnarray}
[\frac{d^{2}}{dr^{2}} - \frac{L(L+1)}{r^{2}} + \frac{2 \mu}{\hbar^{2}}\{ E - V(r)\}]\psi = 0.
\label{eqschroedinger}
\end{eqnarray}
\noindent
Here, $V(r)$ is a central potential that consists of the nuclear and Coulomb parts ($V = V_{\rm N} + V_{\rm C}$; see \ref{sectPotential}), $L$ is the relative orbital angular momentum, $\mu$ is the reduced mass, and $\psi$ represents the resulting radial wave function, which vanishes at the origin. [We will {\it not explicitly} include spin-orbit couplings into $V_{\rm N}$.] 

For the sake of clarity, we will replace  $\psi$ by $\chi$ in the case of scattering problems ($E > 0$) and by $\phi$ in eigen-value problems ($E < 0$).  For scattering states ($E = \hbar^{2} k^{2} / 2\mu$ with the wave number $k$), the radial wave functions $\chi_{L}(k,r)$ behave asymptotically at large distances where $V_{\rm N}$ is negligible as
\begin{eqnarray}
\chi_{L}(k,r) \longrightarrow \frac{i}{2} [H_{L}^{(-)}(kr) - S_{kL} H_{L}^{(+)}(kr)]{\rm e}^{i \delta_{L}^{c}}
\ \ \ \ {\rm as}\ \  \ r\rightarrow\infty,
\label{eqboundary}
\end{eqnarray}
\noindent
where $H_{L}^{(\mp)}(kr) = [ G_{L}(kr) \mp i F_{L}(kr)]$ are the incoming ($-$) and the outgoing (+) Coulomb wave functions expressed in terms of the regular and the irregular Coulomb wave functions $F_{L}(kr)$ and $G_{L}(kr)$, the Coulomb phase shift is denoted by $\delta_{L}^{c}$, and $S_{kL} = {\rm e}^{2 i \delta_{L}}$ is the scattering matrix for the elastic scattering with $\delta_{L}$ being the phase shift by the nuclear potential. Thus, the right-hand side of Eq.\,(\ref{eqboundary}) becomes ${\rm exp}[i(\delta_{L}^{c}+\delta_{L})] [{\rm cos}(\delta_{L}) F_{L}(kr) + {\rm sin}(\delta_{L}) G_{L}(kr)]$.

For bound states, the radial wave functions must vanish at infinity and be normalised:
\begin{eqnarray}
\phi_{nL}(r) \longrightarrow\ \ 0\ \ \ {\rm as}\ \ r\ \rightarrow\ \infty;\ \ \ \ \ \int_{0}^{\infty} |\phi_{nL}(r)|^{2} dr = 1,
\label{eqbound}
\end{eqnarray}
\noindent
where $n$ standing for the radial quantum number has to be chosen appropriately.
\subsubsection{The DWBA cross section for transfer reactions}
\label{sectDWBAcross}
For the transfer reaction $A(a,b)B$, two processes are considered: the stripping ($a=x+b, A+x=B$) and the pickup ($a+x=b, A=B+x$). Of interest in this compilation the former includes (d,n), (d,p) and ($\alpha$,n) and the latter (p,d) and (p,$\alpha$) reactions. We treat them as reactions transferring p, n, $\tau$ ($^{3}$He nucleus) or $t$ ($^{3}$H nucleus). The transfer is specified by the spin $s$, the orbital angular momentum $l$ and the total angular momentum $j$.  The triangular relations ${\bf l} = {\bf j} - {\bf s}$, ${\bf j} = {\bf J}_{B} - {\bf J}_{A}$ and ${\bf s} = {\bf J}_{a} - {\bf J}_{b}$ hold, where $J_{A}, J_{a}, J_{b}$ and $J_{B}$ are the spins of the four participating nuclei. The residual state is usually, but not limited to, the ground state of the nucleus $B$.

If the interaction of the nuclei $A$ and $a$ is described by a potential just like in the elastic scattering, the particle transfer is understood to occur as the result of the residual interaction. The DWBA treats it as a perturbation, and obtains the transition matrix element by sandwiching it between the distorted waves in the initial and the final channels. 

For simplicity, we adopt the zero-range approximation, namely for the interaction between the transferred particle and the core of the projectile (in a stripping) or ejectile (in a pickup). Furthermore, we do not explicitly include spin-orbit couplings in the distorted waves. Then, the differential cross section, with the normalisation of Eq.\,(\ref{eqboundary}), becomes (e.g.\,\cite{Satchlerbook,code})
\begin{eqnarray}
\frac{d\sigma(\theta)}{d\Omega} = \frac{1}{4\pi} C_{\alpha,\beta} \frac{1}{E E_{f}} \frac{k_{f}}{k} \frac{m_{B}^{2}}{m_{A}^{2}} \sum_{slj} \frac{S_{\rm F} D_{0}^{2}}{(2s+1)} \sum_{m} {\Big{\vert}} t(\theta){\Big{\vert}}^2,
\label{dwbadif1}
\end{eqnarray}
\noindent
where the coefficient $C_{\alpha,\beta}$ equals to $(2J_{B}+1)/(2J_{A}+1)$ for a stripping reaction, and to $(2J_{b}+1)/(2J_{a}+1)$ for a pickup reaction. The quantities  $E (\equiv E_{i})$ and $E_{f}$ are the centre-of-mass energies in the entrance and exit channels, $k(\equiv k_{i})$ and $k_{f}$ are the corresponding wave numbers, $m_{A}$ and $m_{B}$ are the masses of the target $A$ and the residue $B$, $S_{\rm F}$ is the spectroscopic factor, and $D_{0}$ measures the strength of the zero-range interaction. 

 Finally, the amplitude $t(\theta)$ is, with the choice of the beam direction as the $z$-axis, given by
\begin{eqnarray}
t(\theta) =
   \sum_{l_{i},l_{f}} c_{l_{f}}^{|m|} P^{|m|}_{l_{f}}({\rm{cos}}\theta)\ T^{if},
\label{dwbadif2}
\end{eqnarray}
\noindent
where the summations run over the relative orbital angular momenta in the entrance and the exit channels, $l_{i}$ and $l_{f}$. The triangular relation ${\bf l} = {\bf l}_{f} - {\bf l}_{i}$ holds, but $l_{i} + l_{f} + l$ must be even. In front of the associated Legendre polynomial $P_{L}^{M}$, the coefficient $c_{L}^{M} = [(2L+1)(L-M)!/(L+M)!]^{1/2}$ is factored out for convenience. The amplitude $T^{if}$ is given by
\begin{eqnarray}
 T^{if}= c_{i,f} \int\chi_{l_{f}}(k_{f},\frac{m_{A}}{m_{B}}r)\frac{\phi_{nl}^{js}(r)}{r}\chi_{l_{i}}(k,r)dr,
\label{dwbadif3}
\end{eqnarray}
\noindent
where 
\begin{eqnarray}
 c_{i,f} = i^{l_{i}-l_{f}}\,<l_{i}l_{f}l|m0m>\,<l_{i}l_{f}l|000> (2l_{i}+1) (2l_{f}+1)^{1/2}/(2l+1)
\end{eqnarray}
\noindent
with $<j_{1}j_{2}J|m_{1}m_{2}M>$ denoting the Clebsch-Gordan coefficients. The radial wave functions, $\chi_{l_{i}}$ and $\chi_{l_{f}}$, are the solutions of Eq.\,(\ref{eqschroedinger}) for the scattering states in the entrance and the exit channels, respectively. The sandwiched $\phi_{nl}^{js}(r)/r$ is the radial form factor related to the stripped or picked-up species  bound in the nucleus $B$ or $A$, respectively. 

The integration of Eq.\,(\ref{dwbadif1}) over the solid angle provides the cross section
\begin{eqnarray}
\sigma_{J_{B}^{\pi}} (E) = C_{\alpha,\beta} \frac{1}{E E_{f}} \frac{k_{f}}{k} \frac{m_{B}^{2}}{m_{A}^{2}} \sum_{slj}  \frac{S_{\rm F}D_{0}^{2}}{2s+1} \sum_{m} \sum_{l_{f}} {\Bigg{\vert}} \sum_{l_{i}}  T^{if} {\Bigg{\vert}}^2.
\label{eqdwbacross}
\end{eqnarray}
A special care must be taken if the entrance or exit channel is composed of identical nuclei (e.g.\,\cite{BW52}). As long as only one set of $slj$ values is considered, the necessary modification can be done by replacing $|t(\theta)|^2$ in Eq.\,(\ref{dwbadif1}) by $|t(\theta)|^2 + |t(\pi-\theta)|^2 + C_{\rm symm} {\rm Re} [t(\theta)t^{*}(\pi-\theta)]$, where $C_{\rm symm}$ depends on $slj(m)$, the spins of the three participating species, and on the even/oddness of the mass number of the identical species \cite{buttle65}. 
\subsubsection{The PM cross section for radiative capture reactions}
\label{sectPMcross}
In the radiative capture reaction $A(a,\gamma)B$, the transition from the initial scattering state $A+a$ forms the nucleus $B$ with accompanying $\gamma$-ray emission. Of interest in this compilation are (p,$\gamma$), (d,$\gamma$) and ($\alpha$,$\gamma$) reactions\footnote{The following formalism (adapted from \cite{Descbook,potenPhD}) is basically an extension of the canonical "Direct Capture" model \cite{rolfspot}. Along a similar vein, a systematic potential model analysis of radiative neutron- and proton-capture reactions to the daughter ground states has been carried out recently \cite{TEXAS10}.}

The PM calculates the transition matrix element between the initial and the final states in a perturbational manner by sandwiching the electromagnetic operators in the long wave-length limit. The consideration of electric dipole (E1), magnetic dipole (M1) and electric quadrupole (E2) operators suffices for the current purpose. The "final state" is either the ground state of the nucleus $B$ if fed directly or, more generally, one of its excited states before the secondary $\gamma$-ray cascade, which may be specified by its spin $J_{f} (\equiv J_{B})$ and parity $\pi_{f}$. Correspondingly, we denote the total angular momentum and the parity of the initial state by $J_{i}$ and $\pi_{i}$. The spin-parity selection rules for the transition between these states  can be expressed by the triangular relation ${\bf J}_{i} = {\bf J}_{f} - {\bf \lambda}$, with $\lambda$ being the multi-polarity (1 for E1 and M1; 2 for E2), and $\pi_{i}\pi_{f} = -$ for E1, and + for M1 and E2.  

The partial cross section to a given final state can be written as 
\begin{eqnarray}
\sigma_{J_{f}^{\pi}}(E) &=& \frac{2J_{f}+1}{(2J_{A}+1)(2J_{a}+1)} \frac{1}{E k}\nonumber\\
&&  \times  \sum_{I_{f},J_{i},l_{i},I_{i}} S_{\rm F} \{ c_1 k_{\gamma}^{3} ( \vert M_{\rm E1} \vert^{2} + \vert M_{\rm M1} \vert^{2}) +  c_2 k_{\gamma}^{5} \vert M_{\rm E2} \vert^{2}  \},
\label{eqpotcross}
\end{eqnarray}
\noindent
where $c_{\lambda} = 4\pi(\lambda+1)(2\lambda+1)/\{\lambda[(2\lambda+1)!!]^{2}\}$, $k_{\gamma}$ is the wave number of the emitted photon, and $S_{\rm F}$ is the spectroscopic factor. The summations run over the channel spin $I_ {i}$, orbital angular momentum $l_{i}$, and $J_{i}$ of the initial state, and over the final channel spin $I_{f}$, provided that the spin-parity selection rules are obeyed.    

The matrix elements consist of the part related to the radial moment, and those related to the internal moments, if any, of the nucleus $A$ or $a$:
\begin{eqnarray}
&& M_{\rm E1}\, = {\cal{M}}_{\rm E1}\nonumber\\
&& M_{\rm M1} = {\cal{M}}_{\rm M1} + {\cal{M}}_{\rm M1}^{{\rm int}}(A) + {\cal{M}}_{\rm M1}^{{\rm int}}(a)\\
&& M_{\rm E2}\, = {\cal{M}}_{\rm E2} \, + \, {\cal{M}}_{\rm E2}^{{\rm int}}\,(A) + {\cal{M}}_{\rm E2}^{{\rm int}}\,(a).\nonumber
\label{eqpotmatrix}
\end{eqnarray}
The radial parts of the E$\lambda$ and M1 matrix elements are given by 
\begin{eqnarray}
&& {\cal{M}}_{{\rm E}\lambda} =  e [Z_A(\frac{m_a}{m_{A}+m_{a}})^{\lambda}+Z_a(\frac{-m_A}{m_{A}+m_{a}})^{\lambda}] \delta_{I_{i}I_{f}} C_{\lambda}^{if} <l_{i} \lambda l_{f}|000> {\cal{I}}_{\lambda}^{if}\nonumber\\ 
&& {\cal{M}}_{\rm M1} = \mu_{\rm N}\, \frac{Z_{A}m_{a}^{2}+Z_{a}m_{A}^{2}}{m_{A}m_{a}(m_{A}+m_{a})}\, \delta_{I_{i}I_{f}}\delta_{l_{i}l_{f}} C_{1}^{if} [l_{i}(l_{i}+1)(2l_{i}+1)]^{1/2}\, {\cal{I}}_{0}^{if},
\label{eqpotEM}
\end{eqnarray}
\noindent
where $\delta_{\kappa\kappa'}$ stands for the Kronecker symbol,
\begin{eqnarray}
 C_{\lambda}^{if} = (-)^{J_{i}+I_{i}+\lambda+l_{f}}\, i^{l_{i}-l_{f}} {\Big{[}} (2J_{i}+1)(2l_{i}+1) {\Big{]}}^{1/2} {\bigg{\{}} \begin{array}{ccc} J_{i}  &  J_{f} & \lambda  \\ l_{f}  & l_{i}  & I_{i} \end{array} {\bigg{\}}},
\label{eqpotC}
\end{eqnarray}
\noindent
and
\begin{eqnarray}
{\cal{I}}_{\nu}^{if} = \int \phi_{n_{f}l_{f}}(r)\, r^{\nu} \chi_{l_{i}}(E,r)\, dr.
\label{eqpotInt}
\end{eqnarray}
\noindent
In the above, the quantity with the curly brackets is the 6$j$ symbol. An additional triangular relation  ${\bf l}_{i} = {\bf l}_{f} - {\bf \lambda}$ holds. $\cal{M}_{{\rm E}\lambda}$ vanishes when $l_{i} + l_{f} + \lambda$ is an odd number, and so does $\cal{M}_{\rm M1}$ when $l_{i} = 0$. The quantum numbers $n_{f}$ and $l_{f}$ have to be appropriately chosen in consideration of the Pauli principle.

The E2 and M1 matrix elements related to the internal moments of the nucleus $A$ are
\begin{eqnarray}
{\cal{M}}_{\rm E2}^{\rm int}(A) = \sqrt{5/4}\, e Q_{2,A}\, \delta_{l_{i}l_{f}}\, D_{2}^{if}\, {\cal{I}}_{0}^{if},\ \ {\cal{M}}_{\rm M1}^{\rm int}(A) = \sqrt{3}\, \mu_{1,A}\, \delta_{l_{i}l_{f}}\,  D_{1}^{if}\, {\cal{I}}_{0}^{if},
\label{eqpotmuA}
\end{eqnarray}
\noindent
where $eQ_{2,A}$ and $\mu_{1,A}$ are the electric quadrupole and the magnetic dipole moments of nucleus $A$, respectively, and  
\begin{eqnarray}
 D_{\lambda}^{if} &=& (-)^{J_{A}+J_{a}-J_{f}-l_{f}} {\Big{[}}(2J_{i}+1)(2J_{A}+1)(2I_{i}+1)(2I_{f}+1) {\Big{]}}^{1/2}\nonumber\\
&& \times {\bigg{\{}} \begin{array}{ccc} J_{i} & \lambda & J_{f} \\ I_{f}  & l_{f}  & I_{i} \end{array} {\bigg{\}}}
{\bigg{\{}} \begin{array}{ccc} J_{A}  &  J_{a} & J_{A} \\ I_{i}  & \lambda  & I_{f} \end{array} {\bigg{\}}}{\big{/}} <J_{A} \lambda J_{A} \vert J_{A} 0 J_{A}>.
\label{eqpotD}
\end{eqnarray}
\noindent
The internal terms related to the partner nucleus $a$ can be obtained by shuffling the suffices $A$ and $a$ in Eqs.\,(\ref{eqpotmuA}-\ref{eqpotD}). For the internal terms, the additional triangular relation is ${\bf I}_{i} = {\bf I}_{f} - {\bf \lambda}$.    

For a reaction between identical nuclides, the even-oddness of $l_{i}$ is to be limited for a given  $I_{i}$ value \cite{BW52}.  
In nuclei composed of the same number of neutrons and protons, E1 transitions are inhibited between isospin-zero states \cite{Greiner}. It is worth noting that the effective charge appearing in Eq.\,(\ref{eqpotEM}) vanishes if the ratios of the masses are replaced by that of the mass numbers, $A_{A}$ and $A_{a}$. The residual contribution owing to the isospin impurity can effectively be taken into account by the inclusion of the proper masses together with the renormalisation factor $S_{\rm F}$  (e.g.\,\cite{BarkerKajino}).
%
\subsection{Selecting the experimental data}
\label{sectSelecting}
The primary ensemble of experimental low-energy cross section data of current interest comprises those included in NACRE and supplementary ones that have become available to the present authors, preferentially, but not limited to, those published in major refereed journals of the field by early 2013. 
Some material that was apparently overlooked by NACRE is also added to the list.
 
 Generally speaking, we take the selected experimental data on cross sections and associated errors at face value since the availability of information required to do otherwise (e.g.\,\cite{Smith}) is often quite limited.  [Attempts of a stricter evaluation can be found in the literature for some specific reactions (e.g.\,[5, 6, 32 - 34].] In some cases, however, we omit from the analysis those data points which deviate very much from other measurements. 

A distinct exception to the above practice concerns some $S$-factors increasing dramatically toward the lowest energy end. The conventional wisdom attributes this observation to the so-called "laboratory screening" effect (Sect.\,\ref{sectScreening}). We simply disregard those parts of the $S$-factor data which exhibit that tendency. The specific selection of low-energy data for fitting and reaction rate calculation is presented in 
 Sect.\,3 for each reaction.

We do not refer in this work to any differential quantities even when they have been measured.
%
\subsection{Fitting procedure}
\label{sectFitting}
The actual computations of the DWBA cross sections have been made with the well-known code DWUCK4 \cite{code} albeit with certain modifications required to meet our goal, whereas a code of our own has been used for the PM cross sections.
 The choice of the form of the nuclear potential differs significantly from the DWBA to the PM analyses, and so does the concrete procedure of parameter fitting. This reflects by and large our preference for purely empirical approaches based on pragmatism to less phenomenological (thus often less flexible) ones.
Later in Sect.\,\ref{sectCritical}, we will briefly discuss the question of the soundness of the present approach.
\subsubsection{The Coulomb and nuclear potentials of choice}
\label{sectPotential}
We adopt the commonly used Coulomb potential 
\begin{eqnarray}
V_{\rm C}(r) &=& \frac{Z_{A}Z_{a}e^{2}}{r}\ \ \ \ \ \ \ \ \ \ \ \ \ \ \ \ \ \ \ \ {\rm for}\ \ \ \ r \ge R_{\rm C}\nonumber\\
&=&\frac{Z_{A}Z_{a} e^{2}}{2R_{\rm C}}{\big{[}}3 - (\frac{r}{R_{\rm C}})^{2}{\big{]}}\ \ \ \ {\rm for}\ \ \ \ r \le R_{\rm C},
\label{eqCoulomb}
\end{eqnarray}
\noindent
which assumes a uniform charge distribution inside the radius $R_{\rm C}$.
The nuclear potential we adopt can be most generally written as a sum of the real Woods-Saxon potential and the surface absorption imaginary part:
\begin{eqnarray}
V_{\rm N}(r)=V_{\rm R} f(x_{\rm R}) + i V_{\rm S} \frac{{\rm d}f}{{\rm d}x}(x_{\rm S}), 
\label{eqnuclpot}
\end{eqnarray}
\noindent
where 
\begin{eqnarray}
f(x_{\kappa}) = [ 1 + {\rm e}^{x_{\kappa}} ]^{-1}\ \ \ \ {\rm and}\ \ \ \ x_{\kappa} = \frac{(r - R_{\kappa})}{a_{\kappa}},
\label{eqWS}
\end{eqnarray}
\noindent
with $\kappa$ referring to the real (R) and surface imaginary (S) terms. 
The procedure adopted to select the potential strengths $V_{\kappa}$, the radius $R_{\kappa}$ and the diffuseness $a_{\kappa}$ depends on the types of reactions, as shall be described in the following.
\subsubsection{DWBA fitting for transfer reactions}
\label{sectDWBAfit}
A cut-off energy for the DWBA fit on the high-energy side is normally set at $E_{\rm cm} \simeq $ 1 MeV. 
 This choice is justified as long as the transfer reactions of the current interest are concerned because their $S$-factors are experimentally known even to relatively low energies, being dominated {\it either} by non-resonant contributions {\it or} by the contribution(s) from relatively broad resonance(s) if any.
This is in a sharp contrast to the cases of radiative capture reactions (see Sect.\,\ref{sectPMfit}).
In exceptional cases the possible contributions from the sub-threshold resonances may have to be considered, however (see Sect.\,\ref{sectSubthreshold}).

If applied to the entrance and exit channels and to the form factor, the potential form given by Eqs.\,(\ref{eqCoulomb}-\ref {eqWS}) introduces clearly too many parameters. Just for the practical purpose of reproducing the measured cross section data, one may, however, reduce the number of parameters drastically without causing much damage. First of all, we generally retain a shallow imaginary part of the nuclear potential only for the entrance channel, which takes into account the weak absorption to the exit channel by particle transfer. 
 Next, we parametrise the radius parameters as
\begin{eqnarray}
R_{\kappa}^{(c)} = r_{\kappa}^{(c)} A_{{\rm t},c}^{1/3},
\label{eqdwbaR}
\end{eqnarray}
\noindent
where the subscript $\kappa$ distinguishes the Coulomb (C), the real Woods-Saxon (R) and the surface imaginary (S) potentials, the superscript $c$ is put for the entrance channel $i$, the form factor $x$ and the exit channel $f$, and $A_{{\rm t},c}$ stands for the mass number of the heaviest nucleus (the "target" or the "core") in the channel $c$. We generally adopt $r_{\kappa}^{(c)}$-values extrapolated from those for the global potentials found elsewhere [35 - 37]. 
 The same source is used for the diffuseness parameter $a_{\kappa}^{(c)}$. Although those values may not adequately apply to very light nuclei, the practice is by and large justified in the spirit of the "equivalent potential". Namely, by re-shuffling the potential depths, quite similar results are obtained with the radius and diffuseness varied within reasonable ranges.  The depth of the (real) potential for the form factor,  $V_{\rm R}^{(x)}$, is determined so as to reproduce the measured binding energy of the particle $x$ in the $B+x$ system for pickup and $A+x$ system for stripping reactions. All in all, we are left in almost all cases only with $V_{\rm R}^{(i)}, V_{\rm S}^{(i)}$, and $V_{\rm R}^{(f)}$ as adjustable potential parameters.
As for the absolute value of the cross section, we treat $S_{\rm F} D_{0}^{2}$ as an adjustable parameter, rather than relying separately on the estimates of spectroscopic factors and of the zero-range interaction strengths found in the literature. 

In case of non-resonant reactions, namely, if no trace of resonances is observed in the energy range of interest, the same potential is used for all the orbital angular momenta $l_{i} (= 0 - 2)$. When a resonance coinciding with  a known level (with spin-parity $J_{\rm R}^{\pi}$) is found in that energy range, the DWBA fit automatically picks an $l_{i}$ value that suits to the formation of the resonance. For the same potential, the (consequently non-resonant) contributions from other waves are negligible. If that $l_{i}$ forms a degenerate resonance with different $J^{\pi}$ values, however, $J_{\rm R}^{\pi}$ must be projected out, especially when another $J^{\pi}$ would lead to a lower $l_{f}$ in the exit channel.
This applies also to the cases in which the data exhibit more than one resonance at close energies.

In general, the optimal values of the adjustable parameters have been derived by applying to the $S$-factors the standard $\chi^{2}$ fit technique. A "fit-by-eye" (linear or logarithmic) is used occasionally, however.

\subsubsection{PM fitting for radiative capture reactions}
\label{sectPMfit}

Radiative capture cross sections have been measured rarely below  $E_{\rm cm} \sim$ 0.1 MeV. This casts doubt upon the notion that the cross sections at the lowest energy range could simply be the tail of a resonance, if at all. Rather, they may well be dominated by non-resonant contributions. Under these circumstances, we are forced to adopt a strategy of parametrisation that is quite different from that we use for transfer reactions. Namely, we assign a real potential for each given set of $l_{i}$ and $J_{i}$. In particular, we try to fit the resonances, if any, in order to deduce the non-resonant contributions simultaneously. As a consequence, the cut-off energy for the fit on the high energy side varies with each reaction and also depends on the model capability.  In order to reduce the number of parameters, we take the same set of radius parameter values for both the initial ($i$) and final ($f$) states. This is in sharp contrast to the procedure used for transfer reactions. In stressing the difference, we set
\begin{eqnarray}
R_{\rm R}^{(i,f)} = R_{\rm C}^{(i,f)} \equiv R_{0}, \ \ \ a_{\rm R}^{(i,f)} \equiv a_{0},\ \ \ {\rm and}\ \ \ R_{0} = r_{0} [ A_{A}^{1/3} + A_{a}^{1/3}].
\label{eqpotR}
\end{eqnarray}
\noindent
Similarly we rewrite $V_{\rm R}^{(c)}$ as $V_{0}^{(c)}$. For the final (bound) state, we determine the depth, $V_{0}^{(f)}$, by matching the binding energy of the particle inserted into the final nucleus $B$. Hence, we are left, for a set of $l_{i}$ and $J_{i}$, with three potential parameters, $V_{0}^{(i)}, r_{0}$ and $a_{0}$, and the 
renormalisation constant, $S_{\rm F}$. 

The measured cross sections (or $S$-factors) reveal in most cases one or more resonances in the energy range of astrophysical interest. The first step we take is to reproduce the excitation energy and width of each resonance (with the spin-parity $J_{\rm R}^{\pi}$) by adjusting the potential parameters. By varying $V_{0}^{(i)}$ and $a_{0}$ with a rather arbitrarily chosen $r_{0}$ and an appropriately chosen $l_{i}$, this can generally be accomplished for not extremely narrow resonances with widths in excess of a few tenths of keV. As the first trial, we can make use of information on the excited levels in the literature. In other words, we mimic the phase-shift analysis for the elastic scattering. If necessary, the widths are altered so as to reproduce the cross section data of our immediate interest. The height of the resonance is then adjusted with $S_{\rm F}$ to match the measured value.  Finally, the non-resonant contributions are calculated with combinations of $J_{i}$ and $l_{i}$ {\it that were not used up for the resonances}. They are added to the resonant contributions with $S_{\rm F}$ values that must be adjusted to reproduce the cross section observed in the lowest energy region without disturbing the fit in the resonance region. 

The optimal values of the adjustable parameters are first derived by applying to the $S$-factors the standard $\chi^{2}$ fit. In many cases, however, an overall iterative re-adjustment becomes due, for which a "fit-by-eye" (linear or logarithmic) technique is often helpful. 
\subsubsection{Nuclear data}
\label{sectNuclearData}
The nuclear mass, $m$, in use in Sects.\,\ref{sectQuantity} and \ref{sectChoice} may be replaced by the measured "atomic mass" less the summed rest mass of the electrons bound to the neutral atom. The electron binding energies need to be considered only in a limited number of very low energy phenomena such as the "laboratory electron screening" (Sect.\,\ref{sectScreening}). As far as the reactions of current interest are concerned, even the neglect of the electron rest mass does not introduce significant errors.
 [It may be worth noting here that the conversion of the mass density to the number density of stellar matter can be most conveniently done with the use of atomic masses \cite{FCZ67}.] The relevant atomic masses are well known \cite{masses}.

Information on the properties of nuclear ground state and excited levels is in most cases available (e.g. in [39 - 41]). 
 The required quantities include the spin, parity and excitation energy for each level.
 For our PM analysis of resonant radiative capture reactions, the total and $\gamma$-widths, and the  $\gamma$-ray branching ratios in the literature are additionally taken as the trial input data for fixing the widths and heights of the resonances.    Nuclear magnetic dipole ($\mu_{1,A}$) and electric quadrupole ($Q_{2,A}$) moments appearing in Eq.\,(\ref{eqpotmuA}) are taken from \cite{moments}.
%
\subsection{The Breit-Wigner formula as a supplement}
\label{sectBreitWigner}
The potential models chosen above are inappropriate for, or fail in, describing certain resonances.  First of all, this is clearly the case  with  very narrow resonances that must be understood in terms of compound nucleus formation, the cross sections near the resonance energy $E_{\rm R}$ being given by the Breit-Wigner formula
\begin{eqnarray}
\sigma_{\rm R}^{(i,f)} = \frac{\pi}{k_{i}^{2}} \omega \frac{\Gamma_{i}\Gamma_{f}}{(E-E_{\rm R})^{2} + (\Gamma/2)^{2}},
\label{eqBreitWigner0}
\end{eqnarray}
\noindent
where $\omega = (2J_{\rm R}+1)/((2J_{A}+1)(2J_{a}+1)$, and $\Gamma_{i}, \Gamma_{f}$ and $\Gamma$ are the entrance partial, exit partial and  total widths, respectively. 

The integrated cross section from Eq.\,(\ref{eqBreitWigner0}) is $(2\pi^{2}/k_{i}^{2})\omega\gamma$. Here, the quantity $\omega\gamma = \Gamma_{i}\Gamma_{f}/\Gamma$ can be experimentally derived in relation to the "thick target yield", and is useful for evaluating the contribution to the reaction rates from a very narrow resonance (Sect.\,\ref{sectNarrow}).

For a relatively broad resonance, the energy dependences of the widths can be approximately introduced such that Eq.\,(\ref{eqBreitWigner0}) may be extended to a wider energy range. Namely, 
 one rewrites, for an entrance channel,
\begin{eqnarray}
\Gamma_{i}(E) = (2 k_{i} a) v_{l_{i}}(E) \gamma_{\rm W}^{2} \theta_{i}^{2}. 
\label{eqBreitWigner1}
\end{eqnarray}
\noindent
in terms of the penetrability $v_{l}(E) = 1 / [F_{l}(ka)^{2} + G_{l}(ka)^{2}]$ (\cite{BW52}) and the reduced width $\theta_{i}^{2} (\leq 1)$ in units of the Wigner limit $\gamma_{\rm W}^{2} = 3\hbar^{2}/(2 \mu a^{2})$,  where $a$ is the channel radius \cite{RRbook}.
 The exit partial width for a transfer reaction can be computed similarly with a suitable shift in energetics.  
For a radiative capture with the multi-polarity $\lambda$, one may take $\Gamma_{\gamma}(E) =  \Gamma_{\gamma}(E_{\rm R}) [(E+q)/(E_{\rm R}+q)]^{2\lambda +1}$ with $q$ being the $Q$-value.
The above formalism will be applied for the evaluation of possible contributions of sub-threshold resonances (Sect.\,\ref{sectSubthreshold}).

Alternatively, the resonance shape and height of our model cross sections  may be made to coincide nearly with those given by Eq.\,(\ref{eqBreitWigner0}) by adjusting the fit parameters. 
This is useful when the dominant (and not very narrow) elastic scattering width is known but, for the reaction, only $\omega\gamma$ is measured.  %
\subsection{Signatures of specific phenomena}
\label{sectSignatures}
In order to warrant a reliable extrapolation to the lowest energy region of the experimentally available $S$-factors, one has to pay special attention to a few phenomena as in the following.
\subsubsection{The laboratory electron screening}
\label{sectScreening}
 In ordinary laboratory experiments, the targets are atomic or molecular. As such, the nuclear charge of a target is screened by the bound electrons in the eyes of the projectile. The cross sections, or more visibly the $S$-factors, get enhanced at very low energies when compared with those in the case of the target being a bare nucleus. That is the so-called laboratory electron screening.\footnote{The screening effect owing to the ionisation (or "free") electrons may have an important effect on reaction rates in stellar environments, particularly at high densities. This problem is not tackled here.} 
 
Since the light species of our interest are most likely fully ionised in stellar interiors, we must question the enhanced segments of the measured data.
A quick estimate of the laboratory screening can be made by shifting the Coulomb barrier by a constant amount $U_{\rm e}$, leading to an approximate enhancement factor of exp($\pi \eta U_{\rm e}/E$). In particular, $U_{\rm e}$ in the adiabatic (as opposed to sudden) approximation can be given as the difference of the atomic binding energy summed for the two separate species and that of the combined "molecular" state with the same number of bound electrons. More advanced dynamical treatments of the screening in the $^{2}$H + d reaction indicate that the adiabatic $U_{\rm e}$ can indeed lead to a good measure for the maximum effect at low energies \cite{screening1, screening2}. On the other hand, various observations appear to indicate that the behaviour of the very low energy $S$-factors depends on the environmental conditions (e.g.\,\cite{BY12}). 

  An indirect approach, the so-called "Trojan horse method" (THM), is meant to avoid the screening effect altogether (e.g.\,\cite{Trojan} for a brief summary).\footnote{Speaking of the indirect methods (\cite{Typel,MU07} for reviews), we note here that the "asymptotic normalisation constant" (ANC) derived from the analysis of suitable experiments gives information on the $S(0)$-value for reaction in a very-loosely bound system.}

Under these circumstances, we disregard in the fit procedure those parts of the measured $S$-factors whose behaviours are "abnormal" and may indicate that screening is in play. This is in contrast to some R-matrix fits that attempt to extract $U_{\rm e}$ simultaneously (e.g.\,\cite{BarkerScreening}). We instead limit ourselves to display in Sect.\,3 an adiabatic screening correction made to the $S$-factors in a few cases just to guide the eye.
\subsubsection{Sub-threshold resonances}
\label{sectSubthreshold}
Another cause for possible enhancements at the lowest energies of the S-factors concerns the "tails" of sub-threshold states (e.g.\,\cite{RRbook}). This problem has been discussed extensively and most often in the framework of R-matrix analyses along with indirect measurements. In general, however, the quantitative estimates of the consequent enhancements of the $S$-factors remain more or less uncertain.

The simple models adopted for this compilation cannot treat the problem in a satisfactory manner. When relevant cross section enhancements are experimentally alluded or theoretically expected, we just explore them by a simple procedure \cite{RRbook} in which the Breit-Wigner formula (Sect.\,\ref{sectBreitWigner}) is applied to the tail distribution of the culprit sub-threshold resonance. The key unknown quantity is $\Gamma_{i}(E)$, whereas $\Gamma_{f}(E)$ can usually be normalised to the experimental value at $E_{\rm R} < 0$. The strength and the slope of that tail distribution may be adjusted by selecting different values for the reduced width $\theta^{2}$ and the channel radius $a$ (see e.g.\,\cite{BE07}). 
If $a$ is parametrised like $R_0$ in Eq.\,(\ref{eqpotR}), $r_{a0}$ may be in the approximate 1.4 $\sim$ 1.9 fm range.
\subsubsection{Interference of resonances}
\label{sectInterference}
In a limited number of cases, two or more resonances with the same $J^{\pi}$ could be attributed to an $l_{i}$ value. 
The $S$-factors in consideration of interference between two resonances (R1 and R2) are given by 
\begin{eqnarray}
S(E) = S_{\rm R1}(E) + S_{\rm R2}(E)+ 2 \sqrt{S_{\rm R1}(E) S_{\rm R2}(E)} {\rm cos}(\delta_{l,R1}(E)-\delta_{l,R2}(E)),
\label{eqInterfere}
\end{eqnarray}
\noindent
where $\delta_{l,R1}$ and $\delta_{l,R2}$ are the respective phase shifts.
We select the overall sign of the interference term by inspecting the observed behaviours of the $S$-factors in the tail regions of the resonances. 
%
\subsection{The rate evaluation}
\label{sectRate}
Following the traditional practice \cite{CF88,NACRE}, we present the thermonuclear reaction rates in the temperature range of $10^6 \lsimeq T \leq 10^{10}$ K. We summarise here the general procedure to evaluate the rates and their uncertainties. Comments on the specifics in individual cases are added in Sect.\,3 along with the results.
\subsubsection{Combining the measured and model $S$-factors}
\label{sectCombining}
 We search for the model parameter values that "best" reproduce the experimental $S$-factor data in a certain range of centre-of-mass energy, $[E_{1}, E_{3}]$, chosen for each reaction (see Sect.\,\ref{sectFitting}). The rates are then computed with the model and observed $S$-factor values in the $[E_{0}, E_{2}]$ and $[E_{2}, E_{4}]$ ranges, respectively, where $E_{0} \approx 0 < E_{1} < E_{2} < E_{3} < E_{4}$. This procedure generally defines the rates labelled as "adopted". In some cases, the lack of experimental data at energies beyond $E_{4}$ makes it necessary to extrapolate the $S$-factors by the model. In certain cases, one has even to resort  to a statistical treatment of compound nucleus reactions in order to compute the rates at very high temperatures (see Sect.\,\ref{sectHighT}). 

The evaluation of the uncertainties in the $[E_{0}, E_{2}]$ range is not always straightforward, and is specific to  each reaction. It depends on how to interpret a scatter (or even a conflict) among the experimental $S$-factors that is more or less large. It also depends on how sensitive the extrapolation to the $[E_{0}, E_{1}]$ range is to the equally good fits with different sets of parameter values. This is of particular concern in the analysis of observed resonances. For transfer reactions with a relatively broad resonance, sets of parameter values that reproduce the data near the resonance as well  as in the "best" fit, but differ significantly in the tail region, are often used to define the "high" and "low" $S$-factor values in the $[E_{0}, E_{2}]$ range.  In other cases, the upper and lower envelopes of the experimental data are used for the fit. The uncertainties regarding a relatively narrow resonance observed in a radiative capture reaction are often evaluated by allowing for 10 - 20 \% errors on the height (or $\gamma$-width) of the resonance.  

The integration over the $[E_{2}, E_{4}]$ range is made with the use of linear fits of the $S$-factor data points of interest in appropriately segmented energy domains. This method, rather than the local linear interpolation adopted in NACRE, avoids unwanted ripples in the computed rates.  
The reaction rates "low" and "high" include the experimental uncertainties in the $[E_{2}, E_{4}]$ range.

\subsubsection{Extremely narrow resonances}
\label{sectNarrow}
In some reactions, resonances with very narrow widths ($\ll 1$ keV) may come into play, to which the potential model is inadequate to apply. With the cross sections unresolved, the experimental information available is the resonance strength $\omega\gamma$ deduced from the thick-target yields in relation to the background cross sections (see Sect.\,\ref{sectBreitWigner}). 
 The additional contribution to the thermonuclear reaction rates (\ref{eqrate1}) from a very narrow resonance is in good approximation given by
\begin{eqnarray}
N_{A}<\sigma v>^{(R)} = 1.54\times10^{11} \hat{\mu}^{-3/2}(\omega \gamma) T_{9}^{-3/2} {\rm exp}[-11.605 E_{\rm R}/ T_{9}],
\label{eqnarrow}
\end{eqnarray}
\noindent
where $\omega\gamma$ and $E_{\rm R}$ are in units of MeV.
\subsubsection{Reaction rates at very high temperatures}
\label{sectHighT}
The estimates of certain reaction rates at temperatures in excess of a few times $10^{9}$ K can get progressively uncertain. This is primarily because the available cross section data may be incomplete for a reaction that may involve a considerable number of compound nuclear resonances at high excitation energies. If that is suspected, the cross sections derived by the state-of-the-art nuclear reaction code TALYS \cite{Talys} may be consulted. This method is clearly more advanced, and can be better controlled, than the extrapolations based on integrated Hauser-Feshbach {\sl rates} as in NACRE.

At high temperatures, the thermal population of the low-lying  states of the {\it targets} may also come into play. The Hauser-Feshbach predictions \cite{Talys} of the ratios of the reaction rates calculated for the thermalised targets relative to those for the ground states imply that the effects are generally small for the systems of interest here.
\subsubsection{Forward and reverse reaction rates in astrophysical environments}
\label{sectastrorates}
All of the reactions of interest in this compilation may be expressed as
\begin{eqnarray}
       n_{1} + n_{2}  \longrightarrow n_{3} + n_{4}\ \ \ \ {\rm\ or}\ \ \ \ \{\,n_{1}\,n_{2}\,n_{3}\,n_{4}\},
\label{eq2body}
\end{eqnarray}
\noindent
where $n_{k}$ refers to the numbers of {\it different} nuclear species $k$. The norms are  $\{1111\}$ for transfer and  $\{1110\}$ for radiative capture reactions that involve four different nuclear species.

The forward and the inverse reaction rates per unit volume are then given by
\begin{eqnarray}
     P_{12} = \rho N_{\rm A} \Lambda_{12 \rightarrow 34}\ \ \ \ P_{34} = \rho N_{\rm A} \Lambda_{34 \rightarrow 12}, 
\label{eqforandback}
\end{eqnarray}
\noindent
where $\rho$ is the matter density. Denoting the number fractions as $Y_{k} \equiv X_{k}/A_{k}$ (with $X_{k}$ being the mass fraction normalised to $\Sigma_{k} X_{k} = 1$), 
\begin{eqnarray}
&&\Lambda_{12 \rightarrow 34} = \frac{N_{\rm A}<\sigma\,v>_{12}}{n_{1}!} Y_{1}^{n_{1}} Y_{2}^{n_{2}} \rho,\\
&&\Lambda_{34 \rightarrow 12} = \frac{N_{\rm A}<\sigma\,v>_{12} \times REV}{n_{3}!} Y_{3}^{n_{3}} Y_{4}^{n_{4}} \rho^{n_{3}+n_{4}-1},
\label{eqreactionrate}
\end{eqnarray}
\noindent
where
\begin{eqnarray}
   REV &\equiv& \frac{x_{1}x_{2}}{x_{3}x_{4}} {\rm exp}{\Big{[}} -\frac{Q}{k_{\rm B}T} {\Big{]}} {\Big{[}}\frac{k_{\rm B}T}{2\pi\hbar^{2}N_{\rm A}^{5/3}}{\Big{]}}^{(3/2)(n_{1} + n_{2}- n_{3}- n_{4})},
\label{eqrev}
\end{eqnarray}
\noindent
with $x_{k} = (g_{k} A_{k}^{3/2})^{n_{k}}/n_{k}!$, where $g_{k}= (2J_{k}+1)G$, $J_{k}$ and $G$ being the ground-state spin and the partition function, and where $Q$ is  the $Q$-value for the forward reaction. 
 The rates of the reverse two-body reactions are given explicitly in Appendix B. 
With these notations, the contributions to the time variation ${\rm d}Y_{k}/{\rm d}t$ of, e.g., the forward reaction are  $- n_{k} \Lambda_{12 \rightarrow 34}$ for $k=1,2$, and $+ n_{k} \Lambda_{12 \rightarrow 34}$ for $k=3,4$.
%
\subsection{Critical assessment}
\label{sectCritical}
Given the purpose of this compilation, we shall not be much devoured by the question of the applicability of the present potential-model approach to each reaction.  Nonetheless, a brief discussion on some issues at stake may be due. A sequence of critical questions may be summarised as follows. Is the reaction a direct process?  If so, is the DWBA or PM applicable?  If so, are the values of the parameters and the goodness of the fit acceptable? 
In many cases, in fact, we work on the answers to these questions in the reversed order.
Namely, if a good fit is achieved with the use of the potential parameter values in "reasonable" ranges, then we answer to the first two questions positively, although there may be many uncomfortable aspects left in the sense of theoretical nuclear physics. 

As mentioned earlier, the potential models (PM and DWBA) are capable of dealing not only with non-resonant reactions at low energies below the Coulomb barrier, but also with reactions forming relatively broad resonances. The present study will indeed reveal (Sect.\,3) that most resonances with their widths as narrow as of the order of 1 keV for radiative captures and of several tens of keV for transfer reactions could still be described by the PM and DWBA, respectively, to a more or less satisfactory extent without much manipulation of the potential parameter values.
One way of judging the "reliability" of the models is through that of the potential parameter values derived from the fit. Whereas the "soundness" of the parameter values can always be checked, their "uniqueness" can hardly be warranted.  For example, one may be able to find many different sets of parameter values that lead to similarly good fits to existing experimental data, which often come with large (or small but conflicting) errors. In order to constrain the spreads to a manageable level, we have limited the number of adjustable parameters (of DWBA in particular).  We stress here that we do not pretend aiming at the construction of a standard (or global) potential of any sort. 

For transfer reactions, we have adopted the zero-range DWBA. Its "validity" in the light of a microscopic cluster model was studied for the $^{13}${C}($\alpha$,n) reaction at $E_{\rm cm} \lsimeq$ 5 MeV  \cite{AD05}. With some clear shortcomings of DWBA set aside, it was expectedly found that the zero-range DWBA could reproduce quite well the shapes (not amplitudes, though) of the $S$-factors of the microscopic model as long as the potential parameters were fit to the same phase-shifts. This gives a sort of support for the present approach to an observed resonance, in which the model parameters are adjusted so as to reproduce the shape {\it and} the height of the resonance. We also note that some numerical experiments have shown that the use of the finite-range DWBA cannot bring notable improvements {\it as far as the empirical fits are concerned}, even for the multi-nucleon transfers considered in the present compilation. Both in $^{9}$Be($\alpha$,n) and $^{11}$B(p,$\alpha$) reactions, for example, the finite-range corrections to $S(E)/S(0)$ {\it for a given potential} do not exceed 10 \% even at $E \simeq E_{\rm R}$. 

In conclusion: Even when the applicability of the present method is stretched too far, our primary aim of a reproduction of experimental cross section data and its extrapolation would be met at least in a sense similar to oft-used polynomial fits. However, large uncertainties in extrapolation would be unavoidable when the experimental data are not enough to constrain the parameter values. This is particularly the case when sub-threshold resonances are involved. Similarly, the non-resonant components may, in some cases, not be constrained well enough as one wishes.  
\section{The results}
\label{sectResults} 
The results for the 34 two-body, exoergic reactions listed in Table \ref{list0} are presented. 
Each subsection discusses 
 1) the available experimental data; 2) the model (PM or DWBA) astrophysical $S$-factor; and 3) the thermonuclear reaction rates. 

The astrophysical $S$-factors are plotted versus the centre-of-mass energy, $E_{\rm cm}$. The solid curve represents the "adopted" model $S$-factors, with the two dashed ones setting their boundaries, "low" and "high".
If not specified otherwise, the $S(0)$-values are those at about  0.5 - 1.0 keV (c.f.\,\cite{Baye00}).
Other compilations in comparison are: [NACRE] (i.e. \cite{NACRE} extrapolation by various methods), [BBN04] (i.e. \cite{Desc04}, R-matrix), [RAD10] (i.e. \cite{TEXAS10}, potential model), and [SUN11] (i.e. \cite{RMP}, various methods).

The reaction rates are given versus $T_9$ in the same mesh as the one used by NACRE, and are graphically compared with those of NACRE (or CF88 \cite{CF88} in two cases).  
 
%
Figure captions may contain  additional information such as on the selection of the experimental data, and on the possible cause of the differences  between the present and previous results.
The origin of these differences is, however, often hard to pin down as the details of the procedures taken by NACRE have meanwhile weathered out.

Additional references are given at the end of the subsections. The lists are, however, not  meant to be exhaustive.
The analytic form of REV [Eq.\,(\ref{eqrev})] is attached to the rate tables. Recall that REV is dimensionless for a transfer reaction and  is in mol/cm$^{3}$ for a capture reaction. 
\clearpage
\begin{table}[ht]
\caption{List of compiled reactions. $Q$ is the $Q$-value.  The first entry of Figures indicates Figure\# that depicts the $S$-factors, and the second the reaction rates in comparison with the previous values. The entries of Tables indicate the Table\# of the reaction rates, the model parameter values (Appendx A) and the reverse reaction rates (Appendix B) in that order. 
 }
\begin{tabular*}{\textwidth}{@{\extracolsep{\fill}} r l r l l l l l l}

\hline

subsection   &     reaction                  &        $Q$ (MeV) &  & Figures      & Tables          &    &                              &   \\
\hline 

$\it{3.1}$ &\reac{2}{H}{p}{\gamma}{3}{He} &+\ \ \,5.493 &\ \ \ &\ \,2,\ \,3 &\ \ref{dpgTab2}, \ref{dpgTab1}, 72  & 
\\

$\it{3.2}$  &\reac{2}{H}{d}{\gamma}{4}{He} &+ 23.847  & &\ \,\ref{ddgFig1},\ \,\ref{ddgFig2}  &\ \ref{ddgTab2}, \ref{ddgTab1}, 72  &
\\

$\it{3.3}$  &\reac{2}{H}{d}{n}{3}{He} &+\ \ \,3.269     & &\ \,\ref{ddnFig1},\ \,\ref{ddnFig2}  &\ \ref{ddnTab2}, \ref{ddnTab1}, 70  &
\\

$\it{3.4}$  &\reac{2}{H}{d}{p}{3}{H} &+\ \ \,4.033   & &\ \,\ref{ddpFig1},\ \,\ref{ddpFig2}  &\ \ref{ddpTab2}, \ref{ddpTab1}, 70  &
\\

$\it{3.5}$  &\reac{2}{H}{\alpha}{\gamma}{6}{Li} &+\ \ \,1.474 & &\ref{dagFig1},\, \ref{dagFig2}  &\ \ref{dagTab2}, \ref{dagTab1}, 72  &
\\

$\it{3.6}$  &\reac{3}{H}{d}{n}{4}{He}    &+ 17.589  & &\ref{tdnFig1}, \ref{tdnFig2}  &\ \ref{tdnTab2}, \ref{tdnTab1}, 70  &
\\

$\it{3.7}$  &\reac{3}{H}{\alpha}{\gamma}{7}{Li} &+\ \ \,2.467 & &\ref{tagFig1}, \ref{tagFig2}  &\ \ref{tagTab2}, \ref{tagTab1}, 72  &
\\

$\it{3.8}$  &\reac{3}{He}{d}{p}{4}{He} &+ 18.353   & &  16-18, \ref{he3dpFig2}  &\ \ref{he3dpTab2}, \ref{he3dpTab1}, 70  &
\\

$\it{3.9}$  &\reac{3}{He}{\tau}{2p}{4}{He} &+ 12.860  & &\ref{he3he3Fig1}, \ref{he3he3Fig2}  & \ref{he3he3Tab2}, \ref{he3he3Tab1},\ \  -  &
\\

$\it{3.10}$  & \reac{3}{He}{\alpha}{\gamma}{7}{Be} &+\ \ \,1.586 & &\ref{he3agFig1}, \ref{he3agFig2}  & \ref{he3agTab2}, \ref{he3agTab1}, 72  &
\\

$\it{3.11}$  &\reac{6}{Li}{p}{\gamma}{7}{Be} &+\ \ \,5.606   & &24-26, 27  & \ref{li6pgTab2}, \ref{li6pgTab1}, 72   &
\\

$\it{3.12}$  &\reac{6}{Li}{p}{\alpha}{3}{He} &+\ \ \,4.020  &  &\ref{li6paFig1}, \ref{li6paFig2}  & \ref{li6paTab2}, \ref{li6paTab1}, 70   &
\\

$\it{3.13}$  &\reac{7}{Li}{p}{\gamma}{8}{Be}(2$\alpha$)$^\dag$  &+ 17.347 & &30-32, 33  & \ref{li7pgTab2}, \ref{li7pgTab1},\ \ -   &
\\

$\it{3.14}$  &\reac{7}{Li}{p}{\alpha}{4}{He} &+ 17.347  & &\ref{li7paFig1}, \ref{li7paFig2}  & \ref{li7paTab2}, \ref{li7paTab1}, 70   &
\\

$\it{3.15}$  &\reac{7}{Li}{\alpha}{\gamma}{11}{B} &+\ \ \,8.665 & &\ref{li7agFig1}, \ref{li7agFig2}  & \ref{li7agTab2}, \ref{li7agTab1}, 74   &
\\

$\it{3.16}$  &\reac{7}{Be}{p}{\gamma}{8}{B} &+\ \ \,0.137 & & 38-40, 41 & \ref{be7pgTab2}, \ref{be7pgTab1}, 73-74   &
\\

$\it{3.17}$  &\reac{7}{Be}{\alpha}{\gamma}{11}{C} &+\ \ \,7.545  & &\ref{be7agFig1}, \ref{be7agFig2}  & \ref{be7agTab2}, \ref{be7agTab1}, 74   &
\\

$\it{3.18}$ &\reac{9}{Be}{p}{\gamma}{10}{B} &+\ \ \,6.586 &  &\ref{be9pgFig1}, \ref{be9pgFig2}  & \ref{be9pgTab2}, \ref{be9pgTab1}, 74   &
\\

$\it{3.19}$ &\reac{9}{Be}{p}{d}{8}{Be}(2$\alpha$)$^\dag$ &+\ \ \,0.651&  &\ref{be9pdFig1}, \ref{be9pdFig2}  & \ref{be9pdTab2}, \ref{be9pdTab1},\ \ -  &
 \\

$\it{3.20}$ &\reac{9}{Be}{p}{\alpha}{6}{Li} &+\ \ \,2.125 &  &\ref{be9paFig1}, \ref{be9paFig2}  & \ref{be9paTab2}, \ref{be9paTab1}, 71  &
 \\

$\it{3.21}$ &\reac{9}{Be}{\alpha}{n}{12}{C} &+\ \ \,5.701 &  &\ref{be9anFig1}, \ref{be9anFig2}  & \ref{be9anTab2}, \ref{be9anTab1}, 71  &
 \\

$\it{3.22}$ &\reac{10}{B}{p}{\gamma}{11}{C} &+\ \ \,8.689 &  &\ref{b10pgFig1}, \ref{b10pgFig2}  & \ref{b10pgTab2}, \ref{b10pgTab1}, 74  &
 \\

$\it{3.23}$ &\reac{10}{B}{p}{\alpha}{7}{Be} &+\ \ \,1.145 &  &\ref{b10paFig1}, \ref{b10paFig2}  & \ref{b10paTab2}, \ref{b10paTab1}, 71  &
 \\

$\it{3.24}$ &\reac{11}{B}{p}{\gamma}{12}{C} &+ 15.957 &  &\ref{b11pgFig1}, \ref{b11pgFig2}  & \ref{b11pgTab2}, \ref{b11pgTab1}, 74  &
\\

$\it{3.25}$ &\reac{11}{B}{p}{\alpha}{8}{Be}(2$\alpha$)$^\dag$ &+\ \ \,8.682 &  &\ref{b11paFig1}, \ref{b11paFig2}  & \ref{b11paTab2}, \ref{b11paTab1},\ \ -  &
\\

$\it{3.26}$ &\reac{11}{B}{\alpha}{n}{14}{N} &+\ \ \,0.158 &  &\ref{b11anFig1}, \ref{b11anFig2}  & \ref{b11anTab2}, \ref{b11anTab1}, 71  &
\\

$\it{3.27}$ &\reac{12}{C}{p}{\gamma}{13}{N}  &+\ \ \,1.943 &  &\ref{c12pgFig1}, \ref{c12pgFig2}  & \ref{c12pgTab2}, \ref{c12pgTab1}, 75  &
\\

$\it{3.28}$ &\reac{12}{C}{\alpha}{\gamma}{16}{O} &+\ \ \,7.162 &  &64-68, 69  & \ref{c12agTab2}, \ref{c12agTab1}, 75  &
\\

$\it{3.29}$ &\reac{13}{C}{p}{\gamma}{14}{N}  &+\ \ \,7.551 &  & 70-71, 72  & \ref{c13pgTab2}, \ref{c13pgTab1}, 75   &
\\

$\it{3.30}$ &\reac{13}{C}{\alpha}{n}{16}{O} &+\ \ \,2.216  &  &73-75, 76  & \ref{c13anTab2}, \ref{c13anTab1}, 71   &
\\

$\it{3.31}$ &\reac{13}{N}{p}{\gamma}{14}{O}  &+\ \ \,4.627 &  &\ref{n13pgFig1}, \ref{n13pgFig2}  & \ref{n13pgTab2}, \ref{n13pgTab1}, 75   &
\\

$\it{3.32}$ &\reac{14}{N}{p}{\gamma}{15}{O}  &+\ \ \,7.297 &  &79-81, 82  & \ref{n14pgTab2}, \ref{n14pgTab1}, 75   &
 \\

$\it{3.33}$ &\reac{15}{N}{p}{\gamma}{16}{O}  &+ 12.127 &  &\ref{n15pgFig1}, \ref{n15pgFig2}  & \ref{n15pgTab2}, \ref{n15pgTab1}, 75   &
\\

$\it{3.34}$ &\reac{15}{N}{p}{\alpha}{12}{C}  &+\ \ \,4.965 &  &\ref{n15paFig1}, \ref{n15paFig2}  & \ref{n15paTab2}, \ref{n15paTab1}, 71  &
\\

\hline
\end{tabular*}

$^\dag Q$ applies to the (2$\alpha$) final state
\label{list0}
\end{table}
\clearpage
\subsection{\reac{2}{H}{p}{\gamma}{3}{He}}
\label{dpgSect}
The experimental data sets referred to in NACRE are GR62 \cite{GR62}, GR63 \cite{GR63}, WA63 \cite{WA63}$^\dag$, BE64 \cite{BE64}$^\dag$, FE65 \cite{FE65}$^\dag$, ST65 \cite{ST65}$^\dag$, GE67 \cite{GE67}, WO67 \cite{WO67}, TI73 \cite{TI73}$^\dag$, SC95a \cite{SC95} and MA97 \cite{MA97}, covering the 0.01 $\lsimeq E_{\rm cm} \lsimeq$ 20 MeV range. \cite{BA70} was superseded by MA97. 
Added are the post-NACRE data sets CA02 \cite{CA02} and BY08a \cite{BY08a}, the former extending the range down to $E_{\rm cm} \simeq$ 0.0025 MeV.
SC95a \cite{SC95} has been corrected by SC96 \cite{SC96} for an error.
[{\footnotesize{$^\dag$from the inverse\reac{3}{He}{\gamma}{p}{2}{H}.}}]

Figure 2 compares the PM and experimental $S$-factors. 
The data in the $E_{\rm cm} \lsimeq$ 2 MeV range are used for the PM fit.
The s- and p-wave captures, leading to M1 and E1 transitions, contribute  predominantly at low energies. 
The adopted parameter values are given in Table \ref{dpgTab1}. 
The resulting  $S(0)$ = 0.21 $\pm$ 0.04 eV\,b. In comparison, $S(0)$ = 0.20 $\pm$ 0.07 eV\,b [NACRE, quadratic polynomial], 0.223 $\pm$ 0.010 eV\,b [BBN04], 0.14 eV\,b [RAD10], and  0.214 $\pm$ 0.017 eV\,b [SUN11, quadratic polynomial].

Table \ref{dpgTab2} gives the reaction rates at 0.001 $\le T_{9} \le$ 10, for which the PM-predicted and the experimental cross sections below and above $E_{\rm cm} \simeq$ 0.1 MeV are used, respectively.
Figure \ref{dpgFig2} compares the present and the NACRE rates.

{\footnotesize See \cite{MA05} for an ${ab\ initio}$ calculation.} 

\begin{figure}[hb]
\centering{
\includegraphics[height=0.50\textheight,width=0.90\textwidth]{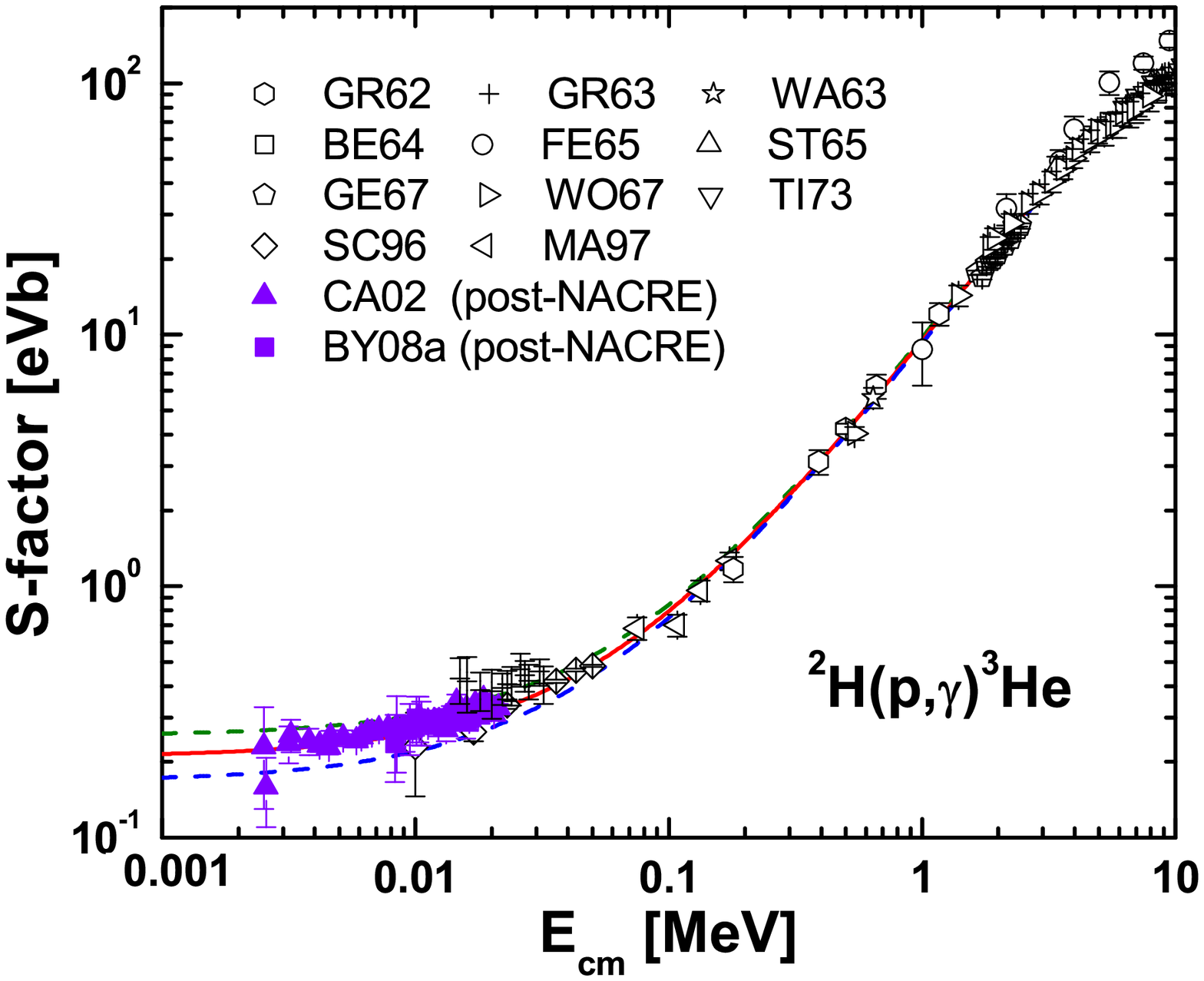}
\vspace{-0.5truecm}
\caption{The $S$-factor for \reac{2}{H}{p}{\gamma}{3}{He}.}
}
\label{dpgFig1}
\end{figure}
\clearpage

\begin{figure}[t]
\centering{
\includegraphics[height=0.33\textheight,width=0.90\textwidth]{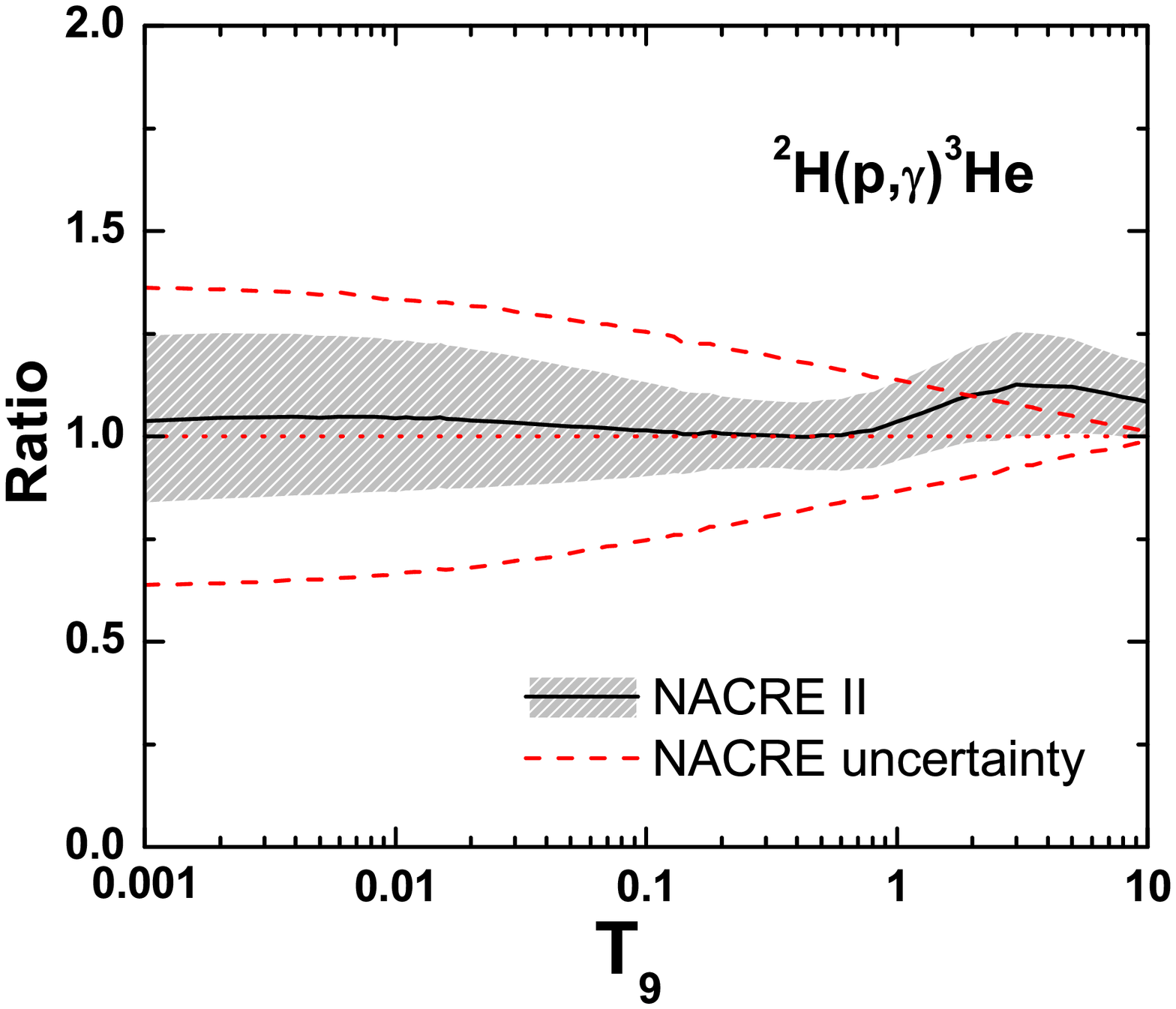}
\vspace{-0.4truecm}
\caption{\reac{2}{H}{p}{\gamma}{3}{He} rates in units of the NACRE (adopt) values.}
\label{dpgFig2}
}
\end{figure}

\begin{table}[hb]
\caption{\reac{2}{H}{p}{\gamma}{3}{He} rates in $\rm{cm^{3}mol^{-1}s^{-1}}$}\scriptsize\rm
\footnotesize{
\begin{tabular*}{\textwidth}{@{\extracolsep{\fill}} l c c c |l c c c}
\hline
$T_{9}$ & adopted & low & high & $T_{9}$ & adopted & low & high \\
\hline
  0.001 & 1.35E$-$11 & 1.09E$-$11 & 1.62E$-$11 &        0.14 & 1.15E+01 & 1.03E+01 & 1.27E+01 \\
  0.002 & 1.87E$-$08 & 1.51E$-$08 & 2.25E$-$08 &        0.15 & 1.33E+01 & 1.20E+01 & 1.47E+01 \\
  0.003 & 6.07E$-$07 & 4.93E$-$07 & 7.27E$-$07 &        0.16 & 1.52E+01 & 1.37E+01 & 1.67E+01 \\
  0.004 & 5.38E$-$06 & 4.38E$-$06 & 6.43E$-$06 &        0.18 & 1.93E+01 & 1.75E+01 & 2.12E+01 \\
  0.005 & 2.52E$-$05 & 2.06E$-$05 & 3.01E$-$05 &        0.2 & 2.37E+01 & 2.16E+01 & 2.60E+01 \\
  0.006 & 8.15E$-$05 & 6.67E$-$05 & 9.71E$-$05 &        0.25 & 3.62E+01 & 3.31E+01 & 3.94E+01 \\
  0.007 & 2.07E$-$04 & 1.70E$-$04 & 2.47E$-$04 &        0.3 & 5.01E+01 & 4.59E+01 & 5.44E+01 \\
  0.008 & 4.47E$-$04 & 3.68E$-$04 & 5.31E$-$04 &        0.35 & 6.52E+01 & 5.98E+01 & 7.08E+01 \\
  0.009 & 8.55E$-$04 & 7.04E$-$04 & 1.01E$-$03 &        0.4 & 8.13E+01 & 7.45E+01 & 8.84E+01 \\
  0.01  & 1.49E$-$03 & 1.23E$-$03 & 1.77E$-$03 &        0.45 & 9.84E+01 & 9.00E+01 & 1.07E+02 \\
  0.011 & 2.43E$-$03 & 2.01E$-$03 & 2.87E$-$03 &        0.5 & 1.16E+02 & 1.06E+02 & 1.27E+02 \\
  0.012 & 3.73E$-$03 & 3.09E$-$03 & 4.40E$-$03 &        0.6 & 1.54E+02 & 1.40E+02 & 1.69E+02 \\
  0.013 & 5.47E$-$03 & 4.54E$-$03 & 6.45E$-$03 &        0.7 & 1.95E+02 & 1.77E+02 & 2.14E+02 \\
  0.014 & 7.73E$-$03 & 6.42E$-$03 & 9.10E$-$03 &        0.8 & 2.38E+02 & 2.16E+02 & 2.61E+02 \\
  0.015 & 1.06E$-$02 & 8.80E$-$03 & 1.24E$-$02 &        0.9 & 2.84E+02 & 2.57E+02 & 3.12E+02 \\
  0.016 & 1.41E$-$02 & 1.17E$-$02 & 1.65E$-$02 &        1. & 3.32E+02 & 2.99E+02 & 3.64E+02 \\
  0.018 & 2.33E$-$02 & 1.95E$-$02 & 2.74E$-$02 &        1.25 & 4.57E+02 & 4.12E+02 & 5.03E+02 \\
  0.02 & 3.60E$-$02 & 3.02E$-$02 & 4.22E$-$02 &        1.5 & 5.91E+02 & 5.31E+02 & 6.52E+02 \\
  0.025 & 8.59E$-$02 & 7.25E$-$02 & 1.00E$-$01 &        1.75 & 7.30E+02 & 6.54E+02 & 8.07E+02 \\
  0.03 & 1.66E$-$01 & 1.41E$-$01 & 1.93E$-$01 &        2. & 8.73E+02 & 7.79E+02 & 9.68E+02 \\
  0.04 & 4.35E$-$01 & 3.73E$-$01 & 5.01E$-$01 &        2.5 & 1.17E+03 & 1.04E+03 & 1.30E+03 \\
  0.05 & 8.62E$-$01 & 7.44E$-$01 & 9.86E$-$01 &        3. & 1.46E+03 & 1.30E+03 & 1.63E+03 \\
  0.06 & 1.45E+00 & 1.26E+00 & 1.65E+00 &        3.5 & 1.76E+03 & 1.56E+03 & 1.97E+03 \\
  0.07 & 2.20E+00 & 1.93E+00 & 2.50E+00 &        4. & 2.07E+03 & 1.84E+03 & 2.30E+03 \\
  0.08 & 3.11E+00 & 2.74E+00 & 3.51E+00 &        5. & 2.67E+03 & 2.39E+03 & 2.95E+03 \\
  0.09 & 4.17E+00 & 3.69E+00 & 4.69E+00 &        6. & 3.27E+03 & 2.95E+03 & 3.60E+03 \\
  0.1 & 5.37E+00 & 4.77E+00 & 6.02E+00 &        7. & 3.86E+03 & 3.50E+03 & 4.23E+03 \\
  0.11 & 6.71E+00 & 5.98E+00 & 7.49E+00 &        8. & 4.45E+03 & 4.05E+03 & 4.86E+03 \\
  0.12 & 8.18E+00 & 7.31E+00 & 9.10E+00 &        9. & 5.02E+03 & 4.58E+03 & 5.48E+03 \\
  0.13 & 9.77E+00 & 8.76E+00 & 1.08E+01 &       10. & 5.59E+03 & 5.10E+03 & 6.08E+03 \\
\hline
\end{tabular*}
\begin{tabular*}{\textwidth}{@{\extracolsep{\fill}} l c }
REV  = 
$1.63 \times 10^{10}T_{9}^{3/2}{\rm exp}(-63.752/T_{9})$
 & \\
\end{tabular*}
}
\label{dpgTab2}
\end{table}
\clearpage
\subsection{\reac{2}{H}{d}{\gamma}{4}{He}}
\label{ddgSect}
The experimental data sets referred to in NACRE are ZU63 \cite{ZU63}, ME69 \cite{ME69}$^\dag$, WI85 \cite{WI85}$^\ddagger$, WE86 \cite{WE86} and BA87 \cite{BA87}$^\ddagger$, covering the 0.025 $\lsimeq E_{\rm cm} \lsimeq$ 5.4 MeV range.
Added is the post-NACRE data point ZH09 \cite{ZH09}$^{\dag\dag}$ at $E_{\rm cm} \simeq$ 0.007 MeV.
[{\footnotesize{$^\dag$re-calculated from the reverse \reac{4}{He}{\gamma}{d}{2}{H} cross sections; $^\ddagger$normalised to the $\sigma_{\rm dp}$ fit of \cite{NACRE}; $^{\dag\dag}$normalised to the extrapolated $\sigma_{\rm dp}$ of \cite{BR90}}}] 

\begin{figure}[hb]
\centering{
\includegraphics[height=0.50\textheight,width=0.9\textwidth]{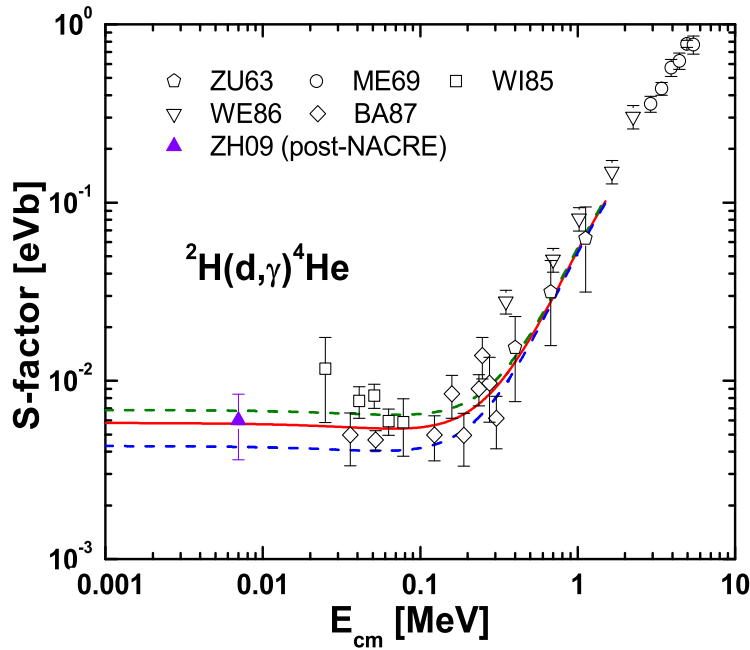}
\vspace{-0.5truecm}
\caption{The $S$-factor for \reac{2}{H}{d}{\gamma}{4}{He}. Both WI85 and WE86 are not considered in the fit because of their inexplicable energy-dependences (cf. \cite{AR11}).}
\label{ddgFig1}
}
\end{figure}

Figure \ref{ddgFig1} compares the PM and experimental $S$-factors.
No clear signature of electron screening has been detected.
The data sets (but WI85 and WE86) in the $E_{\rm cm} \lsimeq$ 2 MeV range are used for the PM fit. The main contribution at low energies is expected to come from the E2 transition resulting from $l_{i}= 2$ (with $I_{i}= 0, 2$), whereas the E1 transition  with $l_{i}= 1$ (with $I_{i}= 1$) is isospin-forbidden.
In  order to reproduce the $S$-factor observed at the lowest energies, a D-state ($l_{f}=2$) admixture in the ground state $^{4}$He (consequently, E2 transition with $l_{i} = 0$) is invoked (e.g. \cite{AS87}).
The adopted parameter values are given in Table \ref{ddgTab1}.
The resulting $S(0)$ = 5.8 $^{+1.0}_{-1.5}$ meV\,b. 

\begin{figure}[t]
\centering{
\includegraphics[height=0.33\textheight,width=0.90\textwidth]{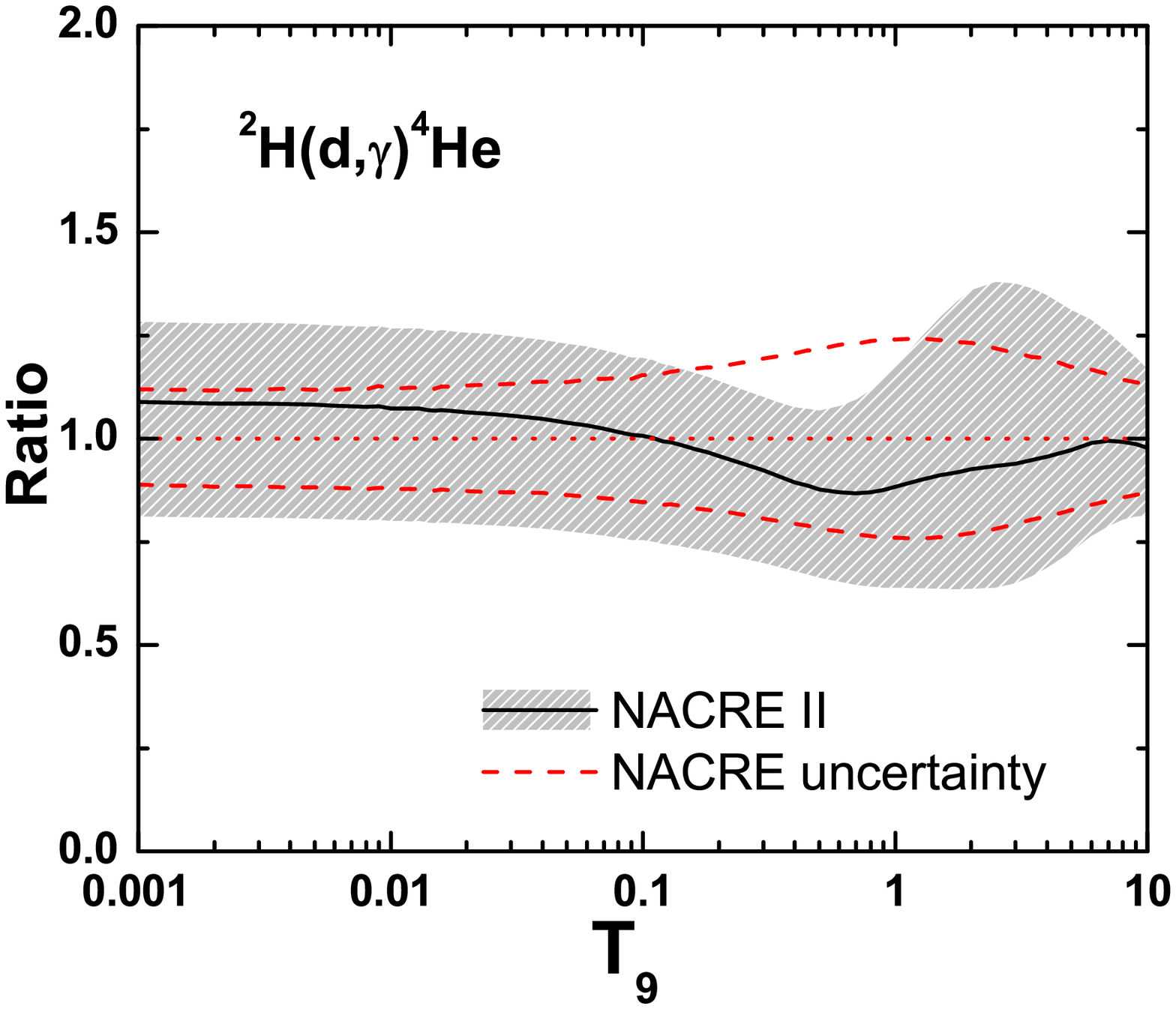}
\vspace{-0.4truecm}
\caption{\reac{2}{H}{d}{\gamma}{4}{He} rates in units of the NACRE (adopt) values.}
\label{ddgFig2}
}
\end{figure}

Table \ref{ddgTab2} gives the reaction rates at $0.001 \le T_{9} \le 10$, for which the PM-predicted and the experimental cross sections below and above $E_{\rm cm} \simeq$ 0.2 MeV are used, respectively.
Figure \ref{ddgFig2} compares the present and the NACRE rates.

{\footnotesize See \cite{AR11} for an ${ab\ initio}$ calculation.}

\clearpage

\begin{table}[hb]
\caption{\reac{2}{H}{d}{\gamma}{4}{He} rates in $\rm{cm^{3}mol^{-1}s^{-1}}$}\scriptsize\rm
\footnotesize{
\begin{tabular*}{\textwidth}{@{\extracolsep{\fill}} l c c c |l c c c}
\hline
$T_{9}$ & adopted & low & high & $T_{9}$ & adopted & low & high \\
\hline
  0.001 & 1.46E$-$15 & 1.08E$-$15 & 1.72E$-$15 &     0.14  & 4.57E$-$02 & 3.41E$-$02 & 5.45E$-$02 \\ 
  0.002 & 6.02E$-$12 & 4.47E$-$12 & 7.12E$-$12 &     0.15  & 5.26E$-$02 & 3.93E$-$02 & 6.27E$-$02 \\ 
  0.003 & 3.30E$-$10 & 2.45E$-$10 & 3.90E$-$10 &     0.16  & 5.98E$-$02 & 4.47E$-$02 & 7.13E$-$02 \\ 
  0.004 & 4.05E$-$09 & 3.01E$-$09 & 4.79E$-$09 &     0.18  & 7.48E$-$02 & 5.60E$-$02 & 8.93E$-$02 \\ 
  0.005 & 2.39E$-$08 & 1.77E$-$08 & 2.83E$-$08 &     0.2 & 9.06E$-$02 & 6.79E$-$02 & 1.08E$-$01 \\ 
  0.006 & 9.19E$-$08 & 6.82E$-$08 & 1.09E$-$07 &     0.25  & 1.32E$-$01 & 9.96E$-$02 & 1.58E$-$01 \\ 
  0.007 & 2.68E$-$07 & 1.99E$-$07 & 3.18E$-$07 &     0.3 & 1.76E$-$01 & 1.33E$-$01 & 2.11E$-$01 \\ 
  0.008 & 6.47E$-$07 & 4.80E$-$07 & 7.66E$-$07 &     0.35  & 2.21E$-$01 & 1.67E$-$01 & 2.66E$-$01 \\ 
  0.009 & 1.36E$-$06 & 1.01E$-$06 & 1.61E$-$06 &     0.4 & 2.67E$-$01 & 2.02E$-$01 & 3.23E$-$01 \\ 
  0.01  & 2.57E$-$06 & 1.90E$-$06 & 3.04E$-$06 &     0.45  & 3.15E$-$01 & 2.37E$-$01 & 3.82E$-$01 \\ 
  0.011 & 4.47E$-$06 & 3.31E$-$06 & 5.29E$-$06 &     0.5 & 3.63E$-$01 & 2.73E$-$01 & 4.44E$-$01 \\ 
  0.012 & 7.28E$-$06 & 5.41E$-$06 & 8.62E$-$06 &     0.6 & 4.65E$-$01 & 3.47E$-$01 & 5.78E$-$01 \\ 
  0.013 & 1.13E$-$05 & 8.36E$-$06 & 1.33E$-$05 &     0.7 & 5.74E$-$01 & 4.25E$-$01 & 7.27E$-$01 \\ 
  0.014 & 1.67E$-$05 & 1.24E$-$05 & 1.98E$-$05 &     0.8 & 6.92E$-$01 & 5.07E$-$01 & 8.92E$-$01 \\ 
  0.015 & 2.38E$-$05 & 1.77E$-$05 & 2.82E$-$05 &     0.9 & 8.18E$-$01 & 5.94E$-$01 & 1.07E+00 \\ 
  0.016 & 3.29E$-$05 & 2.44E$-$05 & 3.90E$-$05 &     1.     & 9.54E$-$01 & 6.86E$-$01 & 1.27E+00 \\ 
  0.018 & 5.83E$-$05 & 4.33E$-$05 & 6.91E$-$05 &     1.25  & 1.33E+00 & 9.38E$-$01 & 1.85E+00 \\ 
  0.02  & 9.52E$-$05 & 7.07E$-$05 & 1.13E$-$04 &     1.5  & 1.76E+00 & 1.22E+00 & 2.51E+00 \\ 
  0.025 & 2.52E$-$04 & 1.87E$-$04 & 2.99E$-$04 &     1.75  & 2.23E+00 & 1.53E+00 & 3.25E+00 \\ 
  0.03  & 5.27E$-$04 & 3.91E$-$04 & 6.25E$-$04 &     2.     & 2.75E+00 & 1.88E+00 & 4.05E+00 \\ 
  0.04  & 1.52E$-$03 & 1.13E$-$03 & 1.80E$-$03 &     2.5  & 3.92E+00 & 2.67E+00 & 5.81E+00 \\ 
  0.05  & 3.19E$-$03 & 2.37E$-$03 & 3.79E$-$03 &     3.     & 5.27E+00 & 3.62E+00 & 7.74E+00 \\ 
  0.06  & 5.57E$-$03 & 4.14E$-$03 & 6.62E$-$03 &     3.5  & 6.80E+00 & 4.76E+00 & 9.81E+00 \\ 
  0.07  & 8.65E$-$03 & 6.44E$-$03 & 1.03E$-$02 &     4.     & 8.50E+00 & 6.07E+00 & 1.20E+01 \\ 
  0.08  & 1.24E$-$02 & 9.22E$-$03 & 1.47E$-$02 &     5.     & 1.23E+01 & 9.17E+00 & 1.67E+01 \\ 
  0.09  & 1.68E$-$02 & 1.25E$-$02 & 1.99E$-$02 &     6.     & 1.66E+01 & 1.28E+01 & 2.17E+01 \\ 
  0.1 & 2.16E$-$02 & 1.61E$-$02 & 2.58E$-$02 &     7.     & 2.12E+01 & 1.67E+01 & 2.69E+01 \\ 
  0.11  & 2.71E$-$02 & 2.02E$-$02 & 3.22E$-$02 &     8.     & 2.58E+01 & 2.08E+01 & 3.20E+01 \\ 
  0.12  & 3.29E$-$02 & 2.45E$-$02 & 3.92E$-$02 &     9.     & 3.04E+01 & 2.49E+01 & 3.70E+01 \\ 
  0.13  & 3.91E$-$02 & 2.92E$-$02 & 4.66E$-$02 &    10.     & 3.48E+01 & 2.88E+01 & 4.18E+01 \\ 
\hline
\end{tabular*}
\begin{tabular*}{\textwidth}{@{\extracolsep{\fill}} l c }
REV  = 
$4.53 \times 10^{10}T_{9}^{3/2}{\rm exp}(-276.74/T_{9})$
 & \\
\end{tabular*}
}
\label{ddgTab2}
\end{table}
\clearpage
\subsection{\reac{2}{H}{d}{n}{3}{He}}
\label{ddnSect}
The experimental data sets referred to in NACRE are SC72 \cite{SC72}, KR87a \cite{KR87a}, BR90 \cite{BR90} and GR95 \cite{GR95}, covering the 0.007 $\lsimeq E_{\rm cm} \lsimeq$ 3.1 MeV range. [82\,-\,85] 
 were apparently superseded by KR87a. 
Added are the post-NACRE data sets HO01 \cite{HO01}, LE06a \cite{LE06a}, BY08b \cite{BY08b} and TU11 \cite{TU11}$^\dag$, extending the range down to $E_{\rm cm} \simeq$ 0.002 MeV.
GA58 \cite{GA58} is also included (cf. \cite{Cyburt2}, however).
 [{\footnotesize{$^\dag$THM via {\rm d($^{3}$He,\,n$^{3}$He})p}}]

\begin{figure}[hb]
\centering{
\includegraphics[height=0.50\textheight,width=0.9\textwidth]{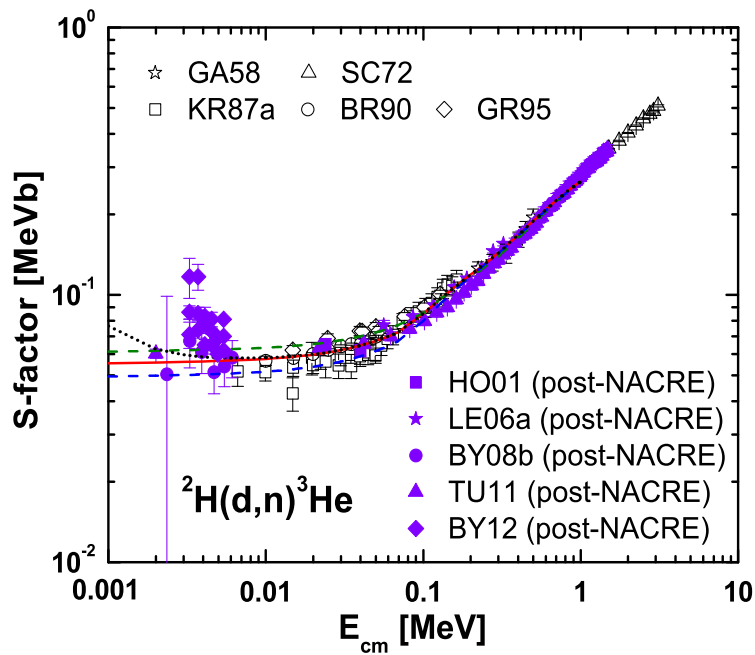}
\vspace{-0.5truecm}
\caption{The $S$-factor for \reac{2}{H}{d}{n}{3}{He}.  The dotted line indicates an adiabatic screening correction ($U_{\rm e}$ = 20.4 eV) to the 'adopt' curve (solid line). A most recent data set BY12 \cite{BY12} revealing in experiments with deuterated metals the extent and scatter of the enhanced $S$-factors at low energies is added just for comparison.}
\label{ddnFig1}
}
\end{figure}

Figure \ref{ddnFig1} compares the DWBA and experimental $S$-factors.
The data in the 0.004 $\lsimeq E_{\rm cm} \lsimeq$ 1 MeV range are used for the DWBA fit.
The symmetry of the entrance channel allows even and odd values of $l_{i}$ for $I_{i}$= 0, 2 and = 1, respectively. 
The adopted parameter values are given in Table \ref{ddnTab1}.
The resulting $S(0)$ = 55.5 $\pm$ 6.0 keV\,b. 
In comparison, $S(0)$ = 55 keV\,b [NACRE, quadratic polynomial], and 52.4 $\pm$ 3.5 keV\,b [BBN04].

\begin{figure}[t]
\centering{
\includegraphics[height=0.33\textheight,width=0.9\textwidth]{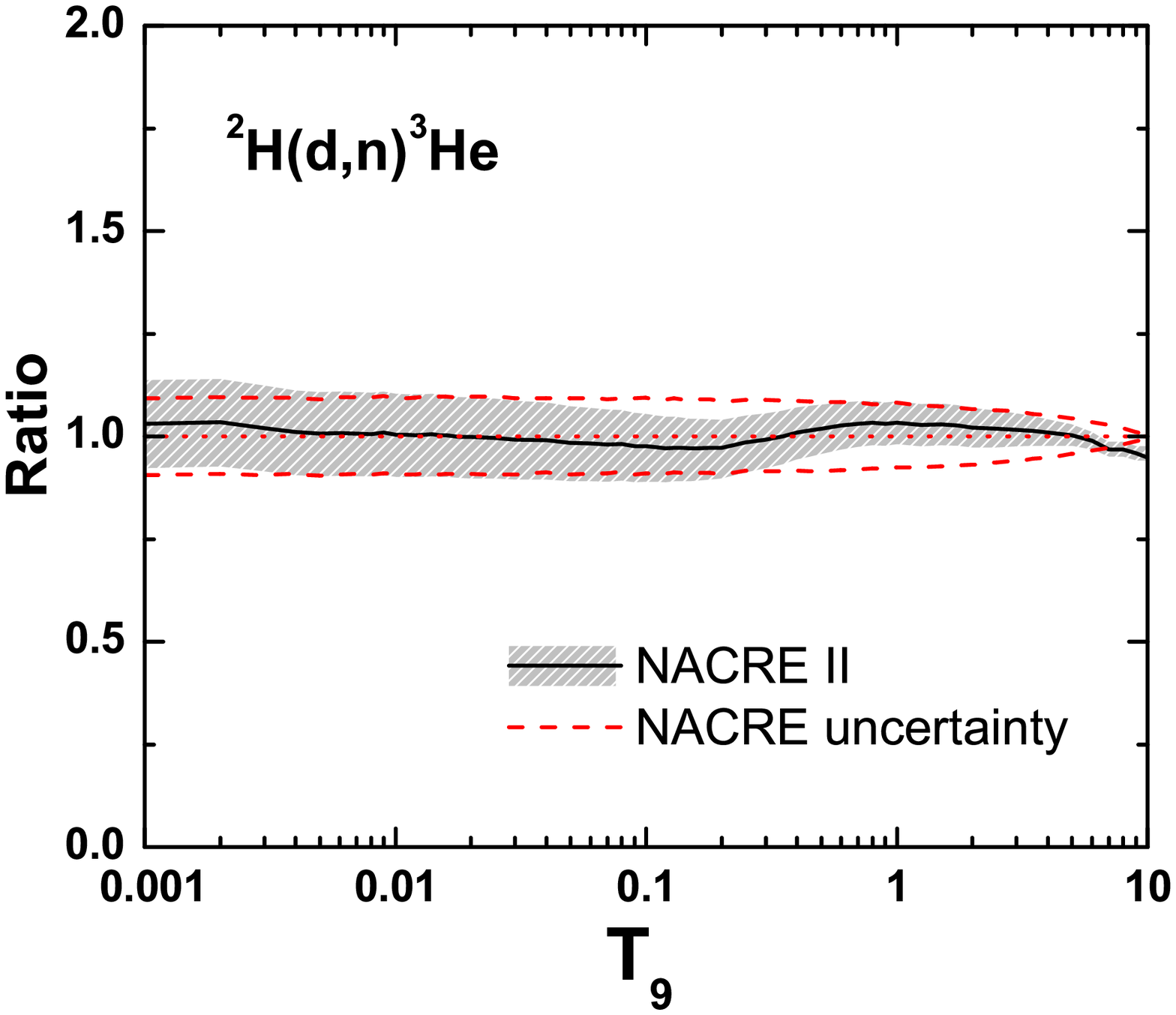}
\vspace{-0.4truecm}
\caption{\reac{2}{H}{d}{n}{3}{He} rates in units of the NACRE (adopt) values.}
\
\label{ddnFig2}
}
\end{figure}

Table \ref{ddnTab2} gives the reaction rates at 0.001 $\le T_{9} \le$ 10, for which the DWBA-predicted and the experimental cross sections below and above $E_{\rm cm} \simeq$ 0.1 MeV are used, respectively.  Figure \ref{ddnFig2} compares the present and the NACRE rates.
 
{\footnotesize See  \cite{AR11} for an ${ab\ initio}$ calculation; \cite{KO90} for a second-order DWBA calculation.} 

\clearpage

\begin{table}[hb]
\caption{\reac{2}{H}{d}{n}{3}{He} rates in $\rm{cm^{3}mol^{-1}s^{-1}}$.} \footnotesize\rm
\begin{tabular*}{\textwidth}{@{\extracolsep{\fill}} l c c c |l c c c}
\hline
$T_{9}$ & adopted & low & high & $T_{9}$ & adopted & low & high \\
\hline
  0.001 & 1.43E$-$08 & 1.28E$-$08 & 1.58E$-$08 &      0.14  & 5.48E+05 & 5.01E+05 & 5.92E+05 \\
  0.002 & 5.95E$-$05 & 5.31E$-$05 & 6.57E$-$05 &      0.15  & 6.38E+05 & 5.83E+05 & 6.88E+05 \\
  0.003 & 3.23E$-$03 & 2.89E$-$03 & 3.57E$-$03 &      0.16  & 7.32E+05 & 6.70E+05 & 7.89E+05 \\
  0.004 & 3.95E$-$02 & 3.53E$-$02 & 4.36E$-$02 &      0.18  & 9.36E+05 & 8.59E+05 & 1.01E+06 \\
  0.005 & 2.34E$-$01 & 2.09E$-$01 & 2.58E$-$01 &      0.2 & 1.16E+06 & 1.06E+06 & 1.24E+06 \\
  0.006 & 9.01E$-$01 & 8.05E$-$01 & 9.95E$-$01 &      0.25  & 1.77E+06 & 1.64E+06 & 1.90E+06 \\
  0.007 & 2.64E+00 & 2.36E+00 & 2.91E+00 &      0.3 & 2.46E+06 & 2.28E+06 & 2.63E+06 \\
  0.008 & 6.38E+00 & 5.70E+00 & 7.04E+00 &      0.35  & 3.20E+06 & 2.97E+06 & 3.41E+06 \\
  0.009 & 1.34E+01 & 1.20E+01 & 1.48E+01 &      0.4 & 3.98E+06 & 3.70E+06 & 4.23E+06 \\
  0.01  & 2.54E+01 & 2.27E+01 & 2.80E+01 &      0.45  & 4.78E+06 & 4.46E+06 & 5.08E+06 \\
  0.011 & 4.44E+01 & 3.97E+01 & 4.89E+01 &      0.5 & 5.59E+06 & 5.23E+06 & 5.93E+06 \\
  0.012 & 7.25E+01 & 6.49E+01 & 7.99E+01 &      0.6 & 7.25E+06 & 6.80E+06 & 7.67E+06 \\
  0.013 & 1.12E+02 & 1.01E+02 & 1.24E+02 &      0.7 & 8.90E+06 & 8.38E+06 & 9.40E+06 \\
  0.014 & 1.67E+02 & 1.49E+02 & 1.84E+02 &      0.8 & 1.05E+07 & 9.95E+06 & 1.11E+07 \\
  0.015 & 2.39E+02 & 2.14E+02 & 2.63E+02 &      0.9 & 1.22E+07 & 1.15E+07 & 1.28E+07 \\
  0.016 & 3.31E+02 & 2.96E+02 & 3.64E+02 &      1.     & 1.37E+07 & 1.30E+07 & 1.45E+07 \\
  0.018 & 5.88E+02 & 5.27E+02 & 6.48E+02 &      1.25  & 1.76E+07 & 1.66E+07 & 1.85E+07 \\
  0.02  & 9.64E+02 & 8.64E+02 & 1.06E+03 &      1.5  & 2.12E+07 & 2.01E+07 & 2.23E+07 \\
  0.025 & 2.58E+03 & 2.31E+03 & 2.84E+03 &      1.75  & 2.46E+07 & 2.34E+07 & 2.59E+07 \\
  0.03  & 5.44E+03 & 4.88E+03 & 5.97E+03 &      2.     & 2.79E+07 & 2.65E+07 & 2.93E+07 \\
  0.04  & 1.60E+04 & 1.44E+04 & 1.75E+04 &      2.5  & 3.39E+07 & 3.23E+07 & 3.55E+07 \\
  0.05  & 3.40E+04 & 3.07E+04 & 3.73E+04 &      3.     & 3.93E+07 & 3.76E+07 & 4.10E+07 \\
  0.06  & 6.04E+04 & 5.45E+04 & 6.60E+04 &      3.5  & 4.43E+07 & 4.25E+07 & 4.60E+07 \\
  0.07  & 9.52E+04 & 8.61E+04 & 1.04E+05 &      4.     & 4.87E+07 & 4.70E+07 & 5.05E+07 \\
  0.08  & 1.38E+05 & 1.25E+05 & 1.51E+05 &      5.     & 5.65E+07 & 5.48E+07 & 5.81E+07 \\
  0.09  & 1.89E+05 & 1.72E+05 & 2.06E+05 &      6.     & 6.26E+07 & 6.10E+07 & 6.42E+07 \\
  0.1 & 2.48E+05 & 2.25E+05 & 2.69E+05 &      7.     & 6.73E+07 & 6.58E+07 & 6.88E+07 \\
  0.11  & 3.13E+05 & 2.85E+05 & 3.40E+05 &      8.     & 7.27E+07 & 7.11E+07 & 7.44E+07 \\
  0.12  & 3.86E+05 & 3.51E+05 & 4.17E+05 &      9.     & 7.72E+07 & 7.54E+07 & 7.89E+07 \\
  0.13  & 4.64E+05 & 4.23E+05 & 5.02E+05 &     10.     & 8.13E+07 & 8.03E+07 & 8.40E+07 \\
\hline
\end{tabular*}
\begin{tabular*}{\textwidth}{@{\extracolsep{\fill}} l c }
REV  = 
$1.73\,{\rm exp}(-37.936/T_{9})$  
 & \\
\end{tabular*}
\label{ddnTab2}
\end{table}
\clearpage\subsection{\reac{2}{H}{d}{p}{3}{H}}
\label{ddpSect}
The experimental data sets referred to in NACRE are SC72 \cite{SC72}, KR87a \cite{KR87a}, BR90 \cite{BR90} and GR95 \cite{GR95}, covering the 0.0016 $\lsimeq E_{\rm cm} \lsimeq$ 3.1 MeV range. 
\cite{MC51,AR54,JA85,WE52} were apparently superseded by KR87a.
Added are the post-NACRE data sets LE06a \cite{LE06a} and TU11 \cite{TU11}$^\dag$. 
[{\footnotesize{$^\dag$from {\rm d($^{3}$He,\,p$^{3}$H})p (THM);
\cite{RI05} (THM) is not considered because of the stated "preliminary" nature of the data.}}]

\begin{figure}[hb]
\centering{
\includegraphics[height=0.50\textheight,width=0.9\textwidth]{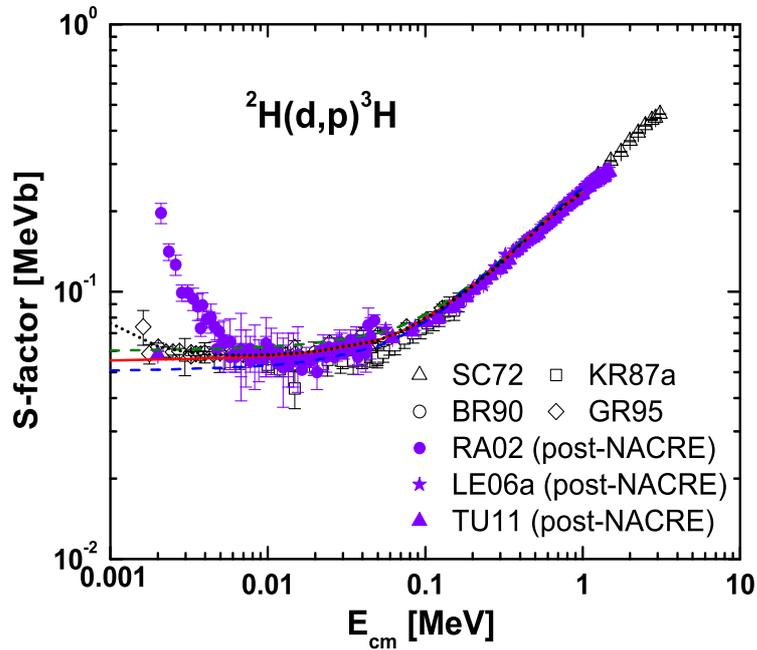}
\vspace{-0.5truecm}
\caption{The $S$-factor for \reac{2}{H}{d}{p}{3}{H}. The dotted line indicates an adiabatic screening correction ($U_{\rm e}$ = 20.4 eV) to the 'adopt' curve (solid line).  RA02 \cite{RA02} with a TaD target is added just to exhibit the hugely enhanced $S$-factors at low energies.}
\label{ddpFig1}
}
\end{figure}

Figure \ref{ddpFig1} compares the DWBA and experimental $S$-factors.
GR95 using a D$_{2}$ gas target appears to indicate a modest electron screening effect, if at all, at the lowest energies. 
In contrast, measurements with deuterated metal targets indicate huge cross section enhancements at low energies (see RA02 \cite{RA02} in Fig.\,\ref{ddpFig1}). 
The data in the 0.004 $\lsimeq E_{\rm cm} \lsimeq$ 1 MeV range are used for the DWBA fit.
The symmetry of the entrance channel allows even and odd values of $l_{i}$ for $I_{i}$= 0, 2 and = 1, respectively. 
The adopted parameter values are given in Table \ref{ddpTab1}.
The present $S(0)$ = 56.2$^{+4.9}_{-4.7}$ keV\,b. 
In comparison, $S(0)$= 56 keV\,b  [NACRE, quadratic polynomial], and 57.1 $\pm$ 0.8 keV\,b [BBN04].

\begin{figure}[t]
\centering{
\includegraphics[height=0.33\textheight,width=0.9\textwidth]{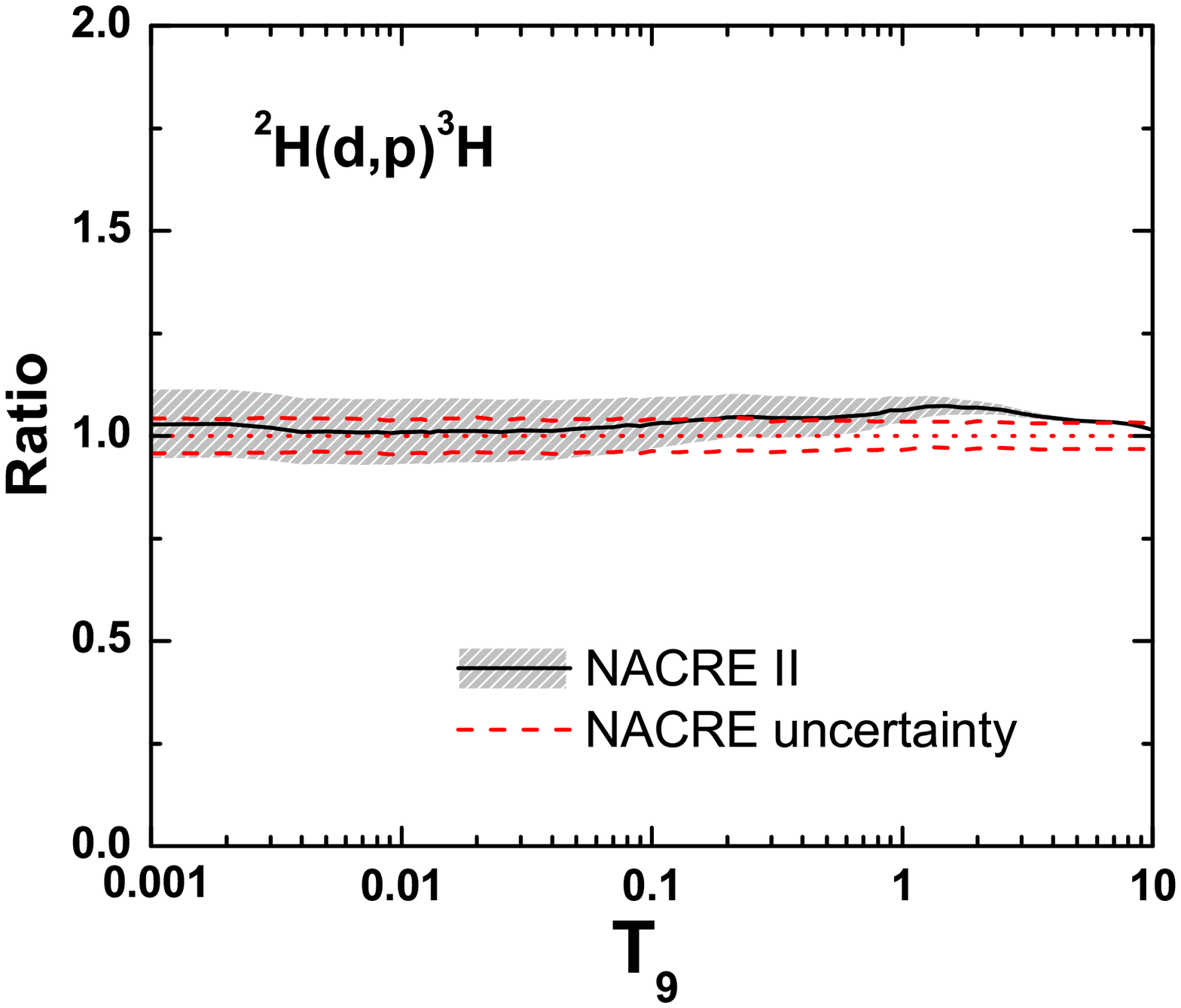}
\vspace{-0.4truecm}
\caption{\reac{2}{H}{d}{p}{3}{H} rates in units of the NACRE (adopt) values.
}
\label{ddpFig2}
}
\end{figure}

Table \ref{ddpTab2} gives the reaction rates at 0.001 $\le T_{9} \le$ 10, for which the DWBA-predicted and experimental cross sections below and above $E_{\rm cm} \simeq$ 0.1 MeV are used, respectively.
Figure \ref{ddpFig2} compares the present and the NACRE rates.

{\footnotesize See the \reac{2}{H}{d}{n}{3}{He} counterpart in Sect.\,\ref{ddnSect}.}

\clearpage

\begin{table}[hb]
\caption{\reac{2}{H}{d}{p}{3}{H} rates in $\rm{cm^{3}mol^{-1}s^{-1}}$}\footnotesize\rm
\begin{tabular*}{\textwidth}{@{\extracolsep{\fill}} l c c c |l c c c}
\hline
$T_{9}$ & adopted & low & high & $T_{9}$ & adopted & low & high \\
\hline
  0.001 & 1.45E$-$08 & 1.33E$-$08 & 1.57E$-$08 &        0.14  & 5.57E+05 & 5.26E+05 & 5.91E+05 \\
  0.002 & 6.02E$-$05 & 5.52E$-$05 & 6.54E$-$05 &        0.15  & 6.47E+05 & 6.12E+05 & 6.87E+05 \\
  0.003 & 3.27E$-$03 & 3.00E$-$03 & 3.55E$-$03 &        0.16  & 7.42E+05 & 7.02E+05 & 7.87E+05 \\
  0.004 & 4.00E$-$02 & 3.67E$-$02 & 4.34E$-$02 &        0.18  & 9.44E+05 & 8.95E+05 & 1.00E+06 \\
  0.005 & 2.36E$-$01 & 2.17E$-$01 & 2.57E$-$01 &        0.2 & 1.16E+06 & 1.10E+06 & 1.23E+06 \\
  0.006 & 9.12E$-$01 & 8.38E$-$01 & 9.89E$-$01 &        0.25  & 1.75E+06 & 1.66E+06 & 1.84E+06 \\
  0.007 & 2.67E+00 & 2.45E+00 & 2.90E+00 &        0.3 & 2.38E+06 & 2.26E+06 & 2.51E+06 \\
  0.008 & 6.46E+00 & 5.94E+00 & 7.00E+00 &        0.35  & 3.04E+06 & 2.89E+06 & 3.20E+06 \\
  0.009 & 1.36E+01 & 1.25E+01 & 1.47E+01 &        0.4 & 3.72E+06 & 3.54E+06 & 3.91E+06 \\
  0.01  & 2.57E+01 & 2.37E+01 & 2.79E+01 &        0.45  & 4.41E+06 & 4.20E+06 & 4.63E+06 \\
  0.011 & 4.49E+01 & 4.13E+01 & 4.87E+01 &        0.5 & 5.11E+06 & 4.87E+06 & 5.36E+06 \\
  0.012 & 7.34E+01 & 6.76E+01 & 7.96E+01 &        0.6 & 6.52E+06 & 6.23E+06 & 6.82E+06 \\
  0.013 & 1.14E+02 & 1.05E+02 & 1.23E+02 &        0.7 & 7.93E+06 & 7.61E+06 & 8.27E+06 \\
  0.014 & 1.69E+02 & 1.56E+02 & 1.83E+02 &        0.8 & 9.34E+06 & 8.99E+06 & 9.71E+06 \\
  0.015 & 2.42E+02 & 2.23E+02 & 2.62E+02 &        0.9 & 1.07E+07 & 1.04E+07 & 1.11E+07 \\
  0.016 & 3.35E+02 & 3.09E+02 & 3.63E+02 &        1.     & 1.21E+07 & 1.17E+07 & 1.25E+07 \\
  0.018 & 5.96E+02 & 5.49E+02 & 6.45E+02 &        1.25  & 1.54E+07 & 1.50E+07 & 1.59E+07 \\
  0.02  & 9.77E+02 & 9.01E+02 & 1.06E+03 &        1.5  & 1.86E+07 & 1.81E+07 & 1.90E+07 \\
  0.025 & 2.62E+03 & 2.41E+03 & 2.83E+03 &        1.75  & 2.15E+07 & 2.11E+07 & 2.19E+07 \\
  0.03  & 5.51E+03 & 5.10E+03 & 5.95E+03 &        2.     & 2.43E+07 & 2.38E+07 & 2.47E+07 \\
  0.04  & 1.62E+04 & 1.50E+04 & 1.74E+04 &        2.5  & 2.94E+07 & 2.89E+07 & 2.98E+07 \\
  0.05  & 3.46E+04 & 3.21E+04 & 3.72E+04 &        3.     & 3.40E+07 & 3.35E+07 & 3.44E+07 \\
  0.06  & 6.14E+04 & 5.71E+04 & 6.59E+04 &        3.5  & 3.82E+07 & 3.77E+07 & 3.86E+07 \\
  0.07  & 9.68E+04 & 9.02E+04 & 1.04E+05 &        4.     & 4.21E+07 & 4.17E+07 & 4.25E+07 \\
  0.08  & 1.41E+05 & 1.31E+05 & 1.50E+05 &        5.     & 4.91E+07 & 4.88E+07 & 4.95E+07 \\
  0.09  & 1.93E+05 & 1.80E+05 & 2.06E+05 &        6.     & 5.54E+07 & 5.51E+07 & 5.58E+07 \\
  0.1 & 2.52E+05 & 2.37E+05 & 2.69E+05 &        7.     & 6.10E+07 & 6.07E+07 & 6.14E+07 \\
  0.11  & 3.19E+05 & 3.00E+05 & 3.40E+05 &        8.     & 6.59E+07 & 6.55E+07 & 6.63E+07 \\
  0.12  & 3.92E+05 & 3.69E+05 & 4.17E+05 &        9.     & 7.00E+07 & 6.96E+07 & 7.03E+07 \\
  0.13  & 4.72E+05 & 4.45E+05 & 5.02E+05 &       10.     & 7.33E+07 & 7.29E+07 & 7.36E+07 \\
\hline
\end{tabular*}
\begin{tabular*}{\textwidth}{@{\extracolsep{\fill}} l c }
REV  = 
$1.73\,{\rm exp}(-46.799/T_{9})$
 & \\
\end{tabular*}
\label{ddpTab2}
\end{table}
\clearpage
\subsection{\reac{2}{H}{\alpha}{\gamma}{6}{Li}}
\label{dagSect}
The experimental data sets referred to in NACRE are RO81 \cite{RO81}, KI91 \cite{KI91}$^\dag$ and MO94 \cite{MO94}, covering the 0.07 $\lsimeq E_{\rm cm} \lsimeq$ 8.3 MeV range. 
No new cross section data are found, but HA10 \cite{HA10}$^\ddagger$ is additionally considered. 
[{\footnotesize{$^\dag$from Coulomb break-up (digital data given in the NACRE home page). $^\ddagger$potential-model analysis after Coulomb break-up experiments.}}] 

\begin{figure}[hb]
\centering{
\includegraphics[height=0.50\textheight,width=0.9\textwidth]{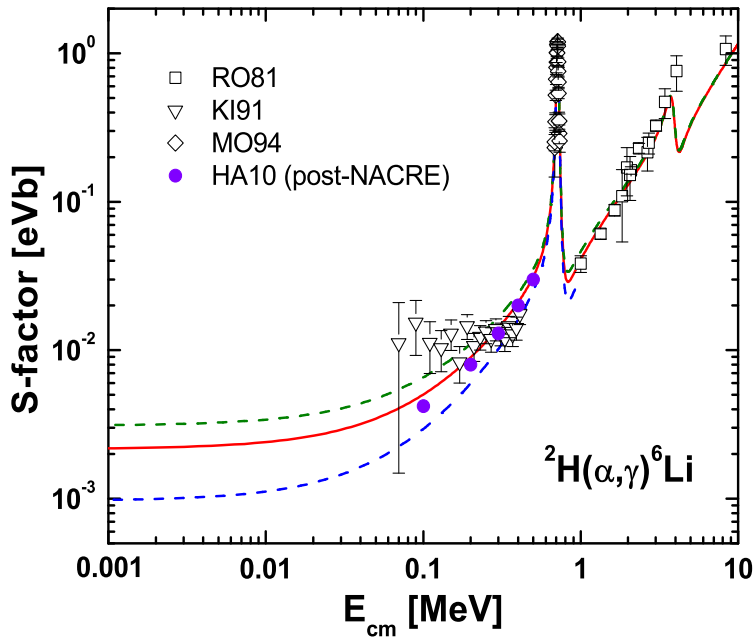}
\vspace{-0.5truecm}
\caption{The $S$-factor for \reac{2}{H}{\alpha}{\gamma}{6}{Li}.
Note that HA10 points between $E_{\rm cm}$ 0.1 and 0.5 are read off from the $S$-factor {\it curve} of the potential model analysis after the Coulomb break-up experiments (see text), which gives $S(0) \simeq 1.9$ meV\,b.   
}
\label{dagFig1}
}
\end{figure}

Figure \ref{dagFig1} compares the PM and experimental $S$-factors.
The data in the $E_{\rm cm} \lsimeq$ 3 MeV range are used for the PM fit. They exhibit the $3^{+}$ and $2^{+}$ resonances at $E_{\rm R} \simeq$  0.71 and 3.89 MeV. 
As discussed in HA10 in detail, Coulomb break-up experiments get progressively unreliable at lower energies because of the larger nuclear break-up contributions to be separated.
Below $E_{\rm cm} \simeq$ 0.7 MeV, therefore, we adopt the model analysis of HA10 to guide the extrapolation (see Fig. \ref{dagFig1}).
 The quasi-orthogonality between the $J_{i}=1^{+}$ scattering state and the $1^{+}$ ground state of $^{6}$Li suppresses M1 transitions (s-wave) at low energies. Isospin-forbidden E1 transitions (p-wave) thus dominate at very low energies, but are surpassed by the E2 transitions (d-wave) at higher energies. 
The adopted parameter values are given in Table \ref{dagTab1}. The present $S(0)$ = $2.2_{-1.2}^{+0.9}$ meV\,b.  

\begin{figure}[t]
\centering{
\includegraphics[height=0.33\textheight,width=0.90\textwidth]{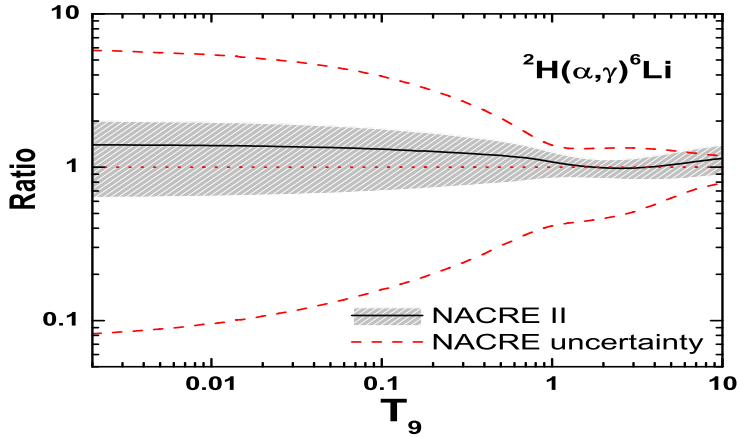}
\vspace{-0.4truecm}
\caption{\reac{2}{H}{\alpha}{\gamma}{6}{Li} rates in units of the NACRE (adopt) values.
Much larger uncertainties in NACRE stem from the adoption of the various estimates of the E1 contributions in early studies, and in particular of KI91 for the upper limit.
}
\label{dagFig2}
}
\end{figure}

Table \ref{dagTab2}  gives the reaction rates at 0.002 $\le T_{9} \le$ 10, for which the PM-predicted and the experimental cross sections below and above $E_{\rm cm} \simeq$ 0.7 MeV are used, respectively. 
Figure \ref{dagFig2} compares the present and the NACRE rates.

{\footnotesize  See  \cite{NO01a} for an ${\it{ab\ initio}}$ calculation}

\begin{table}[hb]
\caption{\reac{2}{H}{\alpha}{\gamma}{6}{Li} rates in $\rm{cm^{3}mol^{-1}s^{-1}}$}\footnotesize\rm
\begin{tabular*}{\textwidth}{@{\extracolsep{\fill}} l c c c | l c c c}
\hline
$T_{9}$ & adopted & low & high & $T_{9}$ & adopted & low & high \\
\hline
  0.002 & 2.98E$-$23 & 1.35E$-$23 & 4.26E$-$23 &      0.15 & 1.13E$-$04 & 6.29E$-$05 & 1.50E$-$04 \\
  0.003 & 4.01E$-$20 & 1.83E$-$20 & 5.72E$-$20 &      0.16 & 1.49E$-$04 & 8.40E$-$05 & 1.99E$-$04 \\
  0.004 & 3.73E$-$18 & 1.71E$-$18 & 5.32E$-$18 &      0.18 & 2.46E$-$04 & 1.41E$-$04 & 3.26E$-$04 \\
  0.005 & 9.34E$-$17 & 4.29E$-$17 & 1.33E$-$16 &      0.2 & 3.80E$-$04 & 2.20E$-$04 & 4.99E$-$04 \\
  0.006 & 1.09E$-$15 & 5.00E$-$16 & 1.54E$-$15 &      0.25 & 9.10E$-$04 & 5.43E$-$04 & 1.18E$-$03 \\
  0.007 & 7.67E$-$15 & 3.55E$-$15 & 1.09E$-$14 &      0.3 & 1.79E$-$03 & 1.09E$-$03 & 2.30E$-$03 \\
  0.008 & 3.85E$-$14 & 1.78E$-$14 & 5.46E$-$14 &      0.35 & 3.09E$-$03 & 1.93E$-$03 & 3.94E$-$03 \\
  0.009 & 1.50E$-$13 & 6.98E$-$14 & 2.13E$-$13 &      0.4 & 4.89E$-$03 & 3.12E$-$03 & 6.16E$-$03 \\
  0.01 & 4.84E$-$13 & 2.26E$-$13 & 6.86E$-$13 &      0.45 & 7.24E$-$03 & 4.70E$-$03 & 9.05E$-$03 \\
  0.011 & 1.35E$-$12 & 6.30E$-$13 & 1.91E$-$12 &      0.5 & 1.02E$-$02 & 6.75E$-$03 & 1.27E$-$02 \\
  0.012 & 3.33E$-$12 & 1.56E$-$12 & 4.71E$-$12 &      0.6 & 1.84E$-$02 & 1.26E$-$02 & 2.24E$-$02 \\
  0.013 & 7.48E$-$12 & 3.52E$-$12 & 1.06E$-$11 &      0.7 & 3.04E$-$02 & 2.15E$-$02 & 3.65E$-$02 \\
  0.014 & 1.55E$-$11 & 7.31E$-$12 & 2.19E$-$11 &      0.8 & 4.78E$-$02 & 3.51E$-$02 & 5.66E$-$02 \\
  0.015 & 3.01E$-$11 & 1.42E$-$11 & 4.25E$-$11 &      0.9 & 7.26E$-$02 & 5.50E$-$02 & 8.48E$-$02 \\
  0.016 & 5.51E$-$11 & 2.61E$-$11 & 7.77E$-$11 &      1. & 1.06E$-$01 & 8.29E$-$02 & 1.23E$-$01 \\
  0.018 & 1.61E$-$10 & 7.65E$-$11 & 2.26E$-$10 &      1.25 & 2.36E$-$01 & 1.93E$-$01 & 2.69E$-$01 \\
  0.02 & 4.04E$-$10 & 1.93E$-$10 & 5.68E$-$10 &      1.5 & 4.28E$-$01 & 3.59E$-$01 & 4.84E$-$01 \\
  0.025 & 2.55E$-$09 & 1.23E$-$09 & 3.58E$-$09 &      1.75 & 6.67E$-$01 & 5.64E$-$01 & 7.53E$-$01 \\
  0.03 & 1.04E$-$08 & 5.07E$-$09 & 1.46E$-$08 &      2. & 9.32E$-$01 & 7.92E$-$01 & 1.05E+00 \\
  0.04 & 8.06E$-$08 & 3.99E$-$08 & 1.12E$-$07 &      2.5 & 1.49E+00 & 1.26E+00 & 1.70E+00 \\
  0.05 & 3.45E$-$07 & 1.74E$-$07 & 4.77E$-$07 &      3. & 2.03E+00 & 1.71E+00 & 2.34E+00 \\
  0.06 & 1.05E$-$06 & 5.34E$-$07 & 1.44E$-$06 &      3.5 & 2.57E+00 & 2.13E+00 & 2.99E+00 \\
  0.07 & 2.54E$-$06 & 1.31E$-$06 & 3.49E$-$06 &      4. & 3.10E+00 & 2.54E+00 & 3.64E+00 \\
  0.08 & 5.29E$-$06 & 2.77E$-$06 & 7.23E$-$06 &      5. & 4.22E+00 & 3.39E+00 & 5.04E+00 \\
  0.09 & 9.85E$-$06 & 5.21E$-$06 & 1.34E$-$05 &      6. & 5.50E+00 & 4.36E+00 & 6.63E+00 \\
  0.1 & 1.68E$-$05 & 8.98E$-$06 & 2.28E$-$05 &      7. & 6.98E+00 & 5.49E+00 & 8.46E+00 \\
  0.11 & 2.69E$-$05 & 1.45E$-$05 & 3.63E$-$05 &      8. & 8.67E+00 & 6.79E+00 & 1.05E+01 \\
  0.12 & 4.08E$-$05 & 2.22E$-$05 & 5.49E$-$05 &      9. & 1.05E+01 & 8.23E+00 & 1.28E+01 \\
  0.13 & 5.92E$-$05 & 3.25E$-$05 & 7.94E$-$05 &     10. & 1.25E+01 & 9.79E+00 & 1.53E+01 \\
  0.14 & 8.29E$-$05 & 4.59E$-$05 & 1.11E$-$04 &            &          &          &          \\
\hline
\end{tabular*}
\begin{tabular*}{\textwidth}{@{\extracolsep{\fill}} l c }
REV  =
$1.53 \times 10^{10}\,T_{9}^{3/2}{\rm exp}(-17.104/T_{9})\,/\,[1.0+2.333{\rm exp}(-25.369/T_{9})]$ 
 & \\
\end{tabular*}
\label{dagTab2}
\end{table}
\clearpage
\subsection{\reac{3}{H}{d}{n}{4}{He}}
\label{tdnSect}
The experimental data sets referred to in NACRE are BR51a \cite{BR51a}, AR52 \cite{AR52}, CO52 \cite{CO52},  AR54 \cite{AR54}, HE55 \cite{HE55}, GA56 \cite{GA56}, BA57 \cite{BA57}, GO61 \cite{GO61}, KO66 \cite{KO66}, MC73 \cite{MC73}, MA75 \cite{MA75}, JA84 \cite{JA84} and BR87a \cite{BR87a}, covering the 0.005 $\lsimeq E_{\rm cm} \lsimeq$ 9.6 MeV range. 
No new cross section data are found.

\begin{figure}[hb]
\centering{
\includegraphics[height=0.50\textheight,width=0.9\textwidth]{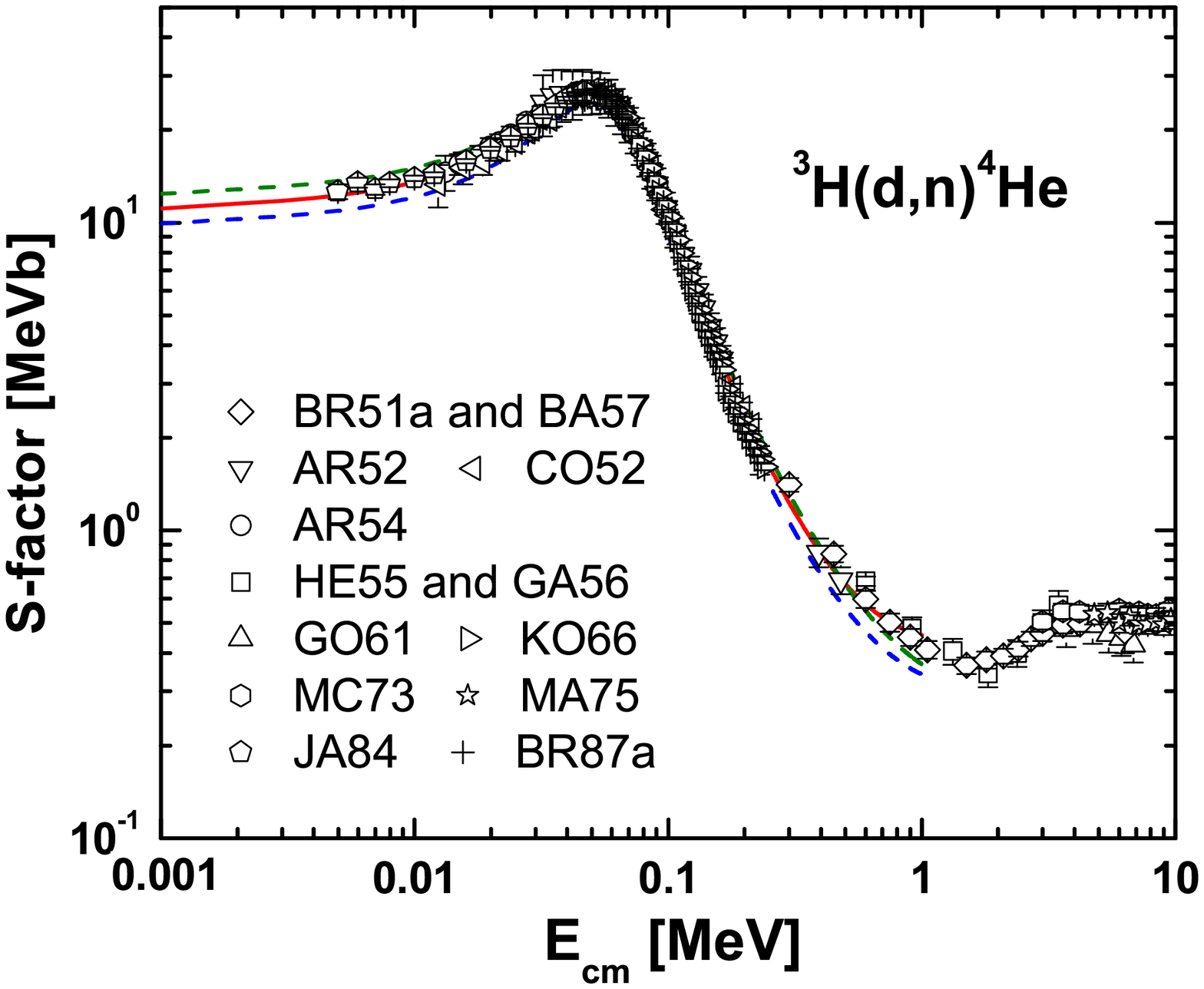}
\vspace{-0.5truecm}
\caption{The $S$-factor for \reac{3}{H}{d}{n}{4}{He}.}
\label{tdnFig1}
}
\end{figure}

Figure \ref{tdnFig1} compares the DWBA and experimental $S$-factors.
The data in the $E_{\rm cm} \lsimeq$  1 MeV range are used for the DWBA fit. They exhibit the $3/2^{+}$ resonance at $E_{\rm R} \simeq$  0.05 MeV.
The adopted parameter values are given in Table \ref{tdnTab1}.
The present $S(0)$ = 11 $\pm$  1 MeV\,b. 
In comparison, $S(0)$ =  11.7 $\pm$ 0.2 MeV\,b [BBN04].

\begin{figure}[t]
\centering{
\includegraphics[height=0.33\textheight,width=0.90\textwidth]{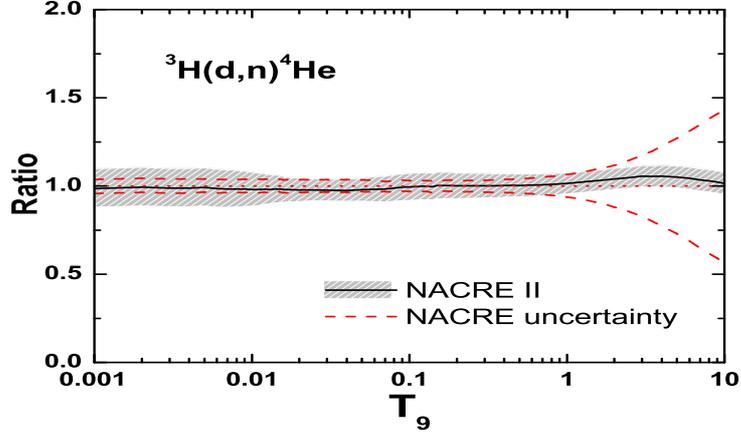}
\vspace{-0.4truecm}
\caption{\reac{3}{H}{d}{n}{4}{He} rates in units of the NACRE (adopt) values. The origin of the NACRE large uncertainties diverging at the highest temperatures is unknown.
}
\label{tdnFig2}
}
\end{figure}

Table \ref{tdnTab2} gives the reaction rates at 0.001 $\le T_{9} \le$ 10, for which the DWBA-predicted and the experimental cross sections below and above $E_{\rm cm} \simeq$ 0.01 MeV are used, respectively.
Figure \ref{tdnFig2} compares the present and the NACRE rates.

{\footnotesize  See  \cite{NA12} for an ${\it{ab\ initio}}$ calculation.}

\clearpage

\begin{table}
\caption{\reac{3}{H}{d}{n}{4}{He} rates in $\rm{cm^{3}mol^{-1}s^{-1}}$.} \footnotesize\rm
\begin{tabular*}{\textwidth}{@{\extracolsep{\fill}} l c c c | l c c c}
\hline
$T_{9}$ & adopted & low & high & $T_{9}$ & adopted & low & high \\
\hline
  0.001 & 1.87E$-$07 & 1.67E$-$07 & 2.09E$-$07 &        0.14  & 1.05E+08 & 9.66E+07 & 1.13E+08 \\  
  0.002 & 1.37E$-$03 & 1.22E$-$03 & 1.53E$-$03 &        0.15  & 1.21E+08 & 1.12E+08 & 1.31E+08 \\  
  0.003 & 9.98E$-$02 & 8.90E$-$02 & 1.11E$-$01 &        0.16  & 1.38E+08 & 1.28E+08 & 1.49E+08 \\  
  0.004 & 1.47E+00 & 1.31E+00 & 1.64E+00 &        0.18  & 1.73E+08 & 1.60E+08 & 1.85E+08 \\  
  0.005 & 9.93E+00 & 8.86E+00 & 1.10E+01 &        0.2 & 2.06E+08 & 1.91E+08 & 2.21E+08 \\  
  0.006 & 4.24E+01 & 3.78E+01 & 4.71E+01 &        0.25  & 2.83E+08 & 2.63E+08 & 3.04E+08 \\  
  0.007 & 1.35E+02 & 1.20E+02 & 1.50E+02 &        0.3 & 3.47E+08 & 3.23E+08 & 3.72E+08 \\  
  0.008 & 3.49E+02 & 3.12E+02 & 3.87E+02 &        0.35  & 3.98E+08 & 3.70E+08 & 4.25E+08 \\  
  0.009 & 7.80E+02 & 7.01E+02 & 8.62E+02 &        0.4 & 4.36E+08 & 4.07E+08 & 4.66E+08 \\  
  0.01  & 1.56E+03 & 1.41E+03 & 1.71E+03 &        0.45  & 4.65E+08 & 4.34E+08 & 4.96E+08 \\  
  0.011 & 2.85E+03 & 2.59E+03 & 3.12E+03 &        0.5 & 4.85E+08 & 4.53E+08 & 5.17E+08 \\  
  0.012 & 4.87E+03 & 4.44E+03 & 5.31E+03 &        0.6 & 5.10E+08 & 4.77E+08 & 5.42E+08 \\  
  0.013 & 7.85E+03 & 7.19E+03 & 8.53E+03 &        0.7 & 5.19E+08 & 4.86E+08 & 5.51E+08 \\  
  0.014 & 1.21E+04 & 1.11E+04 & 1.31E+04 &        0.8 & 5.19E+08 & 4.87E+08 & 5.50E+08 \\  
  0.015 & 1.79E+04 & 1.65E+04 & 1.93E+04 &        0.9 & 5.13E+08 & 4.82E+08 & 5.44E+08 \\  
  0.016 & 2.56E+04 & 2.37E+04 & 2.75E+04 &        1.     & 5.04E+08 & 4.74E+08 & 5.34E+08 \\  
  0.018 & 4.81E+04 & 4.48E+04 & 5.16E+04 &        1.25  & 4.76E+08 & 4.48E+08 & 5.04E+08 \\  
  0.02  & 8.30E+04 & 7.74E+04 & 8.87E+04 &        1.5  & 4.47E+08 & 4.21E+08 & 4.73E+08 \\  
  0.025 & 2.47E+05 & 2.32E+05 & 2.63E+05 &        1.75  & 4.19E+08 & 3.95E+08 & 4.43E+08 \\  
  0.03  & 5.70E+05 & 5.34E+05 & 6.06E+05 &        2.     & 3.94E+08 & 3.71E+08 & 4.17E+08 \\  
  0.04  & 1.93E+06 & 1.81E+06 & 2.06E+06 &        2.5  & 3.52E+08 & 3.32E+08 & 3.72E+08 \\  
  0.05  & 4.62E+06 & 4.30E+06 & 4.93E+06 &        3.     & 3.19E+08 & 3.00E+08 & 3.37E+08 \\  
  0.06  & 8.98E+06 & 8.34E+06 & 9.63E+06 &        3.5  & 2.92E+08 & 2.75E+08 & 3.09E+08 \\  
  0.07  & 1.52E+07 & 1.41E+07 & 1.64E+07 &        4.     & 2.70E+08 & 2.54E+08 & 2.86E+08 \\  
  0.08  & 2.34E+07 & 2.16E+07 & 2.52E+07 &        5.     & 2.37E+08 & 2.23E+08 & 2.52E+08 \\  
  0.09  & 3.35E+07 & 3.09E+07 & 3.61E+07 &        6.     & 2.14E+08 & 2.01E+08 & 2.27E+08 \\  
  0.1 & 4.52E+07 & 4.17E+07 & 4.88E+07 &        7.     & 1.97E+08 & 1.84E+08 & 2.09E+08 \\  
  0.11  & 5.85E+07 & 5.40E+07 & 6.30E+07 &        8.     & 1.83E+08 & 1.72E+08 & 1.95E+08 \\  
  0.12  & 7.30E+07 & 6.73E+07 & 7.86E+07 &        9.     & 1.73E+08 & 1.62E+08 & 1.84E+08 \\  
  0.13  & 8.84E+07 & 8.16E+07 & 9.52E+07 &       10.     & 1.64E+08 & 1.54E+08 & 1.75E+08 \\  
\hline
\end{tabular*}
\begin{tabular*}{\textwidth}{@{\extracolsep{\fill}} l c }
REV  = 
$ 5.54\,{\rm exp}(-204.12/T_{9})$
 & \\
\end{tabular*}
\label{tdnTab2}
\end{table}
\clearpage
\subsection{\reac{3}{H}{\alpha}{\gamma}{7}{Li}}
\label{tagSect}
The experimental data sets referred to in NACRE are  GR61 \cite{GR61}, BU87 \cite{BU87}, and BR94 \cite{BR94}, covering the 0.05 $\lsimeq E_{\rm cm} \lsimeq$ 1.2 MeV range. \cite{SC87X} was rejected. 
Added is the post-NACRE data set TO01 \cite{TO01}$^\dag$.
 [{\footnotesize{$^\dag$from Coulomb break-up}}]

\begin{figure}[hb]
\centering{
\includegraphics[height=0.50\textheight,width=0.90\textwidth]{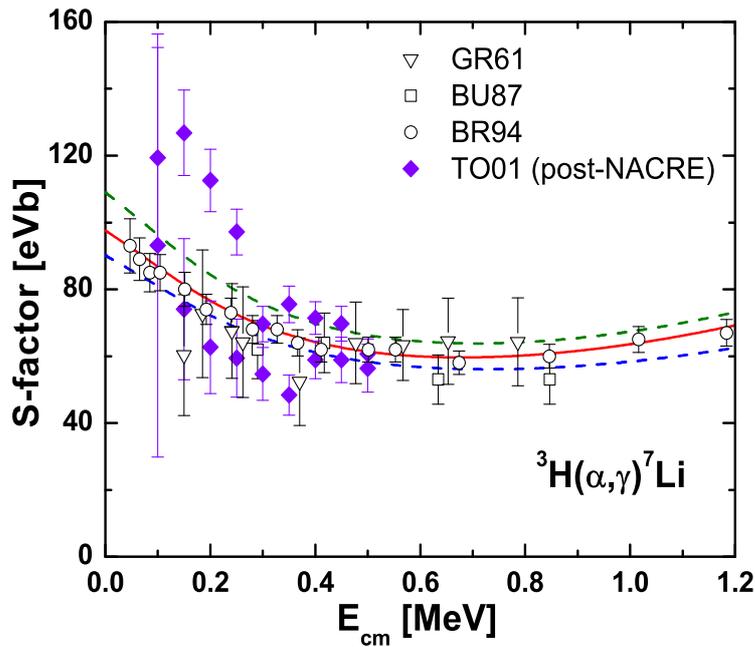}
\vspace{-0.5truecm}
\caption{The $S$-factor for \reac{3}{H}{\alpha}{\gamma}{7}{Li}. 
 TO01 is shown just for comparison and has not been used in the fit.}
\label{tagFig1}
}
\end{figure}

Figure \ref{tagFig1} compares the PM and experimental $S$-factors.
All the data sets but TO01 are used  in the whole $E_{\rm cm}$ range for the PM fit.
The transitions to the ground and the first excited states of $^{7}$Li are considered inclusively. 
The adopted parameter values are given in Table \ref{tagTab1}.
The present $S(0)$ =  98 $_{-8}^{+11}$ eV\,b.
In comparison, $S(0)$ = 100 eV\,b [NACRE, $E$-dependence of \cite{KA86}], and 95 $\pm{5}$ eV\,b [BBN04]. 

\begin{figure}[t]
\centering{
\includegraphics[height=0.33\textheight,width=0.90\textwidth]{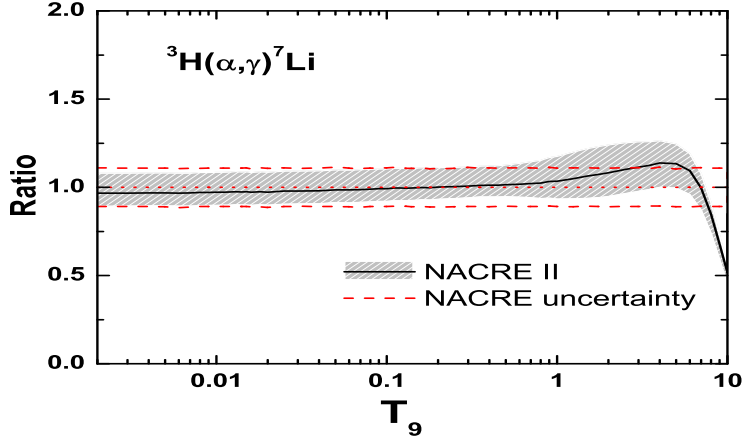}
\vspace{-0.4truecm}
\caption{\reac{3}{H}{\alpha}{\gamma}{7}{Li} rates in units of the NACRE (adopt) values. 
The sharp decrease of the ratio at the highest temperatures results from 
the rapid increase of the $S$-factors at high energies \cite{KA86} adopted by NACRE.
}
\label{tagFig2}
}
\end{figure}

Table \ref{tagTab2} gives the reaction rates at 0.002 $\le T_{9} \le$ 10, for which the PM-predicted cross sections in the  $E_{\rm cm} \lsimeq$ 0.3 MeV and $E_{\rm cm} \gsimeq$ 1.2 MeV ranges, and the experimental ones in the 0.3 $\lsimeq E_{\rm cm} \lsimeq$ 1.2 MeV range are used, respectively.
Figure \ref{tagFig2} compares the present and the NACRE rates.

{\footnotesize  See  \cite{NE11} for an ${\it{ab\ initio}}$ calculation.}

\begin{table}[hb]
\caption{\reac{3}{H}{\alpha}{\gamma}{7}{Li} rates in $\rm{cm^{3}mol^{-1}s^{-1}}$}\footnotesize\rm
\begin{tabular*}{\textwidth}{@{\extracolsep{\fill}} l c c c | l c c c}
\hline
$T_{9}$ & adopted & low & high & $T_{9}$ & adopted & low & high \\
\hline
0.002 & 7.06E$-$21 & 6.52E$-$21 & 7.88E$-$21 &      0.15  & 6.64E$-$01 & 6.18E$-$01 & 7.38E$-$01   \\
0.003 & 1.79E$-$17 & 1.65E$-$17 & 2.00E$-$17 &      0.16  & 8.76E$-$01 & 8.15E$-$01 & 9.73E$-$01   \\
0.004 & 2.48E$-$15 & 2.29E$-$15 & 2.77E$-$15 &      0.18  & 1.43E+00 & 1.33E+00 & 1.58E+00   \\
0.005 & 8.21E$-$14 & 7.59E$-$14 & 9.17E$-$14 &      0.2 & 2.16E+00 & 2.01E+00 & 2.40E+00   \\
0.006 & 1.18E$-$12 & 1.09E$-$12 & 1.32E$-$12 &      0.25  & 4.93E+00 & 4.60E+00 & 5.46E+00   \\
0.007 & 9.87E$-$12 & 9.12E$-$12 & 1.10E$-$11 &      0.3 & 9.14E+00 & 8.54E+00 & 1.01E+01   \\
0.008 & 5.68E$-$11 & 5.25E$-$11 & 6.34E$-$11 &      0.35  & 1.49E+01 & 1.39E+01 & 1.65E+01   \\
0.009 & 2.49E$-$10 & 2.30E$-$10 & 2.77E$-$10 &      0.4 & 2.21E+01 & 2.06E+01 & 2.44E+01   \\
0.01  & 8.85E$-$10 & 8.18E$-$10 & 9.88E$-$10 &      0.45  & 3.07E+01 & 2.86E+01 & 3.40E+01   \\
0.011 & 2.68E$-$09 & 2.48E$-$09 & 2.99E$-$09 &      0.5 & 4.06E+01 & 3.79E+01 & 4.51E+01   \\
0.012 & 7.15E$-$09 & 6.61E$-$09 & 7.98E$-$09 &      0.6 & 6.39E+01 & 5.93E+01 & 7.13E+01   \\
0.013 & 1.72E$-$08 & 1.59E$-$08 & 1.92E$-$08 &      0.7 & 9.12E+01 & 8.41E+01 & 1.02E+02   \\
0.014 & 3.78E$-$08 & 3.49E$-$08 & 4.22E$-$08 &      0.8 & 1.22E+02 & 1.11E+02 & 1.37E+02   \\
0.015 & 7.73E$-$08 & 7.15E$-$08 & 8.62E$-$08 &      0.9 & 1.55E+02 & 1.41E+02 & 1.75E+02   \\
0.016 & 1.49E$-$07 & 1.37E$-$07 & 1.66E$-$07 &      1.     & 1.90E+02 & 1.72E+02 & 2.16E+02   \\
0.018 & 4.72E$-$07 & 4.36E$-$07 & 5.26E$-$07 &      1.25  & 2.85E+02 & 2.54E+02 & 3.26E+02   \\
0.02  & 1.27E$-$06 & 1.18E$-$06 & 1.42E$-$06 &      1.5  & 3.87E+02 & 3.42E+02 & 4.44E+02   \\
0.025 & 9.25E$-$06 & 8.56E$-$06 & 1.03E$-$05 &      1.75  & 4.92E+02 & 4.32E+02 & 5.65E+02   \\
0.03  & 4.17E$-$05 & 3.86E$-$05 & 4.65E$-$05 &      2.     & 5.99E+02 & 5.24E+02 & 6.86E+02   \\
0.04  & 3.70E$-$04 & 3.42E$-$04 & 4.12E$-$04 &      2.5  & 8.15E+02 & 7.13E+02 & 9.28E+02   \\
0.05  & 1.73E$-$03 & 1.60E$-$03 & 1.92E$-$03 &      3.     & 1.03E+03 & 9.04E+02 & 1.17E+03   \\
0.06  & 5.55E$-$03 & 5.14E$-$03 & 6.18E$-$03 &      3.5  & 1.25E+03 & 1.11E+03 & 1.40E+03   \\
0.07  & 1.40E$-$02 & 1.30E$-$02 & 1.56E$-$02 &      4.     & 1.46E+03 & 1.28E+03 & 1.62E+03   \\
0.08  & 3.00E$-$02 & 2.78E$-$02 & 3.34E$-$02 &      5.     & 1.86E+03 & 1.63E+03 & 2.04E+03   \\
0.09  & 5.68E$-$02 & 5.27E$-$02 & 6.32E$-$02 &      6.     & 2.27E+03 & 1.99E+03 & 2.42E+03   \\
0.1 & 9.83E$-$02 & 9.12E$-$02 & 1.09E$-$01 &      7.     & 2.51E+03 & 2.19E+03 & 2.73E+03   \\
0.11  & 1.58E$-$01 & 1.47E$-$01 & 1.76E$-$01 &      8.     & 2.74E+03 & 2.38E+03 & 2.96E+03   \\
0.12  & 2.41E$-$01 & 2.24E$-$01 & 2.68E$-$01 &      9.     & 2.91E+03 & 2.51E+03 & 3.13E+03   \\
0.13  & 3.50E$-$01 & 3.26E$-$01 & 3.90E$-$01 &     10.     & 3.02E+03 & 2.61E+03 & 3.25E+03   \\
0.14  & 4.90E$-$01 & 4.56E$-$01 & 5.45E$-$01 &      &  &  &                \\
\hline
\end{tabular*}
\begin{tabular*}{\textwidth}{@{\extracolsep{\fill}} l c }
REV  = 
$1.11 \times 10^{10}T_{9}^{3/2}{\rm exp}(-28.625/T_{9})\,/\,[1.0+0.5{\rm exp}(-5.543/T_{9})]$
 & \\
\end{tabular*}
\label{tagTab2}
\end{table}
\clearpage
\subsection{\reac{3}{He}{d}{p}{4}{He}}
\label{he3dpSect}
This reaction is not included in NACRE, but is present in CF88. 
The experimental data sets adopted here are  KR87a \cite{KR87a},  SC89 \cite{SC89}, GE99 \cite{GE99}, CO00 \cite{CO00}, AL01 \cite{AL01}, and LA05 \cite{LA05}$^\dag$, covering the 0.005 $\lsimeq E_{\rm cm} \lsimeq$ 1 MeV range. \cite{EN88,PR94} are superseded by AL01.  
[{\footnotesize{$^\dag$from $^{6}$Li($^{3}$He,\,p$\alpha$)$^{4}$He (THM).}}]

\begin{figure}[hb]
\centering{
\includegraphics[height=0.50\textheight,width=0.90\textwidth]{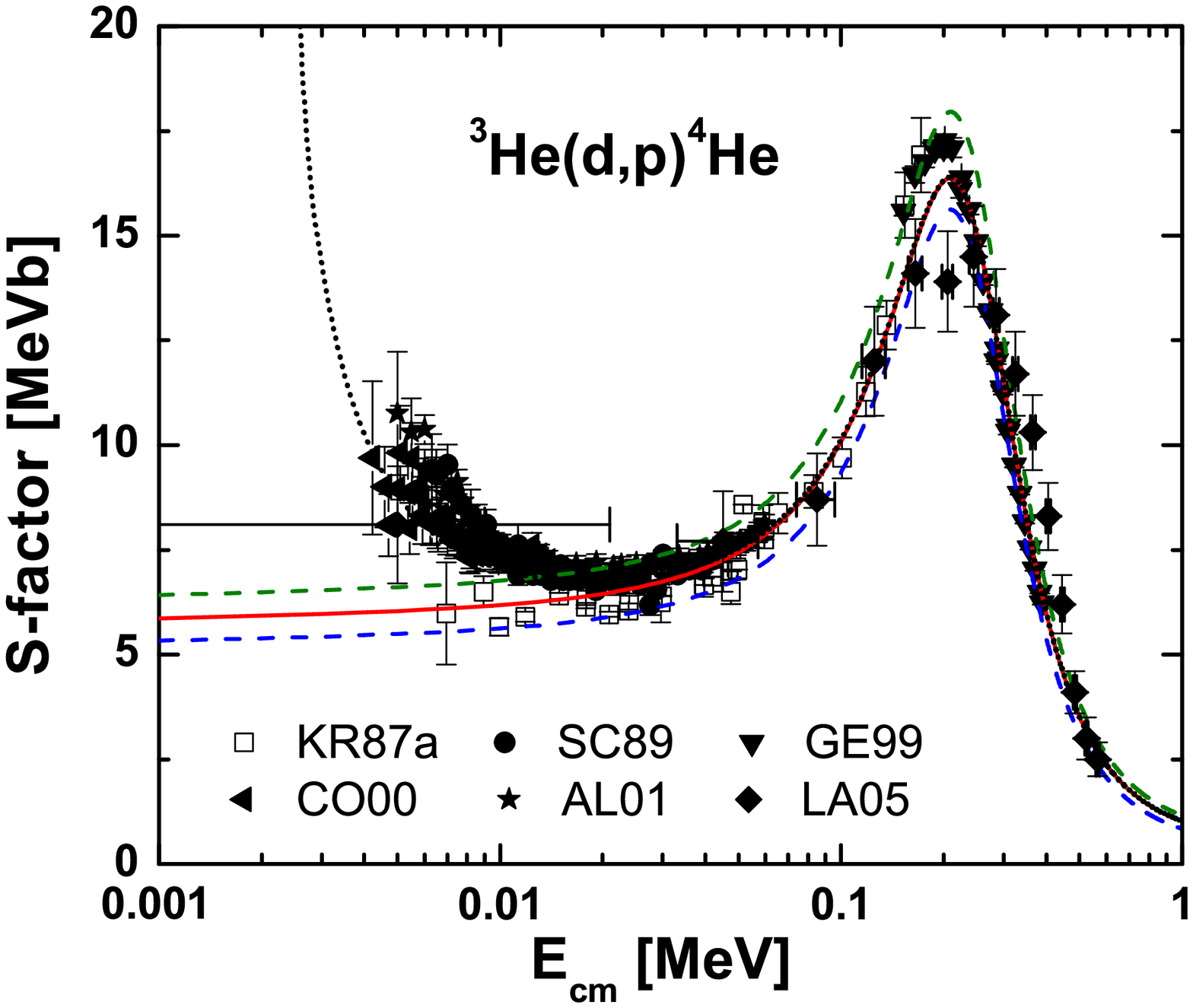}
\vspace{-0.5truecm}
\caption{The $S$-factor for \reac{3}{He}{d}{p}{4}{He}.  The dotted line indicates an adiabatic screening correction ($U_{\rm e}$ = 119 eV) to the 'adopt' curve (solid line). 
\label{he3dpFig1a}
}}
\end{figure}

\begin{figure}[t]
\centering{
\includegraphics[height=0.40\textheight,width=0.90\textwidth]{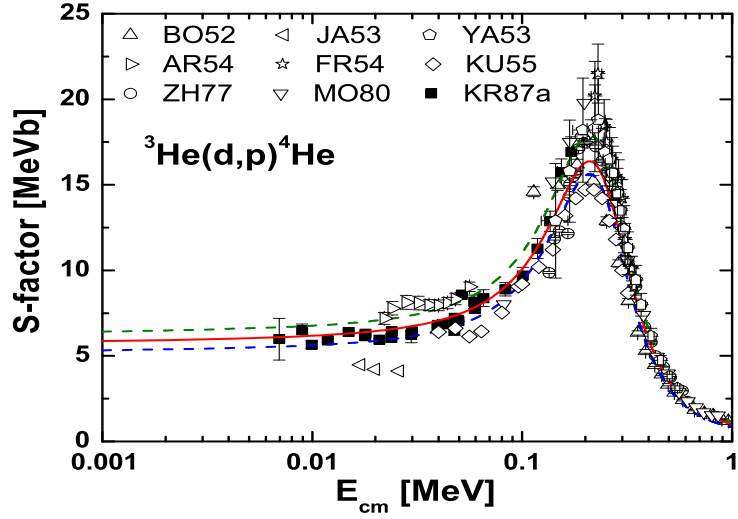}
\vspace{-0.5truecm}
\caption{The pre-CF88 experimental $S$-factor data sets for \reac{3}{He}{d}{p}{4}{He}: BO52 \cite{BO52}, JA53 \cite{JA53}, YA53 \cite{YA53}, AR54 \cite{AR54}, FR54 \cite{FR54}, KU55 \cite{KU55}, ZH77 \cite{ZH77}, MO80 \cite{MO80}, and KR87a \cite{KR87a}. The DWBA curves are added to guide the eye. CF88 relies on KR87a at low energies. See  \cite{GR71,ST60} for data above $E_{\rm cm} \gsimeq$ 1 MeV.
}
\label{he3dpFig1b}
}
\end{figure}

\begin{figure}[t]
\centering{
\includegraphics[height=0.40\textheight,width=0.90\textwidth]{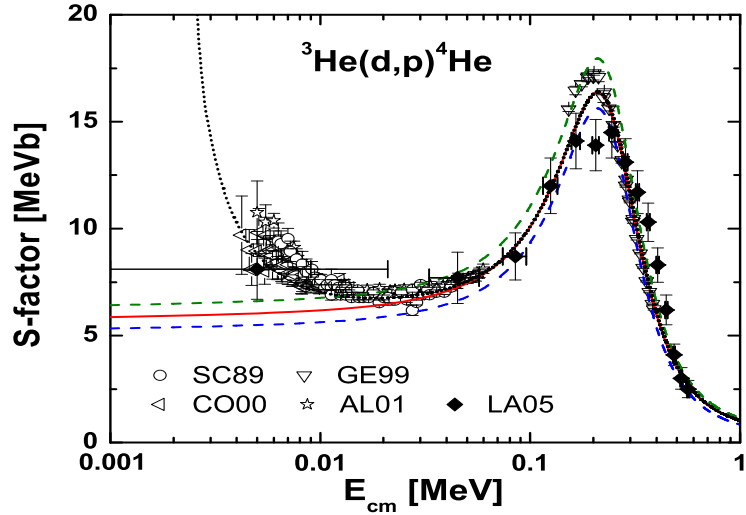}
\vspace{-0.5truecm}
\caption{The post-CF88 experimental $S$-factor data sets for \reac{3}{He}{d}{p}{4}{He}. The DWBAhe3ag
 and screening correction curves are added to guide the eyes. Recall that the LA05 data points are from THM, but one at the lowest energy is not considered in our analysis because of its large uncertainties.
}
\label{he3dpFig1c}
}
\end{figure}

Figure \ref{he3dpFig1a} compares the DWBA and the experimental $S$-factors, whereas Figs.\,\ref{he3dpFig1b} and \ref{he3dpFig1c} summarise the pre- and post-CF88 experimental data sets, respectively.
Many post-CF88 measurements below $E_{\rm cm} \simeq$ 0.02 MeV look highly contaminated by electron screening (see Fig.\,\ref{he3dpFig1c}; also see \cite{BA07}).
Only the data in the 0.02 $\lsimeq$ $E_{\rm cm} \lsimeq$ 1 MeV range are used for the DWBA fit. They exhibit the $3/2^{+}$ resonance at $E_{\rm R} \simeq$  0.21 MeV.
The adopted parameter values are given in Table \ref{he3dpTab1}.
The present $S(0)$ = 5.9 $\pm$ 0.5  MeV\,b.
In comparison, $S(0)$= 5.9 $\pm$ 0.3 MeV\,b [BBN04].

\begin{figure}[t]
\centering{
\includegraphics[height=0.33\textheight,width=0.90\textwidth]{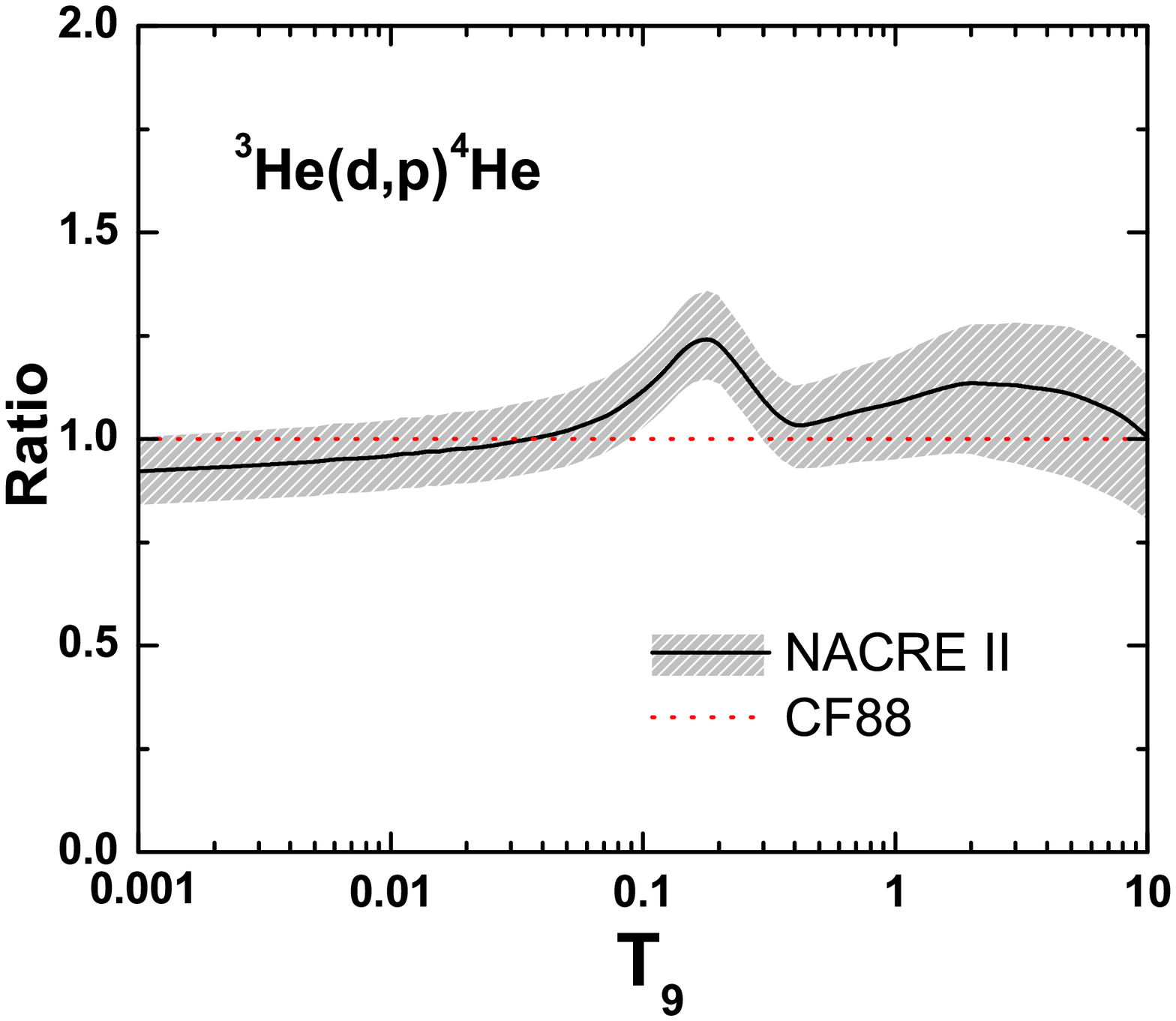}
\vspace{-0.4truecm}
\caption{\reac{3}{He}{d}{p}{4}{He} rates in units of the CF88 values.
}
\label{he3dpFig2}
}
\end{figure}

Table \ref{he3dpTab2} gives the reaction rates at 0.001 $\le T_{9} \le$ 10, for which the DWBA-predicted and the experimental cross sections below and above $E_{\rm cm} \simeq$ 0.05 MeV are used, respectively. 
Figure \ref{he3dpFig2} compares the present and the CF88 rates.

{\footnotesize  See \cite{NA12} for an ${\it{ab\ initio}}$ calculation.}

\clearpage

\clearpage

\begin{table}[hb]
\caption{\reac{3}{He}{d}{p}{4}{He} rates in $\rm{cm^{3}mol^{-1}s^{-1}}$.}\footnotesize\rm
\begin{tabular*}{\textwidth}{@{\extracolsep{\fill}} l c c c | l c c c}
\hline
$T_{9}$ & adopted & low & high & $T_{9}$ & adopted & low & high \\
\hline
  0.001 & 3.54E$-$19 & 3.21E$-$19 & 3.87E$-$19 &        0.14  & 2.88E+05 & 2.64E+05 & 3.15E+05 \\
  0.002 & 6.14E$-$13 & 5.58E$-$13 & 6.72E$-$13 &        0.15  & 3.85E+05 & 3.53E+05 & 4.22E+05 \\
  0.003 & 6.36E$-$10 & 5.78E$-$10 & 6.96E$-$10 &        0.16  & 5.03E+05 & 4.61E+05 & 5.51E+05 \\
  0.004 & 5.02E$-$08 & 4.56E$-$08 & 5.49E$-$08 &        0.18  & 8.07E+05 & 7.40E+05 & 8.84E+05 \\
  0.005 & 1.11E$-$06 & 1.01E$-$06 & 1.22E$-$06 &        0.2 & 1.22E+06 & 1.11E+06 & 1.33E+06 \\
  0.006 & 1.18E$-$05 & 1.07E$-$05 & 1.29E$-$05 &        0.25  & 2.77E+06 & 2.53E+06 & 3.04E+06 \\
  0.007 & 7.74E$-$05 & 7.04E$-$05 & 8.47E$-$05 &        0.3 & 5.19E+06 & 4.71E+06 & 5.69E+06 \\
  0.008 & 3.64E$-$04 & 3.31E$-$04 & 3.99E$-$04 &        0.35  & 8.51E+06 & 7.68E+06 & 9.34E+06 \\
  0.009 & 1.35E$-$03 & 1.23E$-$03 & 1.48E$-$03 &        0.4 & 1.27E+07 & 1.14E+07 & 1.39E+07 \\
  0.01  & 4.15E$-$03 & 3.78E$-$03 & 4.54E$-$03 &        0.45  & 1.76E+07 & 1.57E+07 & 1.93E+07 \\
  0.011 & 1.11E$-$02 & 1.01E$-$02 & 1.21E$-$02 &        0.5 & 2.31E+07 & 2.06E+07 & 2.54E+07 \\
  0.012 & 2.64E$-$02 & 2.40E$-$02 & 2.89E$-$02 &        0.6 & 3.53E+07 & 3.13E+07 & 3.89E+07 \\
  0.013 & 5.74E$-$02 & 5.22E$-$02 & 6.29E$-$02 &        0.7 & 4.83E+07 & 4.26E+07 & 5.33E+07 \\
  0.014 & 1.16E$-$01 & 1.05E$-$01 & 1.26E$-$01 &        0.8 & 6.12E+07 & 5.37E+07 & 6.77E+07 \\
  0.015 & 2.18E$-$01 & 1.98E$-$01 & 2.39E$-$01 &        0.9 & 7.35E+07 & 6.42E+07 & 8.15E+07 \\
  0.016 & 3.89E$-$01 & 3.54E$-$01 & 4.26E$-$01 &        1.     & 8.49E+07 & 7.39E+07 & 9.43E+07 \\
  0.018 & 1.08E+00 & 9.87E$-$01 & 1.19E+00 &        1.25  & 1.09E+08 & 9.40E+07 & 1.21E+08 \\
  0.02  & 2.62E+00 & 2.38E+00 & 2.87E+00 &        1.5  & 1.27E+08 & 1.08E+08 & 1.42E+08 \\
  0.025 & 1.52E+01 & 1.39E+01 & 1.67E+01 &        1.75  & 1.39E+08 & 1.18E+08 & 1.56E+08 \\
  0.03  & 5.82E+01 & 5.30E+01 & 6.37E+01 &        2.     & 1.48E+08 & 1.25E+08 & 1.67E+08 \\
  0.04  & 4.07E+02 & 3.71E+02 & 4.46E+02 &        2.5  & 1.57E+08 & 1.31E+08 & 1.78E+08 \\
  0.05  & 1.62E+03 & 1.48E+03 & 1.77E+03 &        3.     & 1.61E+08 & 1.33E+08 & 1.83E+08 \\
  0.06  & 4.63E+03 & 4.23E+03 & 5.07E+03 &        3.5  & 1.61E+08 & 1.32E+08 & 1.83E+08 \\
  0.07  & 1.07E+04 & 9.80E+03 & 1.17E+04 &        4.     & 1.59E+08 & 1.30E+08 & 1.82E+08 \\
  0.08  & 2.14E+04 & 1.96E+04 & 2.35E+04 &        5.     & 1.53E+08 & 1.24E+08 & 1.76E+08 \\
  0.09  & 3.85E+04 & 3.52E+04 & 4.22E+04 &        6.     & 1.46E+08 & 1.18E+08 & 1.68E+08 \\
  0.1 & 6.38E+04 & 5.85E+04 & 6.99E+04 &        7.     & 1.38E+08 & 1.11E+08 & 1.59E+08 \\
  0.11  & 9.94E+04 & 9.11E+04 & 1.09E+05 &        8.     & 1.31E+08 & 1.05E+08 & 1.51E+08 \\
  0.12  & 1.47E+05 & 1.35E+05 & 1.61E+05 &        9.     & 1.24E+08 & 9.85E+07 & 1.42E+08 \\
  0.13  & 2.09E+05 & 1.92E+05 & 2.29E+05 &       10.     & 1.17E+08 & 9.27E+07 & 1.34E+08 \\
\hline
\end{tabular*}
\begin{tabular*}{\textwidth}{@{\extracolsep{\fill}} l c }
REV  = 
$5.54\,{\rm exp}(-212.99/T_{9})$ 
 & \\
\end{tabular*}
\label{he3dpTab2}
\end{table}
\clearpage
\subsection{\rm $^{3}$He($^{3}$He,\,2p)$^{4}$He}
\label{he3he3Sect}
The experimental data sets referred to in NACRE are WA66 \cite{WA66}, BA67 \cite{BA67}, DW71 \cite{DW71}, DW74 \cite{DW74}, BR87b \cite{BR87b}, KR87b \cite{KR87b} and JU98 \cite{JU98}, covering the 0.02 $\lsimeq E_{\rm cm} \lsimeq$ 12 MeV range.
\cite{AR96} was superseded by JU98.
Added are the post-NACRE data sets BO99 \cite{BO99} and KU04 \cite{KU04}, extending the range down to $E_{\rm cm} \simeq$ 0.016 MeV. 

\begin{figure}[hb]
\centering{
\includegraphics[height=0.50\textheight,width=0.90\textwidth]{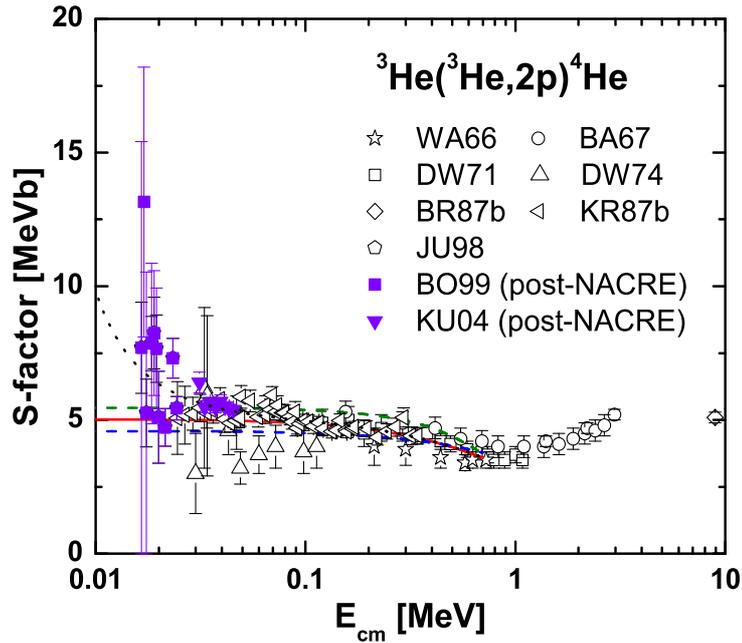}
\vspace{-0.5truecm}
\caption{The $S$-factor for $^{3}$He($^{3}$He,2p)$^{4}$He.  The dotted line indicates an adiabatic screening correction ($U_{\rm e}$ = 241 eV) to the 'adopt' curve (solid line). DW74 is not included in the fit.
}
\label{he3he3Fig1}
}
\end{figure}

Figure \ref{he3he3Fig1} compares the DWBA and experimental $S$-factors.
JU98 and BO99 look contaminated by electron screening at the lowest energies.
The data in the  0.03 $\lsimeq$ $E_{\rm cm} \lsimeq$ 1 MeV range are used for the DWBA fit.
The two-protons in the exit channel are treated as a point-like spinless particle.
The adopted parameter values are given in Table \ref{he3he3Tab1}.
The present $S(0)$ = 5.3 $\pm$ 0.5 MeV\,b.
In comparison, $S(0)$ = 5.18 MeV\,b [NACRE, quadratic polynomial], and 5.21 $\pm$ 0.27 MeV\,b [SUN11, quadratic polynomial].

\begin{figure}[t]
\centering{
\includegraphics[height=0.33\textheight,width=0.90\textwidth]{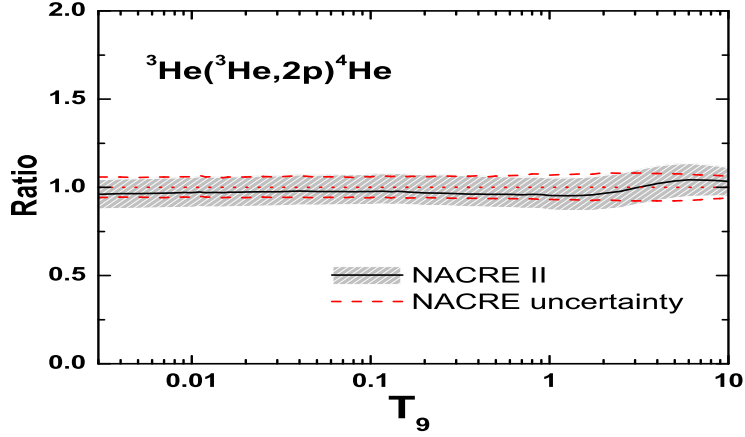}
\vspace{-0.4truecm}
\caption{$^{3}$He($^{3}$He,2p)$^{4}$He rates in units of the NACRE (adopt) values.
}
\label{he3he3Fig2}
}
\end{figure}

Table \ref{he3he3Tab2} gives the reaction rates at 0.003 $\le T_{9} \le$ 10, for which the DWBA-predicted and the experimental cross sections below and above $E_{\rm cm} \simeq$ 0.1 MeV are used, respectively.
Figure \ref{he3he3Fig2} compares the present and the NACRE rates.

{\footnotesize  See   \cite{CS99} for a large scale cluster model calculation; \cite{dwbahe3he3} for a DWBA analysis.}

\begin{table}[hb]
\caption{$^{3}$He($^{3}$He,2p)$^{4}$He rates in units of $\rm{cm^{3}mol^{-1}s^{-1}}$.} \footnotesize\rm
\begin{tabular*}{\textwidth}{@{\extracolsep{\fill}} l c c c | l c c c}
\hline
$T_{9}$ & adopted & low & high & $T_{9}$ & adopted & low & high \\
\hline
  0.003 & 2.80E$-$25 & 2.55E$-$25 & 3.04E$-$25 &        0.15  & 1.74E+01 & 1.61E+01 & 1.92E+01 \\  
  0.004 & 5.55E$-$22 & 5.07E$-$22 & 6.04E$-$22 &        0.16  & 2.71E+01 & 2.51E+01 & 3.00E+01 \\  
  0.005 & 1.22E$-$19 & 1.12E$-$19 & 1.33E$-$19 &        0.18  & 5.96E+01 & 5.52E+01 & 6.59E+01 \\  
  0.006 & 7.48E$-$18 & 6.83E$-$18 & 8.14E$-$18 &        0.2 & 1.17E+02 & 1.08E+02 & 1.30E+02 \\  
  0.007 & 1.99E$-$16 & 1.82E$-$16 & 2.17E$-$16 &        0.25  & 4.50E+02 & 4.16E+02 & 4.98E+02 \\  
  0.008 & 2.98E$-$15 & 2.72E$-$15 & 3.24E$-$15 &        0.3 & 1.24E+03 & 1.15E+03 & 1.38E+03 \\  
  0.009 & 2.93E$-$14 & 2.68E$-$14 & 3.19E$-$14 &        0.35  & 2.79E+03 & 2.58E+03 & 3.09E+03 \\  
  0.01  & 2.09E$-$13 & 1.91E$-$13 & 2.28E$-$13 &        0.4 & 5.40E+03 & 4.99E+03 & 5.98E+03 \\  
  0.011 & 1.17E$-$12 & 1.07E$-$12 & 1.27E$-$12 &        0.45  & 9.42E+03 & 8.70E+03 & 1.04E+04 \\  
  0.012 & 5.34E$-$12 & 4.89E$-$12 & 5.82E$-$12 &        0.5 & 1.52E+04 & 1.40E+04 & 1.68E+04 \\  
  0.013 & 2.08E$-$11 & 1.90E$-$11 & 2.27E$-$11 &        0.6 & 3.30E+04 & 3.04E+04 & 3.64E+04 \\  
  0.014 & 7.08E$-$11 & 6.47E$-$11 & 7.71E$-$11 &        0.7 & 6.10E+04 & 5.61E+04 & 6.73E+04 \\  
  0.015 & 2.15E$-$10 & 1.97E$-$10 & 2.35E$-$10 &        0.8 & 1.01E+05 & 9.24E+04 & 1.11E+05 \\  
  0.016 & 5.95E$-$10 & 5.44E$-$10 & 6.48E$-$10 &        0.9 & 1.53E+05 & 1.40E+05 & 1.69E+05 \\  
  0.018 & 3.59E$-$09 & 3.28E$-$09 & 3.91E$-$09 &        1.     & 2.19E+05 & 2.01E+05 & 2.42E+05 \\  
  0.02  & 1.68E$-$08 & 1.54E$-$08 & 1.84E$-$08 &        1.25  & 4.46E+05 & 4.06E+05 & 4.93E+05 \\  
  0.025 & 3.71E$-$07 & 3.40E$-$07 & 4.05E$-$07 &        1.5  & 7.60E+05 & 6.90E+05 & 8.40E+05 \\  
  0.03  & 3.91E$-$06 & 3.58E$-$06 & 4.27E$-$06 &        1.75  & 1.16E+06 & 1.05E+06 & 1.28E+06 \\  
  0.04  & 1.19E$-$04 & 1.10E$-$04 & 1.31E$-$04 &        2.     & 1.64E+06 & 1.49E+06 & 1.81E+06 \\  
  0.05  & 1.35E$-$03 & 1.24E$-$03 & 1.48E$-$03 &        2.5  & 2.83E+06 & 2.56E+06 & 3.13E+06 \\  
  0.06  & 8.49E$-$03 & 7.82E$-$03 & 9.32E$-$03 &        3.     & 4.30E+06 & 3.89E+06 & 4.74E+06 \\  
  0.07  & 3.68E$-$02 & 3.39E$-$02 & 4.04E$-$02 &        3.5  & 6.01E+06 & 5.43E+06 & 6.60E+06 \\  
  0.08  & 1.23E$-$01 & 1.13E$-$01 & 1.35E$-$01 &        4.     & 7.92E+06 & 7.16E+06 & 8.69E+06 \\  
  0.09  & 3.40E$-$01 & 3.14E$-$01 & 3.74E$-$01 &        5.     & 1.22E+07 & 1.11E+07 & 1.34E+07 \\  
  0.1 & 8.13E$-$01 & 7.52E$-$01 & 8.95E$-$01 &        6.     & 1.70E+07 & 1.54E+07 & 1.85E+07 \\  
  0.11  & 1.74E+00 & 1.61E+00 & 1.92E+00 &        7.     & 2.21E+07 & 2.01E+07 & 2.40E+07 \\  
  0.12  & 3.41E+00 & 3.15E+00 & 3.76E+00 &        8.     & 2.73E+07 & 2.50E+07 & 2.96E+07 \\  
  0.13  & 6.21E+00 & 5.75E+00 & 6.85E+00 &        9.     & 3.27E+07 & 2.99E+07 & 3.53E+07 \\  
  0.14  & 1.07E+01 & 9.86E+00 & 1.18E+01 &       10.     & 3.80E+07 & 3.49E+07 & 4.09E+07 \\  
\hline
\end{tabular*}
\label{he3he3Tab2}
\end{table}
\clearpage
\subsection{\reac{3}{He}{\alpha}{\gamma}{7}{Be}}
\label{he3agSect}
The experimental data sets referred to in NACRE are HO59 \cite{HO59}, PA63 \cite{PA63}$^\dag$, NA69 \cite{NA69}, KR82 \cite{KR82}$^\ddag$, RO83 \cite{RO83}, AL84 \cite{AL84}, OS84 \cite{OS84} and HI88 \cite{HI88}, covering the 0.1 $\lsimeq E_{\rm cm} \lsimeq$ 1.5 MeV range. 
Added are the post-NACRE data sets NA04 \cite{NA04}, BE06 \cite{BE06}, BR07 \cite{BR07}, CO07 \cite{CO07}, GY07 \cite{GY07}, CO08 \cite{CO08} and DI09 \cite{DI09}, extending the range to 0.07 $\lsimeq E_{\rm cm} \lsimeq$ 3 MeV. Most recently, BO13 \cite{BO13} and KO13 \cite{KO13} have become available.
[{\footnotesize{$^\dag$taken in part from NA69; $^\ddag$modified following HI88.}}]

\begin{figure}[hb]
\centering{
\includegraphics[height=0.50\textheight,width=0.90\textwidth]{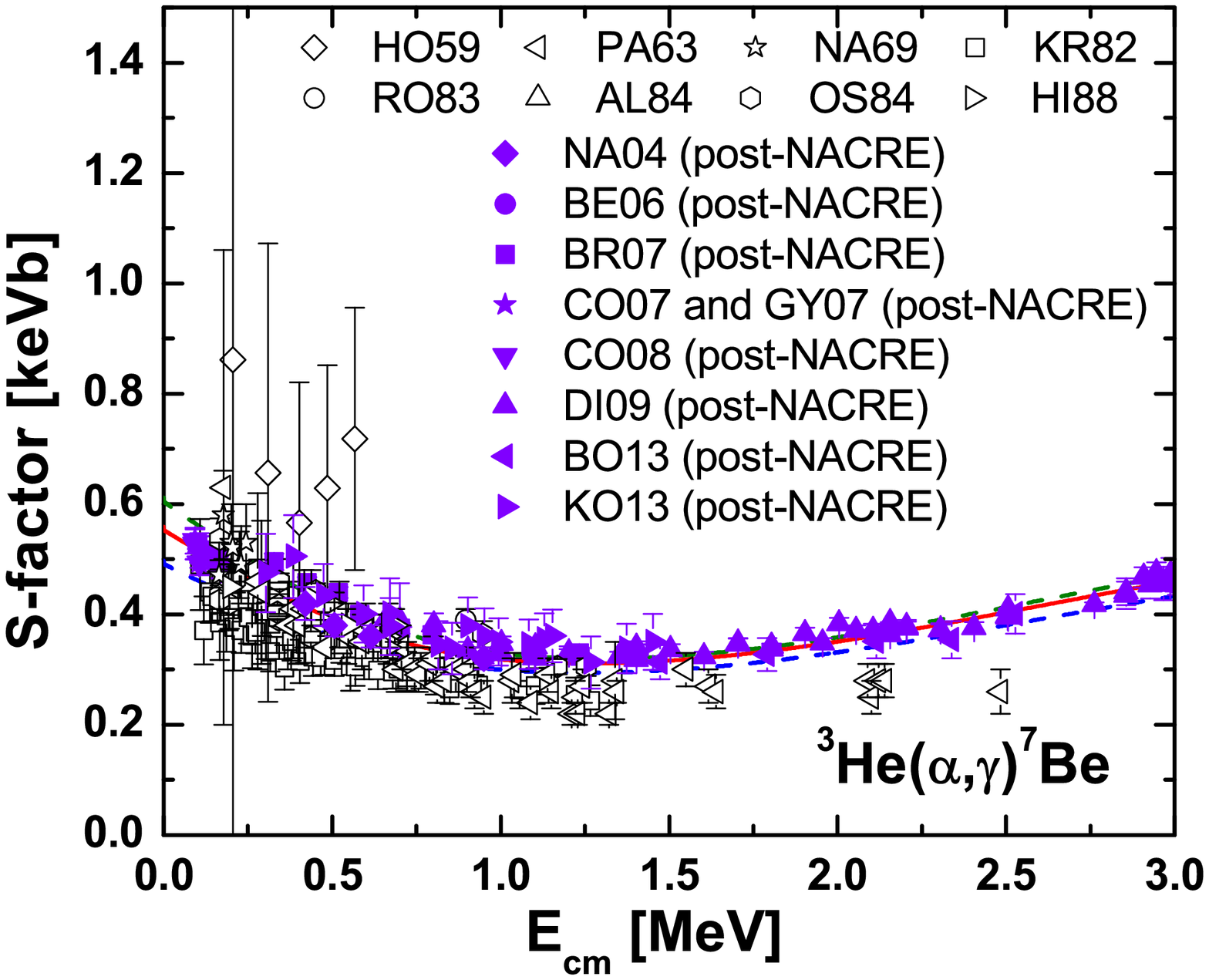}
\vspace{-0.5truecm}
\caption{The $S$-factor for \reac{3}{He}{\alpha}{\gamma}{7}{Be}. PA63 is not included in the fit.
}
\label{he3agFig1}
}
\end{figure}

Figure \ref{he3agFig1} compares the PM and experimental $S$-factors.
All the data sets but PA63 in the whole $E_{\rm cm}$ range are used for the PM fit.
The transitions to the ground and the first excited states of $^{7}$Be are considered inclusively. 
The adopted parameter values are given in Table \ref{he3agTab1}.
The present $S(0)$ = 0.56 $_{-0.07}^{+0.05}$ keV\,b. 
In comparison,  $S(0)$ = 0.54 $\pm$ 0.09 keV\,b [NACRE, $E$-dependence of \cite{KA86}], 0.51 $\pm$ 0.04 [BBN04] keV\,b, and 0.56 $\pm$ 0.04 keV\,b of [SUN11, $E$-dependence of \cite{NO01a}].

\begin{figure}[t]
\centering{
\includegraphics[height=0.33\textheight,width=0.90\textwidth]{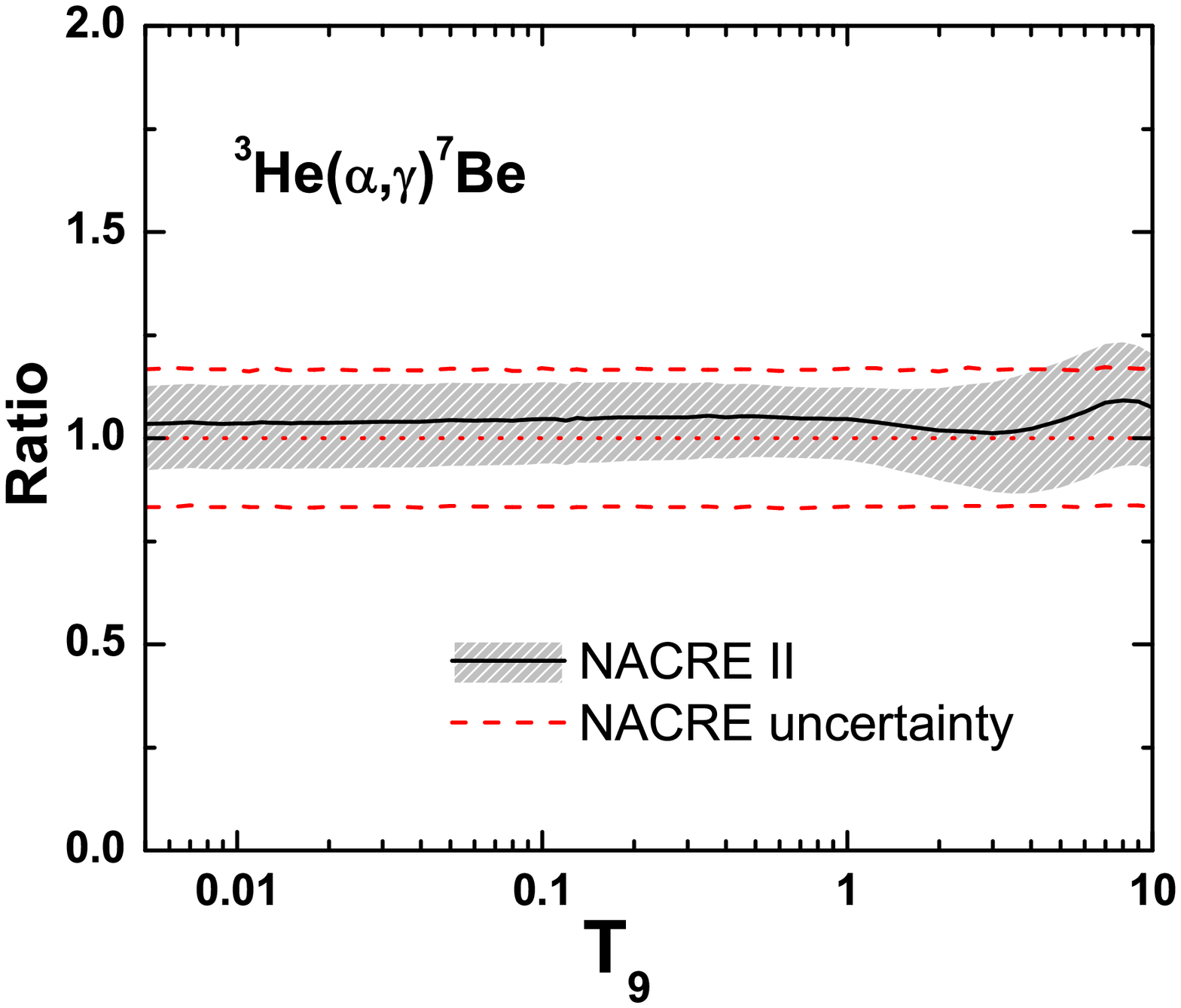}
\vspace{-0.4truecm}
\caption{\reac{3}{He}{\alpha}{\gamma}{7}{Be} rates in units of the NACRE (adopt) values.
}
\label{he3agFig2}
}
\end{figure}

Table \ref{he3agTab2} gives the reaction rates at 0.005 $\le T_{9} \le$ 10, for which the PM-predicted and the experimental cross sections below and above $E_{\rm cm} \simeq$ 0.5 MeV are used, respectively. 
Figure \ref{he3agFig2}  compares the present and the NACRE rates.

{\footnotesize  See  \cite{NE11} for an ${\it{ab\ initio}}$ calculation.}

\begin{table}[hb]
\caption{\reac{3}{He}{\alpha}{\gamma}{7}{Be} rates in $\rm{cm^{3}mol^{-1}s^{-1}}$}\footnotesize\rm
\begin{tabular*}{\textwidth}{@{\extracolsep{\fill}} l c c c | l c c c}
\hline
$T_{9}$ & adopted & low & high & $T_{9}$ & adopted & low & high \\
\hline
0.005 & 5.14E$-$25 & 4.57E$-$25 & 5.62E$-$25 &       0.16  & 9.98E$-$04 & 8.93E$-$04 & 1.08E$-$03  \\
0.006 & 3.79E$-$23 & 3.37E$-$23 & 4.14E$-$23 &       0.18  & 2.28E$-$03 & 2.04E$-$03 & 2.47E$-$03  \\
0.007 & 1.17E$-$21 & 1.04E$-$21 & 1.28E$-$21 &       0.2 & 4.62E$-$03 & 4.14E$-$03 & 5.02E$-$03  \\
0.008 & 1.99E$-$20 & 1.77E$-$20 & 2.17E$-$20 &       0.25  & 1.89E$-$02 & 1.70E$-$02 & 2.05E$-$02  \\
0.009 & 2.17E$-$19 & 1.93E$-$19 & 2.37E$-$19 &       0.3 & 5.49E$-$02 & 4.93E$-$02 & 5.94E$-$02  \\
0.01  & 1.70E$-$18 & 1.51E$-$18 & 1.86E$-$18 &       0.35  & 1.28E$-$01 & 1.15E$-$01 & 1.38E$-$01  \\
0.011 & 1.03E$-$17 & 9.13E$-$18 & 1.12E$-$17 &       0.4 & 2.54E$-$01 & 2.29E$-$01 & 2.74E$-$01  \\
0.012 & 5.03E$-$17 & 4.48E$-$17 & 5.50E$-$17 &       0.45  & 4.54E$-$01 & 4.09E$-$01 & 4.89E$-$01  \\
0.013 & 2.09E$-$16 & 1.86E$-$16 & 2.28E$-$16 &       0.5 & 7.44E$-$01 & 6.72E$-$01 & 8.02E$-$01  \\
0.014 & 7.51E$-$16 & 6.68E$-$16 & 8.20E$-$16 &       0.6 & 1.67E+00 & 1.51E+00 & 1.80E+00  \\
0.015 & 2.40E$-$15 & 2.14E$-$15 & 2.63E$-$15 &       0.7 & 3.16E+00 & 2.86E+00 & 3.39E+00  \\
0.016 & 6.96E$-$15 & 6.20E$-$15 & 7.60E$-$15 &       0.8 & 5.30E+00 & 4.79E+00 & 5.70E+00  \\
0.018 & 4.56E$-$14 & 4.06E$-$14 & 4.98E$-$14 &       0.9 & 8.17E+00 & 7.38E+00 & 8.79E+00  \\
0.02  & 2.30E$-$13 & 2.05E$-$13 & 2.51E$-$13 &       1.     & 1.18E+01 & 1.07E+01 & 1.27E+01  \\
0.025 & 5.86E$-$12 & 5.21E$-$12 & 6.39E$-$12 &       1.25  & 2.44E+01 & 2.19E+01 & 2.64E+01  \\
0.03  & 6.88E$-$11 & 6.12E$-$11 & 7.50E$-$11 &       1.5  & 4.19E+01 & 3.73E+01 & 4.57E+01  \\
0.04  & 2.46E$-$09 & 2.19E$-$09 & 2.69E$-$09 &       1.75  & 6.40E+01 & 5.66E+01 & 7.02E+01  \\
0.05  & 3.11E$-$08 & 2.77E$-$08 & 3.39E$-$08 &       2.     & 9.04E+01 & 7.93E+01 & 9.98E+01  \\
0.06  & 2.14E$-$07 & 1.91E$-$07 & 2.33E$-$07 &       2.5  & 1.54E+02 & 1.34E+02 & 1.72E+02  \\
0.07  & 9.92E$-$07 & 8.85E$-$07 & 1.08E$-$06 &       3.     & 2.32E+02 & 1.98E+02 & 2.61E+02  \\
0.08  & 3.50E$-$06 & 3.13E$-$06 & 3.82E$-$06 &       3.5  & 3.21E+02 & 2.73E+02 & 3.64E+02  \\
0.09  & 1.02E$-$05 & 9.06E$-$06 & 1.11E$-$05 &       4.     & 4.21E+02 & 3.56E+02 & 4.79E+02  \\
0.1 & 2.53E$-$05 & 2.26E$-$05 & 2.76E$-$05 &       5.     & 6.53E+02 & 5.49E+02 & 7.44E+02  \\
0.11  & 5.62E$-$05 & 5.02E$-$05 & 6.11E$-$05 &       6.     & 9.15E+02 & 7.71E+02 & 1.04E+03  \\
0.12  & 1.14E$-$04 & 1.02E$-$04 & 1.24E$-$04 &       7.     & 1.19E+03 & 1.01E+03 & 1.36E+03  \\
0.13  & 2.13E$-$04 & 1.90E$-$04 & 2.31E$-$04 &       8.     & 1.47E+03 & 1.26E+03 & 1.67E+03  \\
0.14  & 3.75E$-$04 & 3.35E$-$04 & 4.07E$-$04 &       9.     & 1.74E+03 & 1.49E+03 & 1.96E+03  \\
0.15  & 6.25E$-$04 & 5.59E$-$04 & 6.80E$-$04 &      10.     & 1.99E+03 & 1.71E+03 & 2.24E+03  \\
\hline
\end{tabular*}
\begin{tabular*}{\textwidth}{@{\extracolsep{\fill}} l c }
REV  = 
$1.11 \times 10^{10}T_{9}^{3/2}{\rm exp}(-18.407/T_{9})\,/\,[1.0+0.5{\,\rm exp}(-4.979/T_{9})]$ 
 & \\
\end{tabular*}
\label{he3agTab2}
\end{table}
\clearpage
\subsection{\reac{6}{Li}{p}{\gamma}{7}{Be}}
\label{li6pgSect}
The experimental data set referred to in NACRE is SW79 \cite{SW79}, covering the 0.14 $\lsimeq E_{\rm cm} \lsimeq$ 1 MeV range.
Added here are BA55 \cite{BA55}, OS83 \cite{OS83},  BR92 \cite{BR92}$^\dag$, PA99 \cite{PA99}, and a most recent "surprise" HE12 \cite{HE12} extending the range down to $E_{\rm cm} \simeq$ 0.03 MeV. \cite{WA56} was rejected because of the unusually low cross section. 
[{\footnotesize{$^\dag$(p,$\gamma_1$) partial cross sections}}]

Figure 24 compares the PM and experimental $S$-factors, where the PM curves are selected from those in Figs.\,25 and 26.
The transitions to the ground and the first excited states of $^{7}$Be are considered inclusively. 
The corresponding parameter values are given in Table \ref{li6pgTab1}.
Note that the HE12 cross sections have been re-normalised by a factor of 1.17 \cite{HE13}.
Waiting for the confirmation of the HE12 data, we retain the uppermost PM curve in Fig.\,25 as the upper limits. 
The present $S(0)$ = 73$^{+56}_{-11}$ eV\,b. 
In comparison,  $S(0)$ = 107 eV\,b [NACRE from \cite{BA80}], and 99.5 eV\,b [RAD10].

Table \ref{li6pgTab2} gives the reaction rates at 0.001 $\le T_{9} \le$ 10. 
The PM $S$-factors below $E_{\rm cm} \simeq$ 0.07, 0.1 and 0.2 MeV in "low", "adopt" and "high" cases, respectively, and above $E_{\rm cm} \simeq$ 1 MeV are used to supplement the experimental data. 
Figure 27  compares the present and the NACRE rates.

{\footnotesize  See  \cite{AR02} for a  cluster model calculation.}

\begin{figure}[hb]
\centering{
\hspace{0.8truein}
\includegraphics[height=0.50\textheight,width=0.90\textwidth]{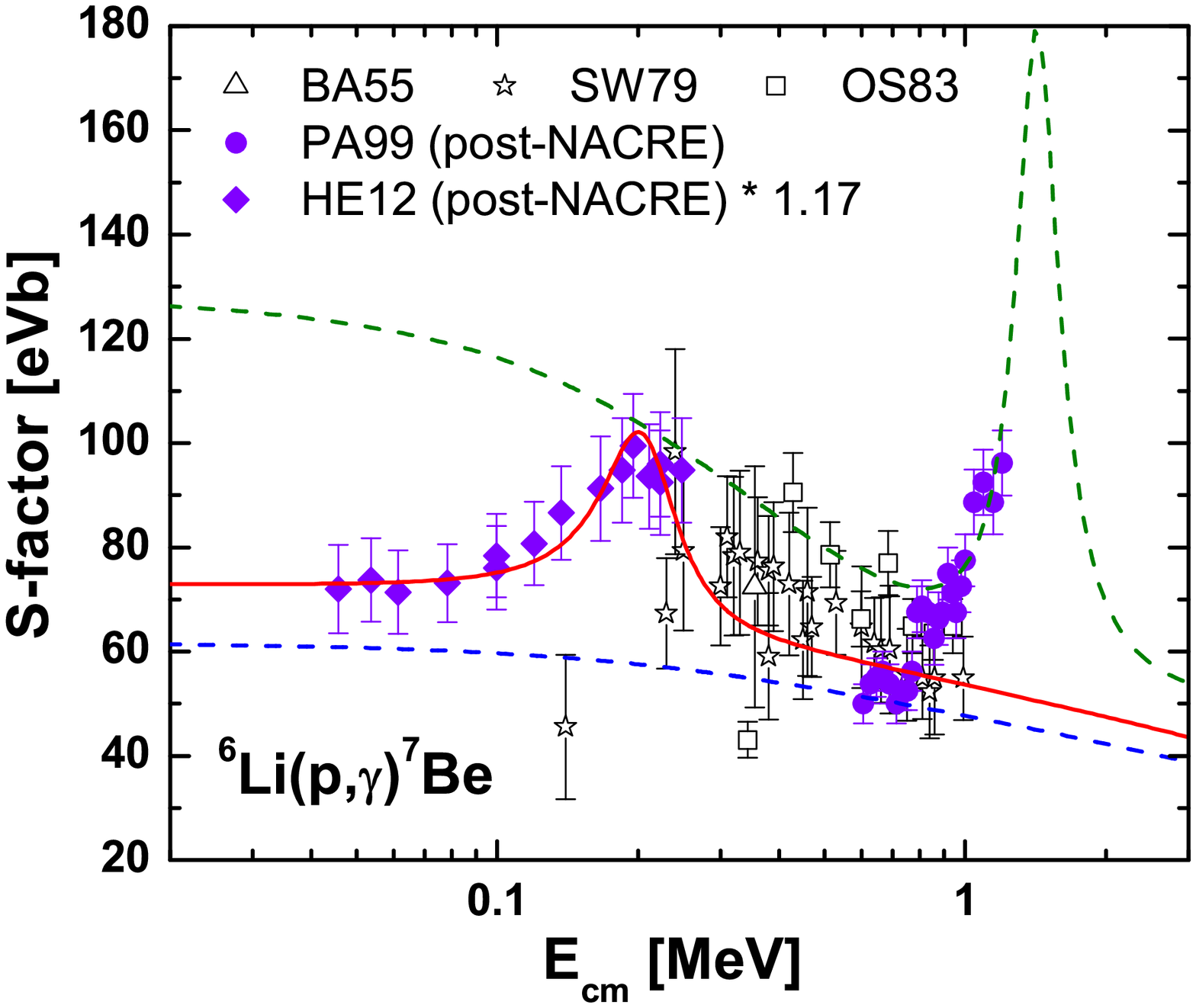}
\vspace{-0.5truecm}
\caption{The $S$-factor for \reac{6}{Li}{p}{\gamma}{7}{Be}.
See Figs.\,25 and 26 for details.}
}
\label{li6pgFig1a}
\end{figure}
\clearpage

\begin{figure}[t]
\centering{
\includegraphics[height=0.4\textheight,width=0.90\textwidth]{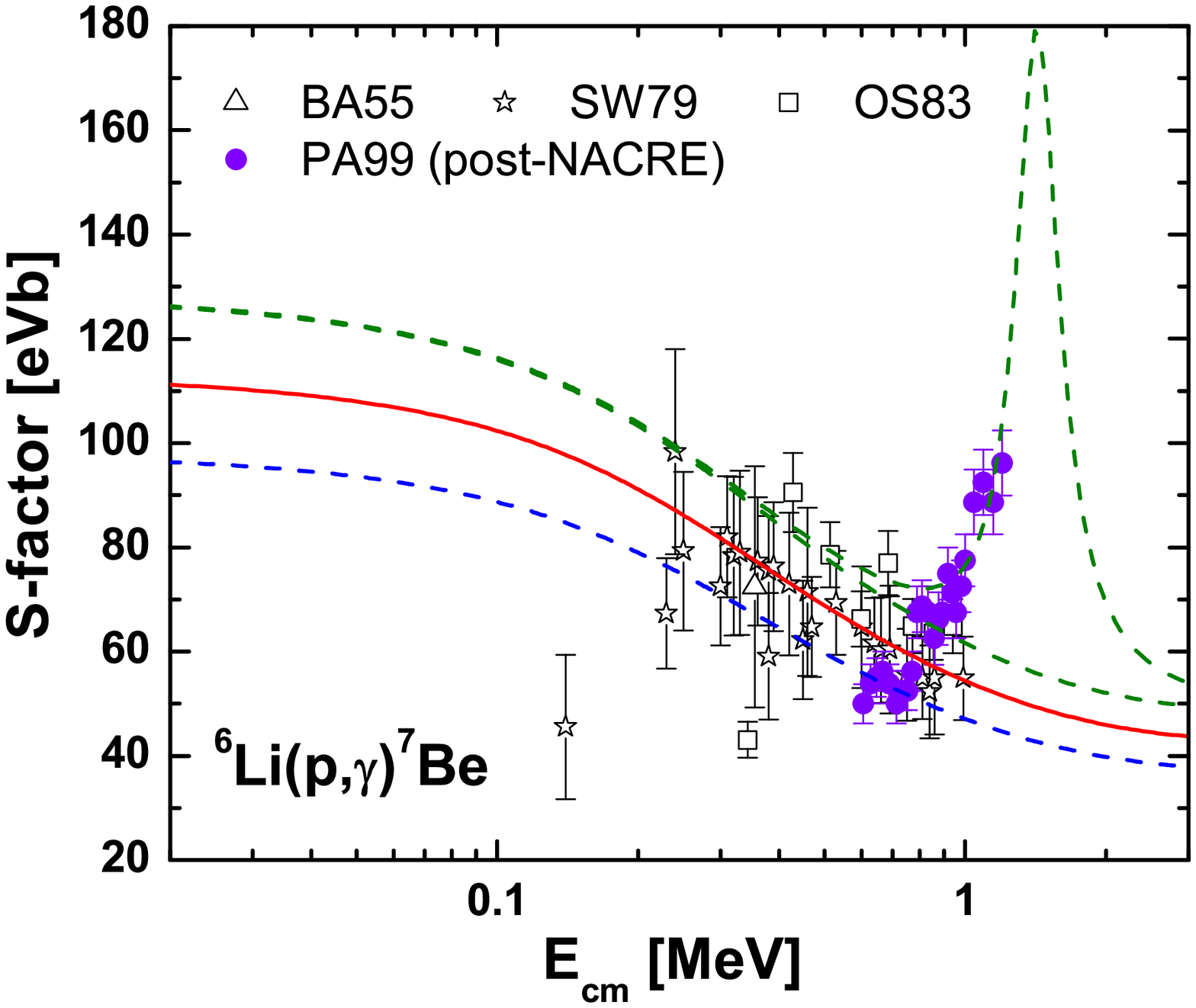}
\vspace{-0.5truecm}
\caption{The $S$-factor for \reac{6}{Li}{p}{\gamma}{7}{Be}. 
The "preliminary" data set PA99 could not be substatiated, but it
 appears to reveal the tail of a 5/2$^{-}$ resonance at $E_{\rm cm} \simeq$ 1.6 MeV that is evident in \reac{6}{Li}{p}{\alpha}{3}{He}. Thus, it is used to set the upper limits.
We note that all the previous theoretical estimates led to d$S/$d$E < 0$ at low energies, and so did an experimental one \cite{PR04}.
Exceptionally, a positive d$S/$d$E$ had been claimed by an experimental study \cite{CE92}.
}
}
\label{li6pgFig1b}
\end{figure}

\begin{figure}[t]
\centering{
\includegraphics[height=0.4\textheight,width=0.90\textwidth]{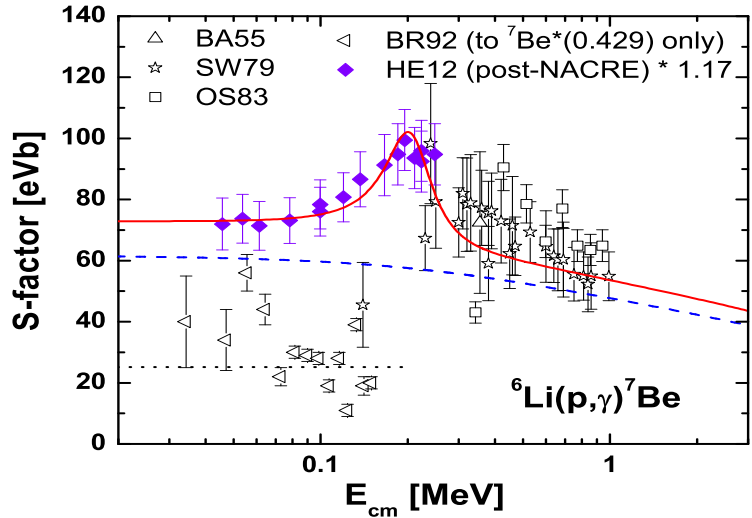}
\vspace{-0.5truecm}
\caption{The $S$-factor for \reac{6}{Li}{p}{\gamma}{7}{Be}.
The upper curve represents an artificial PM fit to the low-energy tail distribution of HE12 (with a re-normalisation factor of 1.17 \cite{HE13}) by {\it assuming} an s-wave (3/2$^{+}$) resonance at $E_{\rm cm} \simeq$ 0.2 MeV.
The dotted line is the mean partial $S$-factor of BR92 for the (p,$\gamma_1$) channel which is known to be about 40 \% of the total.
It is utilised to set the lower asymptotic bound for the $S$-factors at low energies (dashed line).
}
}
\label{li6pgFig1c}
\end{figure}
\clearpage

\begin{figure}[t]
\centering{
\includegraphics[height=0.33\textheight,width=0.90\textwidth]{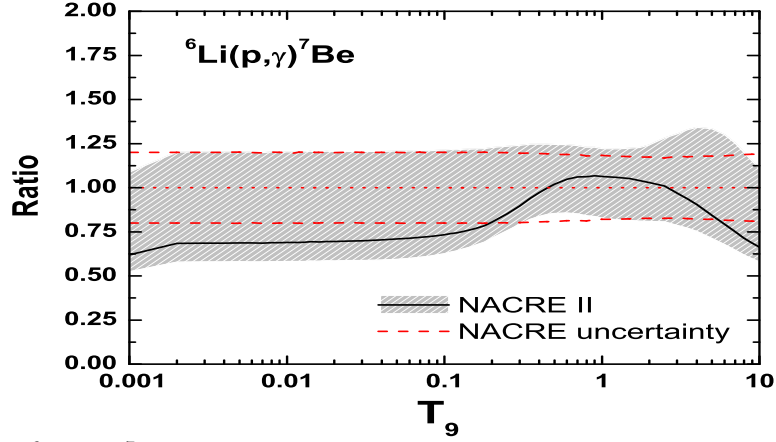}
\vspace{-0.5truecm}
\caption{\reac{6}{Li}{p}{\gamma}{7}{Be} rates in units of the NACRE (adopt) values. The use of HE12 (re-normalised) has expectedly reduced the ratios at low temperatures. The decrease at the highest temperatures reflects the steep increase of the NACRE $S$-factors as extrapolated from \cite{BA80}.}
}
\label{li6pgFig2}
\end{figure}

\begin{table}[hb]
\caption{\reac{6}{Li}{p}{\gamma}{7}{Be} rates in $\rm{cm^{3}mol^{-1}s^{-1}}$}\footnotesize\rm
\begin{tabular*}{\textwidth}{@{\extracolsep{\fill}}  l c c c | l c c c}
\hline
$T_{9}$ & adopted & low & high & $T_{9}$ & adopted & low & high \\
\hline
0.001 & 2.23E$-$29 & 1.89E$-$29 & 3.92E$-$29 &    0.14 & 3.07E$-$01 & 2.67E$-$01 & 4.94E$-$01 \\
0.002 & 5.38E$-$22 & 4.55E$-$22 & 9.45E$-$22 &    0.15 & 4.25E$-$01 & 3.72E$-$01 & 6.79E$-$01 \\
0.003 & 1.91E$-$18 & 1.62E$-$18 & 3.36E$-$18 &    0.16& 5.74E$-$01 & 5.03E$-$01 & 9.08E$-$01 \\
0.004 & 3.28E$-$16 & 2.77E$-$16 & 5.75E$-$16 &    0.18& 9.74E$-$01 & 8.59E$-$01 & 1.51E+00 \\
0.005 & 1.26E$-$14 & 1.07E$-$14 & 2.21E$-$14 &    0.2 & 1.54E+00 & 1.36E+00 & 2.35E+00 \\
0.006 & 2.04E$-$13 & 1.72E$-$13 & 3.57E$-$13 &    0.25 & 3.85E+00 & 3.42E+00 & 5.58E+00 \\
0.007 & 1.87E$-$12 & 1.58E$-$12 & 3.27E$-$12 &    0.3 & 7.73E+00 & 6.82E+00 & 1.07E+01 \\
0.008 & 1.16E$-$11 & 9.82E$-$12 & 2.03E$-$11 &    0.35 & 1.34E+01 & 1.17E+01 & 1.78E+01 \\
0.009 & 5.44E$-$11 & 4.59E$-$11 & 9.49E$-$11 &    0.4 & 2.11E+01 & 1.82E+01 & 2.70E+01 \\
0.01 & 2.05E$-$10 & 1.73E$-$10 & 3.57E$-$10 &    0.45 & 3.07E+01 & 2.60E+01 & 3.83E+01 \\
0.011 & 6.52E$-$10 & 5.50E$-$10 & 1.14E$-$09 &    0.5 & 4.21E+01 & 3.53E+01 & 5.15E+01 \\
0.012 & 1.82E$-$09 & 1.53E$-$09 & 3.16E$-$09 &    0.6 & 7.00E+01 & 5.72E+01 & 8.31E+01 \\
0.013 & 4.53E$-$09 & 3.82E$-$09 & 7.89E$-$09 &    0.7 & 1.03E+02 & 8.30E+01 & 1.21E+02 \\
0.014 & 1.03E$-$08 & 8.71E$-$09 & 1.80E$-$08 &    0.8 & 1.41E+02 & 1.12E+02 & 1.63E+02 \\
0.015 & 2.18E$-$08 & 1.84E$-$08 & 3.80E$-$08 &    0.9 & 1.83E+02 & 1.43E+02 & 2.10E+02 \\
0.016 & 4.32E$-$08 & 3.64E$-$08 & 7.51E$-$08 &    1. & 2.26E+02 & 1.76E+02 & 2.60E+02 \\
0.018 & 1.45E$-$07 & 1.22E$-$07 & 2.51E$-$07 &    1.25 & 3.43E+02 & 2.65E+02 & 3.95E+02 \\
0.02 & 4.09E$-$07 & 3.44E$-$07 & 7.08E$-$07 &    1.5 & 4.64E+02 & 3.59E+02 & 5.40E+02 \\
0.025 & 3.25E$-$06 & 2.74E$-$06 & 5.62E$-$06 &    1.75& 5.85E+02 & 4.56E+02 & 6.92E+02 \\
0.03 & 1.57E$-$05 & 1.32E$-$05 & 2.71E$-$05 &    2. & 7.04E+02 & 5.53E+02 & 8.50E+02 \\
0.04 & 1.55E$-$04 & 1.30E$-$04 & 2.65E$-$04 &    2.5 & 9.32E+02 & 7.45E+02 & 1.19E+03 \\
0.05 & 7.80E$-$04 & 6.54E$-$04 & 1.33E$-$03 &    3. & 1.14E+03 & 9.28E+02 & 1.54E+03 \\
0.06 & 2.66E$-$03 & 2.24E$-$03 & 4.51E$-$03 &    3.5 & 1.34E+03 & 1.10E+03 & 1.90E+03 \\
0.07 & 7.06E$-$03 & 5.95E$-$03 & 1.19E$-$02 &    4. & 1.52E+03 & 1.26E+03 & 2.25E+03 \\
0.08 & 1.58E$-$02 & 1.33E$-$02 & 2.64E$-$02 &    5. & 1.84E+03 & 1.55E+03 & 2.91E+03 \\
0.09 & 3.10E$-$02 & 2.63E$-$02 & 5.15E$-$02 &    6. & 2.11E+03 & 1.80E+03 & 3.46E+03 \\
0.1 & 5.53E$-$02 & 4.73E$-$02 & 9.15E$-$02 &    7. & 2.34E+03 & 2.02E+03 & 3.91E+03 \\
0.11 & 9.18E$-$02 & 7.88E$-$02 & 1.51E$-$01 &    8. & 2.54E+03 & 2.20E+03 & 4.27E+03 \\
0.12 & 1.44E$-$01 & 1.24E$-$01 & 2.34E$-$01 &    9. & 2.71E+03 & 2.36E+03 & 4.56E+03 \\
0.13 & 2.14E$-$01 & 1.85E$-$01 & 3.47E$-$01 &   10. & 2.86E+03 & 2.50E+03 & 4.79E+03 \\
\hline
\end{tabular*}
\begin{tabular*}{\textwidth}{@{\extracolsep{\fill}} l c c }
REV  = 
$1.19 \times 10^{10}T_{9}^{3/2}{\rm exp}(-65.054/T_{9})\,[1.0+2.333{\rm exp}(-25.369/T_{9})]$ \\ 
\hspace{2truecm}
$ /\,[1.0+0.5{\rm exp}(-4.979/T_{9})]$
 & \\
\end{tabular*}
\label{li6pgTab2}
\end{table}
\clearpage\subsection{\reac{6}{Li}{p}{\alpha}{3}{He}}
\label{li6paSect}
The experimental data sets referred to in NACRE are MA56 \cite{MA56}, GE66 \cite{GE66}, FA64 \cite{FA64}, SP71 \cite{SP71}, GO74 \cite{GO74}, LI77 \cite{LI77}, EL79 \cite{EL79}, SH79 \cite{SH79}, KW89 \cite{KW89} and EN92 \cite{EN92}, covering the 0.01 $\lsimeq E_{\rm cm} \lsimeq$ 12 MeV range.
\cite{FI67} was rejected.
Added are the post-NACRE data sets TU03 \cite{TU03}$^\dag$, CR05 \cite{CR05} and CR08 \cite{CR08}.
[{\footnotesize{$^\dag$from  $d(^{6}$Li,$\alpha)^{3}$He (THM)}}] 

\begin{figure}[hb]
\centering{
\includegraphics[height=0.50\textheight,width=0.90\textwidth]{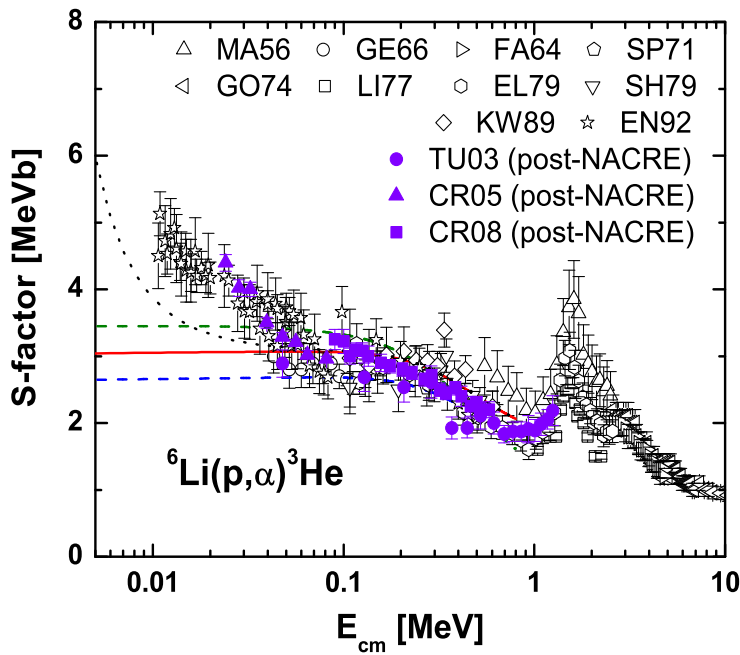}
\vspace{-0.5truecm}
\caption{The $S$-factor for \reac{6}{Li}{p}{\alpha}{3}{He}. The dotted line indicates an adiabatic screening correction ($U_{\rm e}$ = 173 eV) to the solid ('adopted') line.  Of CR05, only the data points obtained with metallic Li are shown.}
\label{li6paFig1}
}
\end{figure}

Figure \ref{li6paFig1} compares the DWBA and experimental $S$-factors.
Some measurements below $E_{\rm cm} \simeq$ 0.05 MeV look contaminated by electron screening (see CR05, and  \cite{LA13}). 
The data in the 0.05 $\lsimeq E_{\rm cm} \lsimeq$ 1 MeV range are used for the DWBA fit.
The adopted parameter values are given in Table \ref{li6paTab1}.
The present  $S(0)$ = 3.1 $\pm$ 0.4  MeV\,b. 
In comparison,  $S(0)$ = 2.97 MeV\,b [NACRE, from KW89].

\begin{figure}[t]
\centering{
\includegraphics[height=0.33\textheight,width=0.90\textwidth]{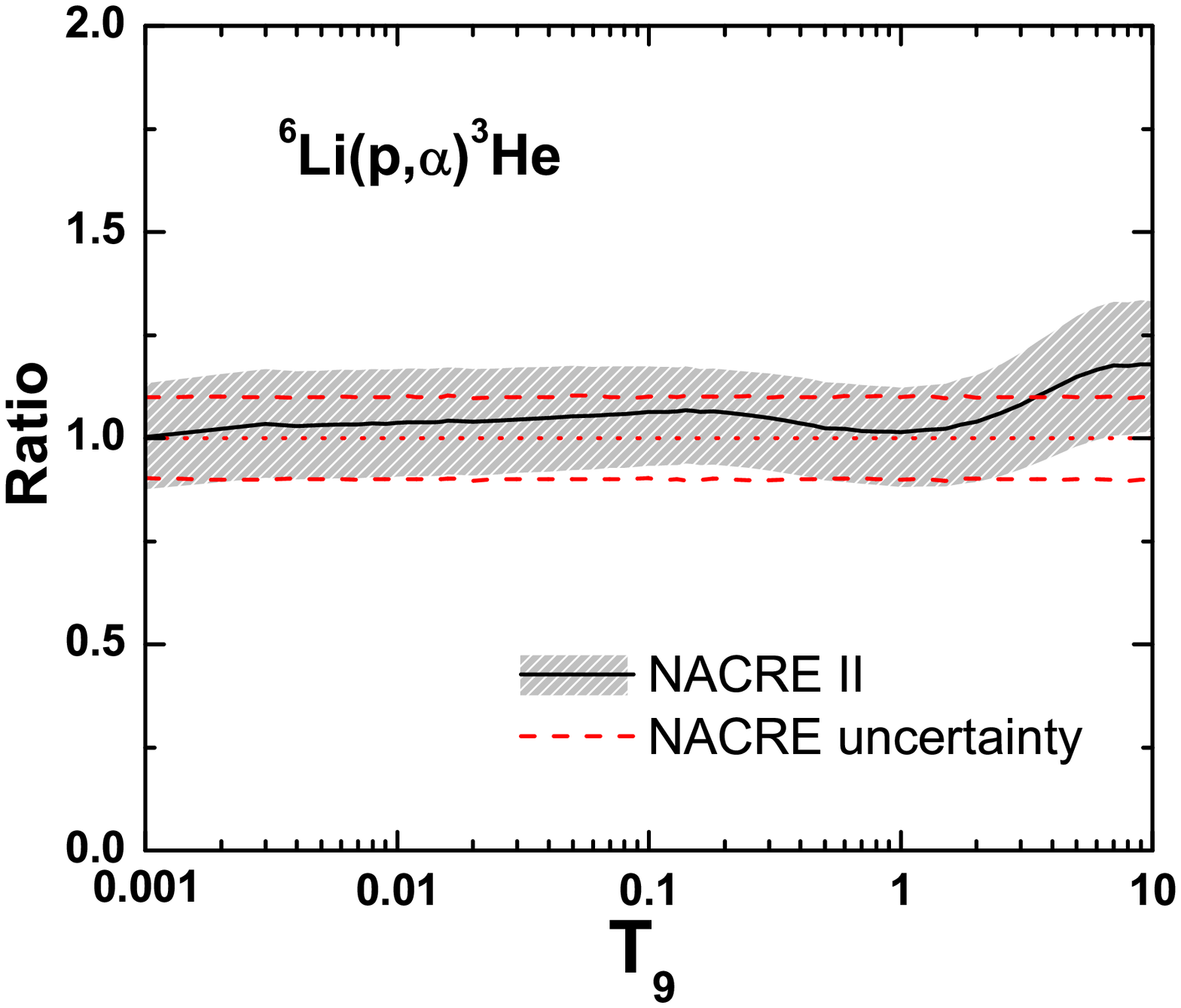}
\vspace{-0.4truecm}
\caption{\reac{6}{Li}{p}{\alpha}{3}{He} rates in units of the NACRE (adopt) values.
}
\label{li6paFig2}
}
\end{figure}

Table \ref{li6paTab2} gives the reaction rates at 0.001 $\le T_{9} \le$ 10, for which the DWBA-predicted and the experimental cross sections below and above $E_{\rm cm} \simeq$ 0.1 MeV are used, respectively.
Figure \ref{li6paFig2} compares the present and the NACRE rates.

{\footnotesize  See  \cite{AR02} for a cluster model calculation.} 

\clearpage

\begin{table}[hb]
\caption{\reac{6}{Li}{p}{\alpha}{3}{He} rates in units of $\rm{cm^{3}mol^{-1}s^{-1}}$.} \footnotesize\rm
\begin{tabular*}{\textwidth}{@{\extracolsep{\fill}}  l c c c | l c c c}
\hline
$T_{9}$ & adopted & low & high & $T_{9}$ & adopted & low & high \\
\hline
  0.001 & 1.00E$-$24 & 8.72E$-$25 & 1.14E$-$24 &          0.14  & 1.26E+04 & 1.10E+04 & 1.39E+04 \\
  0.002 & 2.24E$-$17 & 1.95E$-$17 & 2.54E$-$17 &          0.15  & 1.74E+04 & 1.52E+04 & 1.92E+04 \\
  0.003 & 8.03E$-$14 & 6.99E$-$14 & 9.09E$-$14 &          0.16  & 2.33E+04 & 2.04E+04 & 2.57E+04 \\
  0.004 & 1.37E$-$11 & 1.19E$-$11 & 1.55E$-$11 &          0.18  & 3.90E+04 & 3.41E+04 & 4.29E+04 \\
  0.005 & 5.28E$-$10 & 4.60E$-$10 & 5.98E$-$10 &          0.2 & 6.05E+04 & 5.30E+04 & 6.67E+04 \\
  0.006 & 8.53E$-$09 & 7.42E$-$09 & 9.65E$-$09 &          0.25  & 1.45E+05 & 1.27E+05 & 1.60E+05 \\
  0.007 & 7.83E$-$08 & 6.82E$-$08 & 8.86E$-$08 &          0.3 & 2.78E+05 & 2.43E+05 & 3.07E+05 \\
  0.008 & 4.87E$-$07 & 4.24E$-$07 & 5.50E$-$07 &          0.35  & 4.65E+05 & 4.06E+05 & 5.15E+05 \\
  0.009 & 2.28E$-$06 & 1.98E$-$06 & 2.57E$-$06 &          0.4 & 7.07E+05 & 6.18E+05 & 7.85E+05 \\
  0.01  & 8.58E$-$06 & 7.47E$-$06 & 9.69E$-$06 &          0.45  & 1.00E+06 & 8.76E+05 & 1.12E+06 \\
  0.011 & 2.73E$-$05 & 2.38E$-$05 & 3.09E$-$05 &          0.5 & 1.35E+06 & 1.18E+06 & 1.50E+06 \\
  0.012 & 7.61E$-$05 & 6.63E$-$05 & 8.59E$-$05 &          0.6 & 2.20E+06 & 1.91E+06 & 2.44E+06 \\
  0.013 & 1.90E$-$04 & 1.65E$-$04 & 2.14E$-$04 &          0.7 & 3.22E+06 & 2.80E+06 & 3.58E+06 \\
  0.014 & 4.33E$-$04 & 3.77E$-$04 & 4.89E$-$04 &          0.8 & 4.38E+06 & 3.81E+06 & 4.87E+06 \\
  0.015 & 9.15E$-$04 & 7.97E$-$04 & 1.03E$-$03 &          0.9 & 5.67E+06 & 4.92E+06 & 6.29E+06 \\
  0.016 & 1.81E$-$03 & 1.58E$-$03 & 2.04E$-$03 &          1.     & 7.05E+06 & 6.11E+06 & 7.83E+06 \\
  0.018 & 6.07E$-$03 & 5.29E$-$03 & 6.84E$-$03 &          1.25  & 1.08E+07 & 9.32E+06 & 1.20E+07 \\
  0.02  & 1.71E$-$02 & 1.49E$-$02 & 1.93E$-$02 &          1.5  & 1.48E+07 & 1.28E+07 & 1.65E+07 \\
  0.025 & 1.37E$-$01 & 1.19E$-$01 & 1.54E$-$01 &          1.75  & 1.90E+07 & 1.63E+07 & 2.11E+07 \\
  0.03  & 6.61E$-$01 & 5.76E$-$01 & 7.43E$-$01 &          2.     & 2.33E+07 & 1.99E+07 & 2.59E+07 \\
  0.04  & 6.50E+00 & 5.68E+00 & 7.29E+00 &          2.5  & 3.21E+07 & 2.74E+07 & 3.59E+07 \\
  0.05  & 3.28E+01 & 2.86E+01 & 3.66E+01 &          3.     & 4.13E+07 & 3.51E+07 & 4.63E+07 \\
  0.06  & 1.12E+02 & 9.75E+01 & 1.25E+02 &          3.5  & 5.07E+07 & 4.31E+07 & 5.70E+07 \\
  0.07  & 2.96E+02 & 2.59E+02 & 3.30E+02 &          4.     & 6.01E+07 & 5.11E+07 & 6.78E+07 \\
  0.08  & 6.59E+02 & 5.76E+02 & 7.33E+02 &          5.     & 7.83E+07 & 6.66E+07 & 8.86E+07 \\
  0.09  & 1.29E+03 & 1.13E+03 & 1.44E+03 &          6.     & 9.48E+07 & 8.07E+07 & 1.07E+08 \\
  0.1 & 2.31E+03 & 2.02E+03 & 2.56E+03 &          7.     & 1.09E+08 & 9.30E+07 & 1.24E+08 \\
  0.11  & 3.82E+03 & 3.34E+03 & 4.22E+03 &          8.     & 1.21E+08 & 1.03E+08 & 1.37E+08 \\
  0.12  & 5.95E+03 & 5.20E+03 & 6.57E+03 &          9.     & 1.31E+08 & 1.12E+08 & 1.49E+08 \\
  0.13  & 8.83E+03 & 7.73E+03 & 9.75E+03 &         10.     & 1.39E+08 & 1.19E+08 & 1.58E+08 \\
\hline
\end{tabular*}
\begin{tabular*}{\textwidth}{@{\extracolsep{\fill}} l c }
REV  = 
$ 1.07\,{\rm exp}(-46.648/T_{9})\,[1.0+2.333\,{\rm exp}(-25.369/T_{9})]$ & \\
\end{tabular*}
\label{li6paTab2}
\end{table}
\clearpage
\subsection{\reac{7}{Li}{p}{\gamma}{8}{Be}}
\label{li7pgSect}
The experimental data sets referred to in NACRE are  PE63 \cite{PE63}$^\dag$, RI63 \cite{RI63}$^\dag$, and ZA95a \cite{ZA95a}, covering 0.09 $\lsimeq E_{\rm cm} \lsimeq$ 10 MeV range. 
No new cross section data are found. 
[{\footnotesize{$^\dag$normalised to ZA95a}}] 

\begin{figure}[hb]
\centering{
\includegraphics[height=0.50\textheight,width=0.90\textwidth]{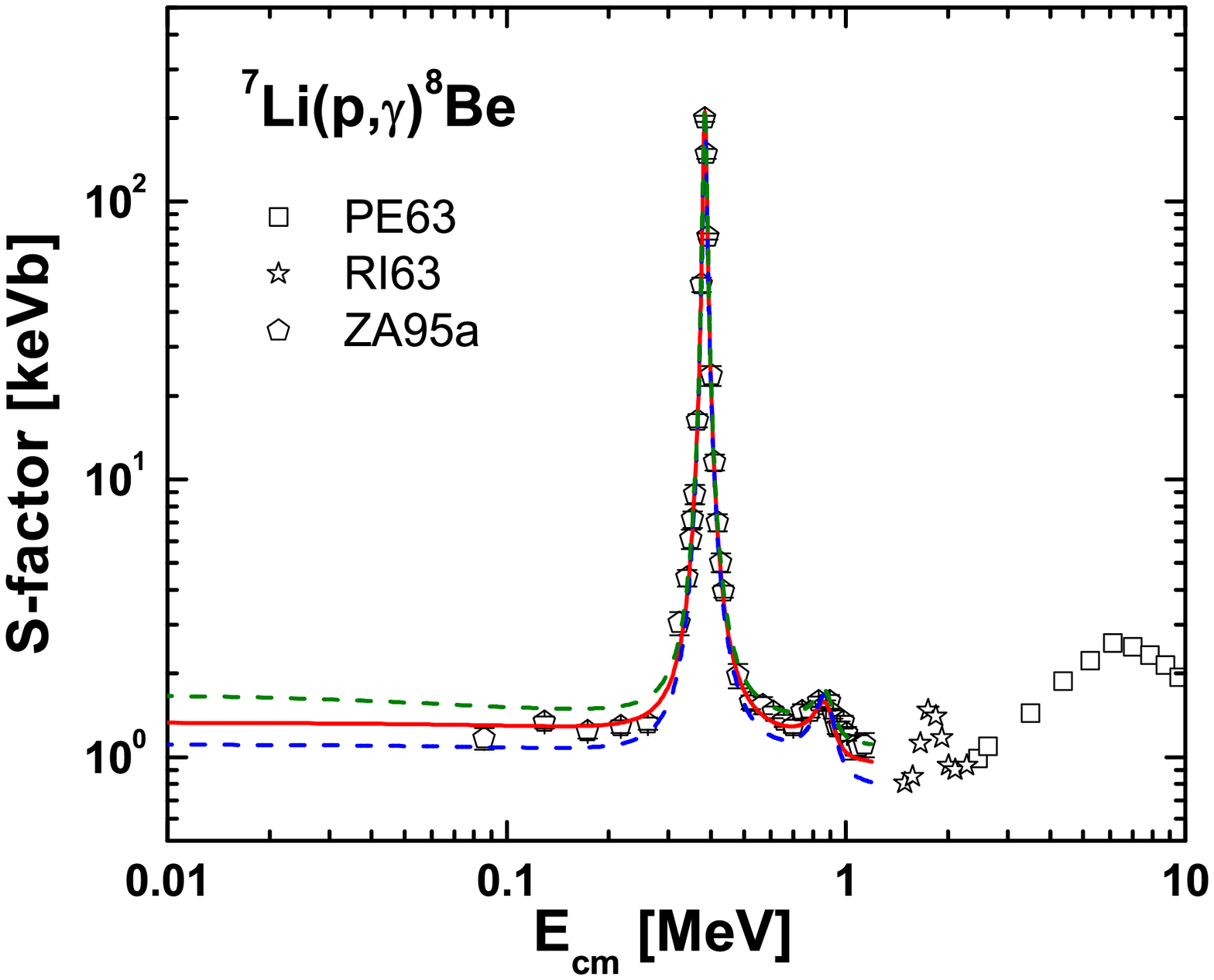}
\vspace{-0.5truecm}
\caption{The $S$-factor  for \reac{7}{Li}{p}{\gamma}{8}{Be}.}
\label{li7pgFig1a}
}
\end{figure}

\begin{figure}[t]
\centering{
\includegraphics[height=0.40\textheight,width=0.90\textwidth]{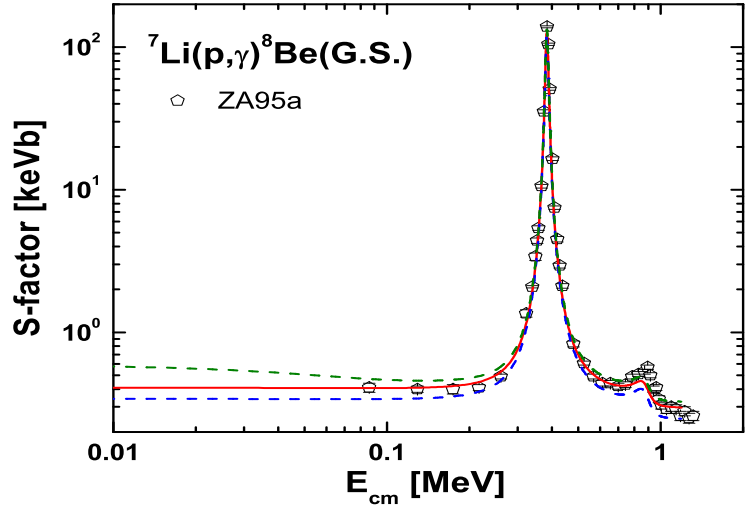}
\vspace{-0.5truecm}
\caption{The partial $S$-factor for the \reac{7}{Li}{p}{\gamma}{8}{Be} transition to the $0^{+}$ ground state. A modest contribution from the sub-threshold "halo-like" $2^{+}$ state at $- 0.63$ MeV with a very large channel radius (\cite{SP99}) has been added to the upper limit.
The M1 transitions via the $1^{+}$ resonances are too weak to explain the observed extent of the p-wave mixture at the low energies. 
[A potential can be found that gives essentially the same shape of the $S$-factors for the s- ($1^{-}$; E1) and p- ($2{+}$; E2) contributions. See Table \ref{li7pgTab1} for an example, which gives about 25\% p-wave non-resonant contributions to the (p, $\gamma_{0}$) channel.] 
}
\label{li7pgFig1b}
}
\end{figure}

\begin{figure}[t]
\centering{
\includegraphics[height=0.40\textheight,width=0.90\textwidth]{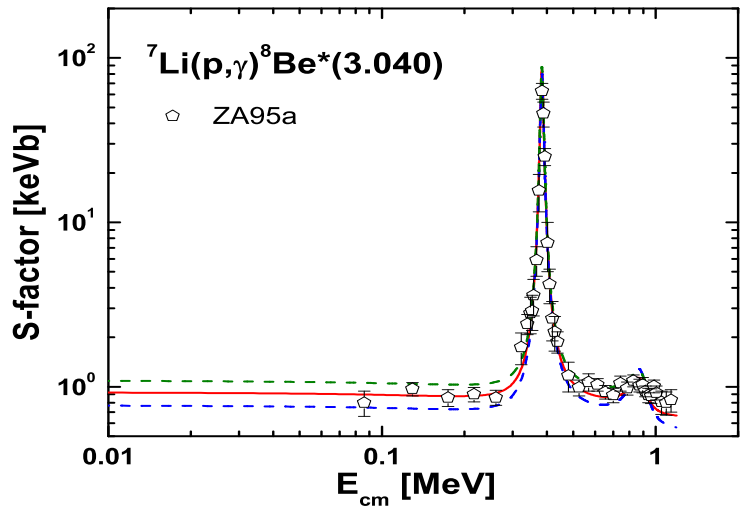}
\vspace{-0.5truecm}
\caption{The partial $S$-factor for the \reac{7}{Li}{p}{\gamma}{8}{Be} transition to the $2^{+}$ first excited state at $E_{\rm x}= 3.040$  MeV.
ZA95a is the difference of the unresolved (p,$\gamma_{0}+\gamma_{1}$) in Fig. \ref{li7pgFig1a} and (p, $\gamma_{0}$) in Fig. \ref{li7pgFig1b}.
}
\label{li7pgFig1c}
}
\end{figure}

Figure \ref{li7pgFig1a} compares the PM and experimental $S$-factors.
At $E_{\rm cm} \lsimeq$ 2 MeV, the partial $S$-factors for the transitions to the ground and the first excited states are used for the PM fit (Figs.\,\ref{li7pgFig1b} and \ref{li7pgFig1c}).
The data exhibit the (interfering) $1^{+}$ resonances at $E_{\rm R} \simeq$ 0.38 and 0.90 MeV. 
Whereas the cross section data at low energies could be well described by the essentially pure s-wave contributions, the observed anisotropies imply a significant mixture of p-wave contributions to (p,$\gamma_0$) cross sections [197\,-\,200]. 
Furthermore, d$S/$d$E < 0$  for the ground- and first-excited-state transitions have been asserted \cite{SP99}. 
Given the difficulty of finding an explanation that is consistent with all these experimental findings in the low-energy range, we allow for relatively large uncertainties in the PM fits. 
The adopted parameter values are given in Table \ref{li7pgTab1}.
The present total $S(0)$ = 1.3$_{-0.2}^{+0.4}$ keV\,b.
In comparison, $S(0)$ = 1.5 $\pm$ 0.2 keV\,b [NACRE, from \cite{CE92}].

\begin{figure}[t]
\centering{
\includegraphics[height=0.33\textheight,width=0.90\textwidth]{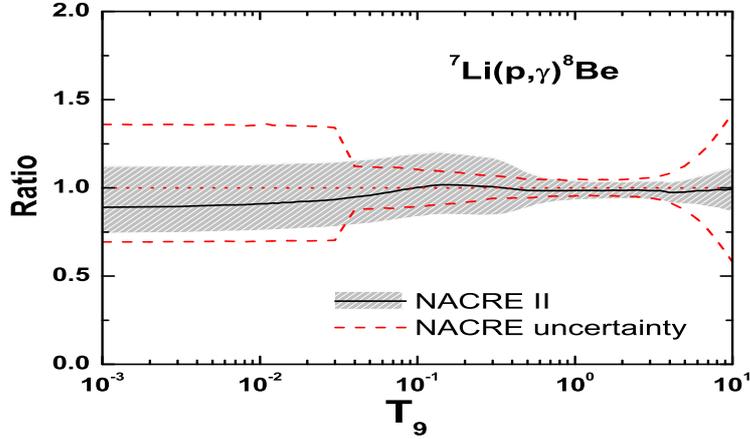}
\vspace{-0.4truecm}
\caption{\reac{7}{Li}{p}{\gamma}{8}{Be} rates in units of the NACRE (adopt) values. The use of PM lowers the upper uncertainty when compared with the NACRE guess.
}
\label{li7pgFig2}
}
\end{figure}

Table \ref{li7pgTab2} gives the reaction rates at 0.001 $\le T_{9} \le$ 10, for which the PM-predicted and the experimental cross sections below and above $E_{\rm cm} \simeq$ 0.3 MeV are used, respectively. 
Figure \ref{li7pgFig2} compares the present and the NACRE rates.

{\footnotesize{See  \cite{SA01} for a potential model analysis.}}

\clearpage

\clearpage

\begin{table}[hb]
\caption{\reac{7}{Li}{p}{\gamma}{8}{Be} rates in $\rm{cm^{3}mol^{-1}s^{-1}}$}\footnotesize\rm
\begin{tabular*}{\textwidth}{@{\extracolsep{\fill}}  l c c c | l c c c}
\hline
$T_{9}$ & adopted & low & high & $T_{9}$ & adopted & low & high \\
\hline
0.001 & 2.50E$-$28 & 2.08E$-$28 & 3.15E$-$28 &      0.14 & 4.79E+00 & 4.00E+00 & 5.66E+00 \\
0.002 & 6.16E$-$21 & 5.14E$-$21 & 7.75E$-$21 &      0.15 & 6.63E+00 & 5.53E+00 & 7.82E+00 \\
0.003 & 2.32E$-$17 & 1.93E$-$17 & 2.91E$-$17 &      0.16 & 8.92E+00 & 7.44E+00 & 1.05E+01 \\
0.004 & 4.12E$-$15 & 3.44E$-$15 & 5.17E$-$15 &      0.18 & 1.50E+01 & 1.25E+01 & 1.77E+01 \\
0.005 & 1.63E$-$13 & 1.36E$-$13 & 2.04E$-$13 &      0.2 & 2.35E+01 & 1.96E+01 & 2.76E+01 \\
0.006 & 2.68E$-$12 & 2.24E$-$12 & 3.36E$-$12 &      0.25 & 5.77E+01 & 4.82E+01 & 6.74E+01 \\
0.007 & 2.50E$-$11 & 2.09E$-$11 & 3.13E$-$11 &      0.3 & 1.17E+02 & 9.81E+01 & 1.36E+02 \\
0.008 & 1.57E$-$10 & 1.31E$-$10 & 1.97E$-$10 &      0.35 & 2.18E+02 & 1.85E+02 & 2.51E+02 \\
0.009 & 7.44E$-$10 & 6.20E$-$10 & 9.28E$-$10 &      0.4 & 3.97E+02 & 3.44E+02 & 4.52E+02 \\
0.01 & 2.83E$-$09 & 2.36E$-$09 & 3.53E$-$09 &      0.45 & 7.10E+02 & 6.28E+02 & 7.94E+02 \\
0.011 & 9.08E$-$09 & 7.57E$-$09 & 1.13E$-$08 &      0.5 & 1.22E+03 & 1.10E+03 & 1.35E+03 \\
0.012 & 2.55E$-$08 & 2.12E$-$08 & 3.17E$-$08 &      0.6 & 3.07E+03 & 2.83E+03 & 3.32E+03 \\
0.013 & 6.40E$-$08 & 5.34E$-$08 & 7.95E$-$08 &      0.7 & 6.22E+03 & 5.80E+03 & 6.66E+03 \\
0.014 & 1.47E$-$07 & 1.22E$-$07 & 1.82E$-$07 &      0.8 & 1.06E+04 & 9.98E+03 & 1.13E+04 \\
0.015 & 3.12E$-$07 & 2.60E$-$07 & 3.87E$-$07 &      0.9 & 1.60E+04 & 1.51E+04 & 1.70E+04 \\
0.016 & 6.21E$-$07 & 5.17E$-$07 & 7.69E$-$07 &      1. & 2.21E+04 & 2.09E+04 & 2.33E+04 \\
0.018 & 2.09E$-$06 & 1.75E$-$06 & 2.59E$-$06 &      1.25 & 3.80E+04 & 3.60E+04 & 4.01E+04 \\
0.02 & 5.96E$-$06 & 4.97E$-$06 & 7.36E$-$06 &      1.5 & 5.27E+04 & 5.00E+04 & 5.54E+04 \\
0.025 & 4.81E$-$05 & 4.01E$-$05 & 5.92E$-$05 &      1.75 & 6.48E+04 & 6.15E+04 & 6.81E+04 \\
0.03 & 2.35E$-$04 & 1.96E$-$04 & 2.89E$-$04 &      2. & 7.42E+04 & 7.05E+04 & 7.79E+04 \\
0.04 & 2.35E$-$03 & 1.96E$-$03 & 2.87E$-$03 &      2.5 & 8.64E+04 & 8.21E+04 & 9.09E+04 \\
0.05 & 1.20E$-$02 & 9.98E$-$03 & 1.45E$-$02 &      3. & 9.29E+04 & 8.80E+04 & 9.78E+04 \\
0.06 & 4.11E$-$02 & 3.43E$-$02 & 4.97E$-$02 &      3.5 & 9.59E+04 & 9.06E+04 & 1.01E+05 \\
0.07 & 1.10E$-$01 & 9.14E$-$02 & 1.32E$-$01 &      4. & 9.71E+04 & 9.13E+04 & 1.03E+05 \\
0.08 & 2.45E$-$01 & 2.05E$-$01 & 2.94E$-$01 &      5. & 9.73E+04 & 9.05E+04 & 1.04E+05 \\
0.09 & 4.84E$-$01 & 4.03E$-$01 & 5.78E$-$01 &      6. & 9.66E+04 & 8.88E+04 & 1.04E+05 \\
0.1 & 8.65E$-$01 & 7.22E$-$01 & 1.03E+00 &      7. & 9.63E+04 & 8.73E+04 & 1.05E+05 \\
0.11 & 1.44E+00 & 1.20E+00 & 1.71E+00 &      8. & 9.66E+04 & 8.65E+04 & 1.07E+05 \\
0.12 & 2.25E+00 & 1.87E+00 & 2.67E+00 &      9. & 9.77E+04 & 8.62E+04 & 1.09E+05 \\
0.13 & 3.35E+00 & 2.79E+00 & 3.96E+00 &     10. & 9.93E+04 & 8.66E+04 & 1.12E+05 \\

\hline
\end{tabular*}
\label{li7pgTab2}
\end{table}
\clearpage
\subsection{\reac{7}{Li}{p}{\alpha}{4}{He}}
\label{li7paSect}
The experimental data sets referred to in NACRE are CA62 \cite{CA62}, MA64 \cite{MA64}, FI67 \cite{FI67}, SP71 \cite{SP71}$^\dag$, RO86 \cite{RO86}, HA89 \cite{HA89}$^{\dag,\ddagger}$, and EN92 \cite{EN92}, covering the 0.011 $\lsimeq E_{\rm cm} \lsimeq$ 10 MeV range.
Added are the post-NACRE data sets LA01 \cite{LA01}$^{\dag\dag}$, CR05 \cite{CR05} and CR09 \cite{CR09}, the first extending the range down to $E_{\rm cm} \simeq$ 0.010 MeV. 
[{\footnotesize{$^\dag$corrected by NACRE for a factor of two; $^\ddagger$relative to $^{6}$Li(p,$\alpha$); $^{\dag\dag}$from d($^{7}$Li, $\alpha\alpha$)n (THM).}}] 

\begin{figure}[hb]
\centering{
\includegraphics[height=0.50\textheight,width=0.90\textwidth]{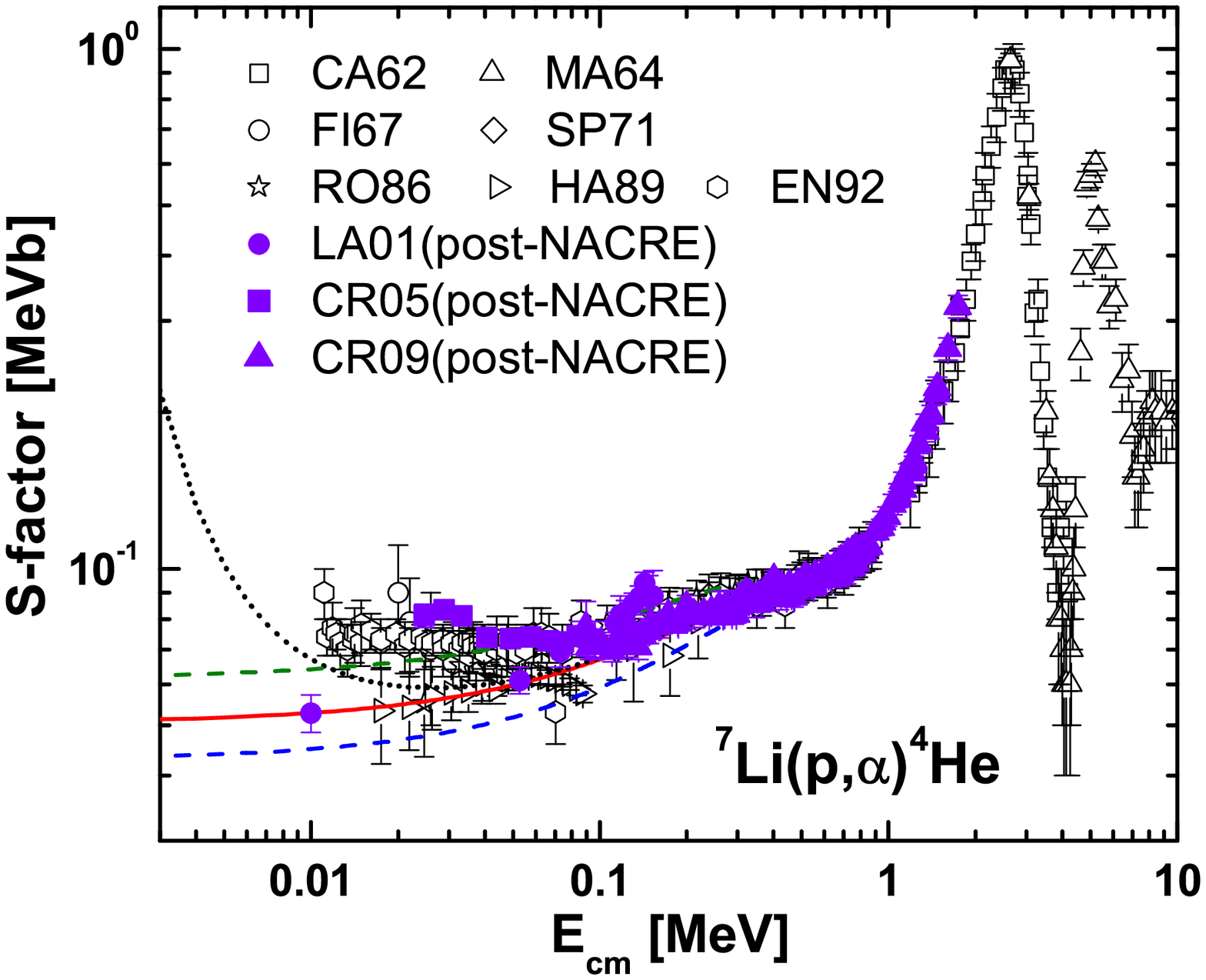}
\vspace{-0.5truecm}
\caption{The $S$-factor for \reac{7}{Li}{p}{\alpha}{4}{He}. The dotted line indicates an adiabatic screening correction ($U_{\rm e}$ = 173 eV) to the 'adopt' curve (solid line).}
\label{li7paFig1}
}
\end{figure}

Figure \ref{li7paFig1} compares the DWBA and experimental $S$-factors. 
Some measurements below $E_{\rm cm} \simeq$ 0.05 MeV look contaminated by electron screening (\cite{CR05} for details).
The data in the 0.05 $\lsimeq$ $E_{\rm cm} \lsimeq$ 1 MeV range (exceptionally, all the LA01 data points by THM extending to the lower energies) are used for the DWBA fit.
The symmetry of the exit channel allows even $l_{f}$, and thus odd $l_{i}$, only.
The adopted parameter values are given in Table \ref{li7paTab1}.
The present  $S(0)$ = 52 $^{+11}_{-8}$ keV\,b.
In comparison,  $S(0)$ = 59.3 keV\,b [NACRE, from EN92], and 67 $\pm$ 4 keV\,b [BBN04].

\begin{figure}[t]
\centering{
\hspace{0.8truein}
\includegraphics[height=0.33\textheight,width=0.90\textwidth]{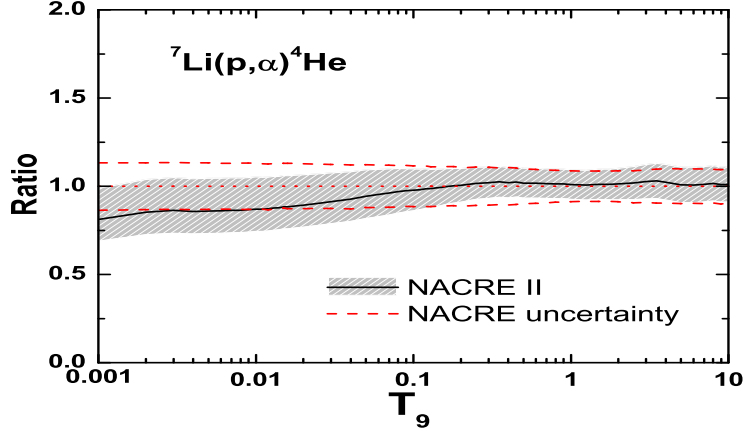}
\vspace{-0.4truecm}
\caption{\reac{7}{Li}{p}{\alpha}{4}{He} rates in units of the NACRE (adopt) values. The adoption of LA01 (THM) reduces the rates at the lowest temperatures.
}
\label{li7paFig2}
}
\end{figure}

Table \ref{li7paTab2} gives the reaction rates at 0.001 $\le T_{9} \le$ 10, for which the DWBA-predicted and the experimental cross sections below and above $E_{\rm cm} \simeq$ 0.05 MeV are used, respectively. 
Figure \ref{li7paFig2} compares the present and the NACRE rates.

{\footnotesize{  See  \cite{RA90,YA95} for DWBA analyses; \cite{PI11} on an invariance of the cross sections extracted from different Trojan horse reactions.}} 

\clearpage

\begin{table}[hb]
\caption{Rates of \reac{7}{Li}{p}{\alpha}{4}{He} in units of $\rm{cm^{3}mol^{-1}s^{-1}}$.} \footnotesize\rm
\begin{tabular*}{\textwidth}{@{\extracolsep{\fill}}  l c c c | l c c c}
\hline
$T_{9}$ & adopted & low & high & $T_{9}$ & adopted & low & high \\
\hline
  0.001 & 9.17E$-$27 & 7.79E$-$27 & 1.12E$-$26 &      0.14 & 2.60E+02 & 2.32E+02 & 2.87E+02 \\
  0.002 & 2.38E$-$19 & 2.02E$-$19 & 2.90E$-$19 &      0.15 & 3.62E+02 & 3.25E+02 & 4.00E+02 \\
  0.003 & 9.06E$-$16 & 7.71E$-$16 & 1.10E$-$15 &      0.16 & 4.92E+02 & 4.42E+02 & 5.42E+02 \\
  0.004 & 1.61E$-$13 & 1.37E$-$13 & 1.96E$-$13 &      0.18 & 8.42E+02 & 7.60E+02 & 9.25E+02 \\
  0.005 & 6.41E$-$12 & 5.45E$-$12 & 7.78E$-$12 &      0.2 & 1.34E+03 & 1.21E+03 & 1.46E+03 \\
  0.006 & 1.06E$-$10 & 9.02E$-$11 & 1.29E$-$10 &      0.25 & 3.35E+03 & 3.05E+03 & 3.66E+03 \\
  0.007 & 9.93E$-$10 & 8.46E$-$10 & 1.20E$-$09 &      0.3 & 6.73E+03 & 6.14E+03 & 7.31E+03 \\
  0.008 & 6.28E$-$09 & 5.36E$-$09 & 7.61E$-$09 &      0.35 & 1.17E+04 & 1.07E+04 & 1.27E+04 \\
  0.009 & 2.98E$-$08 & 2.55E$-$08 & 3.61E$-$08 &      0.4 & 1.84E+04 & 1.68E+04 & 1.99E+04 \\
  0.01 & 1.14E$-$07 & 9.74E$-$08 & 1.38E$-$07 &      0.45 & 2.68E+04 & 2.46E+04 & 2.91E+04 \\
  0.011 & 3.68E$-$07 & 3.15E$-$07 & 4.45E$-$07 &      0.5 & 3.71E+04 & 3.40E+04 & 4.02E+04 \\
  0.012 & 1.04E$-$06 & 8.89E$-$07 & 1.25E$-$06 &      0.6 & 6.31E+04 & 5.78E+04 & 6.83E+04 \\
  0.013 & 2.62E$-$06 & 2.25E$-$06 & 3.16E$-$06 &      0.7 & 9.58E+04 & 8.79E+04 & 1.04E+05 \\
  0.014 & 6.04E$-$06 & 5.18E$-$06 & 7.27E$-$06 &      0.8 & 1.35E+05 & 1.24E+05 & 1.46E+05 \\
  0.015 & 1.29E$-$05 & 1.11E$-$05 & 1.55E$-$05 &      0.9 & 1.79E+05 & 1.65E+05 & 1.94E+05 \\
  0.016 & 2.58E$-$05 & 2.22E$-$05 & 3.10E$-$05 &      1. & 2.29E+05 & 2.10E+05 & 2.48E+05 \\
  0.018 & 8.80E$-$05 & 7.56E$-$05 & 1.05E$-$04 &      1.25 & 3.72E+05 & 3.41E+05 & 4.03E+05 \\
  0.02 & 2.53E$-$04 & 2.17E$-$04 & 3.01E$-$04 &      1.5 & 5.39E+05 & 4.93E+05 & 5.84E+05 \\
  0.025 & 2.09E$-$03 & 1.80E$-$03 & 2.47E$-$03 &      1.75 & 7.25E+05 & 6.63E+05 & 7.88E+05 \\
  0.03 & 1.04E$-$02 & 8.98E$-$03 & 1.23E$-$02 &      2. & 9.31E+05 & 8.49E+05 & 1.01E+06 \\
  0.04 & 1.08E$-$01 & 9.30E$-$02 & 1.26E$-$01 &      2.5 & 1.40E+06 & 1.28E+06 & 1.53E+06 \\
  0.05 & 5.63E$-$01 & 4.88E$-$01 & 6.52E$-$01 &      3. & 1.98E+06 & 1.79E+06 & 2.17E+06 \\
  0.06 & 1.98E+00 & 1.72E+00 & 2.28E+00 &      3.5 & 2.68E+06 & 2.42E+06 & 2.94E+06 \\
  0.07 & 5.40E+00 & 4.71E+00 & 6.17E+00 &      4. & 3.51E+06 & 3.16E+06 & 3.85E+06 \\
  0.08 & 1.23E+01 & 1.08E+01 & 1.40E+01 &      5. & 5.52E+06 & 4.97E+06 & 6.08E+06 \\
  0.09 & 2.47E+01 & 2.16E+01 & 2.78E+01 &      6. & 7.86E+06 & 7.08E+06 & 8.65E+06 \\
  0.1 & 4.48E+01 & 3.94E+01 & 5.03E+01 &      7. & 1.03E+07 & 9.29E+06 & 1.13E+07 \\
  0.11 & 7.53E+01 & 6.66E+01 & 8.42E+01 &      8. & 1.27E+07 & 1.14E+07 & 1.40E+07 \\
  0.12 & 1.19E+02 & 1.06E+02 & 1.33E+02 &      9. & 1.49E+07 & 1.35E+07 & 1.64E+07 \\
  0.13 & 1.79E+02 & 1.60E+02 & 1.99E+02 &     10. & 1.70E+07 & 1.53E+07 & 1.86E+07 \\

\hline
\end{tabular*}
\begin{tabular*}{\textwidth}{@{\extracolsep{\fill}} l c }
REV  = 
$ 4.69\,{\rm exp}(-201.32/T_{9})\,[1.0+0.5\,{\rm exp}(-5.543/T_{9})]$
 & \\
\end{tabular*}
\label{li7paTab2}
\end{table}
\clearpage
\subsection{\reac{7}{Li}{\alpha}{\gamma}{11}{B}}
\label{li7agSect}
No experimental cross section data are found.
NACRE refers to the measured strengths of seven low-lying resonances \cite{PA67,HA84}. They have been supplemented by the data of \cite{GY04}.
The resonance assigned at 1.782 MeV is, however, discarded in the present work for the state was merely "inferred from the radiative capture cross section" (\cite{PA67}).

\begin{figure}[hb]
\centering{
\includegraphics[height=0.50\textheight,width=0.90\textwidth]{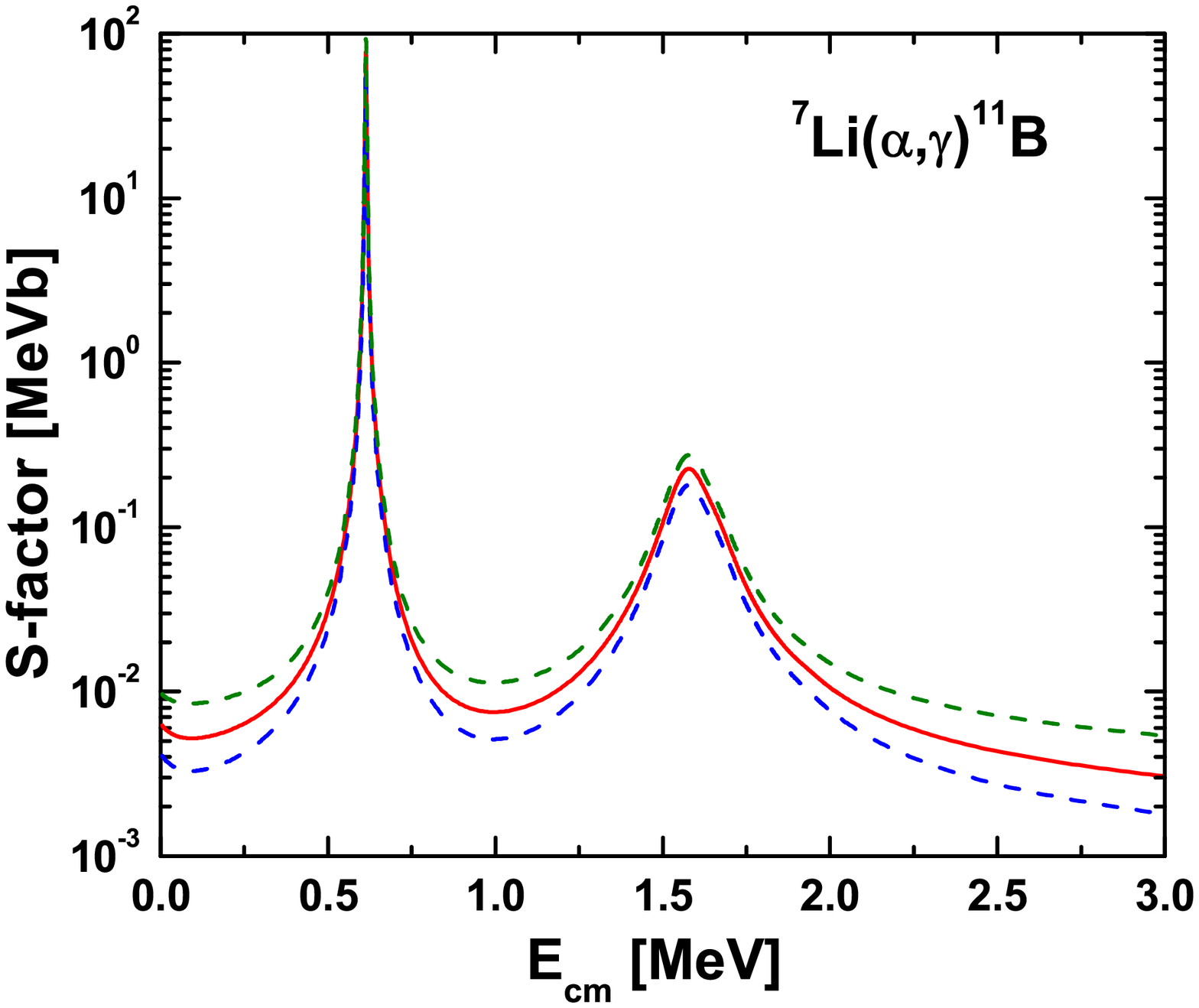}
\vspace{-0.5truecm}
\caption{The $S$-factor for \reac{7}{Li}{\alpha}{\gamma}{11}{B}.}
\label{li7agFig1}
}
\end{figure}

Figure \ref{li7agFig1} presents the PM prediction of the $S$-factors.
The $5/2^{+}, 3/2^{-},$  $5/2^{-}$ and $7/2^{+}$  resonances at $E_{\rm R} \simeq$ 0.61, 1.60, 1.67 and 1.93 MeV \cite{KE12} are considered.
The observed $\gamma$ and total widths are used to adjust the height of each resonance according to Eq.\,(\ref{eqBreitWigner0}).
The transitions to the ground and the first four excited states of $^{11}$B are taken into account inclusively.
The p-wave non-resonant ($J^{\pi} \neq 5/2^{+}$), and the 3/2$^{-}$ sub-threshold resonance (at  $E_{\rm R}= - 0.105$ MeV) contributions are additionally considered.
The adopted parameter values are given in Table \ref{li7agTab1}.
The present  $S$(0.01 MeV) = 6.0$_{-2.0}^{+3.5}$ keV\,b.

\begin{figure}[t]
\centering{
\includegraphics[height=0.33\textheight,width=0.90\textwidth]{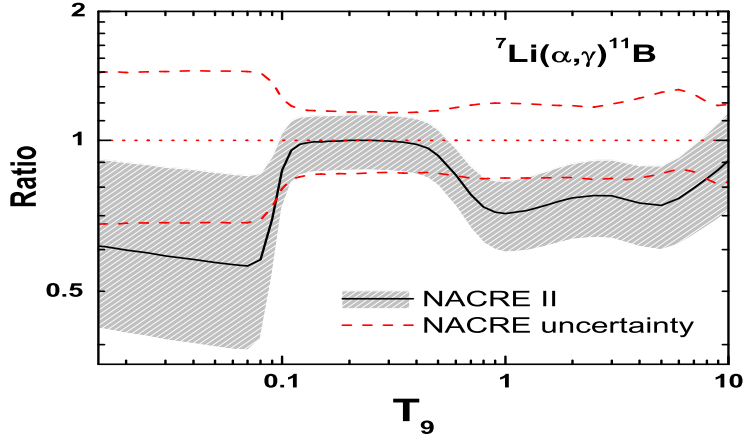}
\vspace{-0.4truecm}
\caption{\reac{7}{Li}{\alpha}{\gamma}{11}{B} rates in units of the NACRE (adopt) values. The suppressed values at the lowest energies reflect the NACRE's large contributions from the tail of a $5/2^{+}$ resonance which is much wider than accepted now \cite{KE12}.
}
\label{li7agFig2}
}
\end{figure}

  Table \ref{li7agTab2} gives the reaction rates at 0.015 $\le T_{9} \le$ 10, for which the PM-predicted cross sections are used.
The very narrow resonances at $E_{\rm R} \simeq$ 0.255 and 0.518 MeV with measured strengths  \cite{HA84,GY04} have also been taken into account. 
 Figure \ref{li7agFig2} compares the present and the NACRE rates.

{\footnotesize  \cite{DE95} for a cluster model calculation.} 

\clearpage

\begin{table}[hb]
\caption{\reac{7}{Li}{\alpha}{\gamma}{11}{B} rates in $\rm{cm^{3}mol^{-1}s^{-1}}$}\footnotesize\rm
\begin{tabular*}{\textwidth}{@{\extracolsep{\fill}}  l c c c | l c c c}
\hline
$T_{9}$ & adopted & low & high & $T_{9}$ & adopted & low & high \\
\hline
   0.015 & 1.73E$-$25 & 1.11E$-$25 & 2.76E$-$25 &    0.35 & 3.43E$-$01 & 2.93E$-$01 & 3.94E$-$01  \\
   0.016 & 8.64E$-$25 & 5.54E$-$25 & 1.38E$-$24 &    0.4 & 8.13E$-$01 & 6.93E$-$01 & 9.34E$-$01  \\
   0.018 & 1.48E$-$23 & 9.49E$-$24 & 2.37E$-$23 &    0.45 & 1.57E+00 & 1.34E+00 & 1.81E+00  \\
   0.02 & 1.71E$-$22 & 1.10E$-$22 & 2.74E$-$22 &    0.5 & 2.66E+00 & 2.26E+00 & 3.07E+00  \\
   0.025 & 2.30E$-$20 & 1.47E$-$20 & 3.69E$-$20 &    0.6 & 6.18E+00 & 5.21E+00 & 7.15E+00  \\
   0.03 & 9.61E$-$19 & 6.12E$-$19 & 1.55E$-$18 &    0.7 & 1.26E+01 & 1.05E+01 & 1.47E+01  \\
   0.04 & 2.20E$-$16 & 1.40E$-$16 & 3.57E$-$16 &    0.8 & 2.40E+01 & 1.98E+01 & 2.83E+01  \\
   0.05 & 1.05E$-$14 & 6.65E$-$15 & 1.70E$-$14 &    0.9 & 4.27E+01 & 3.49E+01 & 5.05E+01  \\
   0.06 & 1.99E$-$13 & 1.26E$-$13 & 3.23E$-$13 &    1. & 7.02E+01 & 5.70E+01 & 8.34E+01  \\
   0.07 & 2.09E$-$12 & 1.33E$-$12 & 3.40E$-$12 &    1.25 & 1.80E+02 & 1.45E+02 & 2.15E+02  \\
   0.08 & 1.57E$-$11 & 1.03E$-$11 & 2.50E$-$11 &    1.5 & 3.37E+02 & 2.70E+02 & 4.04E+02  \\
   0.09 & 1.39E$-$10 & 1.03E$-$10 & 1.95E$-$10 &    1.75 & 5.20E+02 & 4.16E+02 & 6.24E+02  \\
   0.1 & 1.79E$-$09 & 1.46E$-$09 & 2.20E$-$09 &    2. & 7.13E+02 & 5.70E+02 & 8.58E+02  \\
   0.11 & 2.00E$-$08 & 1.68E$-$08 & 2.34E$-$08 &    2.5 & 1.10E+03 & 8.78E+02 & 1.33E+03  \\
   0.12 & 1.60E$-$07 & 1.35E$-$07 & 1.85E$-$07 &    3. & 1.48E+03 & 1.18E+03 & 1.80E+03  \\
   0.13 & 9.33E$-$07 & 7.94E$-$07 & 1.07E$-$06 &    3.5 & 1.87E+03 & 1.48E+03 & 2.27E+03  \\
   0.14 & 4.23E$-$06 & 3.60E$-$06 & 4.86E$-$06 &    4. & 2.25E+03 & 1.78E+03 & 2.75E+03  \\
   0.15 & 1.56E$-$05 & 1.33E$-$05 & 1.79E$-$05 &    5. & 2.97E+03 & 2.34E+03 & 3.67E+03  \\
   0.16 & 4.84E$-$05 & 4.13E$-$05 & 5.56E$-$05 &    6. & 3.59E+03 & 2.81E+03 & 4.48E+03  \\
   0.18 & 3.17E$-$04 & 2.70E$-$04 & 3.64E$-$04 &    7. & 4.10E+03 & 3.19E+03 & 5.17E+03  \\
   0.2 & 1.40E$-$03 & 1.19E$-$03 & 1.61E$-$03 &    8. & 4.50E+03 & 3.47E+03 & 5.73E+03  \\
   0.25 & 1.93E$-$02 & 1.64E$-$02 & 2.22E$-$02 &    9. & 4.79E+03 & 3.68E+03 & 6.17E+03  \\
   0.3 & 1.06E$-$01 & 9.00E$-$02 & 1.21E$-$01 &   10. & 5.00E+03 & 3.81E+03 & 6.50E+03  \\
\hline
\end{tabular*}
\begin{tabular*}{\textwidth}{@{\extracolsep{\fill}} l c }
REV  = 
$ 4.02 \times 10^{10}T_{9}^{3/2}{\rm exp}(-100.56/T_{9})\,[1.0+0.5\,{\rm exp}(-5.543/T_{9})] $\\
\hspace{2truecm}
$/ \,[1.0+0.5\,{\rm exp}(-24.657/T_{9})]$
 & \\
\end{tabular*}
\label{li7agTab2}
\end{table}
\clearpage
\subsection{\reac{7}{Be}{p}{\gamma}{8}{B}}
\label{be7pgSect}
The experimental data sets referred to in NACRE are KA60 \cite{KA60}, PA66 \cite{PA66}, KA69 \cite{KA69}, VA70a \cite{VA70a}, FI83a \cite{FI83a}, FI83b \cite{FI83b} and HA98 \cite{HA98}, covering the 0.12 $\lsimeq E_{\rm cm} \lsimeq$ 9 MeV range. \cite{WI77} was omitted.
Added are the post-NACRE data sets HA99 \cite{HA99}, HA01 \cite{HA01}, ST01 \cite{ST01}, BA03 \cite{BA03}, JU03 \cite{JU03}, SC06 \cite{SC06}$^\dag$ and JU10 \cite{JU10}, the second extending the range down to  $E_{\rm cm} \simeq$ 0.11 MeV.
[{\footnotesize{$^\dag$from $^{8}$B Coulomb break-up.
The indirect Coulomb break-up experiments prior to SC06, starting from \cite{MO94x}, are not considered here (see the review \cite{TY08} for references).}}]

\begin{figure}[hb]
\centering{
\hspace{0.8truein}
\includegraphics[height=0.50\textheight,width=0.90\textwidth]{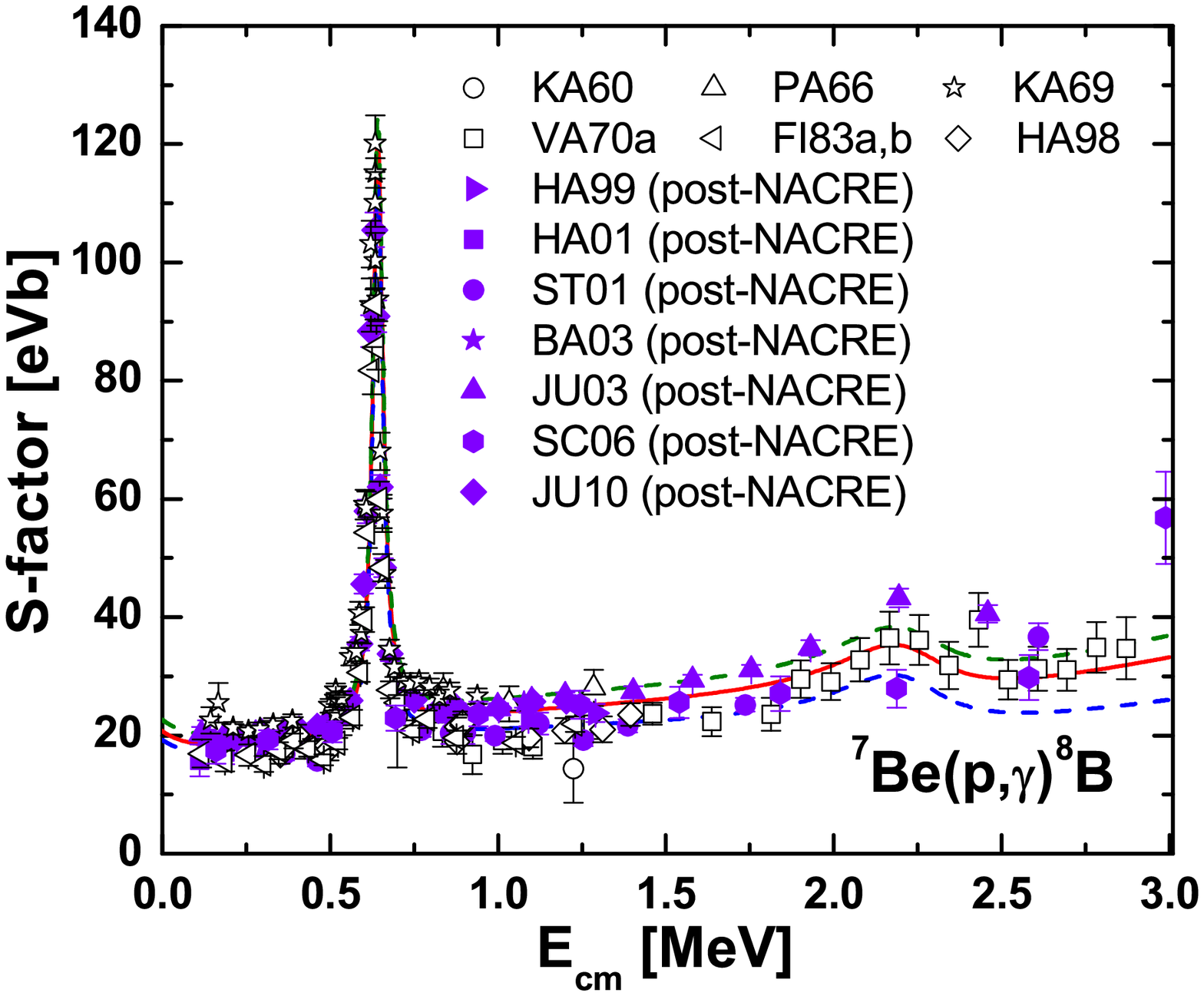}
\vspace{-0.5truecm}
\caption{The $S$-factor for \reac{7}{Be}{p}{\gamma}{8}{B}.}
\label{be7pgFig1a}
}
\end{figure}

\begin{figure}[t]
\centering{
\hspace{0.8truein}
\includegraphics[height=0.450\textheight,width=0.90\textwidth]{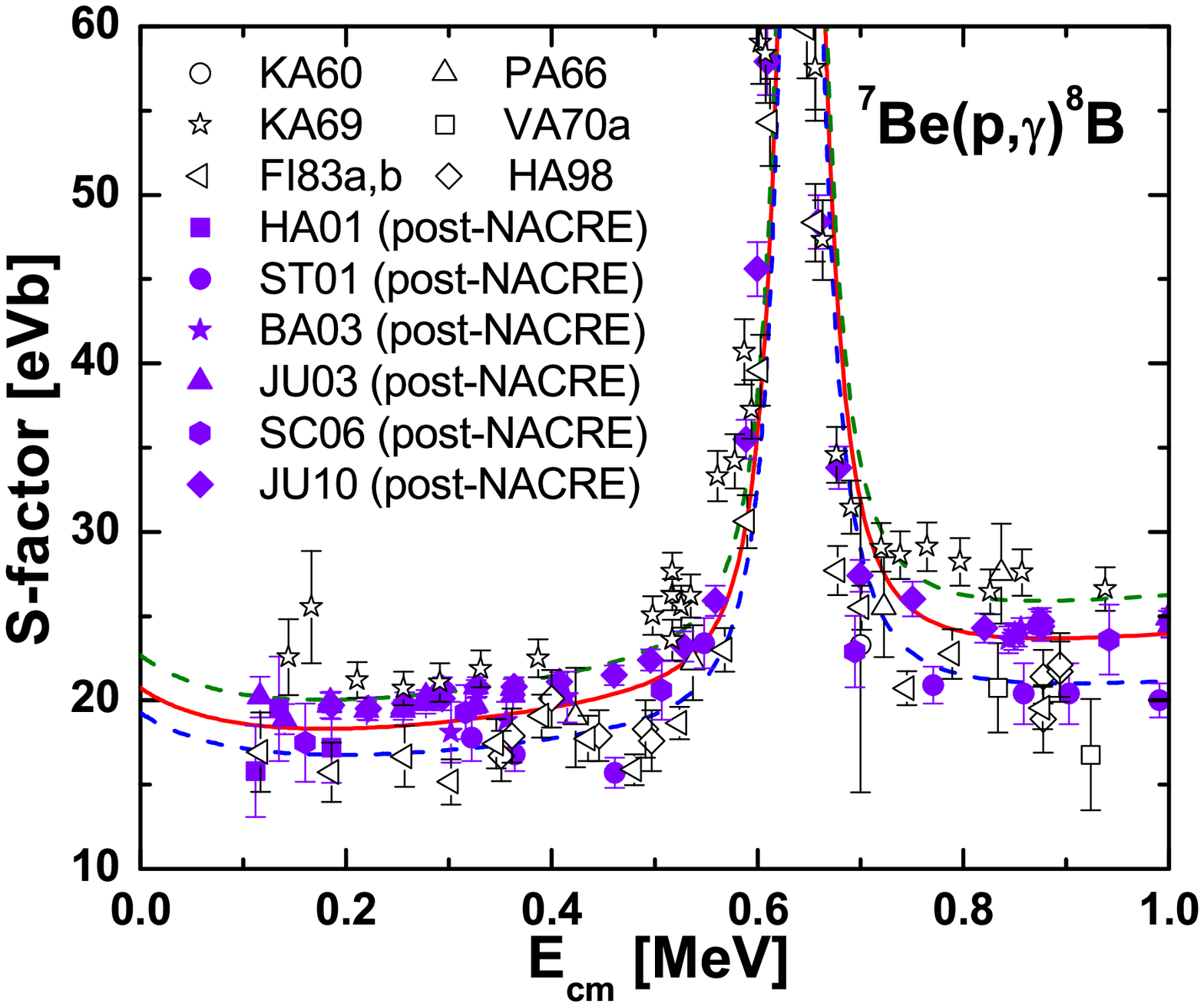}
\vspace{-0.5truecm}
\caption{The $S$-factor for \reac{7}{Be}{p}{\gamma}{8}{B} in the lowest energy range.}
\label{be7pgFig1b}
}
\end{figure}

\begin{figure}[t]
\centering{
\hspace{0.8truein}
\includegraphics[height=0.450\textheight,width=0.90\textwidth]{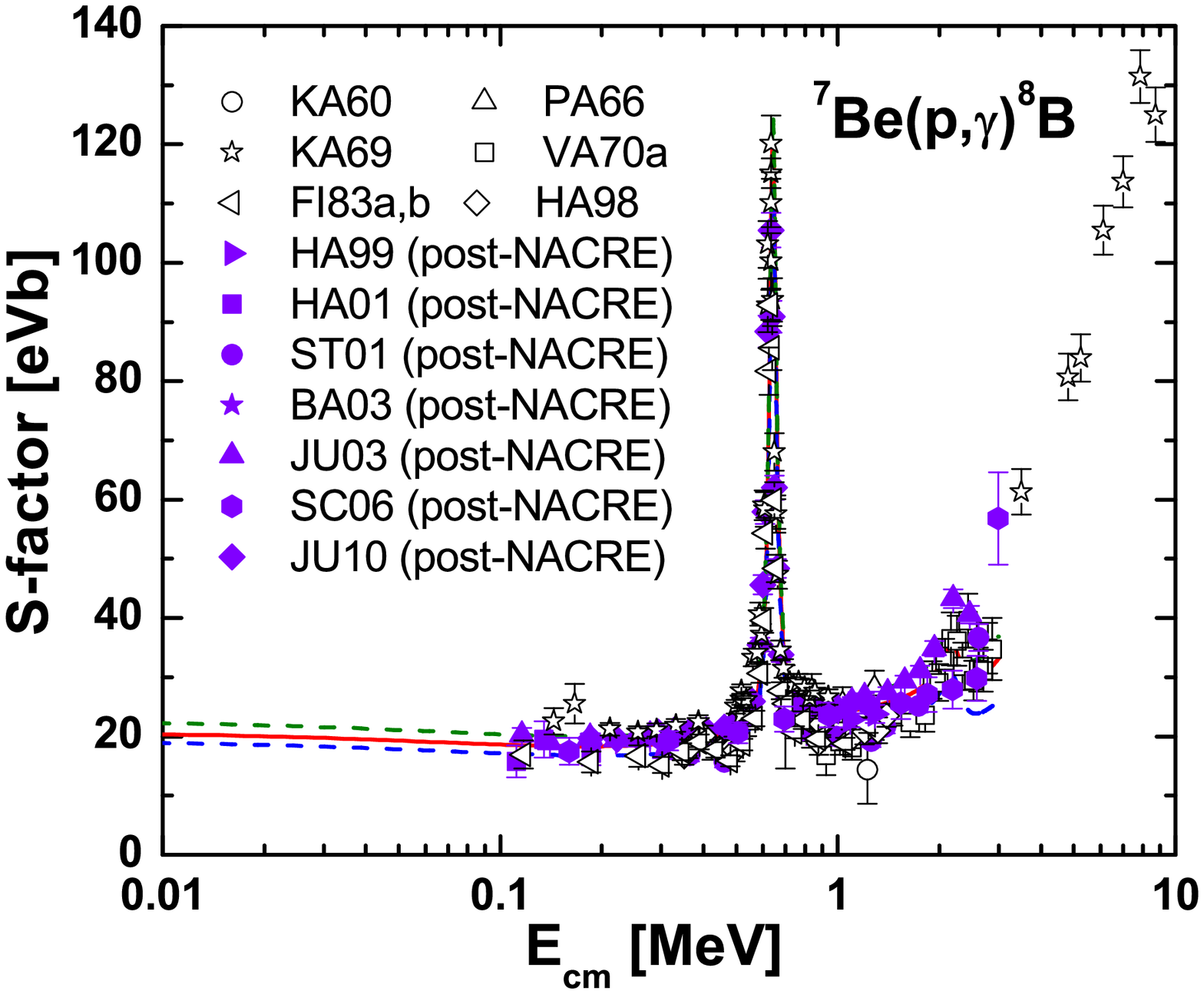}
\vspace{-0.5truecm}
\caption{The $S$-factor for \reac{7}{Be}{p}{\gamma}{8}{B} in the whole $E_{\rm cm}$ range  (logarithmic scale).}
\label{be7pgFig1c}
}
\end{figure}

Figure \ref{be7pgFig1a} compares the PM and experimental $S$-factors. It is supplemented by Figs.\,\ref{be7pgFig1b} and \ref{be7pgFig1c}.
The data in the $E_{\rm cm} \lsimeq$ 3 MeV range are used for the PM fit. They exhibit the $1^{+}$ and $3^{+}$ resonances at $E_{\rm R} \simeq$  0.64  and 2.18 MeV.
The adopted parameter values are given in Table \ref{be7pgTab1}.
The present $S(0)$ = $20.8^{+1.9}_{-1.5}$ eV\,b.
In comparison,  $S(0)$ = 21 $\pm$ 2 eV\,b [NACRE, $E$-dependence of \cite{DE94}], 20.8 $\pm$ 2.1 eV\,b [AD11, direct measurements only; $E$-dependence of \cite{DE04}], and 19.4 eV\,b [RAD10], whereas an indirect derivation by the ANC method leads to $S(0)$ = 18.0 $\pm$ 1.9 eV\,b \cite{TA06}.

\begin{figure}[t]
\centering{
\includegraphics[height=0.33\textheight,width=0.90\textwidth]{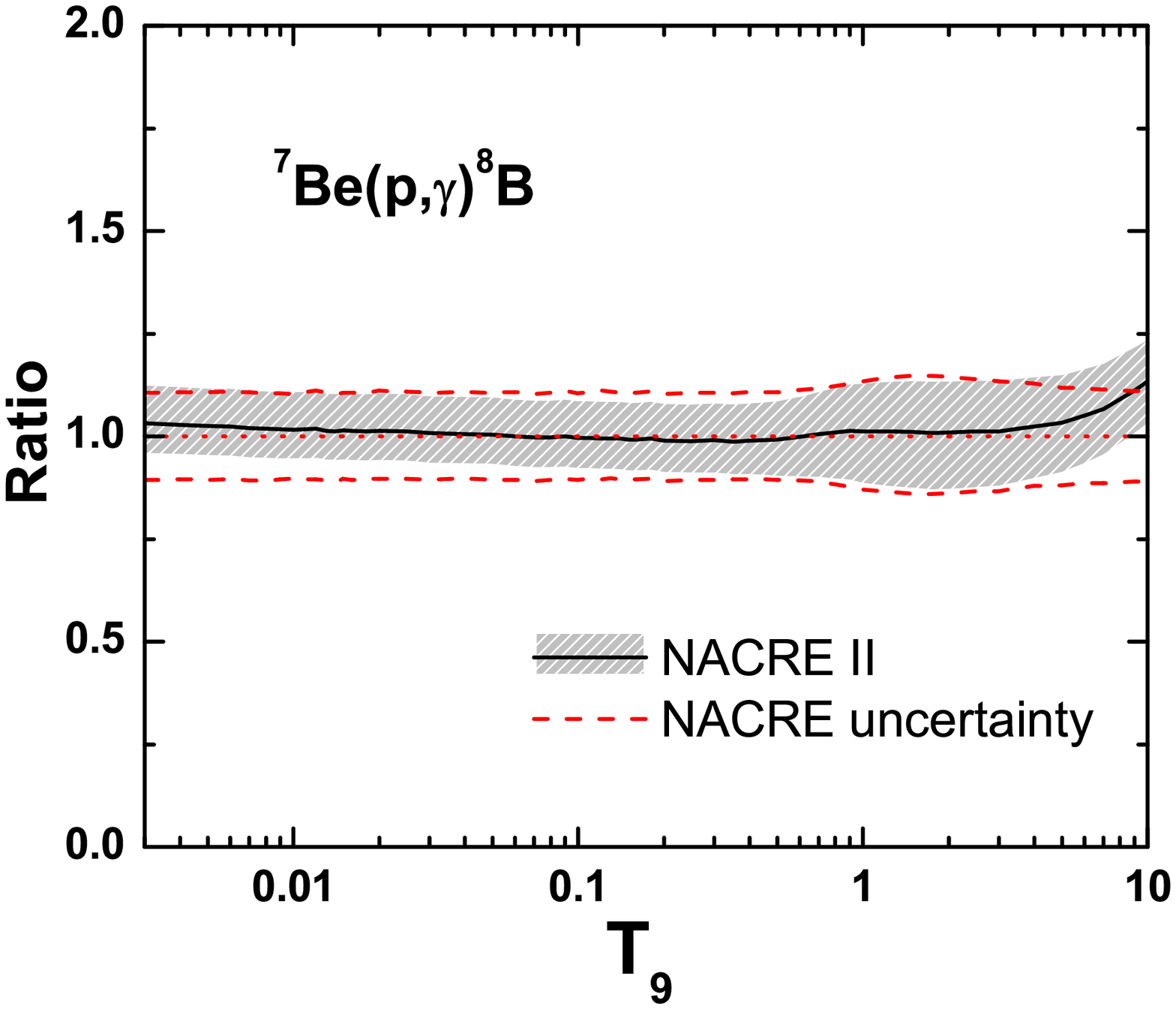}
\vspace{-0.4truecm}
\caption{\reac{7}{Be}{p}{\gamma}{8}{B} rates in units of the NACRE (adopt) values.
}
\label{be7pgFig2}
}
\end{figure}

Table \ref{be7pgTab2} gives the reaction rates at 0.003 $\le T_{9} \le$ 10, for which the PM-predicted and the experimental cross sections below and above $E_{\rm cm} \simeq$ 0.3 MeV are used, respectively.
Figure \ref{be7pgFig2} compares the present and the NACRE rates.

{\footnotesize  See  \cite{NA11} for an ${\it{ab\ initio}}$ calculation.}

\clearpage

\clearpage

\begin{table}[hb]
\caption{\reac{7}{Be}{p}{\gamma}{8}{B} rates in $\rm{cm^{3}mol^{-1}s^{-1}}$}\footnotesize\rm
\begin{tabular*}{\textwidth}{@{\extracolsep{\fill}}  l c c c | l c c c}
\hline
$T_{9}$ & adopted & low & high & $T_{9}$ & adopted & low & high \\
\hline
   0.003 &   1.65E$-$24 &   1.53E$-$24 &   1.80E$-$24 &    0.15  &   3.61E$-$03 &   3.33E$-$03 &   3.95E$-$03 \\ 
   0.004 &   9.08E$-$22 &   8.42E$-$22 &   9.92E$-$22 &    0.16  &   5.21E$-$03 &   4.80E$-$03 &   5.70E$-$03 \\  
   0.005 &   8.02E$-$20 &   7.44E$-$20 &   8.77E$-$20 &    0.18  &   9.94E$-$03 &   9.15E$-$03 &   1.09E$-$02 \\ 
   0.006 &   2.44E$-$18 &   2.26E$-$18 &   2.66E$-$18 &    0.2 &   1.73E$-$02 &   1.59E$-$02 &   1.89E$-$02 \\ 
   0.007 &   3.71E$-$17 &   3.44E$-$17 &   4.06E$-$17 &    0.25  &   5.23E$-$02 &   4.81E$-$02 &   5.72E$-$02 \\ 
   0.008 &   3.50E$-$16 &   3.25E$-$16 &   3.83E$-$16 &    0.3 &   1.21E$-$01 &   1.11E$-$01 &   1.32E$-$01 \\ 
   0.009 &   2.33E$-$15 &   2.16E$-$15 &   2.55E$-$15 &    0.35  &   2.35E$-$01 &   2.15E$-$01 &   2.57E$-$01 \\ 
   0.01  &   1.19E$-$14 &   1.10E$-$14 &   1.30E$-$14 &    0.4 &   4.06E$-$01 &   3.71E$-$01 &   4.44E$-$01 \\ 
   0.011 &   4.94E$-$14 &   4.58E$-$14 &   5.40E$-$14 &    0.45  &   6.44E$-$01 &   5.87E$-$01 &   7.06E$-$01 \\ 
   0.012 &   1.74E$-$13 &   1.61E$-$13 &   1.90E$-$13 &    0.5 &   9.58E$-$01 &   8.71E$-$01 &   1.05E+00 \\ 
   0.013 &   5.36E$-$13 &   4.97E$-$13 &   5.86E$-$13 &    0.6 &   1.86E+00 &   1.67E+00 &   2.05E+00 \\ 
   0.014 &   1.48E$-$12 &   1.37E$-$12 &   1.62E$-$12 &    0.7 &   3.18E+00 &   2.85E+00 &   3.52E+00 \\ 
   0.015 &   3.71E$-$12 &   3.44E$-$12 &   4.06E$-$12 &    0.8 &   5.03E+00 &   4.45E+00 &   5.58E+00 \\ 
   0.016 &   8.61E$-$12 &   7.98E$-$12 &   9.41E$-$12 &    0.9 &   7.46E+00 &   6.55E+00 &   8.31E+00 \\ 
   0.018 &   3.81E$-$11 &   3.53E$-$11 &   4.16E$-$11 &    1.     &   1.05E+01 &   9.19E+00 &   1.18E+01 \\ 
   0.02  &   1.37E$-$10 &   1.27E$-$10 &   1.49E$-$10 &    1.25  &   2.10E+01 &   1.81E+01 &   2.35E+01 \\ 
   0.025 &   1.76E$-$09 &   1.63E$-$09 &   1.92E$-$09 &    1.5  &   3.49E+01 &   3.01E+01 &   3.93E+01 \\ 
   0.03  &   1.23E$-$08 &   1.14E$-$08 &   1.34E$-$08 &    1.75  &   5.14E+01 &   4.42E+01 &   5.79E+01 \\ 
   0.04  &   2.06E$-$07 &   1.91E$-$07 &   2.25E$-$07 &    2.     &   6.97E+01 &   6.00E+01 &   7.86E+01 \\ 
   0.05  &   1.52E$-$06 &   1.40E$-$06 &   1.66E$-$06 &    2.5  &   1.09E+02 &   9.43E+01 &   1.23E+02 \\ 
   0.06  &   6.90E$-$06 &   6.38E$-$06 &   7.55E$-$06 &    3.     &   1.51E+02 &   1.31E+02 &   1.70E+02 \\ 
   0.07  &   2.30E$-$05 &   2.13E$-$05 &   2.52E$-$05 &    3.5  &   1.93E+02 &   1.68E+02 &   2.17E+02 \\ 
   0.08  &   6.20E$-$05 &   5.73E$-$05 &   6.78E$-$05 &    4.     &   2.37E+02 &   2.08E+02 &   2.66E+02 \\ 
   0.09  &   1.43E$-$04 &   1.32E$-$04 &   1.56E$-$04 &    5.     &   3.31E+02 &   2.91E+02 &   3.69E+02 \\ 
   0.1 &   2.93E$-$04 &   2.70E$-$04 &   3.20E$-$04 &    6.     &   4.33E+02 &   3.84E+02 &   4.81E+02 \\ 
   0.11  &   5.47E$-$04 &   5.05E$-$04 &   5.98E$-$04 &    7.     &   5.45E+02 &   4.87E+02 &   6.03E+02 \\ 
   0.12  &   9.49E$-$04 &   8.76E$-$04 &   1.04E$-$03 &    8.     &   6.68E+02 &   6.00E+02 &   7.36E+02 \\ 
   0.13  &   1.55E$-$03 &   1.43E$-$03 &   1.70E$-$03 &    9.     &   8.02E+02 &   7.24E+02 &   8.79E+02 \\ 
   0.14  &   2.42E$-$03 &   2.23E$-$03 &   2.64E$-$03 &   10.     &   9.43E+02 &   8.56E+02 &   1.03E+03 \\ 
\hline
\end{tabular*}
\begin{tabular*}{\textwidth}{@{\extracolsep{\fill}} l c }
REV  = 
 $1.31 \times 10^{10}T_{9}^{3/2}{\rm exp}(-1.596/T_{9})\,[1.0+0.5\,{\rm exp}(-4.979/T_{9})] $\\
\hspace{2truecm}
$/\,[1.0+0.6\,{\rm exp}(-8.982/T_{9})+1.4\,{\rm exp}(-26.924/T_{9})]$
 & \\
\end{tabular*}
\label{be7pgTab2}
\end{table}
\clearpage
\subsection{\reac{7}{Be}{\alpha}{\gamma}{11}{C}}
\label{be7agSect}
No experimental cross section data are found.
NACRE refers to the measured strengths for the two lowest-lying resonances \cite{HA84}.

\begin{figure}[hb]
\includegraphics[height=0.50\textheight,width=0.90\textwidth]{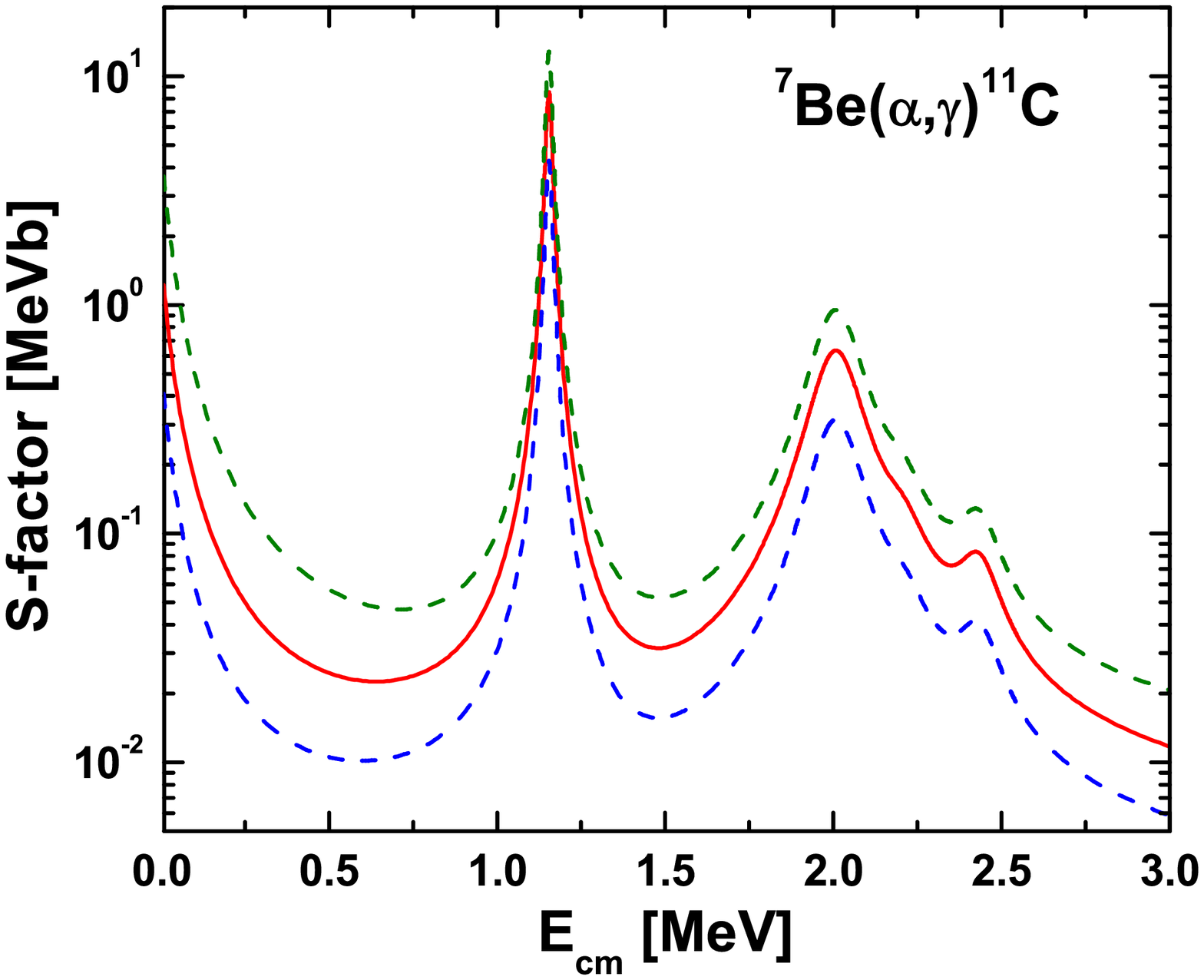}
\vspace{-0.5truecm}
\caption{The $S$-factor for \reac{7}{Be}{\alpha}{\gamma}{11}{C}.}
\label{be7agFig1}
\end{figure}

Figure \ref{be7agFig1} presents the PM $S$-factors.
The $5/2^{+}, 3/2^{-}, 5/2^{-}$ and $7/2^{+}$ resonances at $E_{\rm R} \simeq$  1.16, 2.11, 2.24 and 2.54 MeV \cite{KE12} have been considered.
The unknown $\gamma$-widths are obtained by scaling in transition energies from those known in $^{11}$B, but by allowing for large uncertainties. Along with the known total widths, they set the heights of the resonances through Eq.\,(\ref{eqBreitWigner0}).
The $7/2^{-}$ resonance at  $E_{\rm R} \simeq$ 2.43 MeV, which does not have its counterpart in $^{11}$B, is included in the evaluation of the higher limit with the assumed $\gamma$-width of 1 eV. 
The transitions to the ground and the first three excited states of $^{11}$C are considered inclusively.
Non-resonant contributions are calculated with the same potential parameter values as those for the mirror \reac{7}{Li}{\alpha}{\gamma}{11}{B} 
reaction. 
Significant contributions from the $3/2^{+}$ sub-threshold state at $E_{\rm R} \simeq  -0.044$ MeV are added (cf. \cite{DE95}).
The adopted parameter values are given in Table \ref{be7agTab1}.
The present  $S$(0.01 MeV) = 1.2$_{-0.8}^{+2.3}$ MeV\,b.

\begin{figure}[t]
\includegraphics[height=0.33\textheight,width=0.90\textwidth]{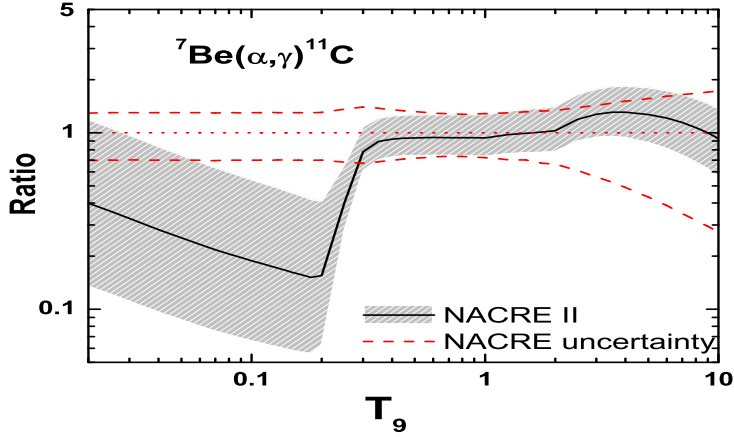}
\vspace{-0.4truecm}
\caption{\reac{7}{Be}{\alpha}{\gamma}{11}{C} rates in units of the NACRE (adopt) values. The reduction of the ratios at low temperatures reflect the NACRE apparent reliance on a cluster-model calculation \cite{DE95} that predicts a much wider $5/2^{+}$ resonance than observed.
}
\label{be7agFig2}
\end{figure}

Table \ref{be7agTab2} gives the reaction rates at 0.02 $\le T_{9} \le$ 10. Figure \ref{be7agFig2} compares the present and the NACRE rates.
The very narrow resonances at $E_{\rm R} \simeq $  0.560 and 0.877 MeV with measured strengths \cite{HA84} are also taken into account. 

{\footnotesize See \cite{DE95} for a cluster model calculation; \cite{BU88} for a potential model analysis.}

\clearpage

\begin{table}[hb]
\caption{\reac{7}{Be}{\alpha}{\gamma}{11}{C} rates in $\rm{cm^{3}mol^{-1}s^{-1}}$}\footnotesize\rm
\begin{tabular*}{\textwidth}{@{\extracolsep{\fill}}  l c c c | l c c c}
\hline
$T_{9}$ & adopted & low & high & $T_{9}$ & adopted & low & high \\
\hline
   0.02 & 4.63E$-$27 & 1.57E$-$27 & 1.38E$-$26 &    0.45 & 2.17E$-$02 & 1.72E$-$02 & 2.92E$-$02  \\
   0.025 & 1.56E$-$24 & 5.30E$-$25 & 4.63E$-$24 &    0.5 & 7.87E$-$02 & 6.25E$-$02 & 1.06E$-$01  \\
   0.03 & 1.30E$-$22 & 4.42E$-$23 & 3.85E$-$22 &    0.6 & 5.31E$-$01 & 4.22E$-$01 & 7.15E$-$01  \\
   0.04 & 7.90E$-$20 & 2.70E$-$20 & 2.33E$-$19 &    0.7 & 2.05E+00 & 1.62E+00 & 2.76E+00  \\
   0.05 & 7.33E$-$18 & 2.52E$-$18 & 2.15E$-$17 &    0.8 & 5.63E+00 & 4.46E+00 & 7.59E+00  \\
   0.06 & 2.28E$-$16 & 7.86E$-$17 & 6.66E$-$16 &    0.9 & 1.25E+01 & 9.85E+00 & 1.68E+01  \\
   0.07 & 3.49E$-$15 & 1.21E$-$15 & 1.02E$-$14 &    1. & 2.37E+01 & 1.87E+01 & 3.21E+01  \\
   0.08 & 3.29E$-$14 & 1.15E$-$14 & 9.55E$-$14 &    1.25 & 7.81E+01 & 6.11E+01 & 1.06E+02  \\
   0.09 & 2.18E$-$13 & 7.63E$-$14 & 6.29E$-$13 &    1.5 & 1.77E+02 & 1.37E+02 & 2.41E+02  \\
   0.1 & 1.10E$-$12 & 3.88E$-$13 & 3.17E$-$12 &    1.75 & 3.20E+02 & 2.46E+02 & 4.38E+02  \\
   0.11 & 4.53E$-$12 & 1.60E$-$12 & 1.30E$-$11 &    2. & 4.97E+02 & 3.80E+02 & 6.82E+02  \\
   0.12 & 1.58E$-$11 & 5.61E$-$12 & 4.50E$-$11 &    2.5 & 9.09E+02 & 6.85E+02 & 1.25E+03  \\
   0.13 & 4.80E$-$11 & 1.71E$-$11 & 1.36E$-$10 &    3. & 1.34E+03 & 9.92E+02 & 1.85E+03  \\
   0.14 & 1.31E$-$10 & 4.69E$-$11 & 3.69E$-$10 &    3.5 & 1.74E+03 & 1.27E+03 & 2.43E+03  \\
   0.15 & 3.24E$-$10 & 1.17E$-$10 & 9.11E$-$10 &    4. & 2.11E+03 & 1.51E+03 & 2.95E+03  \\
   0.16 & 7.41E$-$10 & 2.69E$-$10 & 2.08E$-$09 &    5. & 2.73E+03 & 1.90E+03 & 3.87E+03  \\
   0.18 & 3.23E$-$09 & 1.20E$-$09 & 8.93E$-$09 &    6. & 3.22E+03 & 2.17E+03 & 4.61E+03  \\
   0.2 & 1.23E$-$08 & 4.98E$-$09 & 3.23E$-$08 &    7. & 3.60E+03 & 2.35E+03 & 5.20E+03  \\
   0.25 & 6.37E$-$07 & 4.49E$-$07 & 1.04E$-$06 &    8. & 3.89E+03 & 2.48E+03 & 5.66E+03  \\
   0.3 & 2.99E$-$05 & 2.34E$-$05 & 4.13E$-$05 &    9. & 4.10E+03 & 2.56E+03 & 6.01E+03  \\
   0.35 & 5.11E$-$04 & 4.05E$-$04 & 6.93E$-$04 &   10. & 4.24E+03 & 2.61E+03 & 6.26E+03  \\
   0.4 & 4.25E$-$03 & 3.37E$-$03 & 5.73E$-$03 &      \\
\hline
\end{tabular*}
\begin{tabular*}{\textwidth}{@{\extracolsep{\fill}} l c }
REV  = 
$4.02 \times 10^{10}T_{9}^{3/2}{\rm exp}(-87.555/T_{9})\,[1.0+0.5\,{\rm exp}(-4.979/T_{9})]$ \\
\hspace{2truecm}
$ /\,[1.0+0.5\,{\rm exp}(-23.210/T_{9})]$
 & \\
\end{tabular*}
\label{be7agTab2}
\end{table}
\clearpage
\subsection{\reac{9}{Be}{p}{\gamma}{10}{B}}
\label{be9pgSect}
The experimental data sets referred to in NACRE are ME59 \cite{ME59}, HO64 \cite{HO64}, AU75 \cite{AU75}, PA89 \cite{PA89} and ZA95b \cite{ZA95b}, covering the 0.07 $\lsimeq E_{\rm cm} \lsimeq$ 1.6 MeV range.
Among them, ME59 was omitted, and HO64, AU75 and PA89 were superseded by ZA95b. In the present work, all sets but ME59 are considered.
No new cross section data are found.

\begin{figure}[hb]
\centering{
\includegraphics[height=0.50\textheight,width=0.90\textwidth]{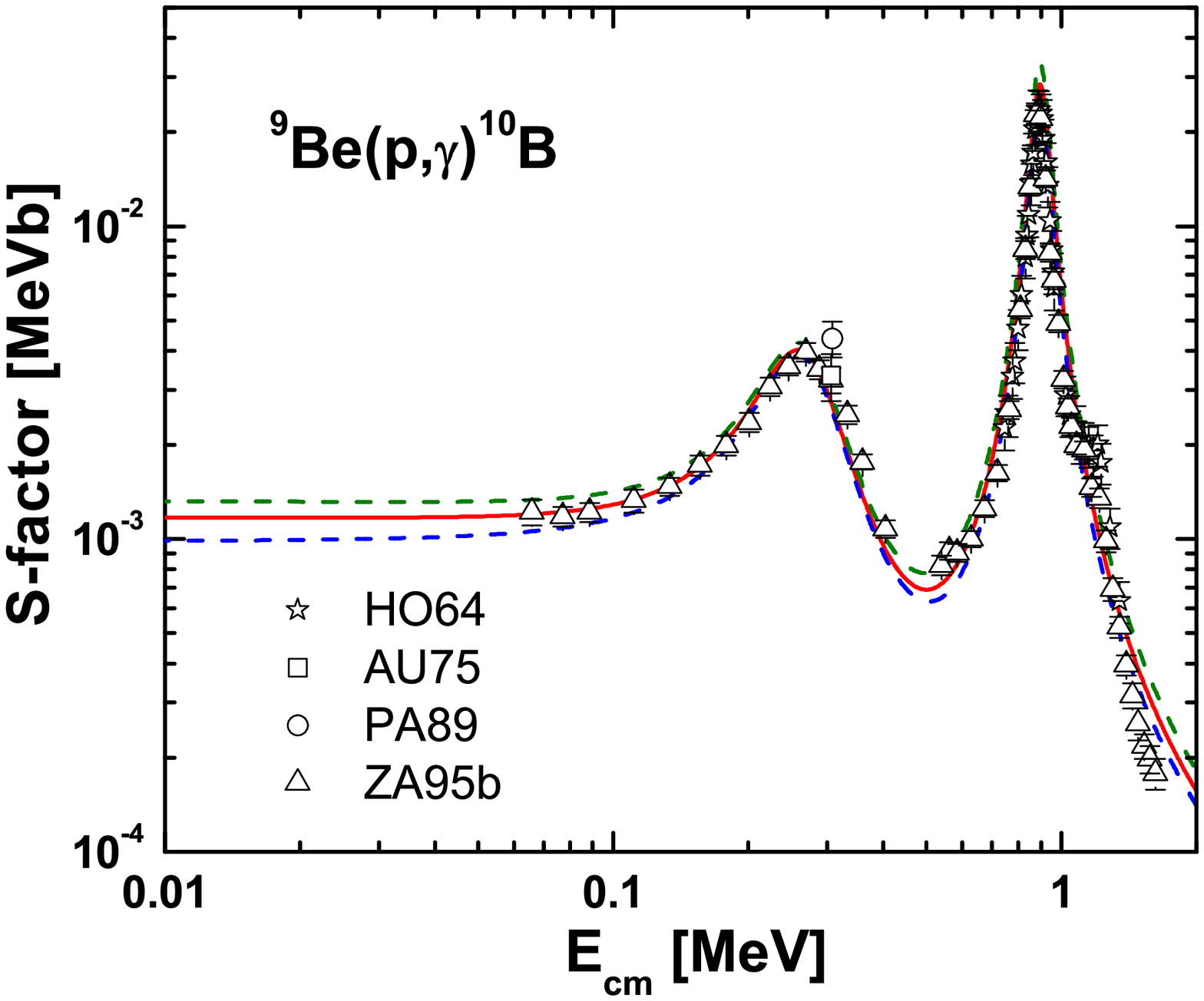}
\vspace{-0.5truecm}
\caption{The $S$-factor for \reac{9}{Be}{p}{\gamma}{10}{B}.}
\label{be9pgFig1}
}
\end{figure}

Figure \ref{be9pgFig1} compares the PM and experimental $S$-factors.
The data in the $E_{\rm cm} \lsimeq$ 1 MeV range are used for the PM fit. They exhibit the $1^{-}$ and $2^{+}$ resonances at $E_{\rm R} \simeq$  0.29 and ß0.89 MeV.
The transitions to the ground and the first three excited states of $^{10}$B are considered inclusively.
The adopted parameter values are given in Table \ref{be9pgTab1}.
The present $S(0)$ = 1.2$_{-0.2}^{+0.1}$  keV\,b.
In comparison, $S(0)$ =   1 keV\,b [NACRE from \cite{ZA95b}], and 1.05 keV\,b [RAD10].

\begin{figure}[t]
\centering{
\includegraphics[height=0.33\textheight,width=0.90\textwidth]{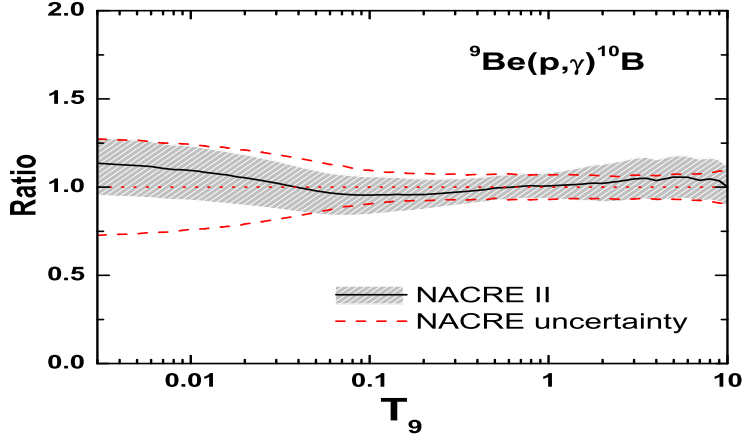}
\vspace{-0.4truecm}
\caption{\reac{9}{Be}{p}{\gamma}{10}{B} rates in units of the NACRE (adopt) values. The use of PM limits the uncertainties when compared with NACRE.  
}
\label{be9pgFig2}
}
\end{figure}

Table \ref{be9pgTab2} gives the reaction rates at 0.003 $\le T_{9} \le $ 10, for which the PM-predicted and the experimental cross sections below and above $E_{\rm cm} \simeq$ 0.1 MeV are used, respectively.
The narrow 0$^{+}$ resonance at $E_{\rm R} =$  0.974 MeV (with known $\omega\gamma$ \cite{ZA95b}) is additionally considered, contributing to the rates marginally (up to 4 \% at $T_{9} > 3$).
In the $E_{\rm cm} \gsimeq$ 1.62 MeV range, the $S$-factors are taken to be constant for simplicity. 
The impact of this assumption is very weak, as the contribution of that energy range to the rates amounts to about 5 \%  only, even at $T_{9}$ = 10.
Figure \ref{be9pgFig2} compares the present and the NACRE rates.

{\footnotesize  See  \cite{BA02} for an R-matrix fit.}

\clearpage

\begin{table}
\caption{\reac{9}{Be}{p}{\gamma}{10}{B} rates in $\rm{cm^{3}mol^{-1}s^{-1}}$}\footnotesize\rm
\begin{tabular*}{\textwidth}{@{\extracolsep{\fill}}  l c c c | l c c c}
\hline
$T_{9}$ & adopted & low & high & $T_{9}$ & adopted & low & high \\
\hline
0.003 & 4.64E$-$23 & 4.42E$-$23 & 5.09E$-$23 &         0.15  & 2.06E$-$01 & 1.92E$-$01 & 2.25E$-$01   \\ 
0.004 & 2.73E$-$20 & 2.60E$-$20 & 3.00E$-$20 &         0.16  & 3.05E$-$01 & 2.84E$-$01 & 3.32E$-$01   \\
0.005 & 2.54E$-$18 & 2.41E$-$18 & 2.78E$-$18 &         0.18  & 6.13E$-$01 & 5.70E$-$01 & 6.65E$-$01   \\
0.006 & 8.00E$-$17 & 7.61E$-$17 & 8.78E$-$17 &         0.2 & 1.13E+00 & 1.05E+00 & 1.22E+00   \\
0.007 & 1.26E$-$15 & 1.19E$-$15 & 1.38E$-$15 &         0.25  & 3.93E+00 & 3.65E+00 & 4.23E+00   \\
0.008 & 1.21E$-$14 & 1.15E$-$14 & 1.33E$-$14 &         0.3 & 1.04E+01 & 9.65E+00 & 1.11E+01   \\
0.009 & 8.25E$-$14 & 7.84E$-$14 & 9.05E$-$14 &         0.35  & 2.25E+01 & 2.10E+01 & 2.42E+01   \\
0.01  & 4.29E$-$13 & 4.08E$-$13 & 4.71E$-$13 &         0.4 & 4.22E+01 & 3.93E+01 & 4.52E+01   \\
0.011 & 1.81E$-$12 & 1.72E$-$12 & 1.99E$-$12 &         0.45  & 7.07E+01 & 6.57E+01 & 7.57E+01   \\
0.012 & 6.47E$-$12 & 6.15E$-$12 & 7.10E$-$12 &         0.5 & 1.08E+02 & 1.01E+02 & 1.16E+02   \\
0.013 & 2.02E$-$11 & 1.92E$-$11 & 2.22E$-$11 &         0.6 & 2.11E+02 & 1.96E+02 & 2.26E+02   \\
0.014 & 5.64E$-$11 & 5.35E$-$11 & 6.18E$-$11 &         0.7 & 3.45E+02 & 3.21E+02 & 3.69E+02   \\
0.015 & 1.43E$-$10 & 1.36E$-$10 & 1.57E$-$10 &         0.8 & 5.08E+02 & 4.72E+02 & 5.43E+02   \\
0.016 & 3.35E$-$10 & 3.18E$-$10 & 3.68E$-$10 &         0.9 & 6.98E+02 & 6.48E+02 & 7.49E+02   \\
0.018 & 1.51E$-$09 & 1.43E$-$09 & 1.66E$-$09 &         1.     & 9.24E+02 & 8.56E+02 & 9.93E+02   \\
0.02  & 5.50E$-$09 & 5.22E$-$09 & 6.04E$-$09 &         1.25  & 1.71E+03 & 1.57E+03 & 1.85E+03   \\
0.025 & 7.32E$-$08 & 6.94E$-$08 & 8.04E$-$08 &         1.5  & 2.93E+03 & 2.65E+03 & 3.20E+03   \\
0.03  & 5.24E$-$07 & 4.96E$-$07 & 5.75E$-$07 &         1.75  & 4.62E+03 & 4.16E+03 & 5.09E+03   \\
0.04  & 9.15E$-$06 & 8.65E$-$06 & 1.00E$-$05 &         2.     & 6.73E+03 & 6.01E+03 & 7.45E+03   \\
0.05  & 6.94E$-$05 & 6.55E$-$05 & 7.62E$-$05 &         2.5  & 1.16E+04 & 1.03E+04 & 1.29E+04   \\
0.06  & 3.24E$-$04 & 3.06E$-$04 & 3.56E$-$04 &         3.     & 1.65E+04 & 1.46E+04 & 1.85E+04   \\
0.07  & 1.11E$-$03 & 1.04E$-$03 & 1.22E$-$03 &         3.5  & 2.09E+04 & 1.85E+04 & 2.34E+04   \\
0.08  & 3.05E$-$03 & 2.86E$-$03 & 3.34E$-$03 &         4.     & 2.45E+04 & 2.16E+04 & 2.74E+04   \\
0.09  & 7.16E$-$03 & 6.73E$-$03 & 7.86E$-$03 &         5.     & 2.94E+04 & 2.59E+04 & 3.29E+04   \\
0.1 & 1.50E$-$02 & 1.40E$-$02 & 1.64E$-$02 &         6.     & 3.18E+04 & 2.80E+04 & 3.56E+04   \\
0.11  & 2.85E$-$02 & 2.67E$-$02 & 3.13E$-$02 &         7.     & 3.27E+04 & 2.88E+04 & 3.66E+04   \\
0.12  & 5.06E$-$02 & 4.73E$-$02 & 5.54E$-$02 &         8.     & 3.26E+04 & 2.88E+04 & 3.65E+04   \\
0.13  & 8.47E$-$02 & 7.90E$-$02 & 9.25E$-$02 &         9.     & 3.20E+04 & 2.82E+04 & 3.58E+04   \\
0.14  & 1.35E$-$01 & 1.26E$-$01 & 1.47E$-$01 &        10.     & 3.10E+04 & 2.74E+04 & 3.47E+04   \\
\hline
\end{tabular*}
\begin{tabular*}{\textwidth}{@{\extracolsep{\fill}} l c }
REV  = 
 $9.74 \times 10^{9}T_{9}^{3/2}{\rm exp}(-76.429/T_{9}) $ \\
\hspace{1truecm}
$\times\,[1.0+0.5\,{\rm exp}(-19.543/T_{9})+1.5\,{\rm exp}(-28.193/T_{9})] $ \\
\hspace{1truecm}
$/\,[1.0+0.429\,{\rm exp}(-8.336/T_{9}) + 0.143\,{\rm exp}(-20.194/T_{9}) + 0.429\,{\rm exp}(-24.997/T_{9})]  $\\
 & \\
\end{tabular*}
\label{be9pgTab2}
\end{table}
\clearpage\subsection{\reac{9}{Be}{p}{d}{8}{Be}}
\label{be9pdSect}
The experimental data sets referred to in NACRE are NE51 \cite{NE51}, WE56 \cite{WE56}, HU72 \cite{HU72}, SI73 \cite{SI73} and ZA97 \cite{ZA97}, covering the 0.02 $\lsimeq E_{\rm cm} \lsimeq$ 10 MeV range.
No new cross section data are found. (For angular distribution in the 0.07 $\lsimeq E_{\rm cm} \lsimeq$0.29 MeV range, see \cite{BR98}.)
The enhancements of the $S$-factors below $E_{\rm cm}$ = 0.03 $\sim$ 0.04 MeV are likely caused by electron screening as it is the case with \reac{9}{Be}{p}{\alpha}{4}{He}, for which THM has been applied (see Sect.\,\ref{be9paSect}).

\begin{figure}[hb]
\centering{
\includegraphics[height=0.50\textheight,width=0.90\textwidth]{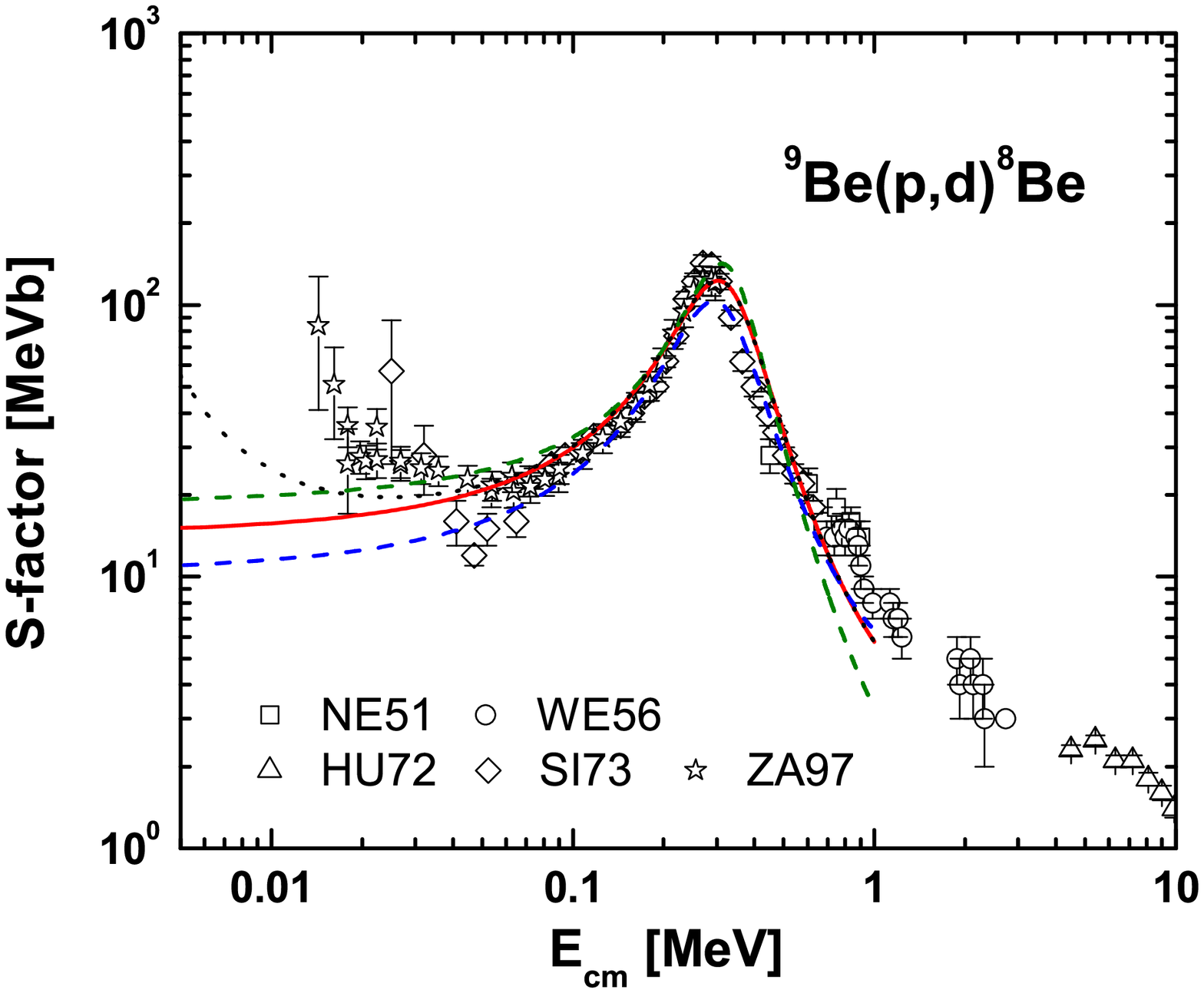}
\vspace{-0.5truecm}
\caption{The $S$-factor for \reac{9}{Be}{p}{d}{8}{Be}.  The dotted line indicates an adiabatic screening correction ($U_{\rm e}$ = 264 eV) to the 'adopt' curve (solid line).}
\label{be9pdFig1}
}
\end{figure}

Figure \ref{be9pdFig1} compares the DWBA and experimental $S$-factors.
The data in the 0.05 $\lsimeq E_{\rm cm} \lsimeq$ 1 MeV range are used for the DWBA  fit. They exhibit the $1^{-}$ resonance at $E_{\rm R} \simeq$  0.29 MeV.
The possible contribution from a $(4)^{-}$ sub-threshold state at $E_{\rm R} \simeq  - 0.026$  MeV  appears to be minor (see ZA97 and \cite{BR98}).
The adopted parameter values are given in Table \ref{be9pdTab1}.
The present $S(0)$ = 15 $\pm$ 4 MeV\,b.
In comparison,  $S(0)$ = 17 $_{-7}^{+25}$ MeV\,b [NACRE, from SI73].

\begin{figure}[t]
\centering{
\includegraphics[height=0.33\textheight,width=0.90\textwidth]{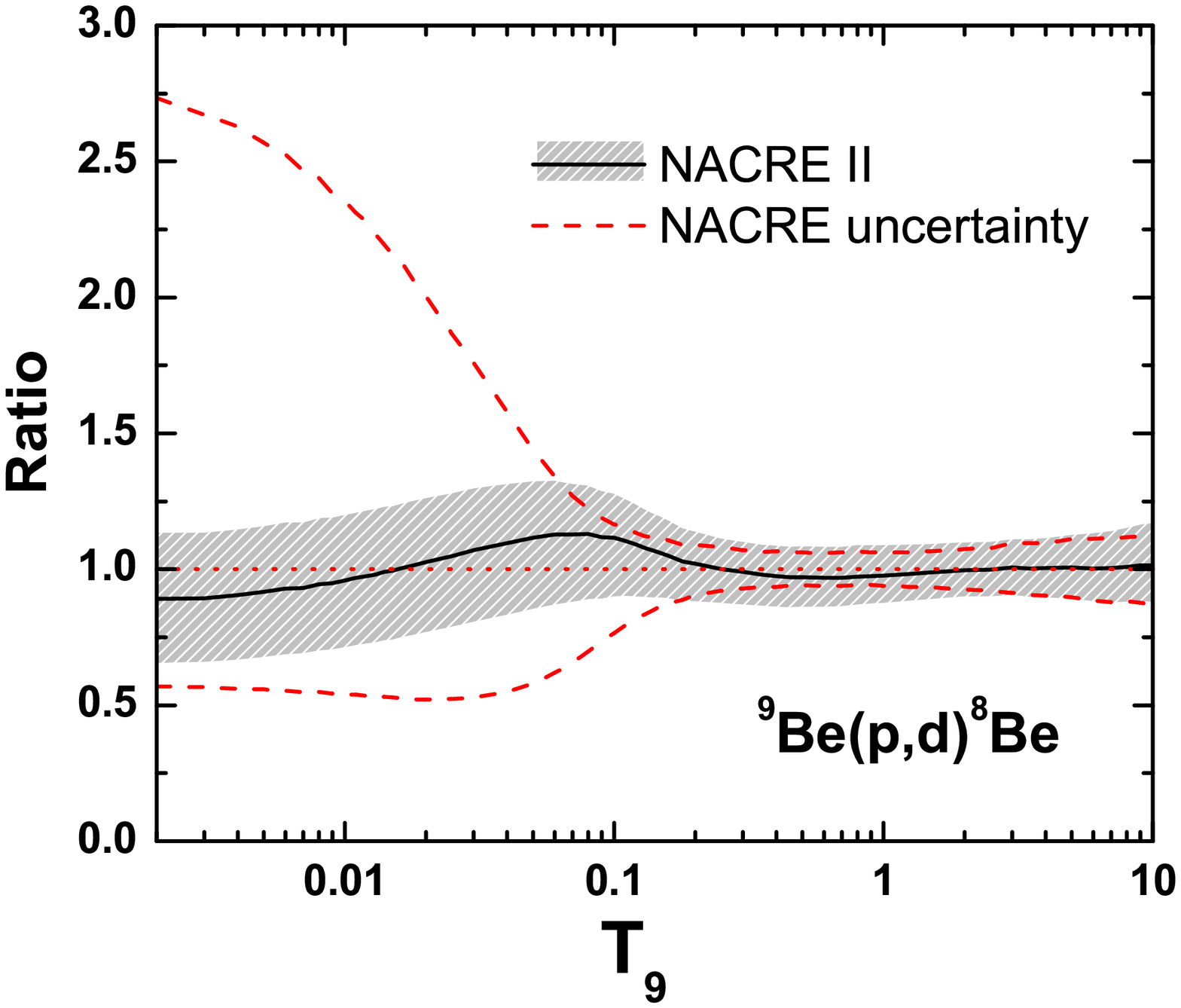}
\vspace{-0.4truecm}
\caption{\reac{9}{Be}{p}{d}{8}{Be} rates in units of the NACRE (adopt) values. The exclusion for the DWBA fit of the experimental data in the $E_{\rm cm} \lsimeq$ 0.05 MeV range (see text) lowers the upper NACRE uncertainty.
}
\label{be9pdFig2}
}
\end{figure}

Table \ref{be9pdTab2} gives the reaction rates at $0.002 \le T_{9} \le 10$, for which the DWBA-predicted and the experimental cross sections below and above $E_{\rm cm} \simeq$ 0.07 MeV are used, respectively. 
Figure \ref{be9pdFig2} compares the present and the NACRE rates.

{\footnotesize  See  \cite{RA91} for a DWBA analysis.}

\clearpage

\begin{table}[hb]
\caption{\reac{9}{Be}{p}{d}{8}{Be} rates in units of $\rm{cm^{3}mol^{-1}s^{-1}}$.} \footnotesize\rm
\begin{tabular*}{\textwidth}{@{\extracolsep{\fill}}  l c c c | l c c c}
\hline
$T_{9}$ & adopted & low & high & $T_{9}$ & adopted & low & high \\
\hline
  0.002 & 2.38E$-$23 & 1.74E$-$23 & 3.04E$-$23 &            0.15  & 4.45E+03 & 3.74E+03 & 5.03E+03 \\
  0.003 & 5.95E$-$19 & 4.36E$-$19 & 7.60E$-$19 &            0.16  & 6.67E+03 & 5.64E+03 & 7.52E+03 \\
  0.004 & 3.55E$-$16 & 2.60E$-$16 & 4.52E$-$16 &            0.18  & 1.38E+04 & 1.18E+04 & 1.55E+04 \\
  0.005 & 3.33E$-$14 & 2.44E$-$14 & 4.22E$-$14 &            0.2 & 2.60E+04 & 2.24E+04 & 2.92E+04 \\
  0.006 & 1.06E$-$12 & 7.78E$-$13 & 1.34E$-$12 &            0.25  & 9.63E+04 & 8.38E+04 & 1.07E+05 \\
  0.007 & 1.68E$-$11 & 1.23E$-$11 & 2.12E$-$11 &            0.3 & 2.68E+05 & 2.35E+05 & 3.00E+05 \\
  0.008 & 1.64E$-$10 & 1.20E$-$10 & 2.06E$-$10 &            0.35  & 6.10E+05 & 5.36E+05 & 6.82E+05 \\
  0.009 & 1.12E$-$09 & 8.26E$-$10 & 1.41E$-$09 &            0.4 & 1.19E+06 & 1.05E+06 & 1.33E+06 \\
  0.01  & 5.87E$-$09 & 4.33E$-$09 & 7.38E$-$09 &            0.45  & 2.06E+06 & 1.81E+06 & 2.31E+06 \\
  0.011 & 2.50E$-$08 & 1.85E$-$08 & 3.13E$-$08 &            0.5 & 3.24E+06 & 2.86E+06 & 3.64E+06 \\
  0.012 & 8.99E$-$08 & 6.65E$-$08 & 1.13E$-$07 &            0.6 & 6.57E+06 & 5.82E+06 & 7.39E+06 \\
  0.013 & 2.83E$-$07 & 2.09E$-$07 & 3.53E$-$07 &            0.7 & 1.10E+07 & 9.79E+06 & 1.24E+07 \\
  0.014 & 7.94E$-$07 & 5.88E$-$07 & 9.90E$-$07 &            0.8 & 1.64E+07 & 1.45E+07 & 1.84E+07 \\
  0.015 & 2.03E$-$06 & 1.50E$-$06 & 2.53E$-$06 &            0.9 & 2.22E+07 & 1.98E+07 & 2.49E+07 \\
  0.016 & 4.78E$-$06 & 3.55E$-$06 & 5.94E$-$06 &            1.     & 2.84E+07 & 2.54E+07 & 3.18E+07 \\
  0.018 & 2.18E$-$05 & 1.62E$-$05 & 2.70E$-$05 &            1.25  & 4.41E+07 & 3.95E+07 & 4.92E+07 \\
  0.02  & 8.05E$-$05 & 5.99E$-$05 & 9.94E$-$05 &            1.5  & 5.88E+07 & 5.27E+07 & 6.54E+07 \\
  0.025 & 1.10E$-$03 & 8.23E$-$04 & 1.35E$-$03 &            1.75  & 7.19E+07 & 6.46E+07 & 7.98E+07 \\
  0.03  & 8.09E$-$03 & 6.07E$-$03 & 9.86E$-$03 &            2.     & 8.34E+07 & 7.49E+07 & 9.24E+07 \\
  0.04  & 1.48E$-$01 & 1.12E$-$01 & 1.78E$-$01 &            2.5  & 1.02E+08 & 9.14E+07 & 1.13E+08 \\
  0.05  & 1.17E+00 & 8.91E$-$01 & 1.39E+00 &            3.     & 1.16E+08 & 1.04E+08 & 1.28E+08 \\
  0.06  & 5.68E+00 & 4.36E+00 & 6.70E+00 &            3.5  & 1.26E+08 & 1.13E+08 & 1.41E+08 \\
  0.07  & 2.01E+01 & 1.56E+01 & 2.35E+01 &            4.     & 1.34E+08 & 1.19E+08 & 1.50E+08 \\
  0.08  & 5.70E+01 & 4.47E+01 & 6.62E+01 &            5.     & 1.46E+08 & 1.29E+08 & 1.64E+08 \\
  0.09  & 1.38E+02 & 1.09E+02 & 1.59E+02 &            6.     & 1.54E+08 & 1.34E+08 & 1.74E+08 \\
  0.1 & 2.95E+02 & 2.37E+02 & 3.39E+02 &            7.     & 1.59E+08 & 1.38E+08 & 1.81E+08 \\
  0.11  & 5.74E+02 & 4.66E+02 & 6.57E+02 &            8.     & 1.63E+08 & 1.41E+08 & 1.87E+08 \\
  0.12  & 1.04E+03 & 8.52E+02 & 1.18E+03 &            9.     & 1.65E+08 & 1.43E+08 & 1.91E+08 \\
  0.13  & 1.77E+03 & 1.46E+03 & 2.01E+03 &           10.     & 1.67E+08 & 1.44E+08 & 1.94E+08 \\
  0.14  & 2.86E+03 & 2.39E+03 & 3.24E+03 &                &     &    &         \\
\hline
\end{tabular*}
\label{be9pdTab2}
\end{table}
\clearpage\subsection{\reac{9}{Be}{p}{\alpha}{6}{Li}}
\label{be9paSect}
The experimental data sets referred to in NACRE are$^{*}$: NE51 \cite{NE51}$^\dag$, DA52 \cite{DA52}$^\ddag$, MA59 \cite{MA59}$^\ddag$, BL63 \cite{BL63}$^{\dag\dag}$, YA64 \cite{YA64}, MO65 \cite{MO65}$^\dag$, SI73 \cite{SI73} and ZA97 \cite{ZA97}, covering the 0.014 $\lsimeq E_{\rm cm} \lsimeq$ 10 MeV range.
Added is the post-NACRE data set WE08 \cite{WE08}$^\ddag,^{\ddag\ddag}$, extending the $E_{\rm cm}$ range down to $\simeq$ 0.012 MeV. 
[{\footnotesize{$^{*}$(p,$\alpha_0)$, i.e. to the ground state of $^{6}$Li with the $Q$-value of 2.125 MeV, if not marked otherwise;  
$^\dag$(re-)normalised (see NACRE);
$^\ddag$(p,$\alpha_2)$, i.e. to the second excited state at 3.562 MeV) with MA59 normalised to DA52;
$^{\dag\dag}$(p,$\alpha_1)$, i.e. to the first excited state at 2.186 MeV;  $^{\ddag\ddag}$from $^{2}$H($^{9}$Be,\,$^{6}$Li$\alpha$)n (THM).}}] 

\begin{figure}[hb]
\centering{
\includegraphics[height=0.50\textheight,width=0.90\textwidth]{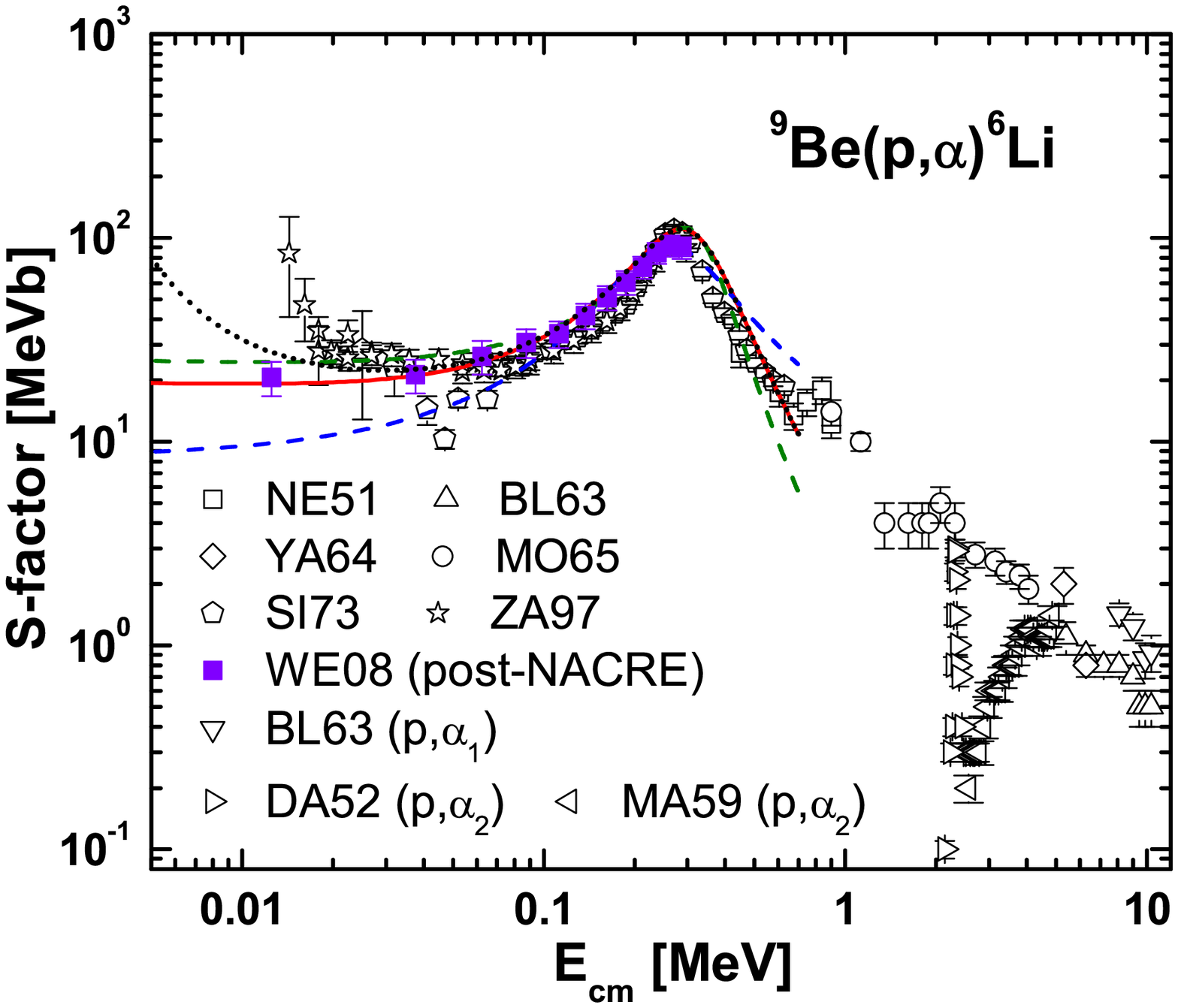}
\vspace{-0.5truecm}
\caption{The $S$-factor for \reac{9}{Be}{p}{\alpha}{6}{Li}.  The dotted line indicates an adiabatic screening correction ($U_{\rm e}$ = 264 eV) to the 'adopt' curve (solid line).}
\label{be9paFig1}
}
\end{figure}

Figure \ref{be9paFig1} compares the DWBA and experimental $S$-factors.
As inferred from WE08 using THM, the enhancements observed in other data sets below $E_{\rm cm} \simeq $ 0.03 - 0.04 MeV are likely caused by electron screening. The contribution from a sub-threshold resonance (SI73) appears to be minor (ZA97; see also \cite{BR98}).
The data in the 0.05 $\lsimeq E_{\rm cm} \lsimeq$ 1 MeV range, and all the WE08 data, are used for the DWBA fit. They exhibit the $1^{-}$ resonance at $E_{\rm R} \simeq$ 0.29 MeV.
The adopted parameter values are given in Table \ref{be9paTab1}.
The present  $S(0)$ = 21 $_{-13}^{+5}$ MeV\,b.
In comparison,  $S(0)$ = 17 $_{-7}^{+25}$ MeV\,b [NACRE, from SI73].

\begin{figure}[t]
\centering{
\hspace{0.8truein}
\includegraphics[height=0.33\textheight,width=0.90\textwidth]{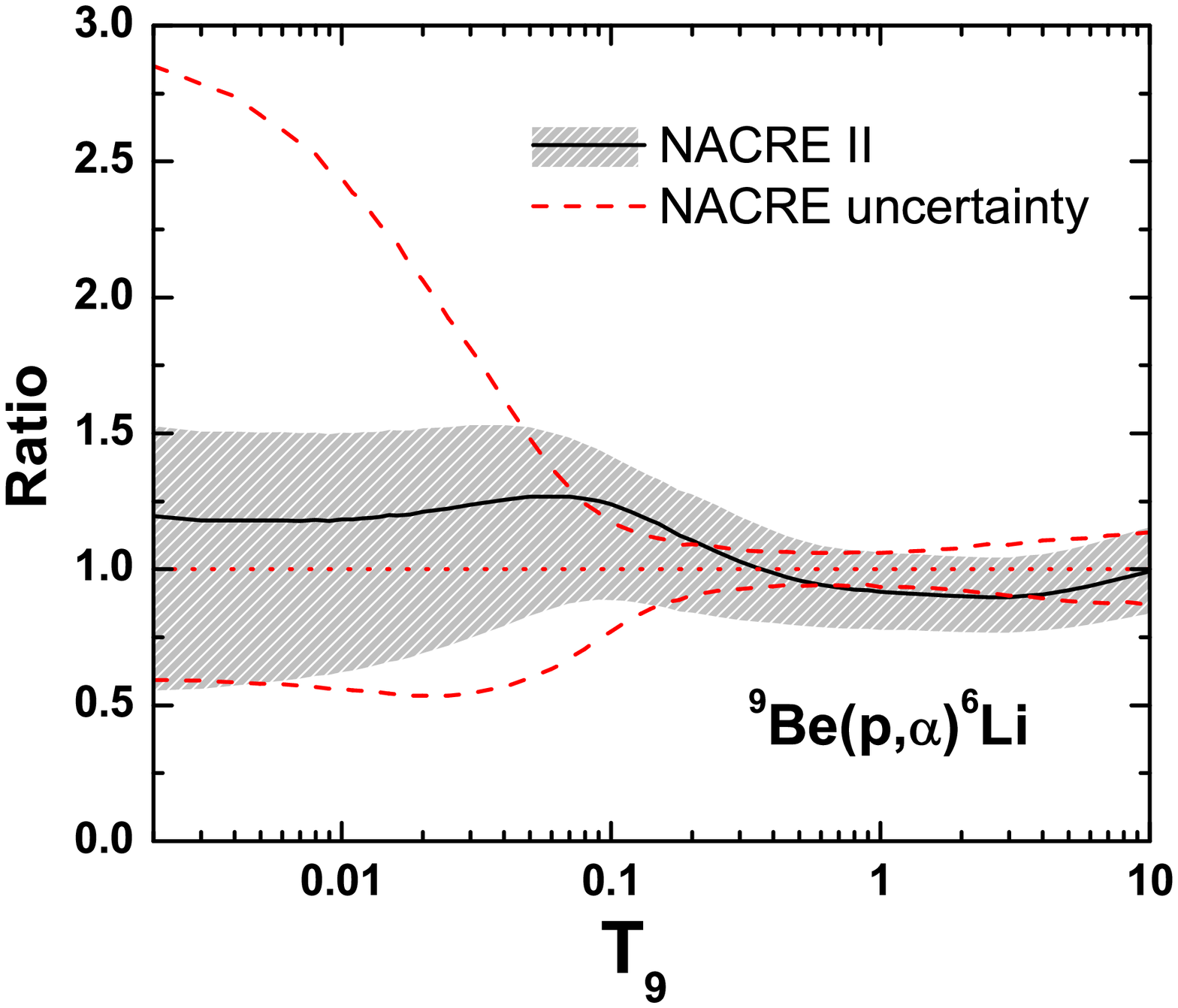}
\vspace{-0.4truecm}
\caption{\reac{9}{Be}{p}{\alpha}{6}{Li} rates in units of the NACRE (adopt) values. The exclusion for the DWBA fit of the experimental data sets in the $E_{\rm cm} \lsimeq$ 0.05 MeV range other than WE08 by THM (see text)  lowers the upper NACRE uncertainty.
}
\label{be9paFig2}
}
\end{figure}

Table \ref{be9paTab2} gives the reaction rates at $0.002 \le T_{9} \le 10$, for which the DWBA-predicted and the experimental cross sections below and above $E_{\rm cm} \simeq$ 0.07 MeV are used, respectively. 
Figure \ref{be9paFig2} compares the present and the NACRE rates.

{\footnotesize  See  BL63  for a DWBA analysis (also see \cite{RA91}).}

\clearpage

\begin{table}[hb]
\caption{\reac{9}{Be}{p}{\alpha}{6}{Li} rates in units of $\rm{cm^{3}mol^{-1}s^{-1}}$.} \footnotesize\rm
\begin{tabular*}{\textwidth}{@{\extracolsep{\fill}}  l c c c | l c c c}
\hline
$T_{9}$ & adopted & low & high & $T_{9}$ & adopted & low & high \\
\hline
  0.002 & 3.06E$-$23 & 1.41E$-$23 & 3.92E$-$23 &      0.15 & 4.88E+03 & 3.59E+03 & 5.59E+03 \\
  0.003 & 7.54E$-$19 & 3.55E$-$19 & 9.66E$-$19 &      0.16 & 7.30E+03 & 5.40E+03 & 8.38E+03 \\
  0.004 & 4.44E$-$16 & 2.13E$-$16 & 5.68E$-$16 &      0.18 & 1.49E+04 & 1.12E+04 & 1.72E+04 \\
  0.005 & 4.11E$-$14 & 2.01E$-$14 & 5.27E$-$14 &      0.2 & 2.79E+04 & 2.10E+04 & 3.23E+04 \\
  0.006 & 1.30E$-$12 & 6.44E$-$13 & 1.66E$-$12 &      0.25 & 9.92E+04 & 7.64E+04 & 1.16E+05 \\
  0.007 & 2.04E$-$11 & 1.03E$-$11 & 2.60E$-$11 &      0.3 & 2.64E+05 & 2.08E+05 & 3.08E+05 \\
  0.008 & 1.97E$-$10 & 1.01E$-$10 & 2.52E$-$10 &      0.35 & 5.77E+05 & 4.60E+05 & 6.72E+05 \\
  0.009 & 1.34E$-$09 & 6.93E$-$10 & 1.71E$-$09 &      0.4 & 1.09E+06 & 8.76E+05 & 1.26E+06 \\
  0.01 & 7.00E$-$09 & 3.65E$-$09 & 8.91E$-$09 &      0.45 & 1.82E+06 & 1.49E+06 & 2.12E+06 \\
  0.011 & 2.96E$-$08 & 1.56E$-$08 & 3.77E$-$08 &      0.5 & 2.81E+06 & 2.31E+06 & 3.26E+06 \\
  0.012 & 1.06E$-$07 & 5.64E$-$08 & 1.35E$-$07 &      0.6 & 5.51E+06 & 4.57E+06 & 6.39E+06 \\
  0.013 & 3.32E$-$07 & 1.78E$-$07 & 4.21E$-$07 &      0.7 & 9.06E+06 & 7.56E+06 & 1.05E+07 \\
  0.014 & 9.28E$-$07 & 5.02E$-$07 & 1.18E$-$06 &      0.8 & 1.32E+07 & 1.11E+07 & 1.54E+07 \\
  0.015 & 2.36E$-$06 & 1.29E$-$06 & 2.99E$-$06 &      0.9 & 1.78E+07 & 1.50E+07 & 2.07E+07 \\
  0.016 & 5.55E$-$06 & 3.05E$-$06 & 7.01E$-$06 &      1. & 2.27E+07 & 1.91E+07 & 2.63E+07 \\
  0.018 & 2.51E$-$05 & 1.40E$-$05 & 3.17E$-$05 &      1.25 & 3.49E+07 & 2.95E+07 & 4.07E+07 \\
  0.02 & 9.22E$-$05 & 5.22E$-$05 & 1.16E$-$04 &      1.5 & 4.67E+07 & 3.96E+07 & 5.45E+07 \\
  0.025 & 1.25E$-$03 & 7.28E$-$04 & 1.56E$-$03 &      1.75 & 5.76E+07 & 4.89E+07 & 6.73E+07 \\
  0.03 & 9.09E$-$03 & 5.45E$-$03 & 1.13E$-$02 &      2. & 6.75E+07 & 5.73E+07 & 7.89E+07 \\
  0.04 & 1.64E$-$01 & 1.03E$-$01 & 2.01E$-$01 &      2.5 & 8.46E+07 & 7.19E+07 & 9.89E+07 \\
  0.05 & 1.29E+00 & 8.40E$-$01 & 1.56E+00 &      3. & 9.86E+07 & 8.37E+07 & 1.15E+08 \\
  0.06 & 6.25E+00 & 4.19E+00 & 7.44E+00 &      3.5 & 1.10E+08 & 9.35E+07 & 1.29E+08 \\
  0.07 & 2.21E+01 & 1.51E+01 & 2.59E+01 &      4. & 1.20E+08 & 1.02E+08 & 1.40E+08 \\
  0.08 & 6.26E+01 & 4.36E+01 & 7.29E+01 &      5. & 1.35E+08 & 1.14E+08 & 1.57E+08 \\
  0.09 & 1.51E+02 & 1.07E+02 & 1.75E+02 &      6. & 1.45E+08 & 1.23E+08 & 1.70E+08 \\
  0.1 & 3.25E+02 & 2.31E+02 & 3.73E+02 &      7. & 1.54E+08 & 1.29E+08 & 1.80E+08 \\
  0.11 & 6.33E+02 & 4.54E+02 & 7.25E+02 &      8. & 1.60E+08 & 1.34E+08 & 1.87E+08 \\
  0.12 & 1.14E+03 & 8.28E+02 & 1.31E+03 &      9. & 1.65E+08 & 1.38E+08 & 1.93E+08 \\
  0.13 & 1.95E+03 & 1.42E+03 & 2.23E+03 &     10. & 1.69E+08 & 1.41E+08 & 1.97E+08 \\
  0.14 & 3.15E+03 & 2.30E+03 & 3.60E+03 &  \\
\hline
\end{tabular*}
\begin{tabular*}{\textwidth}{@{\extracolsep{\fill}} l c }
REV  = 
$ 0.618\,{\rm exp}(-24.660/T_{9})\,/\,[1.0+2.333\,{\rm exp}(-25.369/T_{9})] $ \\
\hspace{2truecm}
$[1.0+0.5\,{\rm exp}(-19.543/T_{9})\,+\,1.5\,{\rm exp}(-28.193/T_{9})] $  & \\
\end{tabular*}
\label{be9paTab2}
\end{table}
\clearpage\subsection{\reac{9}{Be}{\alpha}{n}{12}{C}}
\label{be9anSect}
The experimental data sets referred to in NACRE are GI65 \cite{GI65}$^\dag$, VA70b \cite{VA70b}, GE75 \cite{GE75}, SC92 \cite{SC92}, WR94 \cite{WR94} and KU96 \cite{KU96}, covering the 0.02 $\lsimeq E_{\rm cm} \lsimeq$ 10 MeV range.
No new cross section data are found.
[{\footnotesize{$^\dag$measured neutrons include those by the break-up reaction \reac{9}{Be}{\alpha}{n\alpha}{8}{Be}}.}] 

\begin{figure}[hb]
\centering{
\includegraphics[height=0.50\textheight,width=0.90\textwidth]{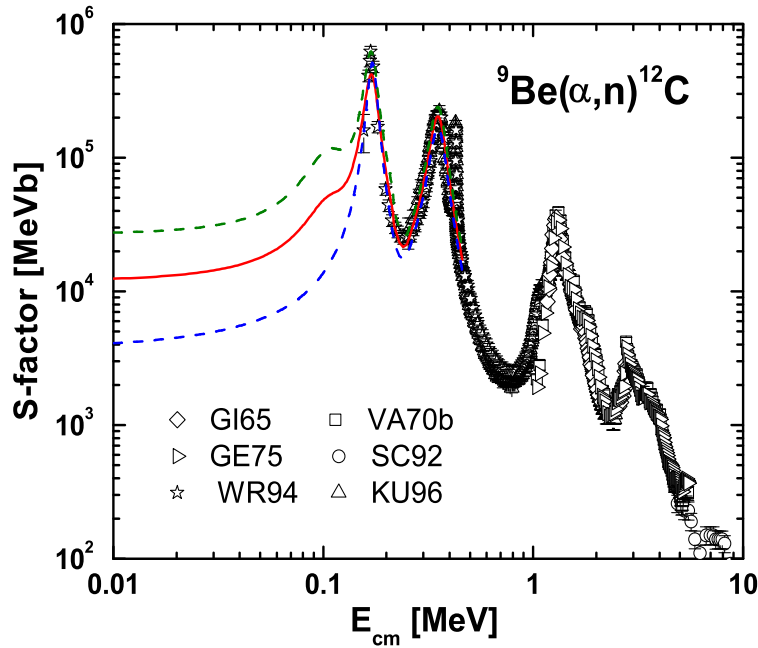}
\vspace{-0.5truecm}
\caption{The $S$-factor for \reac{9}{Be}{\alpha}{n}{12}{C}. GI65 data above $E_{\rm cm} \simeq$ 3 MeV are not used because of the possible contamination of break-up neutrons.}
\label{be9anFig1}
}
\end{figure}

Figure \ref{be9anFig1} compares model and experimental $S$-factors.
The data in the $E_{\rm cm} \lsimeq$ 0.4 MeV range are used for the DWBA fit. They exhibit the $5/2^{-}$ and $1/2^{+}$ resonances at $E_{\rm R} \simeq$  0.17 and 0.35 MeV. 
The likely contribution to the low-energy $S$-factors from the $7/2^{-}$ resonance at 0.10 MeV is estimated by Eqs.\,(\ref{eqBreitWigner0}) and (\ref{eqBreitWigner1}) with $\theta_\alpha^{2} = 0.1$ along with $\Gamma_{\rm cm}(E_{\rm R})$ from \cite{isotopes} and the $(\alpha, n_0)$ and $(\alpha, n_1)$ branching ratios from \cite{CI80}. For the upper limit, $\theta_\alpha^{2} = 0.3$ is used so as to avoid conflicts with the data at higher energies. 
Furthermore, the same $\theta_\alpha$ values are used to evaluate the possible contributions from the 3/2$^{-}$ (s-wave) sub-threshold resonance at $- 0.751$ MeV. For the lower limit, those contributions are assumed to be negligibly small.
The adopted parameter values are given in Table \ref{be9anTab1}.
The present $S$(0.01 MeV) = 11.5 $_{-7.4}^{+16} \times 10^{3}$ MeV\,b. 

Table \ref{be9anTab2} gives the reaction rates at $0.018 \le T_{9} \le 10$, for which the predicted and the experimental cross sections below and above $E_{\rm cm} \simeq$ 0.165 MeV are used, respectively. 

\begin{figure}[t]
\centering{
\includegraphics[height=0.33\textheight,width=0.90\textwidth]{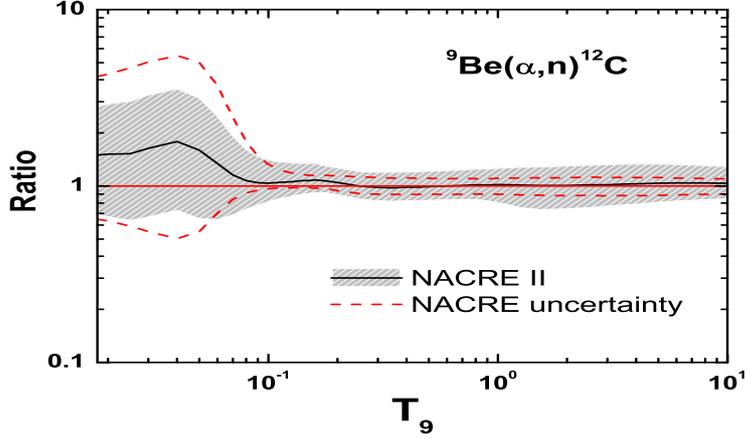}
\vspace{-0.4truecm}
\caption{\reac{9}{Be}{\alpha}{n}{12}{C} rates in units of the NACRE (adopt) values. The reduced upper limits at low temperatures result from a constraint imposed on the possible maximum contribution from the $7/2^{-}$ resonance (see text). The uncertainties at highest temperatures are slightly larger than in NACRE because of the less selective choice of the experimental data.  
}
\label{be9anFig2}
}
\end{figure}

Figure \ref{be9anFig2} compares the present and the NACRE rates.

\clearpage

\begin{table}[hb]
\caption{\reac{9}{Be}{\alpha}{n}{12}{C} rates in units of $\rm{cm^{3}mol^{-1}s^{-1}}$.} \footnotesize\rm
\begin{tabular*}{\textwidth}{@{\extracolsep{\fill}}  l c c c | l c c c}
\hline
$T_{9}$ & adopted & low & high & $T_{9}$ & adopted & low & high \\
\hline
   0.018 & 5.95E$-$25 & 2.73E$-$25 & 1.12E$-$24 &    0.4 & 1.22E+01 & 1.02E+01 & 1.49E+01  \\
   0.02 & 1.40E$-$23 & 6.17E$-$24 & 2.70E$-$23 &    0.45 & 3.30E+01 & 2.76E+01 & 4.04E+01  \\
   0.025 & 8.04E$-$21 & 3.35E$-$21 & 1.59E$-$20 &    0.5 & 7.37E+01 & 6.15E+01 & 9.03E+01  \\
   0.03 & 1.15E$-$18 & 4.69E$-$19 & 2.31E$-$18 &    0.6 & 2.49E+02 & 2.07E+02 & 3.05E+02  \\
   0.04 & 1.74E$-$15 & 7.06E$-$16 & 3.46E$-$15 &    0.7 & 6.05E+02 & 5.03E+02 & 7.42E+02  \\
   0.05 & 2.84E$-$13 & 1.16E$-$13 & 5.62E$-$13 &    0.8 & 1.23E+03 & 1.01E+03 & 1.51E+03  \\
   0.06 & 1.46E$-$11 & 7.00E$-$12 & 2.71E$-$11 &    0.9 & 2.27E+03 & 1.84E+03 & 2.80E+03  \\
   0.07 & 3.96E$-$10 & 2.35E$-$10 & 6.56E$-$10 &    1. & 4.09E+03 & 3.24E+03 & 5.07E+03  \\
   0.08 & 6.33E$-$09 & 4.36E$-$09 & 9.51E$-$09 &    1.25 & 1.76E+04 & 1.31E+04 & 2.22E+04  \\
   0.09 & 6.28E$-$08 & 4.70E$-$08 & 8.82E$-$08 &    1.5 & 6.51E+04 & 4.75E+04 & 8.32E+04  \\
   0.1 & 4.19E$-$07 & 3.30E$-$07 & 5.64E$-$07 &    1.75 & 1.87E+05 & 1.37E+05 & 2.40E+05  \\
   0.11 & 2.04E$-$06 & 1.66E$-$06 & 2.66E$-$06 &    2. & 4.28E+05 & 3.15E+05 & 5.52E+05  \\
   0.12 & 7.80E$-$06 & 6.45E$-$06 & 9.95E$-$06 &    2.5 & 1.40E+06 & 1.04E+06 & 1.81E+06  \\
   0.13 & 2.46E$-$05 & 2.06E$-$05 & 3.08E$-$05 &    3. & 3.11E+06 & 2.35E+06 & 4.04E+06  \\
   0.14 & 6.66E$-$05 & 5.64E$-$05 & 8.23E$-$05 &    3.5 & 5.53E+06 & 4.21E+06 & 7.17E+06  \\
   0.15 & 1.60E$-$04 & 1.37E$-$04 & 1.97E$-$04 &    4. & 8.57E+06 & 6.57E+06 & 1.11E+07  \\
   0.16 & 3.53E$-$04 & 3.03E$-$04 & 4.30E$-$04 &    5. & 1.62E+07 & 1.26E+07 & 2.08E+07  \\
   0.18 & 1.42E$-$03 & 1.23E$-$03 & 1.70E$-$03 &    6. & 2.54E+07 & 2.00E+07 & 3.23E+07  \\
   0.2 & 4.98E$-$03 & 4.28E$-$03 & 5.94E$-$03 &    7. & 3.57E+07 & 2.84E+07 & 4.51E+07  \\
   0.25 & 7.67E$-$02 & 6.49E$-$02 & 9.25E$-$02 &    8. & 4.66E+07 & 3.74E+07 & 5.85E+07  \\
   0.3 & 6.65E$-$01 & 5.58E$-$01 & 8.09E$-$01 &    9. & 5.76E+07 & 4.66E+07 & 7.19E+07  \\
   0.35 & 3.44E+00 & 2.88E+00 & 4.20E+00 &   10. & 6.85E+07 & 5.56E+07 & 8.50E+07  \\
\hline
\end{tabular*}
\begin{tabular*}{\textwidth}{@{\extracolsep{\fill}} l c }
REV  = 
$ 10.3\,{\rm exp}(-66.163/T_{9}) \,[1.0+0.5\,{\rm exp}(-19.543/T_{9})
+1.5\,{\rm exp}(-28.193/T_{9})] $ 
 & \\
\end{tabular*}
\label{be9anTab2}
\end{table}
\clearpage\subsection{\reac{10}{B}{p}{\gamma}{11}{C}}
\label{b10pgSect}
The experimental data sets referred to in NACRE are KU70 \cite{KU70} and WI83 \cite{WI83}, covering the 0.09 $\lsimeq E_{\rm cm} \lsimeq$ 13 MeV range.
Added is the post-NACRE data set TO03 \cite{TO03}.

\begin{figure}[hb]
\centering{
\includegraphics[height=0.50\textheight,width=0.90\textwidth]{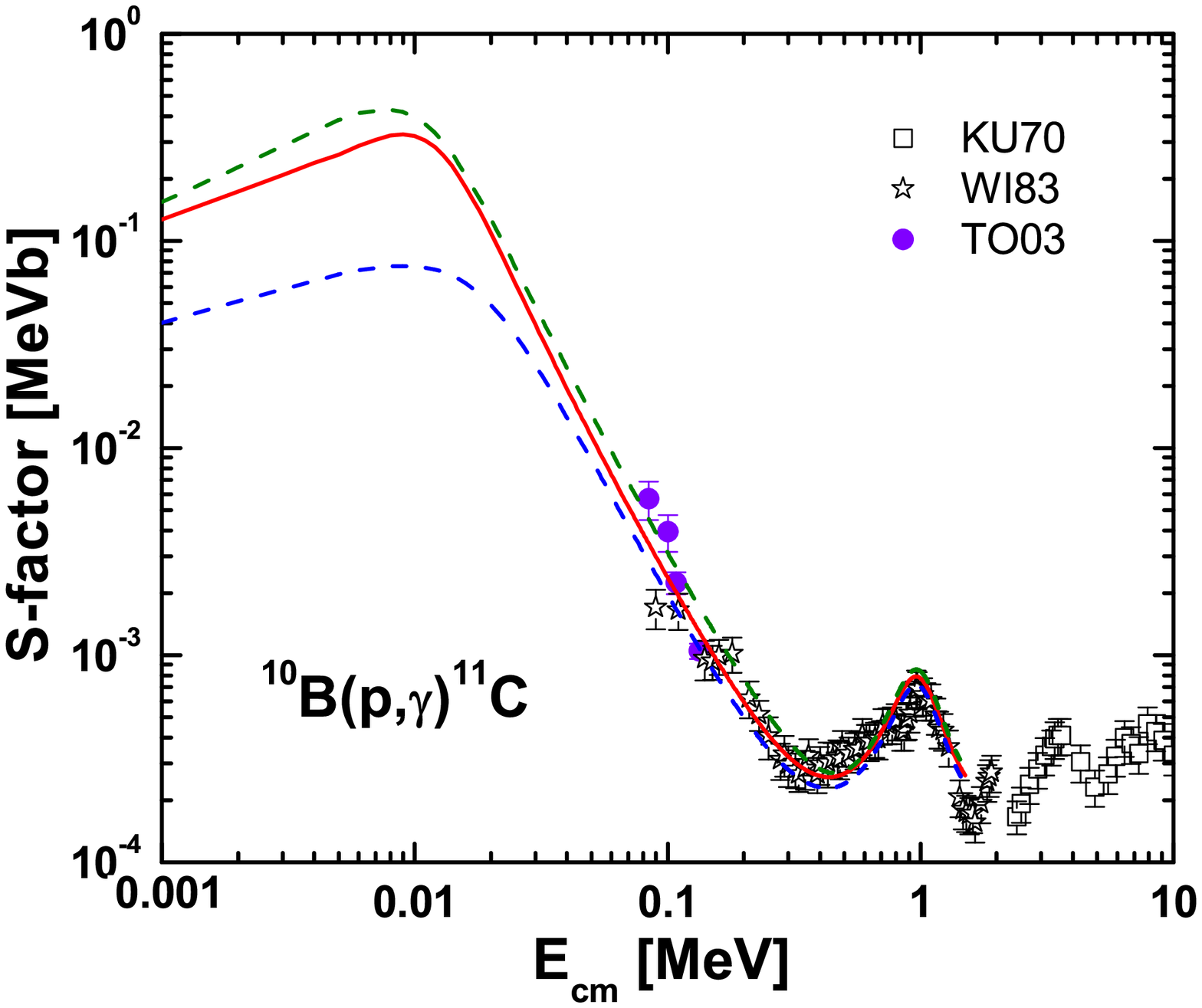}
\vspace{-0.5truecm}
\caption{The $S$-factor for \reac{10}{B}{p}{\gamma}{11}{C}.}
\label{b10pgFig1}
}
\end{figure}

Figure \ref{b10pgFig1} compares the model and experimental $S$-factors.
The data for $E_{\rm cm} \lsimeq$ 1.0 MeV are used for the PM fit. They exhibit the $3/2^{-}$ resonance at $E_{\rm R} \simeq$  0.96 MeV. The transitions to the ground and the first five excited states of $^{11}$C are considered inclusively.
The $S$-factor data below $E_{\rm cm} \simeq$ 0.2 MeV are thought to reflect the tail of the $5/2^{+}$ resonance expected at $E_{\rm cm} \simeq$ 0.010 MeV. 
The extrapolation to the resonance peak region is performed in parallel to the \reac{10}{B}{p}{\alpha}{7}{Be} reaction, the $S$-factors of which are known to lower energies than in the (p,$\gamma$) case. 
In practice, we use the Breit-Wigner formula, Eqs.\,(\ref{eqBreitWigner0}) and (\ref{eqBreitWigner1}),  with resonance energy and width that well reproduce the DWBA result. 
At $E_{\rm cm} \lsimeq$ 0.005 MeV, a correction is made by directly scaling the DWBA results.
No significant sub-threshold contributions (e.g. from the $7/2^{+}$ resonance at $E_{\rm R} \simeq -0.034$ MeV) to the $S$-factors in the low-energy range are concordant with the experimental data at $E_{\rm cm} \gsimeq$ 0.1 MeV.
The adopted parameter values are given in Table \ref{b10pgTab1}.
The present $S(0.001 MeV) = 0.13^{+0.03}_{-0.09}$ MeV\,b.

\begin{figure}[t]
\centering{
\includegraphics[height=0.33\textheight,width=0.90\textwidth]{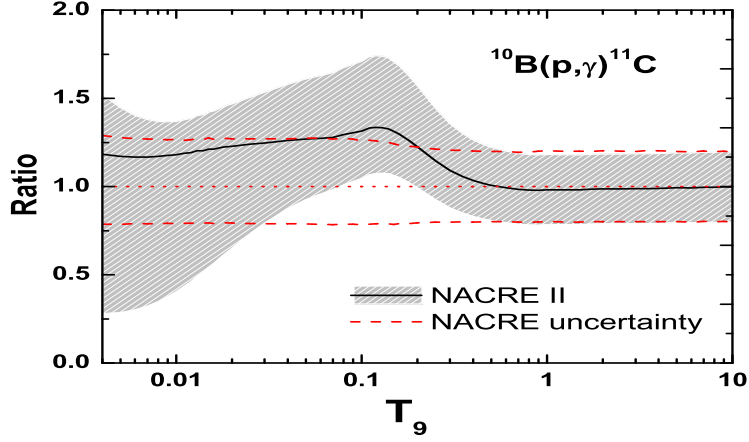}
\vspace{-0.4truecm}
\caption{\reac{10}{B}{p}{\gamma}{11}{C} rates in units of the NACRE (adopt) values. As for the large uncertainties at the lowest temperatures, see the caption to Fig.\,\ref{b10paFig2}.
}
\label{b10pgFig2}
}
\end{figure}

Table \ref{b10pgTab2} gives the reaction rates at $0.004 \le T_{9} \le 10$, for which the PM-predicted and the experimental cross sections below and above $E_{\rm cm} \simeq$ 0.1 MeV are used, respectively. 
Figure \ref{b10pgFig2} compares the present and the NACRE rates.

\clearpage

\begin{table}[hb]
\caption{\reac{10}{B}{p}{\gamma}{11}{C} rates in $\rm{cm^{3}mol^{-1}s^{-1}}$}\scriptsize\rm
\footnotesize{
\begin{tabular*}{\textwidth}{@{\extracolsep{\fill}} l c c c |l c c c}
\hline
$T_{9}$ & adopted & low & high & $T_{9}$ & adopted & low & high \\
\hline
  0.004 & 1.71E$-$22 & 4.05E$-$23 & 2.21E$-$22 &     0.16 & 2.60E$-$02 & 2.10E$-$02 & 3.37E$-$02 \\
  0.005 & 3.29E$-$20 & 8.02E$-$21 & 4.07E$-$20 &     0.18 & 4.77E$-$02 & 3.87E$-$02 & 6.16E$-$02 \\
  0.006 & 1.72E$-$18 & 4.42E$-$19 & 2.06E$-$18 &     0.2 & 8.03E$-$02 & 6.50E$-$02 & 1.03E$-$01 \\
  0.007 & 3.87E$-$17 & 1.07E$-$17 & 4.56E$-$17 &     0.25 & 2.27E$-$01 & 1.83E$-$01 & 2.86E$-$01 \\
  0.008 & 4.89E$-$16 & 1.45E$-$16 & 5.71E$-$16 &     0.3 & 4.99E$-$01 & 4.02E$-$01 & 6.20E$-$01 \\
  0.009 & 4.07E$-$15 & 1.30E$-$15 & 4.73E$-$15 &     0.35 & 9.29E$-$01 & 7.47E$-$01 & 1.14E+00 \\
  0.01 & 2.48E$-$14 & 8.47E$-$15 & 2.87E$-$14 &     0.4 & 1.54E+00 & 1.24E+00 & 1.88E+00 \\
  0.011 & 1.19E$-$13 & 4.32E$-$14 & 1.38E$-$13 &     0.45 & 2.35E+00 & 1.89E+00 & 2.86E+00 \\
  0.012 & 4.69E$-$13 & 1.81E$-$13 & 5.45E$-$13 &     0.5 & 3.38E+00 & 2.71E+00 & 4.11E+00 \\
  0.013 & 1.59E$-$12 & 6.47E$-$13 & 1.85E$-$12 &     0.6 & 6.18E+00 & 4.95E+00 & 7.47E+00 \\
  0.014 & 4.74E$-$12 & 2.03E$-$12 & 5.54E$-$12 &     0.7 & 1.01E+01 & 8.10E+00 & 1.22E+01 \\
  0.015 & 1.27E$-$11 & 5.70E$-$12 & 1.49E$-$11 &     0.8 & 1.55E+01 & 1.24E+01 & 1.87E+01 \\
  0.016 & 3.13E$-$11 & 1.46E$-$11 & 3.68E$-$11 &     0.9 & 2.25E+01 & 1.80E+01 & 2.71E+01 \\
  0.018 & 1.52E$-$10 & 7.59E$-$11 & 1.80E$-$10 &     1. & 3.15E+01 & 2.52E+01 & 3.79E+01 \\
  0.02 & 5.86E$-$10 & 3.10E$-$10 & 7.00E$-$10 &     1.25 & 6.41E+01 & 5.12E+01 & 7.70E+01 \\
  0.025 & 8.50E$-$09 & 5.02E$-$09 & 1.03E$-$08 &     1.5 & 1.13E+02 & 9.02E+01 & 1.36E+02 \\
  0.03 & 6.37E$-$08 & 4.05E$-$08 & 7.84E$-$08 &     1.75 & 1.78E+02 & 1.42E+02 & 2.14E+02 \\
  0.04 & 1.15E$-$06 & 7.98E$-$07 & 1.44E$-$06 &     2. & 2.58E+02 & 2.07E+02 & 3.10E+02 \\
  0.05 & 8.72E$-$06 & 6.36E$-$06 & 1.11E$-$05 &     2.5 & 4.52E+02 & 3.61E+02 & 5.42E+02 \\
  0.06 & 4.00E$-$05 & 3.01E$-$05 & 5.12E$-$05 &     3. & 6.70E+02 & 5.36E+02 & 8.04E+02 \\
  0.07 & 1.33E$-$04 & 1.02E$-$04 & 1.71E$-$04 &     3.5 & 8.94E+02 & 7.16E+02 & 1.07E+03 \\
  0.08 & 3.55E$-$04 & 2.76E$-$04 & 4.58E$-$04 &     4. & 1.12E+03 & 8.94E+02 & 1.34E+03 \\
  0.09 & 8.07E$-$04 & 6.36E$-$04 & 1.05E$-$03 &     5. & 1.55E+03 & 1.24E+03 & 1.86E+03 \\
  0.1 & 1.63E$-$03 & 1.30E$-$03 & 2.12E$-$03 &     6. & 1.96E+03 & 1.57E+03 & 2.36E+03 \\
  0.11 & 3.00E$-$03 & 2.40E$-$03 & 3.91E$-$03 &     7. & 2.37E+03 & 1.90E+03 & 2.84E+03 \\
  0.12 & 5.11E$-$03 & 4.11E$-$03 & 6.67E$-$03 &     8. & 2.78E+03 & 2.22E+03 & 3.33E+03 \\
  0.13 & 8.21E$-$03 & 6.62E$-$03 & 1.07E$-$02 &     9. & 3.18E+03 & 2.55E+03 & 3.82E+03 \\
  0.14 & 1.25E$-$02 & 1.01E$-$02 & 1.63E$-$02 &    10. & 3.59E+03 & 2.88E+03 & 4.31E+03 \\
  0.15 & 1.84E$-$02 & 1.49E$-$02 & 2.39E$-$02 &     \\

\hline
\end{tabular*}
\begin{tabular*}{\textwidth}{@{\extracolsep{\fill}} l c }
REV  = 
 $3.03 \times 10^{10}T_{9}^{3/2}{\rm exp}(-100.84/T_{9})\,/\,[1.0+0.5\,{\rm exp}(-23.210/T_{9})] $ \\
\hspace{1truecm}
$\times\,[1.0+0.429\,{\rm exp}(-8.336/T_{9})+0.143\,{\rm exp}(-20.194/T_{9})+0.429\,{\rm exp}(-24.997/T_{9})] $ 
 & \\
\end{tabular*}
}
\label{b10pgTab2}
\end{table}
\clearpage

\subsection{\reac{10}{B}{p}{\alpha}{7}{Be}}
\label{b10paSect}
The experimental data sets referred to in NACRE are BU50 \cite{BU50}, BR51b \cite{BR51b}, JE64 \cite{JE64}$^\dag$, SZ72 \cite{SZ72}, RO79 \cite{RO79}, YO91 \cite{YO91}, KN93 \cite{KN93} and AN93 \cite{AN93}, covering the 0.02 $\lsimeq E_{\rm cm} \lsimeq$ 6.4 MeV range.
Added are the post-NACRE data sets LA10 \cite{LA10}$^\ddag$ and PU10 \cite{PU10}$^\ddag$, extending the range down to $E_{\rm cm} \simeq$ 0.004 MeV. \cite{LA07a}$^\ddag$ is superseded by LA10.
Some data in the lowest energy range may possibly be contaminated  by electron screening.  
[{\footnotesize{$^\dag$(p,$\alpha_1$) also; $^\ddag$from $^{2}$H\,($^{10}$B, $\alpha^{7}$Be)\,n (THM).
}}] 

\begin{figure}[hb]
\centering{
\includegraphics[height=0.50\textheight,width=0.90\textwidth]{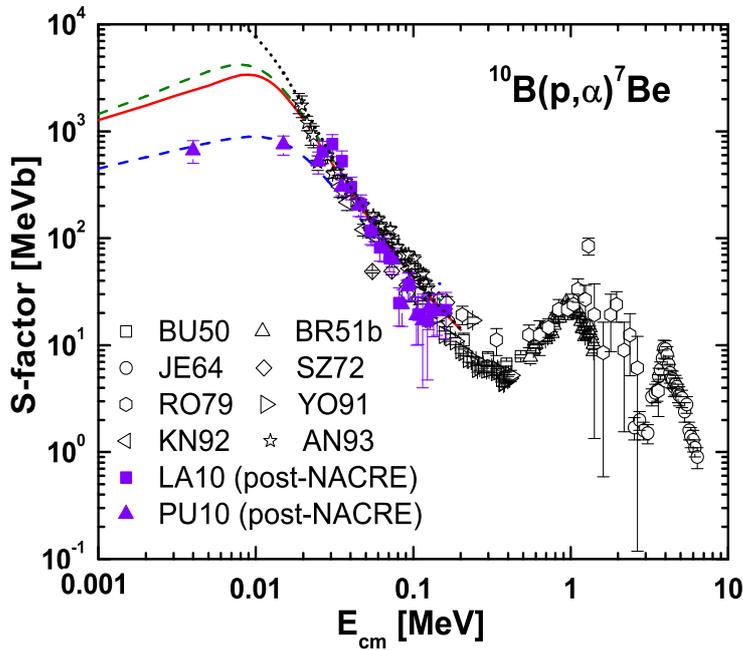}
\vspace{-0.5truecm}
\caption{The $S$-factor for \reac{10}{B}{p}{\alpha}{7}{Be}. The dotted line indicates an adiabatic screening correction ($U_{\rm e}$ = 348 eV) to the "adopt" curve (solid line). The recent data \cite{FR12}, not shown here, from the reverse \reac{7}{Be}{\alpha}{p}{10}{B} reaction imply (but with large uncertainties) less structured $S$-factors in the 0.3 $\lsimeq E_{\rm cm} \lsimeq$ 1.1 MeV range, and some fine structures in the 2.3 $\lsimeq E_{\rm cm} \lsimeq$ 6.5 MeV range. 
}
\label{b10paFig1}
}
\end{figure}

Figure \ref{b10paFig1} compares the DWBA and experimental $S$-factors.
The data in the $E_{\rm cm} \lsimeq$ 0.1 MeV range are used for the DWBA fit. 
They appear to exhibit the $5/2^{+}$ resonance, or its tail, expected at $E_{\rm cm} \simeq$ 0.010 MeV.
The "low" curve results from the fit to PU10, albeit this data set (which is "still under study") requires a width of about 30 keV, twice as much as those obtained for the "adopt" and "high" curves here and by other methods.   
The adopted parameter values are given in Table \ref{b10paTab1}.
The present $S(0.001) = 1.3^{+0.2}_{-0.9} \times 10^{3}$ MeV\,b.

\begin{figure}[t]
\centering{
\includegraphics[height=0.33\textheight,width=0.90\textwidth]{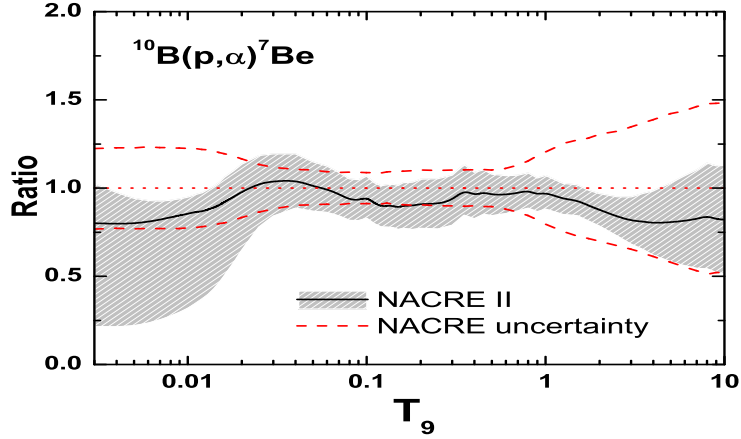}
\vspace{-0.4truecm}
\caption{\reac{10}{B}{p}{\alpha}{7}{Be} rates in units of the NACRE (adopt) values. The consideration of PU10 (THM) explains the large uncertainty at the lowest temperatures. What is seen at high temperatures may be the consequence of the different interpolation/integration techniques of high-energy data.
}
\label{b10paFig2}
}
\end{figure}

Table \ref{b10paTab2} gives the reaction rates at $0.003 \le T_{9} \le 10$, for which the DWBA-predicted and the experimental cross sections below and above $E_{\rm cm} \simeq$ 0.02 MeV are used, respectively. 
Figure \ref{b10paFig2} compares the present and the NACRE rates.

{\footnotesize{See \cite{RA96} for a DWBA analysis.
}} 
\clearpage

\begin{table}[hb]
\caption{\reac{10}{B}{p}{\alpha}{7}{Be} rates in $\rm{cm^{3}mol^{-1}s^{-1}}$}\scriptsize\rm
\footnotesize{
\begin{tabular*}{\textwidth}{@{\extracolsep{\fill}} l c c c |l c c c}
\hline
$T_{9}$ & adopted & low & high & $T_{9}$ & adopted & low & high \\
\hline
  0.003 & 1.00E$-$21 & 2.68E$-$22 & 1.28E$-$21 &     0.15 & 3.17E+02 & 2.73E+02 & 3.61E+02 \\
  0.004 & 1.78E$-$18 & 4.75E$-$19 & 2.17E$-$18 &     0.16 & 4.52E+02 & 3.88E+02 & 5.17E+02 \\
  0.005 & 3.43E$-$16 & 9.47E$-$17 & 4.04E$-$16 &     0.18 & 8.43E+02 & 7.17E+02 & 9.68E+02 \\
  0.006 & 1.80E$-$14 & 5.25E$-$15 & 2.07E$-$14 &     0.2 & 1.43E+03 & 1.21E+03 & 1.65E+03 \\
  0.007 & 4.11E$-$13 & 1.27E$-$13 & 4.64E$-$13 &     0.25 & 4.04E+03 & 3.41E+03 & 4.67E+03 \\
  0.008 & 5.26E$-$12 & 1.73E$-$12 & 5.87E$-$12 &     0.3 & 8.79E+03 & 7.45E+03 & 1.01E+04 \\
  0.009 & 4.43E$-$11 & 1.55E$-$11 & 4.92E$-$11 &     0.35 & 1.62E+04 & 1.39E+04 & 1.86E+04 \\
  0.01 & 2.72E$-$10 & 1.02E$-$10 & 3.03E$-$10 &     0.4 & 2.69E+04 & 2.32E+04 & 3.06E+04 \\
  0.011 & 1.31E$-$09 & 5.22E$-$10 & 1.46E$-$09 &     0.45 & 4.13E+04 & 3.59E+04 & 4.66E+04 \\
  0.012 & 5.22E$-$09 & 2.22E$-$09 & 5.86E$-$09 &     0.5 & 5.99E+04 & 5.26E+04 & 6.72E+04 \\
  0.013 & 1.78E$-$08 & 8.05E$-$09 & 2.01E$-$08 &     0.6 & 1.13E+05 & 1.00E+05 & 1.25E+05 \\
  0.014 & 5.32E$-$08 & 2.57E$-$08 & 6.05E$-$08 &     0.7 & 1.92E+05 & 1.72E+05 & 2.13E+05 \\
  0.015 & 1.44E$-$07 & 7.37E$-$08 & 1.64E$-$07 &     0.8 & 3.09E+05 & 2.77E+05 & 3.40E+05 \\ 
 0.016 & 3.54E$-$07 & 1.93E$-$07 & 4.07E$-$07 &     0.9 & 4.73E+05 & 4.26E+05 & 5.20E+05 \\
  0.018 & 1.74E$-$06 & 1.05E$-$06 & 2.01E$-$06 &     1. & 6.98E+05 & 6.30E+05 & 7.66E+05 \\
  0.02 & 6.79E$-$06 & 4.49E$-$06 & 7.88E$-$06 &     1.25 & 1.60E+06 & 1.45E+06 & 1.76E+06 \\
  0.025 & 1.02E$-$04 & 7.76E$-$05 & 1.18E$-$04 &     1.5 & 3.12E+06 & 2.83E+06 & 3.41E+06 \\
  0.03 & 7.90E$-$04 & 6.39E$-$04 & 9.08E$-$04 &     1.75 & 5.32E+06 & 4.80E+06 & 5.83E+06 \\
  0.04 & 1.49E$-$02 & 1.27E$-$02 & 1.71E$-$02 &     2. & 8.19E+06 & 7.33E+06 & 9.05E+06 \\
  0.05 & 1.18E$-$01 & 1.02E$-$01 & 1.34E$-$01 &     2.5 & 1.57E+07 & 1.36E+07 & 1.78E+07 \\
  0.06 & 5.66E$-$01 & 4.91E$-$01 & 6.41E$-$01 &     3. & 2.50E+07 & 2.09E+07 & 2.91E+07 \\
  0.07 & 1.96E+00 & 1.71E+00 & 2.21E+00 &     3.5 & 3.53E+07 & 2.83E+07 & 4.22E+07 \\
  0.08 & 5.41E+00 & 4.73E+00 & 6.09E+00 &     4. & 4.59E+07 & 3.55E+07 & 5.64E+07 \\
  0.09 & 1.27E+01 & 1.11E+01 & 1.43E+01 &     5. & 6.68E+07 & 4.83E+07 & 8.54E+07 \\
  0.10 & 2.62E+01 & 2.29E+01 & 2.94E+01 &     6. & 8.55E+07 & 5.87E+07 & 1.12E+08 \\
  0.11 & 4.90E+01 & 4.28E+01 & 5.53E+01 &     7. & 1.01E+08 & 6.71E+07 & 1.36E+08 \\
  0.12 & 8.50E+01 & 7.39E+01 & 9.60E+01 &     8. & 1.15E+08 & 7.38E+07 & 1.55E+08 \\
  0.13 & 1.38E+02 & 1.20E+02 & 1.57E+02 &     9. & 1.25E+08 & 7.92E+07 & 1.71E+08 \\
  0.14 & 2.14E+02 & 1.85E+02 & 2.43E+02 &    10. & 1.34E+08 & 8.36E+07 & 1.84E+08 \\

\hline
\end{tabular*}
\begin{tabular*}{\textwidth}{@{\extracolsep{\fill}} l c }
REV  = 
$ 0.754\,{\rm exp}(-13.285/T_{9})\,/\,[1.0+0.5\,{\rm exp}(-4.979/T_{9})] $ \\
\hspace{1truecm}
$ \times\,[1.0+0.429\,{\rm exp}(-8.336/T_{9})+0.143\,{\rm exp}(-20.194/T_{9})+0.429\,{\rm exp}(-24.997/T_{9})]$ \\
 & \\
\end{tabular*}
}
\label{b10paTab2}
\end{table}
\clearpage

\subsection{\reac{11}{B}{p}{\gamma}{12}{C}}
\label{b11pgSect}
The experimental data sets referred to in NACRE are HU53 \cite{HU53}, AL64 \cite{AL64} and SE65 \cite{SE65}, covering the 0.3 $\lsimeq E_{\rm cm} \lsimeq$ 9 MeV range.
Added is the post-NACRE data set KE00 \cite{KE00}, exploring the 0.07 $\lsimeq E_{\rm cm} \lsimeq$ 0.09 MeV range.

\begin{figure}[hb]
\centering{
\includegraphics[height=0.50\textheight,width=0.90\textwidth]{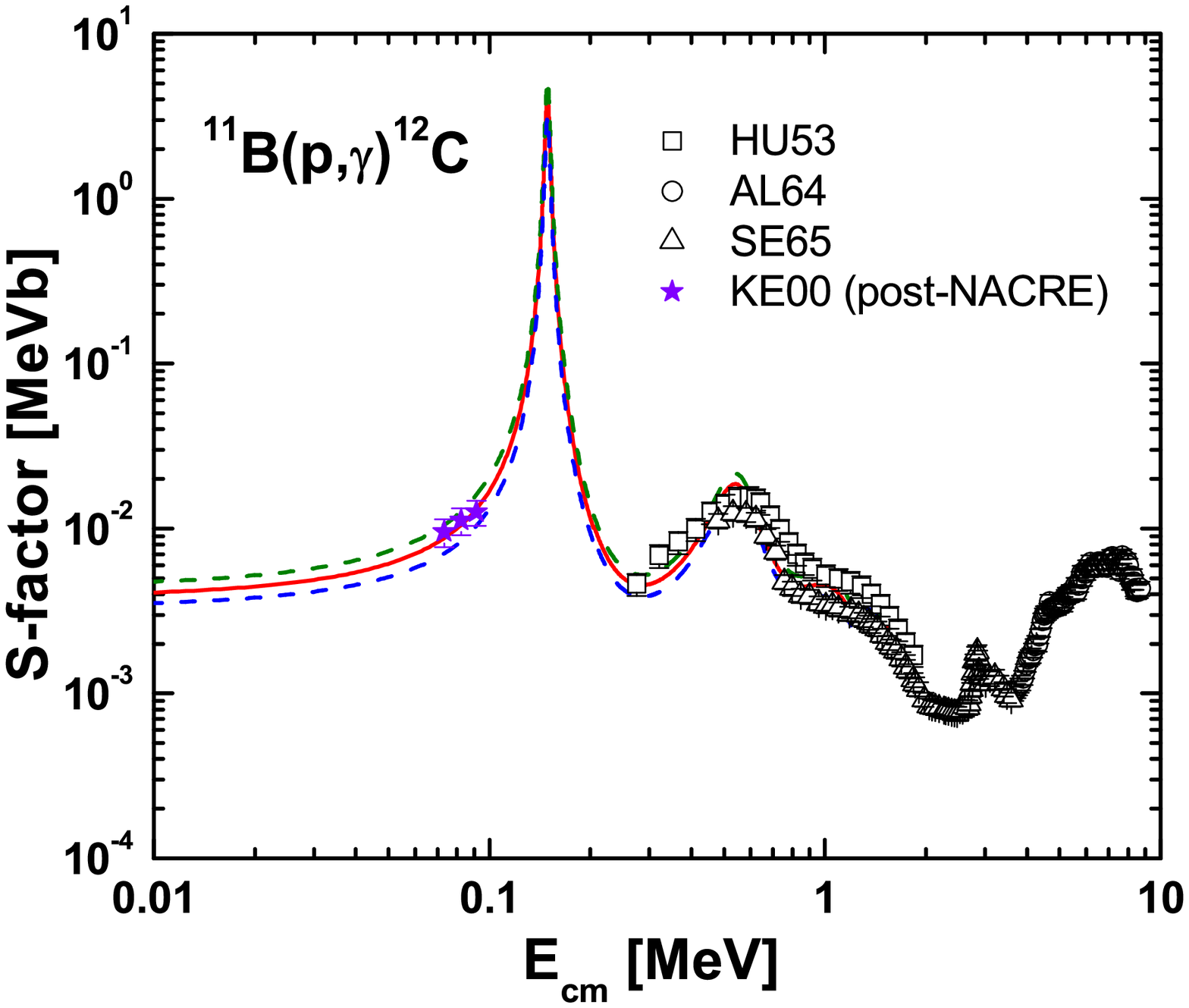}
\vspace{-0.5truecm}
\caption{The $S$-factor for \reac{11}{B}{p}{\gamma}{12}{C}.}
\label{b11pgFig1}
}
\end{figure}

Figure \ref{b11pgFig1} compares the PM and experimental $S$-factors.
The data in the $E_{\rm cm} \lsimeq$ 0.7 MeV range are used for the PM fit. They suggest a $2^{+}$ resonance at $E_{\rm R} \simeq$  0.15 MeV, and  exhibit the $2^{-}$ resonance$^{\dag}$ at $E_{\rm R} \simeq$ 0.61 MeV.
For the former, the Breit-Wigner formula is used to fix its height, and the PM  extrapolation then nicely goes through the KE00 data points. The transitions to the ground and the first excited states of $^{12}$C are considered inclusively.
The adopted parameter values are given in Table \ref{b11pgTab1}.
The present $S(0) = 4.3^{+0.8}_{-0.6}$ keV\,b. In comparison, $S(0)$ = 3.3 $\pm$ 0.5 keV\,b [NACRE, from \cite{CE92}]. 
[{\footnotesize{$^{\dag}$This state at $E_{\rm x} \simeq$ 16.57 MeV decays to the ground state 100 \% via an M2 transition \cite{isotopes}. Whereas this mode is not included in the formalism given in Sect.\,\ref{sectPMcross}, the {\it shape} of the corresponding $S$-factor is determined essentially by the scattering state. It is therefore nearly independent of the multipolarities when the $Q$-value is as high as in the current case.}}]

\begin{figure}[t]
\centering{
\includegraphics[height=0.33\textheight,width=0.90\textwidth]{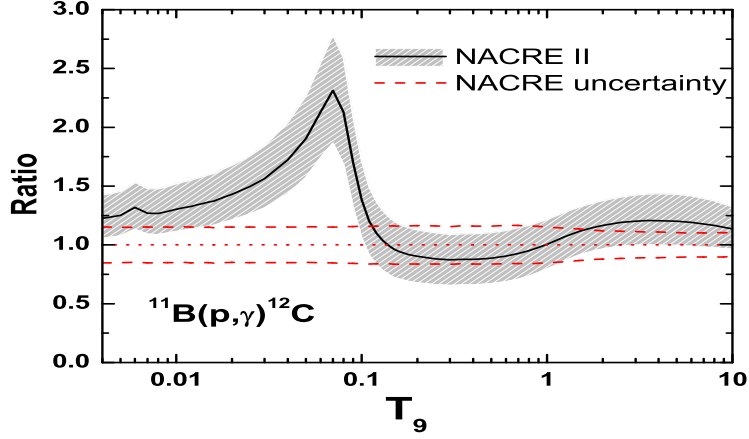}
\vspace{-0.4truecm}
\caption{\reac{11}{B}{p}{\gamma}{12}{C} rates in units of the NACRE (adopt) values. The peculiar increase of the ratio at $T_{9} \lsimeq$ 0.1 reflects the improved S-factor shape of the low-energy tail of the $2^{+}$ resonance.  
}
\label{b11pgFig2}
}
\end{figure}

Table \ref{b11pgTab2} gives the reaction rates at $0.004 \le T_{9} \le 10$, for which the PM-predicted and the experimental cross sections below and above $E_{\rm cm} \simeq$ 0.2 MeV are used, respectively. 
Figure \ref{b11pgFig2} compares the present and the NACRE rates.

\clearpage

\begin{table}[hb]
\caption{\reac{11}{B}{p}{\gamma}{12}{C} rates in $\rm{cm^{3}mol^{-1}s^{-1}}$}\scriptsize\rm
\footnotesize{
\begin{tabular*}{\textwidth}{@{\extracolsep{\fill}} l c c c |l c c c}
\hline
$T_{9}$ & adopted & low & high & $T_{9}$ & adopted & low & high \\
\hline
  0.004 & 1.78E$-$24 & 1.53E$-$24 & 2.06E$-$24 &           0.16  & 1.95E+00 & 1.47E+00 & 2.44E+00 \\
  0.005 & 3.65E$-$22 & 3.15E$-$22 & 4.25E$-$22 &           0.18  & 5.29E+00 & 3.98E+00 & 6.61E+00 \\
  0.006 & 2.12E$-$20 & 1.83E$-$20 & 2.47E$-$20 &           0.2 & 1.16E+01 & 8.75E+00 & 1.45E+01 \\
  0.007 & 5.43E$-$19 & 4.67E$-$19 & 6.33E$-$19 &           0.25  & 4.63E+01 & 3.48E+01 & 5.78E+01 \\
  0.008 & 7.89E$-$18 & 6.77E$-$18 & 9.19E$-$18 &           0.3 & 1.12E+02 & 8.40E+01 & 1.39E+02 \\
  0.009 & 7.56E$-$17 & 6.49E$-$17 & 8.82E$-$17 &           0.35  & 2.03E+02 & 1.53E+02 & 2.54E+02 \\
  0.01  & 5.30E$-$16 & 4.54E$-$16 & 6.18E$-$16 &           0.4 & 3.12E+02 & 2.35E+02 & 3.89E+02 \\
  0.011 & 2.91E$-$15 & 2.49E$-$15 & 3.39E$-$15 &           0.45  & 4.29E+02 & 3.24E+02 & 5.35E+02 \\
  0.012 & 1.31E$-$14 & 1.12E$-$14 & 1.53E$-$14 &           0.5 & 5.49E+02 & 4.15E+02 & 6.83E+02 \\
  0.013 & 5.05E$-$14 & 4.32E$-$14 & 5.90E$-$14 &           0.6 & 7.83E+02 & 5.96E+02 & 9.71E+02 \\
  0.014 & 1.70E$-$13 & 1.45E$-$13 & 1.99E$-$13 &           0.7 & 1.01E+03 & 7.75E+02 & 1.25E+03 \\
  0.015 & 5.14E$-$13 & 4.39E$-$13 & 6.02E$-$13 &           0.8 & 1.24E+03 & 9.63E+02 & 1.53E+03 \\
  0.016 & 1.41E$-$12 & 1.20E$-$12 & 1.65E$-$12 &           0.9 & 1.50E+03 & 1.17E+03 & 1.82E+03 \\
  0.018 & 8.45E$-$12 & 7.19E$-$12 & 9.90E$-$12 &           1.     & 1.78E+03 & 1.41E+03 & 2.15E+03 \\
  0.02  & 3.95E$-$11 & 3.35E$-$11 & 4.63E$-$11 &           1.25  & 2.64E+03 & 2.14E+03 & 3.16E+03 \\
  0.025 & 8.70E$-$10 & 7.35E$-$10 & 1.02E$-$09 &           1.5  & 3.72E+03 & 3.05E+03 & 4.40E+03 \\
  0.03  & 9.26E$-$09 & 7.79E$-$09 & 1.09E$-$08 &           1.75  & 4.94E+03 & 4.08E+03 & 5.83E+03 \\
  0.04  & 2.96E$-$07 & 2.47E$-$07 & 3.51E$-$07 &           2.     & 6.24E+03 & 5.18E+03 & 7.36E+03 \\
  0.05  & 3.58E$-$06 & 2.96E$-$06 & 4.26E$-$06 &           2.5  & 8.87E+03 & 7.36E+03 & 1.05E+04 \\
  0.06  & 2.49E$-$05 & 2.04E$-$05 & 2.98E$-$05 &           3.     & 1.13E+04 & 9.40E+03 & 1.34E+04 \\
  0.07  & 1.28E$-$04 & 1.03E$-$04 & 1.54E$-$04 &           3.5  & 1.35E+04 & 1.12E+04 & 1.60E+04 \\
  0.08  & 5.79E$-$04 & 4.58E$-$04 & 7.04E$-$04 &           4.     & 1.54E+04 & 1.28E+04 & 1.83E+04 \\
  0.09  & 2.50E$-$03 & 1.94E$-$03 & 3.07E$-$03 &           5.     & 1.85E+04 & 1.53E+04 & 2.20E+04 \\
  0.1 & 9.84E$-$03 & 7.54E$-$03 & 1.22E$-$02 &           6.     & 2.09E+04 & 1.73E+04 & 2.48E+04 \\
  0.11  & 3.34E$-$02 & 2.54E$-$02 & 4.15E$-$02 &           7.     & 2.28E+04 & 1.90E+04 & 2.70E+04 \\
  0.12  & 9.67E$-$02 & 7.32E$-$02 & 1.20E$-$01 &           8.     & 2.47E+04 & 2.07E+04 & 2.91E+04 \\
  0.13  & 2.42E$-$01 & 1.83E$-$01 & 3.02E$-$01 &           9.     & 2.67E+04 & 2.25E+04 & 3.13E+04 \\
  0.14  & 5.36E$-$01 & 4.04E$-$01 & 6.68E$-$01 &          10.     & 2.89E+04 & 2.45E+04 & 3.36E+04 \\
  0.15  & 1.07E+00 & 8.05E$-$01 & 1.33E+00 &               &        &        &          \\

\hline
\end{tabular*}
\begin{tabular*}{\textwidth}{@{\extracolsep{\fill}} l c }
REV  = 
$ 7.02 \times 10^{10}T_{9}^{3/2}{\rm exp}(-185.18/T_{9})\,[1.0+0.5\,{\rm exp}(-24.657/T_{9})] $
 & \\
\end{tabular*}
}
\label{b11pgTab2}
\end{table}
\clearpage

\subsection{\reac{11}{B}{p}{\alpha}{8}{Be}}
\label{b11paSect}
The experimental data sets referred to in NACRE are SE65 \cite{SE65}, DA79 \cite{DA79}, BE87 \cite{BE87}, and AN93 \cite{AN93}, covering the 0.017 $\lsimeq E_{\rm cm} \lsimeq$ 3.5 MeV range.
For the \reac{11}{B}{p}{\alpha_1}{8}{Be}($E_{\rm x} \simeq$  3.04 MeV) channel, which is overwhelming, no new cross section data are  found. 
The data below $E_{\rm cm} \simeq$ 0.04 MeV appear to be contaminated by electron screening, which seems to be supported by the recent  data for the (p,$\alpha_0$) channel \cite{LA12}$^\dag$.   
[{\footnotesize{$^\dag$from $^{2}$H($^{11}$B,\,$\alpha_0${$^{8}$Be)n} (THM).}}]

\begin{figure}[hb]
\centering{
\includegraphics[height=0.50\textheight,width=0.90\textwidth]{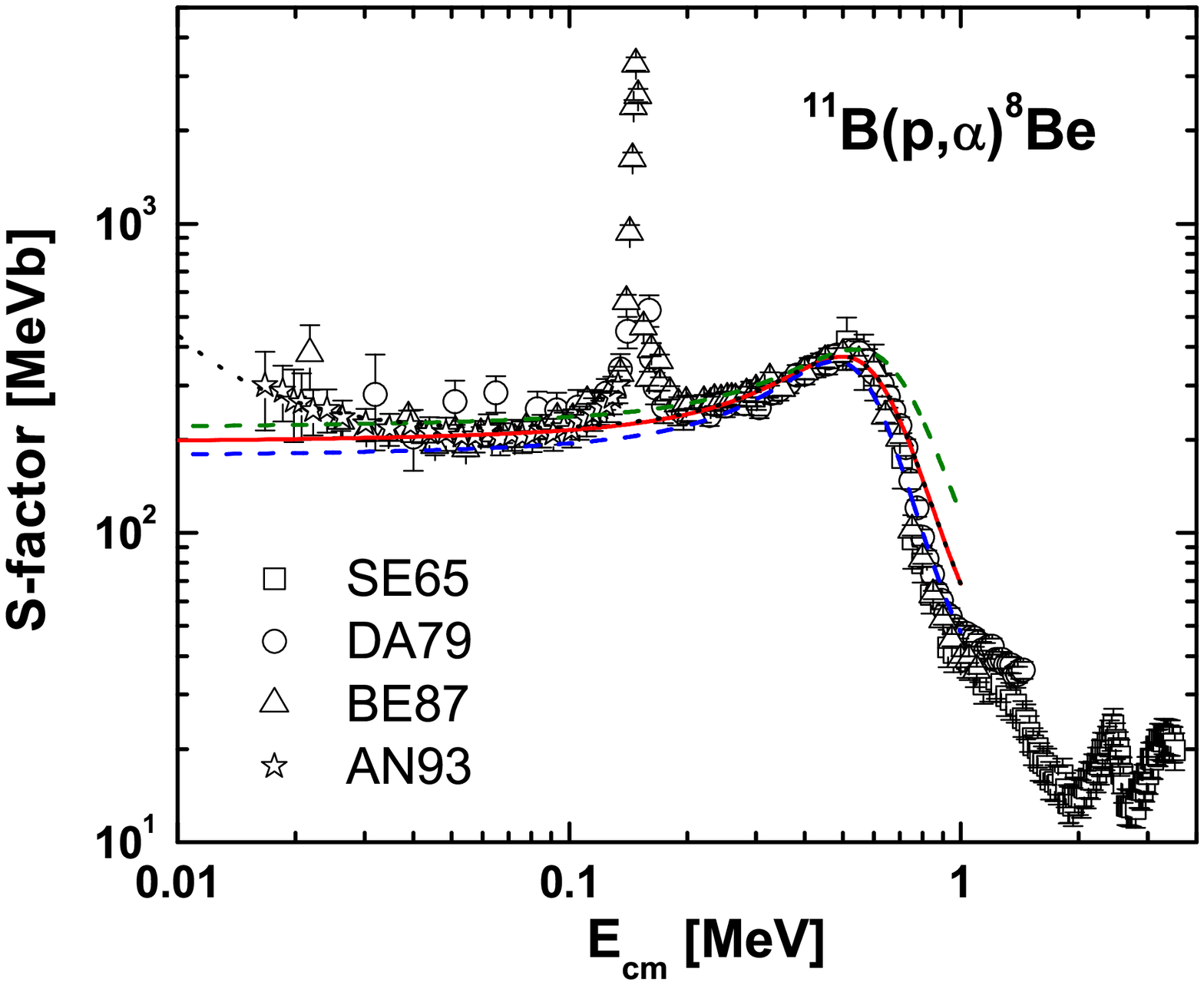}
\vspace{-0.5truecm}
\caption{The $S$-factor for \reac{11}{B}{p}{\alpha}{8}{Be}. The dotted line indicates an adiabatic screening correction ($U_{\rm e}$ = 348 eV) to the 'adopt' curve (solid line).}
\label{b11paFig1}
}
\end{figure}

Figure \ref{b11paFig1} compares the DWBA and experimental $S$-factors.
The data in the 0.04 $\lsimeq E_{\rm cm} \lsimeq$ 1 MeV range are used for the DWBA fit. They exhibit the $2^{-}$ resonance at $E_{\rm R} \simeq$  0.61 MeV.
The 0.1 $\lsimeq E_{\rm cm} \lsimeq$ 0.2 MeV range containing the narrow  $2^{+}$ resonance at $E_{\rm R} \simeq$ 0.15 MeV is excluded in the fitting procedure.
The adopted parameter values are given in Table \ref{b11paTab1}.
The present $S(0)$ = 210$^{+20}_{-30}$  MeV\,b. In comparison, $S(0)$ = 187 $\pm$ 30 MeV\,b [NACRE, from AN93]. 

\begin{figure}[t]
\centering{
\hspace{0.8truein}
\includegraphics[height=0.33\textheight,width=0.90\textwidth]{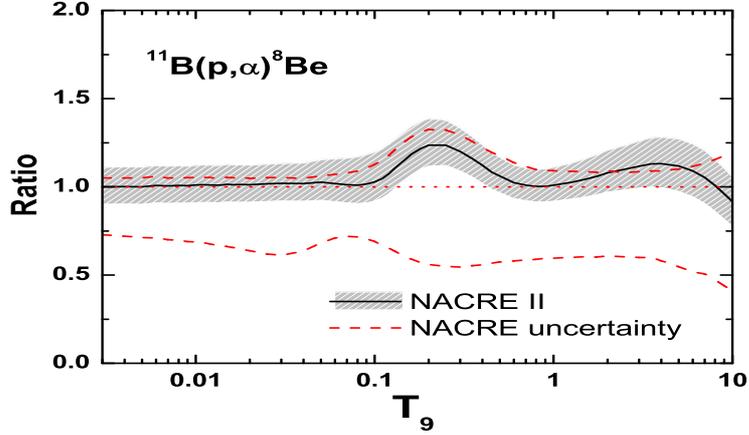}
\vspace{-0.4truecm}
\caption{\reac{11}{B}{p}{\alpha}{8}{Be} rates in units of the NACRE (adopt) values. The origin of the large (and asymmetric) uncertainty given in NACRE for the lower limits is unclear.  
}
\label{b11paFig2}
}
\end{figure}

Table \ref{b11paTab2} gives the reaction rates at $0.003 \le T_{9} \le 10$, for which the DWBA-predicted and the experimental cross sections below and above $E_{\rm cm} \simeq$ 0.05 MeV are used, respectively.
The narrow resonance at $E_{\rm R} \simeq$ 0.15 MeV with measured strength  \cite{DA79,SE61,AN74} is also taken into account. 
Figure \ref{b11paFig2} compares the present and the NACRE rates.

{\footnotesize{See \cite{RA96} for a DWBA analysis.
}} 

\clearpage

\begin{table}[hb]
\caption{\reac{11}{B}{p}{\alpha}{8}{Be} rates  in unit is $\rm{cm^{3}mol^{-1}s^{-1}}$.} \footnotesize\rm
\begin{tabular*}{\textwidth}{@{\extracolsep{\fill}}  l c c c | l c c c}
%
\hline
$T_{9}$ & adopted & low & high & $T_{9}$ & adopted & low & high \\
\hline
  0.003 & 4.90E$-$23 & 4.41E$-$23 & 5.45E$-$23 &     0.15  & 2.22E+03 & 1.99E+03 & 2.52E+03 \\
  0.004 & 8.69E$-$20 & 7.83E$-$20 & 9.67E$-$20 &     0.16  & 3.69E+03 & 3.33E+03 & 4.19E+03 \\
  0.005 & 1.77E$-$17 & 1.59E$-$17 & 1.97E$-$17 &     0.18  & 8.96E+03 & 8.08E+03 & 1.01E+04 \\
  0.006 & 1.02E$-$15 & 9.17E$-$16 & 1.13E$-$15 &     0.2 & 1.87E+04 & 1.69E+04 & 2.09E+04 \\
  0.007 & 2.58E$-$14 & 2.33E$-$14 & 2.87E$-$14 &     0.25  & 7.35E+04 & 6.67E+04 & 8.21E+04 \\
  0.008 & 3.72E$-$13 & 3.35E$-$13 & 4.13E$-$13 &     0.3 & 1.92E+05 & 1.74E+05 & 2.13E+05 \\
  0.009 & 3.53E$-$12 & 3.18E$-$12 & 3.93E$-$12 &     0.35  & 3.97E+05 & 3.60E+05 & 4.40E+05 \\
  0.01  & 2.45E$-$11 & 2.21E$-$11 & 2.73E$-$11 &     0.4 & 7.13E+05 & 6.48E+05 & 7.89E+05 \\
  0.011 & 1.34E$-$10 & 1.20E$-$10 & 1.48E$-$10 &     0.45  & 1.17E+06 & 1.06E+06 & 1.29E+06 \\
  0.012 & 5.98E$-$10 & 5.38E$-$10 & 6.64E$-$10 &     0.5 & 1.80E+06 & 1.64E+06 & 1.99E+06 \\
  0.013 & 2.28E$-$09 & 2.05E$-$09 & 2.53E$-$09 &     0.6 & 3.75E+06 & 3.41E+06 & 4.12E+06 \\
  0.014 & 7.62E$-$09 & 6.87E$-$09 & 8.47E$-$09 &     0.7 & 6.82E+06 & 6.22E+06 & 7.49E+06 \\
  0.015 & 2.28E$-$08 & 2.06E$-$08 & 2.53E$-$08 &     0.8 & 1.12E+07 & 1.02E+07 & 1.23E+07 \\
  0.016 & 6.22E$-$08 & 5.60E$-$08 & 6.90E$-$08 &     0.9 & 1.70E+07 & 1.55E+07 & 1.87E+07 \\
  0.018 & 3.66E$-$07 & 3.30E$-$07 & 4.06E$-$07 &     1.     & 2.42E+07 & 2.20E+07 & 2.67E+07 \\
  0.02  & 1.68E$-$06 & 1.51E$-$06 & 1.87E$-$06 &     1.25  & 4.71E+07 & 4.27E+07 & 5.21E+07 \\
  0.025 & 3.55E$-$05 & 3.20E$-$05 & 3.94E$-$05 &     1.5  & 7.43E+07 & 6.71E+07 & 8.25E+07 \\
  0.03  & 3.62E$-$04 & 3.26E$-$04 & 4.02E$-$04 &     1.75  & 1.03E+08 & 9.26E+07 & 1.15E+08 \\
  0.04  & 1.06E$-$02 & 9.54E$-$03 & 1.18E$-$02 &     2.     & 1.31E+08 & 1.17E+08 & 1.46E+08 \\
  0.05  & 1.16E$-$01 & 1.04E$-$01 & 1.30E$-$01 &     2.5  & 1.80E+08 & 1.61E+08 & 2.03E+08 \\
  0.06  & 7.15E$-$01 & 6.43E$-$01 & 8.14E$-$01 &     3.     & 2.20E+08 & 1.95E+08 & 2.48E+08 \\
  0.07  & 3.06E+00 & 2.74E+00 & 3.51E+00 &     3.5  & 2.51E+08 & 2.22E+08 & 2.84E+08 \\
  0.08  & 1.02E+01 & 9.13E+00 & 1.18E+01 &     4.     & 2.75E+08 & 2.42E+08 & 3.12E+08 \\
  0.09  & 2.86E+01 & 2.56E+01 & 3.33E+01 &     5.     & 3.08E+08 & 2.69E+08 & 3.51E+08 \\
  0.1 & 7.13E+01 & 6.37E+01 & 8.31E+01 &     6.     & 3.30E+08 & 2.86E+08 & 3.77E+08 \\
  0.11  & 1.62E+02 & 1.45E+02 & 1.89E+02 &     7.     & 3.46E+08 & 2.97E+08 & 3.96E+08 \\
  0.12  & 3.43E+02 & 3.06E+02 & 3.96E+02 &     8.     & 3.57E+08 & 3.04E+08 & 4.10E+08 \\
  0.13  & 6.78E+02 & 6.07E+02 & 7.80E+02 &     9.     & 3.65E+08 & 3.09E+08 & 4.20E+08 \\
  0.14  & 1.26E+03 & 1.13E+03 & 1.44E+03 &    10.     & 3.70E+08 & 3.13E+08 & 4.28E+08 \\
\hline
\end{tabular*}
\label{b11paTab2}
\end{table}
\clearpage
\subsection{\reac{11}{B}{\alpha}{n}{14}{N}}
\label{b11anSect}
This reaction is not included in NACRE, but is present in CF88. 
The experimental data sets adopted here are VA75 \cite{VA75}, and WA91 \cite{WA91}, covering the 0.4 $\lsimeq E_{\rm cm} \lsimeq$ 6 MeV range.
\cite{NI77} is superseded by WA91.

\begin{figure}[hb]
\centering{
\includegraphics[height=0.50\textheight,width=0.90\textwidth]{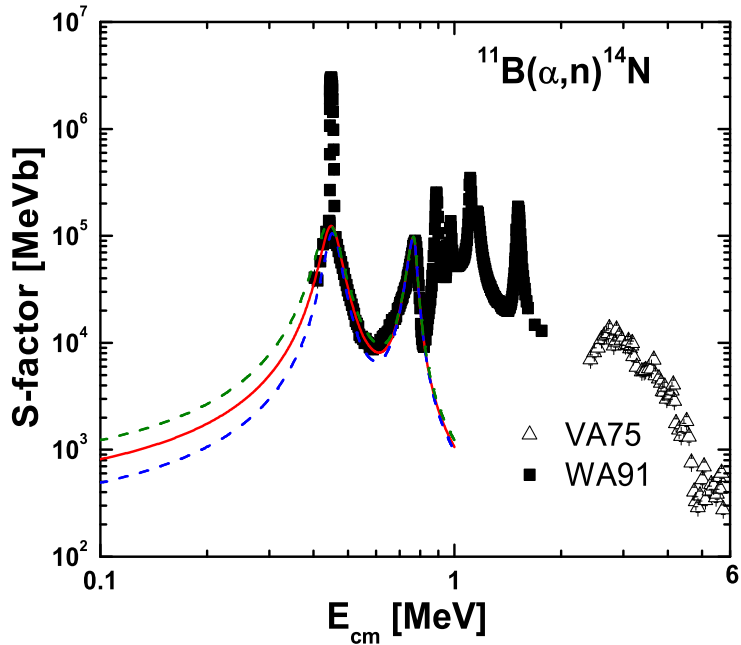}
\vspace{-0.5truecm}
\caption{The $S$-factor for \reac{11}{B}{\alpha}{n}{14}{N}. Note that the very narrow peak ($7/2^{-}$), not used for the fit, is shown overlapping with the broader $1/2^{+}$ resonance at  $E_{\rm cm} \simeq$ 0.44 MeV. 
}
\label{b11anFig1}
}
\end{figure}

Figure \ref{b11anFig1} compares the DWBA and experimental $S$-factors.
The data in the $E_{\rm cm} \lsimeq$ 0.8 MeV range are used for the DWBA fit. They exhibit the $1/2^{+}$ and $3/2^{+}$ resonances at $E_{\rm R} \simeq$ 0.45 and 0.77 MeV, the former being almost degenerate in energy with the narrow $7/2^{-}$ resonance.   
The adopted parameter values are given in Table \ref{b11anTab1}.
The present $S$(0.001 MeV) = $7.5^{+3.7}_{-3.1} \times 10^{2}$ MeV\,b.

\begin{figure}[t]
\centering{
\includegraphics[height=0.33\textheight,width=0.90\textwidth]{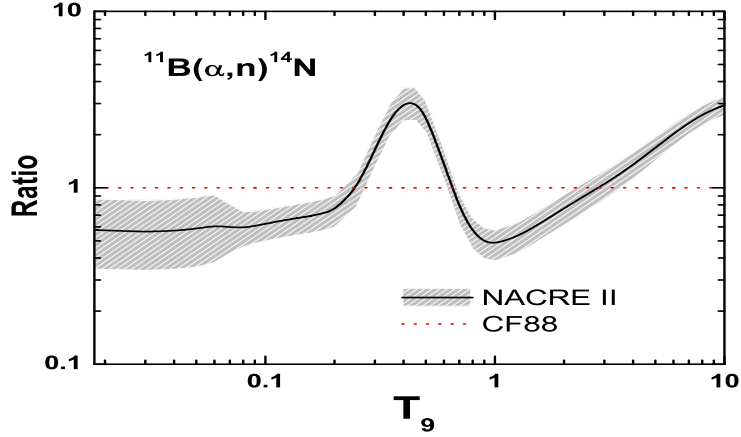}
\vspace{-0.4truecm}
\caption{\reac{11}{B}{\alpha}{n}{14}{N} rates in units of the CF88 values. The difference from CF88 stems mostly from the inclusion of the new data set WA91.
}
\label{b11anFig2}
}
\end{figure}

Table \ref{b11anTab2} gives the reaction rates at $0.03 \le T_{9} \le 10$, for which the DWBA-predicted and the experimental cross sections below and above $E_{\rm cm} \simeq$ 0.4 MeV are used, respectively. 
In addition to the one at $E_{\rm cm} \simeq$ 0.45 MeV  mentioned above, the very narrow $1/2^{-}$ resonance at $E_{\rm R}\simeq$ 0.30 MeV with measured strength \cite{WA91} is also taken into account. 
Figure \ref{b11anFig2} compares the present and the CF88 rates.

\clearpage

\begin{table}[hb]
\caption{Rates of \reac{11}{B}{\alpha}{n}{14}{N} in units of $\rm{cm^{3}mol^{-1}s^{-1}}$.} \footnotesize\rm
\begin{tabular*}{\textwidth}{@{\extracolsep{\fill}}  l c c c | l c c c}
\hline
$T_{9}$ & adopted & low & high & $T_{9}$ & adopted & low & high \\
\hline
  0.03  & 2.75E$-$26 & 1.56E$-$26 & 4.25E$-$26 &          0.5    & 7.07E$-$01 & 5.44E$-$01 & 8.82E$-$01 \\
  0.04  & 1.04E$-$22 & 5.96E$-$23 & 1.62E$-$22 &          0.6    & 3.67E+00 & 2.84E+00 & 4.56E+00 \\ 
  0.05  & 3.77E$-$20 & 2.16E$-$20 & 5.85E$-$20 &          0.7    & 1.44E+01 & 1.12E+01 & 1.78E+01 \\
  0.06  & 3.65E$-$18 & 2.14E$-$18 & 5.59E$-$18 &          0.8    & 4.95E+01 & 3.87E+01 & 6.07E+01 \\
  0.07  & 2.82E$-$16 & 1.97E$-$16 & 3.85E$-$16 &          0.9    & 1.50E+02 & 1.18E+02 & 1.84E+02 \\
  0.08  & 2.32E$-$14 & 1.80E$-$14 & 2.88E$-$14 &          1.     & 4.02E+02 & 3.17E+02 & 4.89E+02 \\
  0.09  & 9.45E$-$13 & 7.49E$-$13 & 1.15E$-$12 &          1.25   & 2.79E+03 & 2.21E+03 & 3.38E+03 \\
  0.1   & 1.89E$-$11 & 1.50E$-$11 & 2.28E$-$11 &          1.5    & 1.12E+04 & 8.87E+03 & 1.36E+04 \\
  0.11  & 2.18E$-$10 & 1.74E$-$10 & 2.62E$-$10 &          1.75   & 3.15E+04 & 2.50E+04 & 3.81E+04 \\
  0.12  & 1.66E$-$09 & 1.32E$-$09 & 1.99E$-$09 &          2.     & 7.02E+04 & 5.57E+04 & 8.50E+04 \\
  0.13  & 9.16E$-$09 & 7.32E$-$09 & 1.10E$-$08 &          2.5    & 2.26E+05 & 1.79E+05 & 2.74E+05 \\ 
  0.14  & 3.94E$-$08 & 3.14E$-$08 & 4.73E$-$08 &          3.     & 5.16E+05 & 4.07E+05 & 6.29E+05 \\
  0.15  & 1.38E$-$07 & 1.10E$-$07 & 1.66E$-$07 &          3.5    & 9.71E+05 & 7.59E+05 & 1.19E+06 \\
  0.16  & 4.14E$-$07 & 3.30E$-$07 & 4.98E$-$07 &          4.     & 1.62E+06 & 1.25E+06 & 2.00E+06 \\
  0.18  & 2.55E$-$06 & 2.03E$-$06 & 3.07E$-$06 &          5.     & 3.55E+06 & 2.71E+06 & 4.46E+06 \\
  0.2   & 1.10E$-$05 & 8.74E$-$06 & 1.33E$-$05 &          6.     & 6.35E+06 & 4.78E+06 & 8.08E+06 \\
  0.25  & 1.88E$-$04 & 1.47E$-$04 & 2.31E$-$04 &          7.     & 9.88E+06 & 7.36E+06 & 1.27E+07 \\
  0.3   & 2.00E$-$03 & 1.55E$-$03 & 2.49E$-$03 &          8.     & 1.39E+07 & 1.03E+07 & 1.79E+07 \\
  0.35  & 1.46E$-$02 & 1.12E$-$02 & 1.82E$-$02 &          9.     & 1.81E+07 & 1.33E+07 & 2.35E+07 \\ 
  0.4   & 7.10E$-$02 & 5.46E$-$02 & 8.87E$-$02 &         10.     & 2.24E+07 & 1.64E+07 & 2.91E+07 \\ 
  0.45  & 2.51E$-$01 & 1.93E$-$01 & 3.14E$-$01 \\

\hline
\end{tabular*}
\begin{tabular*}{\textwidth}{@{\extracolsep{\fill}} l c }
REV  = 
$ 3.67\,{\rm exp}(-1.835/T_{9})\,[1.0+0.5\,{\rm exp}(-24.657/T_{9})] /\,[1.0+0.333\,{\rm exp}(-26.840/T_{9})] $
 & \\
\end{tabular*}
\label{b11anTab2}
\end{table}
\clearpage
\subsection{\reac{12}{C}{p}{\gamma}{13}{N}}
\label{c12pgSect}
The experimental data sets referred to in NACRE are BA50 \cite{BA50}, HA50 \cite{HA50}, LA57a \cite{LA57a}, VO63 \cite{VO63}$^\dag$ and RO74a \cite{RO74a}$^\ddag$, covering the 0.07 $\lsimeq E_{\rm cm} \lsimeq$ 2.3 MeV range. \cite{HE60a} was apparently superseded by VO63.
Added is the post-NACRE data set BU08 \cite{BU08}.
[{\footnotesize{$^\dag$$S$-factors re-calculated in the present work from the original cross section table; $^\ddag$from d$\sigma$/d$\Omega$ at 0 and 90 degrees, used here in the $E_{\rm cm} \gsimeq$ 0.63 MeV range.}}] 

\begin{figure}[hb]
\centering{
\includegraphics[height=0.50\textheight,width=0.90\textwidth]{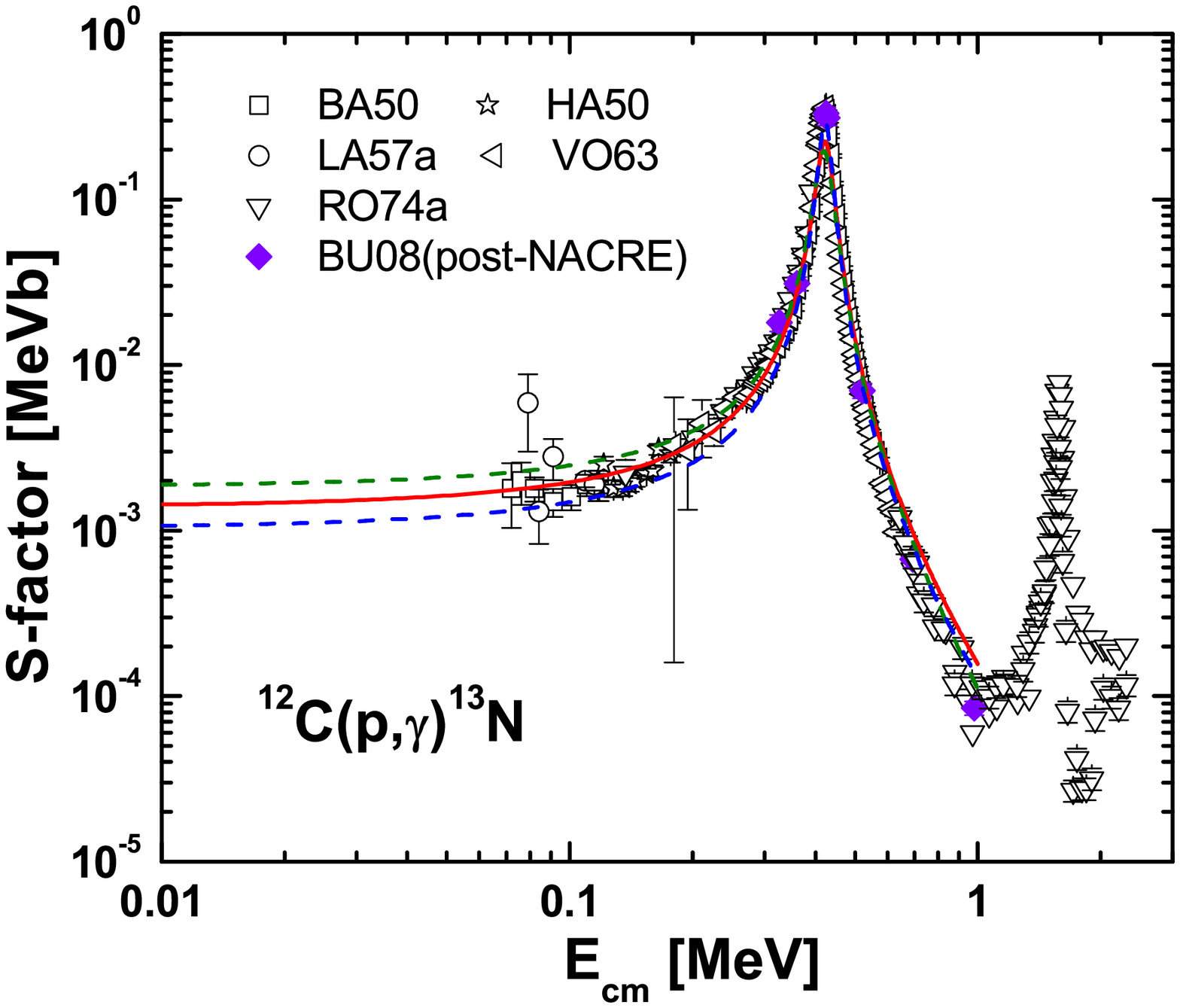}
\vspace{-0.5truecm}
\caption{The $S$-factor for \reac{12}{C}{p}{\gamma}{13}{N}. See the footnotes to VO63 and RO74a in the text.
}
\label{c12pgFig1}
}
\end{figure}

Figure \ref{c12pgFig1} compares the PM and experimental $S$-factors.
The data in the $E_{\rm cm} \lsimeq$ 1 MeV range are used for the PM fit. They exhibit the $1/2^{+}$ resonance at $E_{\rm R} \simeq$  0.42 MeV.
The adopted parameter values are given in Table \ref{c12pgTab1}.
The present $S(0)$ = 1.4 $\pm$ 0.5 keV\,b.
In comparison,  $S(0)$ = 1.45 $\pm$ 0.20 keV\,b [NACRE from \cite{RO74a}], and  2.35 keV\,b [RAD10].

\begin{figure}[t]
\centering{
\includegraphics[height=0.33\textheight,width=0.90\textwidth]{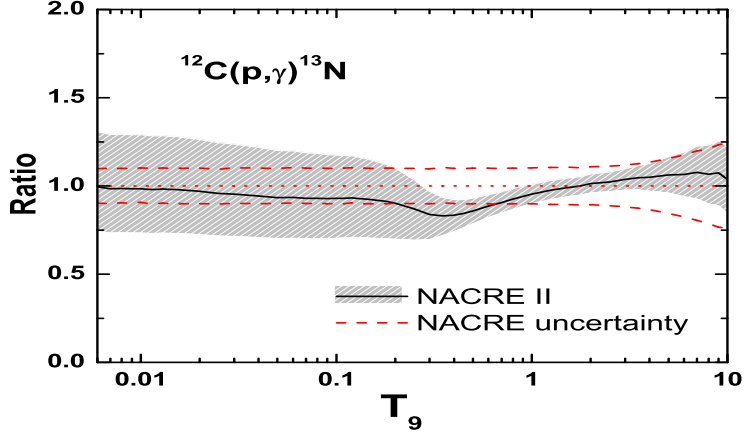}
\vspace{-0.4truecm}
\caption{\reac{12}{C}{p}{\gamma}{13}{N} rates in units of the NACRE (adopt) values. The  NACRE small uncertainty at low temperatures may not be warranted.
}
\label{c12pgFig2}
}
\end{figure}

Table \ref{c12pgTab2} gives the reaction rates at $0.006 \le T_{9} \le 10$, for which the PM-predicted and the experimental cross sections  below and above $E_{\rm cm} \simeq$ 0.2 MeV are used, respectively. 
Figure \ref{c12pgFig2} compares the present and the NACRE rates.

{\footnotesize{See \cite{DU97} for  a cluster model calculation; \cite{LA85} for a potential model fit.
}}

\clearpage

\begin{table}[hb]
\caption{\reac{12}{C}{p}{\gamma}{13}{N} rates in $\rm{cm^{3}mol^{-1}s^{-1}}$}\footnotesize\rm
\begin{tabular*}{\textwidth}{@{\extracolsep{\fill}}  l c c c | l c c c}
\hline
$T_{9}$ & adopted & low & high & $T_{9}$ & adopted & low & high \\
\hline
  0.006 & 1.21E$-$24 & 9.04E$-$25 & 1.60E$-$24 &       0.18 & 3.35E$-$03 & 2.57E$-$03 & 4.12E$-$03 \\
  0.007 & 4.75E$-$23 & 3.54E$-$23 & 6.24E$-$23 &       0.2 & 7.81E$-$03 & 6.04E$-$03 & 9.54E$-$03 \\
  0.008 & 9.85E$-$22 & 7.34E$-$22 & 1.29E$-$21 &       0.25 & 4.78E$-$02 & 3.81E$-$02 & 5.69E$-$02 \\
  0.009 & 1.28E$-$20 & 9.53E$-$21 & 1.68E$-$20 &       0.3 & 2.33E$-$01 & 1.94E$-$01 & 2.69E$-$01 \\
  0.01 & 1.16E$-$19 & 8.66E$-$20 & 1.52E$-$19 &       0.35 & 9.55E$-$01 & 8.26E$-$01 & 1.08E+00 \\
  0.011 & 7.99E$-$19 & 5.96E$-$19 & 1.05E$-$18 &       0.4 & 3.18E+00 & 2.83E+00 & 3.52E+00 \\
  0.012 & 4.40E$-$18 & 3.28E$-$18 & 5.77E$-$18 &       0.45 & 8.62E+00 & 7.79E+00 & 9.41E+00 \\
  0.013 & 2.02E$-$17 & 1.51E$-$17 & 2.65E$-$17 &       0.5 & 1.96E+01 & 1.79E+01 & 2.12E+01 \\
  0.014 & 8.01E$-$17 & 5.97E$-$17 & 1.05E$-$16 &       0.6 & 6.79E+01 & 6.30E+01 & 7.29E+01 \\
  0.015 & 2.79E$-$16 & 2.08E$-$16 & 3.65E$-$16 &       0.7 & 1.64E+02 & 1.53E+02 & 1.75E+02 \\
  0.016 & 8.75E$-$16 & 6.53E$-$16 & 1.14E$-$15 &       0.8 & 3.13E+02 & 2.93E+02 & 3.33E+02 \\
  0.018 & 6.60E$-$15 & 4.93E$-$15 & 8.62E$-$15 &       0.9 & 5.10E+02 & 4.79E+02 & 5.41E+02 \\
  0.02 & 3.76E$-$14 & 2.81E$-$14 & 4.90E$-$14 &       1. & 7.43E+02 & 7.01E+02 & 7.88E+02 \\
  0.025 & 1.23E$-$12 & 9.17E$-$13 & 1.59E$-$12 &       1.25 & 1.41E+03 & 1.33E+03 & 1.49E+03 \\
  0.03 & 1.74E$-$11 & 1.31E$-$11 & 2.26E$-$11 &       1.5 & 2.06E+03 & 1.95E+03 & 2.17E+03 \\
  0.04 & 8.35E$-$10 & 6.26E$-$10 & 1.08E$-$09 &       1.75 & 2.62E+03 & 2.48E+03 & 2.76E+03 \\
  0.05 & 1.31E$-$08 & 9.82E$-$09 & 1.68E$-$08 &       2. & 3.06E+03 & 2.90E+03 & 3.22E+03 \\
  0.06 & 1.07E$-$07 & 8.02E$-$08 & 1.37E$-$07 &       2.5 & 3.64E+03 & 3.45E+03 & 3.84E+03 \\
  0.07 & 5.70E$-$07 & 4.30E$-$07 & 7.29E$-$07 &       3. & 3.94E+03 & 3.71E+03 & 4.18E+03 \\
  0.08 & 2.28E$-$06 & 1.72E$-$06 & 2.90E$-$06 &       3.5 & 4.08E+03 & 3.81E+03 & 4.35E+03 \\
  0.09 & 7.36E$-$06 & 5.57E$-$06 & 9.34E$-$06 &       4. & 4.13E+03 & 3.82E+03 & 4.44E+03 \\
  0.1 & 2.03E$-$05 & 1.53E$-$05 & 2.56E$-$05 &       5. & 4.12E+03 & 3.73E+03 & 4.52E+03 \\
  0.11 & 4.92E$-$05 & 3.73E$-$05 & 6.20E$-$05 &       6. & 4.06E+03 & 3.58E+03 & 4.54E+03 \\
  0.12 & 1.08E$-$04 & 8.21E$-$05 & 1.36E$-$04 &       7. & 3.97E+03 & 3.42E+03 & 4.52E+03 \\
  0.13 & 2.19E$-$04 & 1.67E$-$04 & 2.75E$-$04 &       8. & 3.88E+03 & 3.27E+03 & 4.48E+03 \\
  0.14 & 4.16E$-$04 & 3.17E$-$04 & 5.19E$-$04 &       9. & 3.78E+03 & 3.12E+03 & 4.43E+03 \\
  0.15 & 7.46E$-$04 & 5.69E$-$04 & 9.28E$-$04 &      10. & 3.68E+03 & 2.99E+03 & 4.35E+03 \\
  0.16 & 1.28E$-$03 & 9.77E$-$04 & 1.58E$-$03 &                   \\
\hline
\end{tabular*}
\begin{tabular*}{\textwidth}{@{\extracolsep{\fill}} l c }
REV  = 
$ 8.85 \times 10^{9}T_{9}^{3/2}{\rm exp}(-22.554/T_{9})\,/\,[1.0+{\rm exp}(-27.445/T_{9})]$   \\
 & \\
\end{tabular*}
\label{c12pgTab2}
\end{table}
\clearpage
\subsection{\reac{12}{C}{\alpha}{\gamma}{16}{O}}
\label{c12agSect}
The experimental data sets referred to in NACRE are DY74 \cite{DY74}$^\dag$, RE87 \cite{RE87}$^{\dag,\dag\dag}$, KR88 \cite{KR88}$^\dag$ and OU96 \cite{OU96}$^{\dag}$, covering the 0.94 $\lsimeq E_{\rm cm} \lsimeq$ 3 MeV range. Added are the post-NACRE data sets RO99 \cite{RO99}$^{\dag}$, GI01 \cite{GI01}$^\dag$, KU01 \cite{KU01}$^{\dag}$, FE04 \cite{FE04}$^{\dag}$, SC05 \cite{SC05}$^{\ddag}$, AS06 \cite{AS06}$^{\dag}$, MA09 \cite{MA09}$^{\dag}$, SC11 \cite{SC11}$^{\dag\dag}$ and PL12 \cite{PL12}$^{\dag,\ddag,\dag\dag}$, extending the range to 0.89 $\lsimeq E_{\rm cm} \lsimeq$ 5 MeV.
KE82 \cite{KE82}$^{\dag\dag}$, KU02 \cite{KU02}$^{\dag\dag}$, and MA06 \cite{MA06}$^{\dag\dag}$ are also considered.
[{\footnotesize{$^\dag$to g.s.; $^{\dag\dag}$cascade; $^{\ddag}$total.}}]

\begin{figure}[hb]
\centering{
\includegraphics[height=0.50\textheight,width=0.90\textwidth]{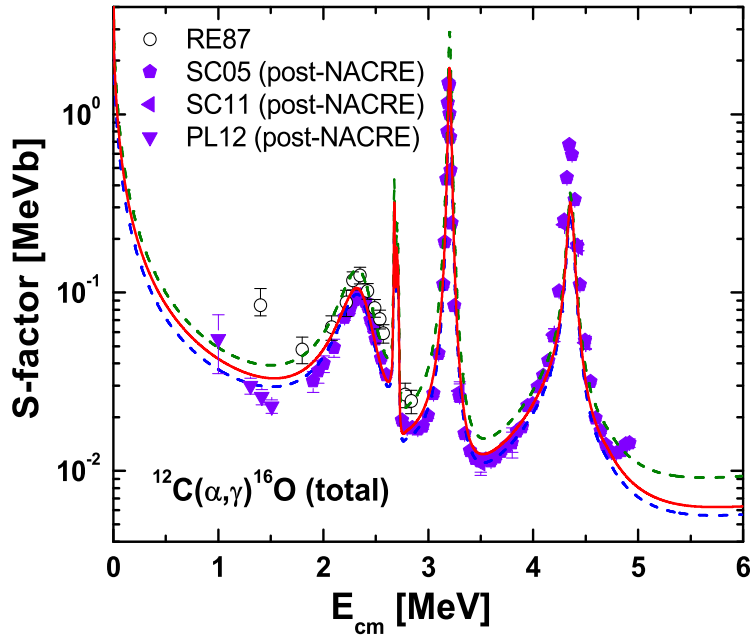}
\vspace{-0.5truecm}
\caption{The $S$-factor for \reac{12}{C}{\alpha}{\gamma}{16}{O}. RE87 and SC11 refer to the sums of the partial $S$-factors with regard to the final states (for RE87, allow for some errors owing to the reading of the graphs).
}
\label{c12agFig1a}
}
\end{figure}

\begin{figure}[t]
\centering{
\includegraphics[height=0.450\textheight,width=0.9\textwidth]{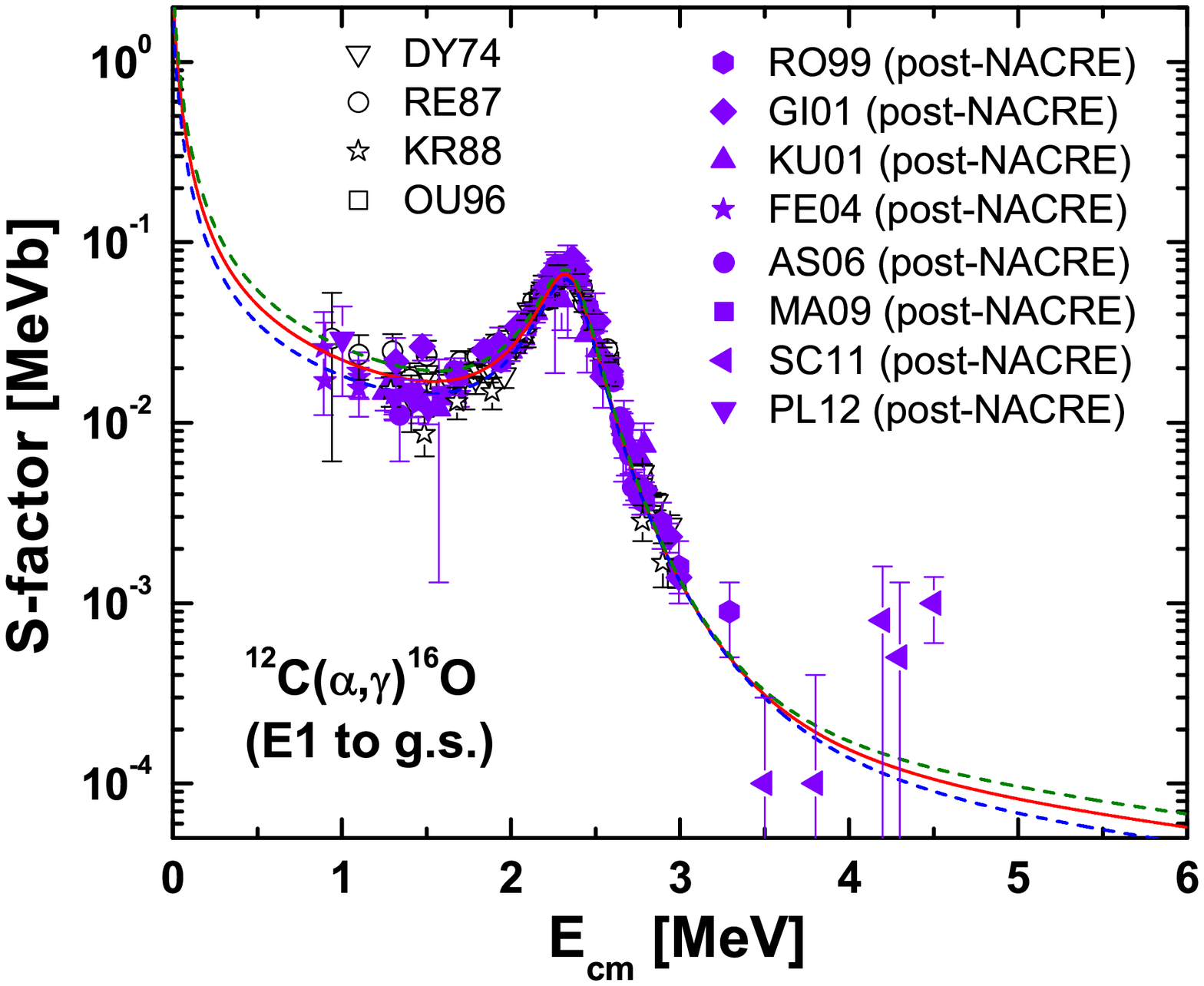}
\vspace{-0.5truecm}
\caption{The E1 $S$-factor for the \reac{12}{C}{\alpha}{\gamma}{16}{O} transition to the ground state. 
}
\label{c12agFig1b}
}
\end{figure}

\begin{figure}[t]
\centering{
\includegraphics[height=0.450\textheight,width=0.9\textwidth]{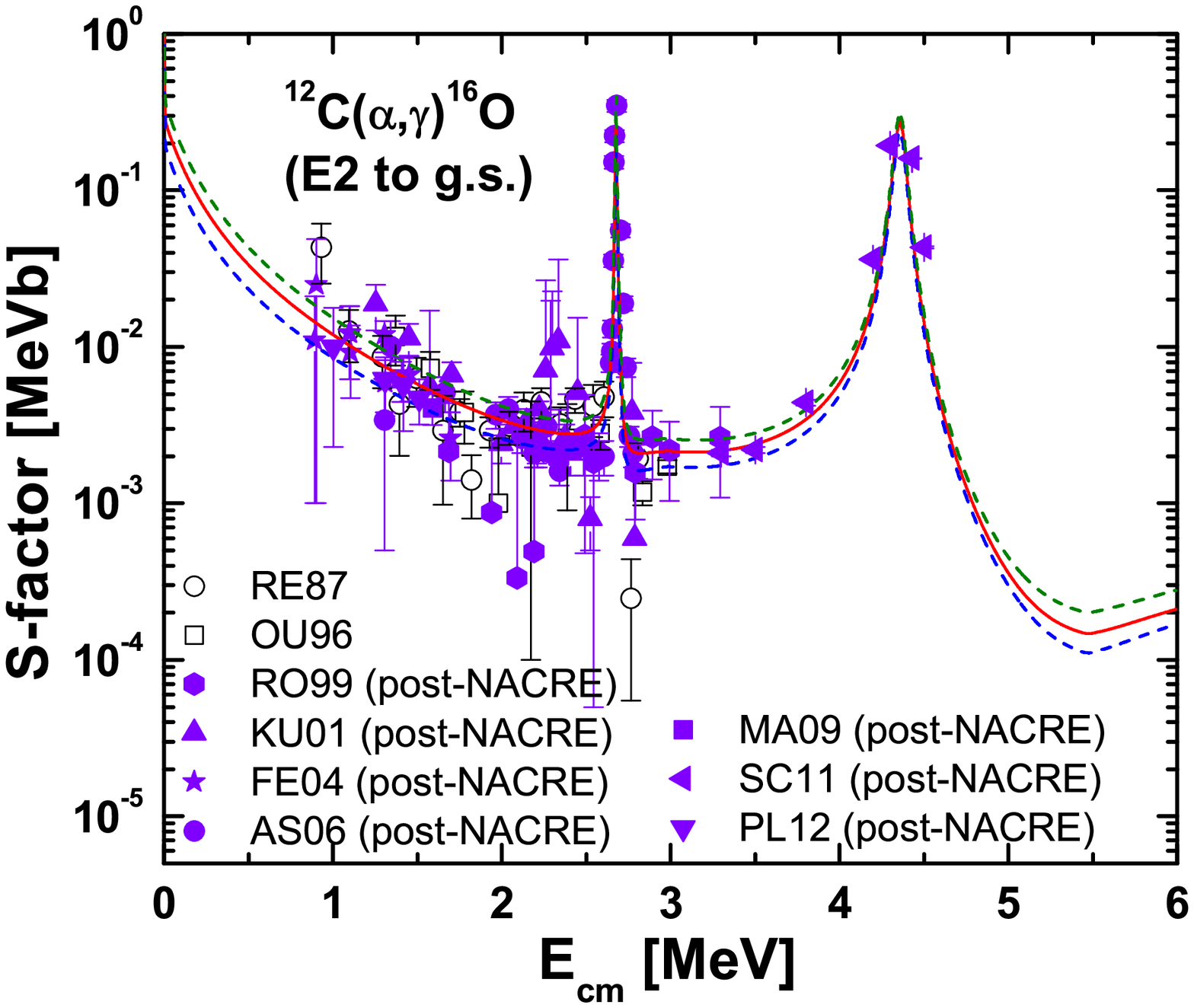}
\vspace{-0.5truecm}
\caption{The E2 $S$-factor for the \reac{12}{C}{\alpha}{\gamma}{16}{O} transition to the ground state. 
}
\label{c12agFig1c}
}
\end{figure}

\begin{figure}[t]
\centering{
\includegraphics[height=0.450\textheight,width=0.9\textwidth]{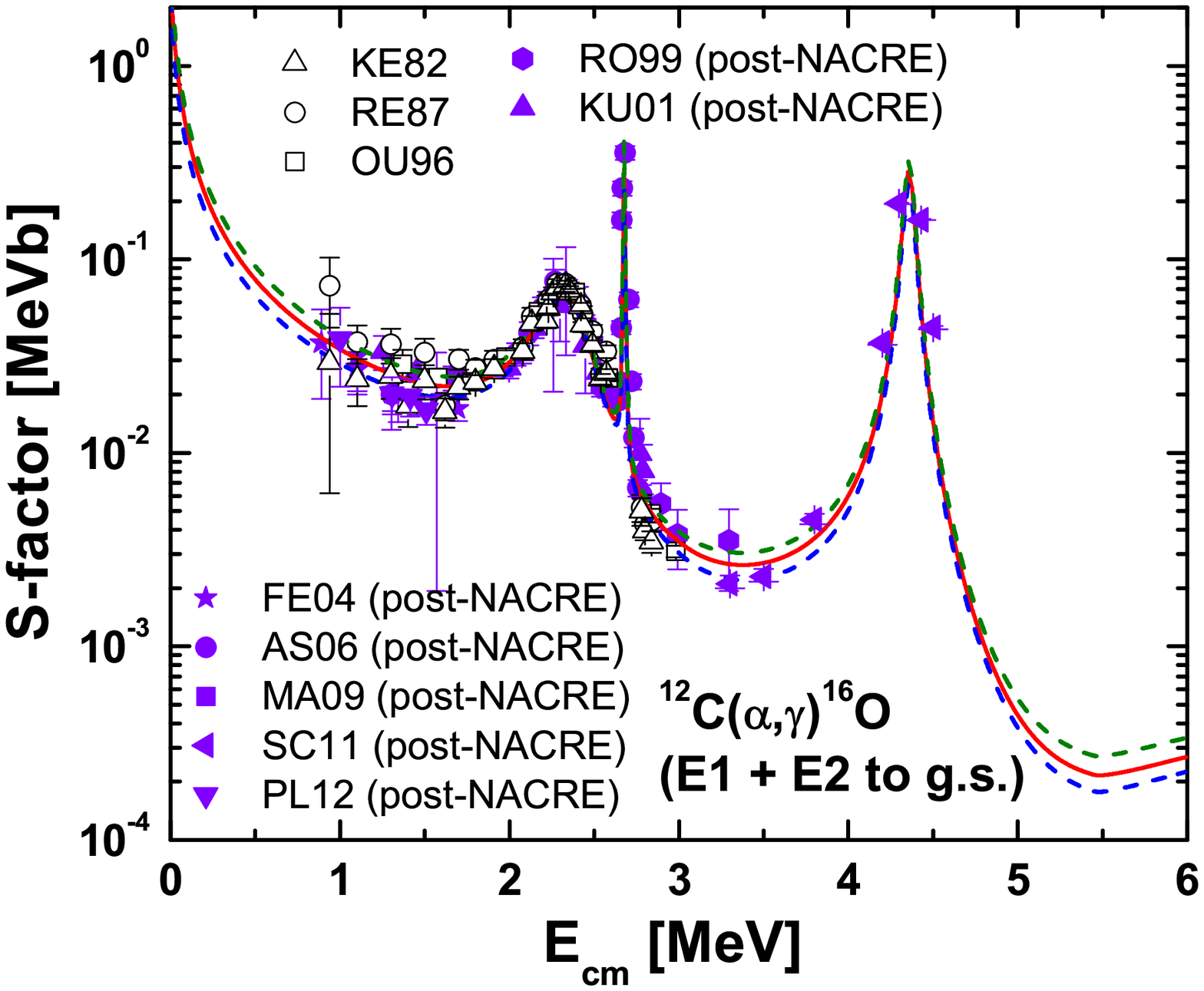}
\vspace{-0.5truecm}
\caption{The $S$-factor for the \reac{12}{C}{\alpha}{\gamma}{16}{O} E1 and E2 transitions to the ground state. 
}
\label{c12agFig1d}
}
\end{figure}

\begin{figure}[t]
\centering{
\includegraphics[height=0.450\textheight,width=0.9\textwidth]{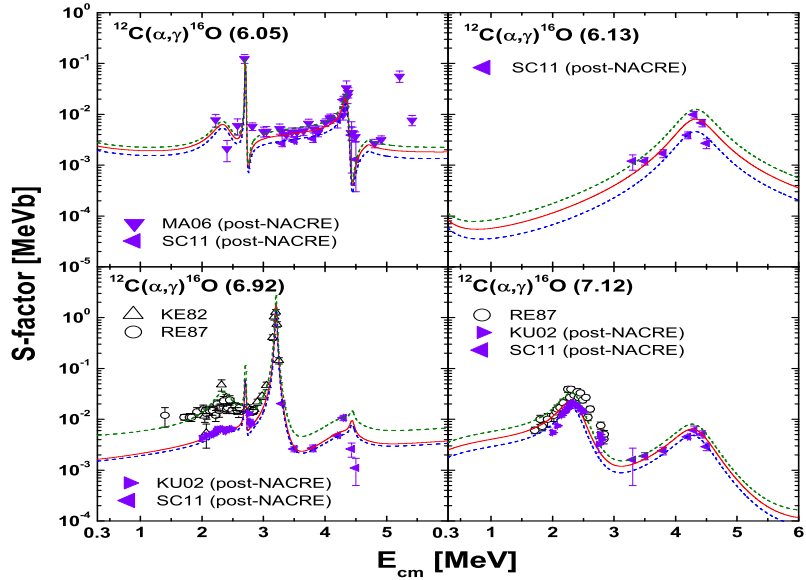}
\vspace{-0.5truecm}
\caption{The "cascade" $S$-factors for the \reac{12}{C}{\alpha}{\gamma}{16}{O}$^{*}$ transitions to the 0$^{+}$, 3$^{-}$, 2$^{+}$ and 1$^{-}$ excited states at $E_{\rm x}$ = 6.05, 6.13, 6.92 and 7.12 MeV. The present $S$-factors sum up to 6.6$_{-2.0}^{+5.3}$ and 7.3$_{-1.3}^{+8.2}$ keV\,b at $E_{\rm cm} \simeq$ 1.0 and 1.5 MeV, to be compared with 16 $\pm$ 8.5 and 7.0 $\pm$ 1.5 keV\,b of PL12 \cite{PL12}, respectively.
}
\label{c12agFig1e}
}
\end{figure}

Figure \ref{c12agFig1a} compares the PM and experimental $S$-factors.
The partial $S$-factors for $E_{\rm cm} \lsimeq$ 4.5 MeV of the transitions to the ground and the four excited states  are used for the PM fit (Figs.\,\ref{c12agFig1b}-\ref{c12agFig1e}). The $1^{-}$, $2^{+}$, $3^{-}$ and $4^{+}$ resonances at $E_{\rm R} \simeq$ 2.42, 2.68 and 4.36, 4.43, and 3.19 MeV are seen. 
Possible contributions from the $1^{-}$ and $2^{+}$ sub-threshold states at $E_{\rm R} \simeq -0.045$ and $-0.245$ MeV (and their interference to the respective resonances at 2.42 and 2.68 MeV) are also considered.
The adopted parameter values are given in Table \ref{c12agTab1}.
The present $S_{\rm E1}$(0.3 MeV) = $80 \pm 18$ keV\,b, $S_{\rm E2}$(0.3 MeV) = $61 \pm 19$ keV\,b, and $S_{\rm casc}$(0.3 MeV) = $6.5_{-2.2}^{+4.7}$ keV\,b, leading to the total $S$(0.3 MeV) = $ 148 \pm 27$ keV\,b. 
These ranges overlap with such previous estimates as [NACRE], \cite{BU06}, and \cite{SC12}.

\begin{figure}[t]
\centering{
\includegraphics[height=0.33\textheight,width=0.90\textwidth]{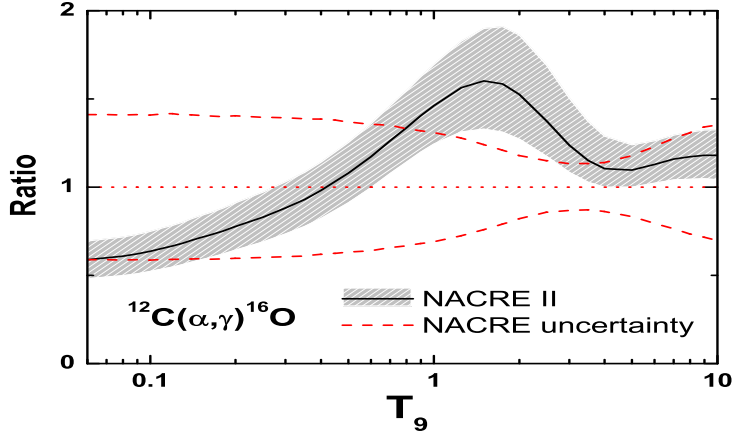}
\vspace{-0.4truecm}
\caption{\reac{12}{C}{\alpha}{\gamma}{16}{O} rates in units of the NACRE (adopt) values. The reduction at low temperatures owes mainly to the NACRE choice of a high $S_{\rm E2}$(0.3)-value (120 $\pm$ 60 keV b) taken from an R-matrix analysis \cite{BarkerKajino} of the early data. The hump at $T_{9} \simeq$ 2 results from the present inclusion of the cascade transitions. 
}
\label{c12agFig2}
}
\end{figure}

Table \ref{c12agTab2} gives the reaction rates at 0.06 $\le T_{9} \le$ 10, for which the PM-predicted and the experimental cross sections below and above $E_{\rm cm} \simeq$ 2 MeV are used, respectively. 
Several very narrow resonances in the 5.28 $\lsimeq E_{\rm cm} \lsimeq$ 5.93 MeV range with measured strengths \cite{TI93} are also taken into account.
Figure \ref{c12agFig2} compares the present and the NACRE rates.

{\footnotesize  See \cite{GA12} for a critical review. See also
 [50, 317, 321\,-\,326] 
 for indirect measurements (such as the $^{16}$N $\beta$-delayed $\alpha$-spectrum, the transfer reaction and the scattering).
}
\vspace{-1.0truecm}
\clearpage

\clearpage

\clearpage

\begin{table}[hb]
\caption{\reac{12}{C}{\alpha}{\gamma}{16}{O} rates in $\rm{cm^{3}mol^{-1}s^{-1}}$}\footnotesize\rm
\begin{tabular*}{\textwidth}{@{\extracolsep{\fill}}  l c c c | l c c c}
\hline
$T_{9}$ & adopted & low & high & $T_{9}$ & adopted & low & high \\
\hline
  0.06 & 6.00E$-$26 & 4.91E$-$26 & 7.09E$-$26 &     0.6 & 3.23E$-$08 & 2.74E$-$08 & 3.74E$-$08 \\
  0.07 & 2.98E$-$24 & 2.44E$-$24 & 3.52E$-$24 &     0.7 & 1.77E$-$07 & 1.51E$-$07 & 2.05E$-$07 \\
  0.08 & 7.43E$-$23 & 6.09E$-$23 & 8.75E$-$23 &     0.8 & 7.16E$-$07 & 6.12E$-$07 & 8.29E$-$07 \\
  0.09 & 1.12E$-$21 & 9.18E$-$22 & 1.32E$-$21 &     0.9 & 2.33E$-$06 & 1.99E$-$06 & 2.70E$-$06 \\
  0.1 & 1.15E$-$20 & 9.46E$-$21 & 1.36E$-$20 &     1. & 6.45E$-$06 & 5.49E$-$06 & 7.48E$-$06 \\
  0.11 & 8.81E$-$20 & 7.24E$-$20 & 1.04E$-$19 &     1.25 & 4.99E$-$05 & 4.19E$-$05 & 5.83E$-$05 \\
  0.12 & 5.32E$-$19 & 4.38E$-$19 & 6.26E$-$19 &     1.5 & 2.45E$-$04 & 2.03E$-$04 & 2.91E$-$04 \\
  0.13 & 2.65E$-$18 & 2.18E$-$18 & 3.12E$-$18 &     1.75 & 9.13E$-$04 & 7.56E$-$04 & 1.10E$-$03 \\
  0.14 & 1.13E$-$17 & 9.27E$-$18 & 1.32E$-$17 &     2. & 2.81E$-$03 & 2.34E$-$03 & 3.43E$-$03 \\
  0.15 & 4.18E$-$17 & 3.44E$-$17 & 4.91E$-$17 &     2.5 & 1.75E$-$02 & 1.49E$-$02 & 2.15E$-$02 \\
  0.16 & 1.38E$-$16 & 1.14E$-$16 & 1.63E$-$16 &     3. & 7.19E$-$02 & 6.30E$-$02 & 8.75E$-$02 \\
  0.18 & 1.15E$-$15 & 9.49E$-$16 & 1.35E$-$15 &     3.5 & 2.23E$-$01 & 2.00E$-$01 & 2.66E$-$01 \\
  0.2 & 7.08E$-$15 & 5.86E$-$15 & 8.30E$-$15 &     4. & 5.77E$-$01 & 5.22E$-$01 & 6.73E$-$01 \\
  0.25 & 2.67E$-$13 & 2.21E$-$13 & 3.12E$-$13 &     5. & 2.75E+00 & 2.51E+00 & 3.11E+00 \\
  0.3 & 4.19E$-$12 & 3.49E$-$12 & 4.89E$-$12 &     6. & 9.63E+00 & 8.72E+00 & 1.08E+01 \\
  0.35 & 3.75E$-$11 & 3.13E$-$11 & 4.37E$-$11 &     7. & 2.60E+01 & 2.33E+01 & 2.90E+01 \\
  0.4 & 2.27E$-$10 & 1.90E$-$10 & 2.64E$-$10 &     8. & 5.66E+01 & 5.05E+01 & 6.33E+01 \\
  0.45 & 1.03E$-$09 & 8.69E$-$10 & 1.20E$-$09 &     9. & 1.05E+02 & 9.29E+01 & 1.17E+02 \\
  0.5 & 3.80E$-$09 & 3.21E$-$09 & 4.41E$-$09 &    10. & 1.71E+02 & 1.51E+02 & 1.92E+02 \\
\hline
\end{tabular*}
\begin{tabular*}{\textwidth}{@{\extracolsep{\fill}} l c }
REV  = 
$5.14 \times 10^{10}T_{9}^{3/2}\,{\rm exp}(-83.114/T_{9}) $
 & \\
\end{tabular*}
\label{c12agTab2}
\end{table}
\clearpage
\subsection{\reac{13}{C}{p}{\gamma}{14}{N}}
\label{c13pgSect}
The experimental data sets referred to in NACRE are HE60a \cite{HE60a}$^\dag$, HE61 \cite{HE61}$^\dag$ and KI94 \cite{KI94}$^{\dag,}$$^{\dag}$$^{\dag}$, covering the 0.1 $\lsimeq E_{\rm cm} \lsimeq$ 0.9 MeV range.
Added is the post-NACRE data set GE10 \cite{GE10}$^{\dag,}$$^{\ddagger}$$^{\ddagger}$.
In the present work, HE60a is superseded by VO63 \cite{VO63}$^\dag$ (see Sect.\,\ref{c12pgSect}). 
The data point WO52 \cite{WO52}$^{\ddagger}$ is added at $E_{\rm cm}$ = 0.12 MeV. 
[{\footnotesize{$^\dag$to  g.s., $^{\dag\dag}$cascades, $^{\ddagger}$total, $^{\ddagger\ddagger}$in reverse kinematics}}]

\begin{figure}[hb]
\centering{
\includegraphics[height=0.50\textheight,width=0.90\textwidth]{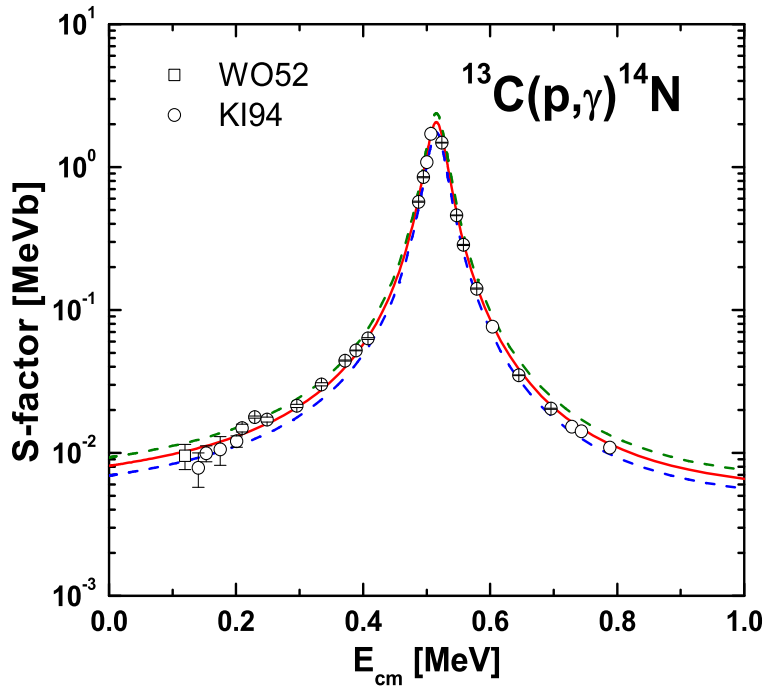}
\vspace{-0.5truecm}
\caption{The $S$-factor for \reac{13}{C}{p}{\gamma}{14}{N}.  KI94 refers to the sum of the partial contributions (allow for some errors in reading the graphs). The PM results also correspond to the sum of the partial $S$-factors shown in Fig. \ref{c13pgFig1b}. Note that in the NACRE $S$-factors for the "total" (inclusive) reaction are from KI94 and 1.18 times the HE60a and HE61 values for the ground-state transition.  
}
\label{c13pgFig1a}
}
\end{figure}

\begin{figure}[t]
\centering{
\includegraphics[height=0.9\textheight,width=0.90\textwidth]{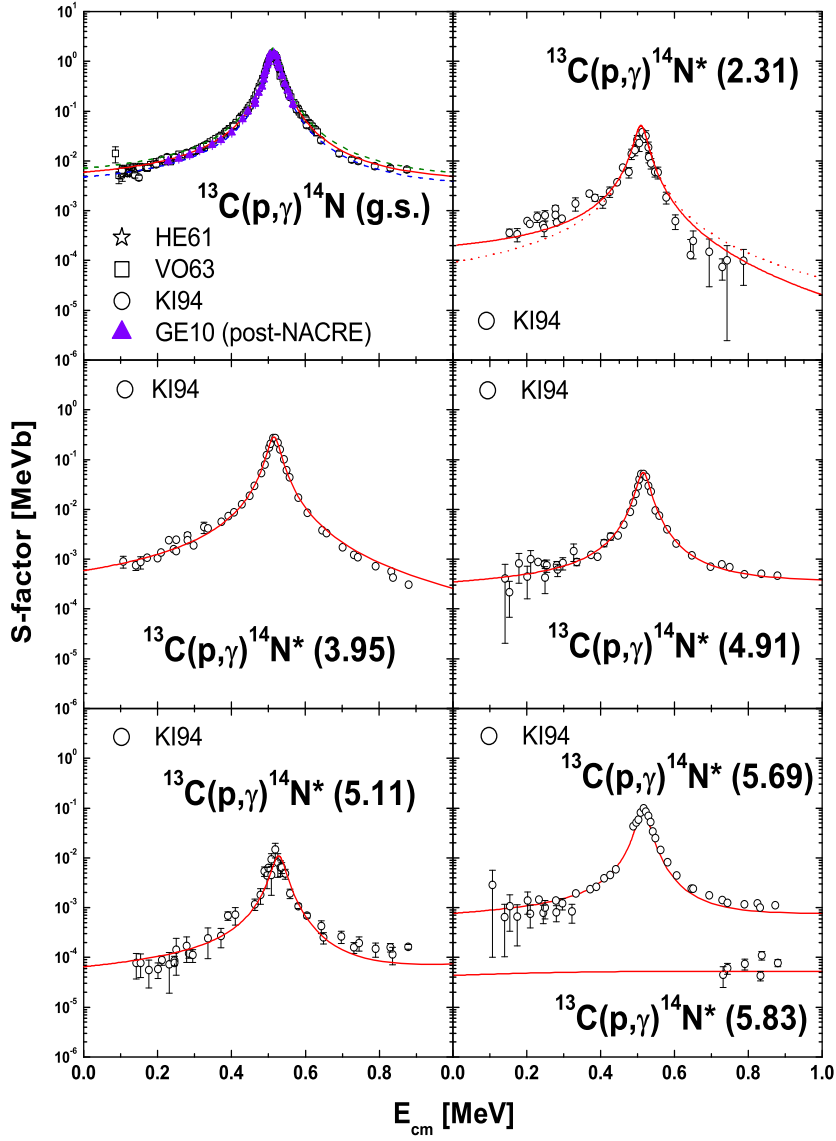}
\vspace{-0.5truecm}
\caption{The partial $S$-factors for the \reac{13}{C}{p}{\gamma}{14}{N} transitions to the $1^{+}$ ground state, and to the $0^{+}, 1^{+}, 0^{-}, 2^{-}, 1^{-}$ and $3^{-}$ excited states at $E_{\rm x}$ = 2.31, 3.95, 4.91, 5.11, 5.69 and 5.83  MeV. For the transitions to the first excited (2.31 MeV) state, the $1^{-}$ sub-threshold state at $E_{\rm R} = -1.86$ MeV and the $1^{-}$ resonance at 0.51 MeV interfere. The dotted line represents the case without the interference. The effect of the interference is negligible for the ground-state transition.  Not used in the fit , the data from the recent measurements in reverse kinematics GE10 \cite{GE10} are added for comparison.
}
\label{c13pgFig1b}
}
\end{figure}

Figure \ref{c13pgFig1a} compares the PM and experimental $S$-factors.
The partial $S$-factors in the $E_{\rm cm} \lsimeq$ 0.8 MeV range for the transitions to the ground and the six  excited states are used for the PM fit (Fig.\,\ref{c13pgFig1b}). The data exhibit the  $1^{-}$ resonance at $E_{\rm R} \simeq$  0.51  MeV. 
The GE10 data were not used in the PM fit because of their systematic deviation from the other data.
The adopted parameter values are given in Table \ref{c13pgTab1}. The present $S(0)$ = $8.1^{+1.2}_{-1.1}$ keV\,b. In comparison, $S(0)$ = 7.0 $\pm$ 1.5 keV\,b [NACRE, R-matrix plus a background], and 6.22 keV\,b [RAD10].

\begin{figure}[t]
\centering{
\includegraphics[height=0.33\textheight,width=0.90\textwidth]{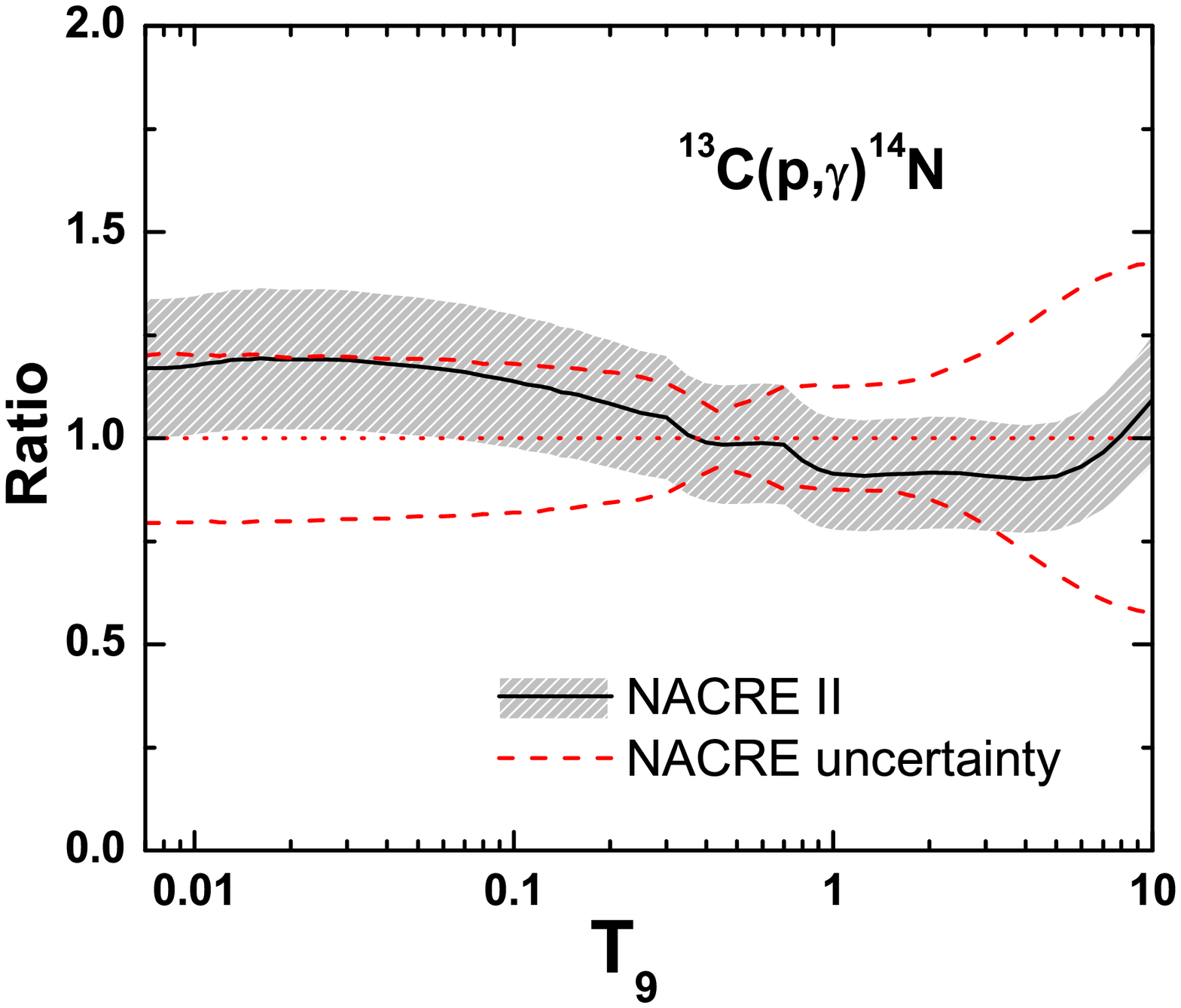}
\vspace{-0.4truecm}
\caption{\reac{13}{C}{p}{\gamma}{14}{N} rates in units of the NACRE (adopt) values. 
}
\label{c13pgFig2}
}
\end{figure}

Table \ref{c13pgTab2} gives the reaction rates at 0.007 $\le T_{9} \le$ 10, for which the PM-predicted cross sections and the measured resonance strengths in the 0.415 $\lsimeq E_{\rm cm} \lsimeq$ 2.881 MeV range 
 ([328, 331\,-\,336]) 
 are used. 
Figure \ref{c13pgFig2} compares the present and the NACRE rates.

{\footnotesize  See \cite{MU02} for an R-matrix fit.
}

\clearpage

\clearpage

\begin{table}[hb]
\caption{\reac{13}{C}{p}{\gamma}{14}{N} rates in $\rm{cm^{3}mol^{-1}s^{-1}}$}\footnotesize\rm
\begin{tabular*}{\textwidth}{@{\extracolsep{\fill}}  l c c c | l c c c}
\hline
$T_{9}$ & adopted & low & high & $T_{9}$ & adopted & low & high \\
\hline
  0.007 & 2.34E$-$22 & 2.00E$-$22 & 2.68E$-$22 &      0.18  & 1.42E$-$02 & 1.21E$-$02 & 1.63E$-$02 \\
  0.008 & 4.86E$-$21 & 4.15E$-$21 & 5.57E$-$21 &      0.2 & 3.18E$-$02 & 2.71E$-$02 & 3.64E$-$02 \\
  0.009 & 6.32E$-$20 & 5.39E$-$20 & 7.24E$-$20 &      0.25  & 1.63E$-$01 & 1.40E$-$01 & 1.87E$-$01 \\
  0.01  & 5.76E$-$19 & 4.92E$-$19 & 6.60E$-$19 &      0.3 & 6.10E$-$01 & 5.21E$-$01 & 6.99E$-$01 \\
  0.011 & 3.98E$-$18 & 3.40E$-$18 & 4.57E$-$18 &      0.35  & 1.98E+00 & 1.69E+00 & 2.26E+00 \\
  0.012 & 2.20E$-$17 & 1.88E$-$17 & 2.52E$-$17 &      0.4 & 5.97E+00 & 5.09E+00 & 6.86E+00 \\
  0.013 & 1.02E$-$16 & 8.67E$-$17 & 1.16E$-$16 &      0.45  & 1.66E+01 & 1.41E+01 & 1.91E+01 \\
  0.014 & 4.02E$-$16 & 3.44E$-$16 & 4.61E$-$16 &      0.5 & 4.14E+01 & 3.52E+01 & 4.76E+01 \\
  0.015 & 1.41E$-$15 & 1.20E$-$15 & 1.61E$-$15 &      0.6 & 1.80E+02 & 1.53E+02 & 2.07E+02 \\
  0.016 & 4.41E$-$15 & 3.76E$-$15 & 5.05E$-$15 &      0.7 & 5.32E+02 & 4.52E+02 & 6.13E+02 \\
  0.018 & 3.32E$-$14 & 2.84E$-$14 & 3.81E$-$14 &      0.8 & 1.20E+03 & 1.02E+03 & 1.38E+03 \\
  0.02  & 1.89E$-$13 & 1.62E$-$13 & 2.17E$-$13 &      0.9 & 2.24E+03 & 1.90E+03 & 2.59E+03 \\
  0.025 & 6.17E$-$12 & 5.27E$-$12 & 7.07E$-$12 &      1.     & 3.66E+03 & 3.11E+03 & 4.22E+03 \\
  0.03  & 8.77E$-$11 & 7.50E$-$11 & 1.01E$-$10 &      1.25  & 8.53E+03 & 7.24E+03 & 9.83E+03 \\
  0.04  & 4.18E$-$09 & 3.57E$-$09 & 4.79E$-$09 &      1.5  & 1.44E+04 & 1.22E+04 & 1.66E+04 \\
  0.05  & 6.49E$-$08 & 5.55E$-$08 & 7.44E$-$08 &      1.75  & 2.04E+04 & 1.73E+04 & 2.35E+04 \\
  0.06  & 5.25E$-$07 & 4.49E$-$07 & 6.01E$-$07 &      2.     & 2.58E+04 & 2.19E+04 & 2.98E+04 \\
  0.07  & 2.78E$-$06 & 2.38E$-$06 & 3.19E$-$06 &      2.5  & 3.46E+04 & 2.94E+04 & 3.98E+04 \\
  0.08  & 1.10E$-$05 & 9.42E$-$06 & 1.26E$-$05 &      3.     & 4.06E+04 & 3.45E+04 & 4.67E+04 \\
  0.09  & 3.52E$-$05 & 3.01E$-$05 & 4.04E$-$05 &      3.5  & 4.46E+04 & 3.80E+04 & 5.13E+04 \\
  0.1 & 9.60E$-$05 & 8.20E$-$05 & 1.10E$-$04 &      4.     & 4.74E+04 & 4.03E+04 & 5.45E+04 \\
  0.11  & 2.31E$-$04 & 1.97E$-$04 & 2.64E$-$04 &      5.     & 5.10E+04 & 4.34E+04 & 5.85E+04 \\
  0.12  & 5.01E$-$04 & 4.28E$-$04 & 5.74E$-$04 &      6.     & 5.35E+04 & 4.56E+04 & 6.13E+04 \\
  0.13  & 1.00E$-$03 & 8.58E$-$04 & 1.15E$-$03 &      7.     & 5.55E+04 & 4.73E+04 & 6.36E+04 \\
  0.14  & 1.88E$-$03 & 1.61E$-$03 & 2.15E$-$03 &      8.     & 5.71E+04 & 4.87E+04 & 6.55E+04 \\
  0.15  & 3.33E$-$03 & 2.84E$-$03 & 3.81E$-$03 &      9.     & 5.84E+04 & 4.97E+04 & 6.70E+04 \\
  0.16  & 5.61E$-$03 & 4.80E$-$03 & 6.43E$-$03 &     10.     & 5.93E+04 & 5.05E+04 & 6.80E+04 \\
\hline
\end{tabular*}
\begin{tabular*}{\textwidth}{@{\extracolsep{\fill}} l c }
REV  = 
$1.19 \times 10^{10}T_{9}^{3/2}\,{\rm exp}(-87.624/T_{9})\,/\,[1.0+0.333\,{\rm exp}(-26.840/T_{9})] $
 & \\
\end{tabular*}
\label{c13pgTab2}
\end{table}
\clearpage
\subsection{\reac{13}{C}{\alpha}{n}{16}{O}}
\label{c13anSect}
The experimental cross section data sets referred to in NACRE are SE67 \cite{SE67}, DA68 \cite{DA68}, BA73 \cite{BA73}, DR93 \cite{DR93} and BR93 \cite{BR93}, covering the 0.28 $\lsimeq E_{\rm cm} \lsimeq$ 4.5 MeV range.
Added are the post-NACRE data sets HA05 \cite{HA05} and HE08 \cite{HE08}, the former extending the range up to $E_{\rm cm} \simeq$ 6 MeV.

\begin{figure}[hb]
\centering{
\includegraphics[height=0.50\textheight,width=0.90\textwidth]{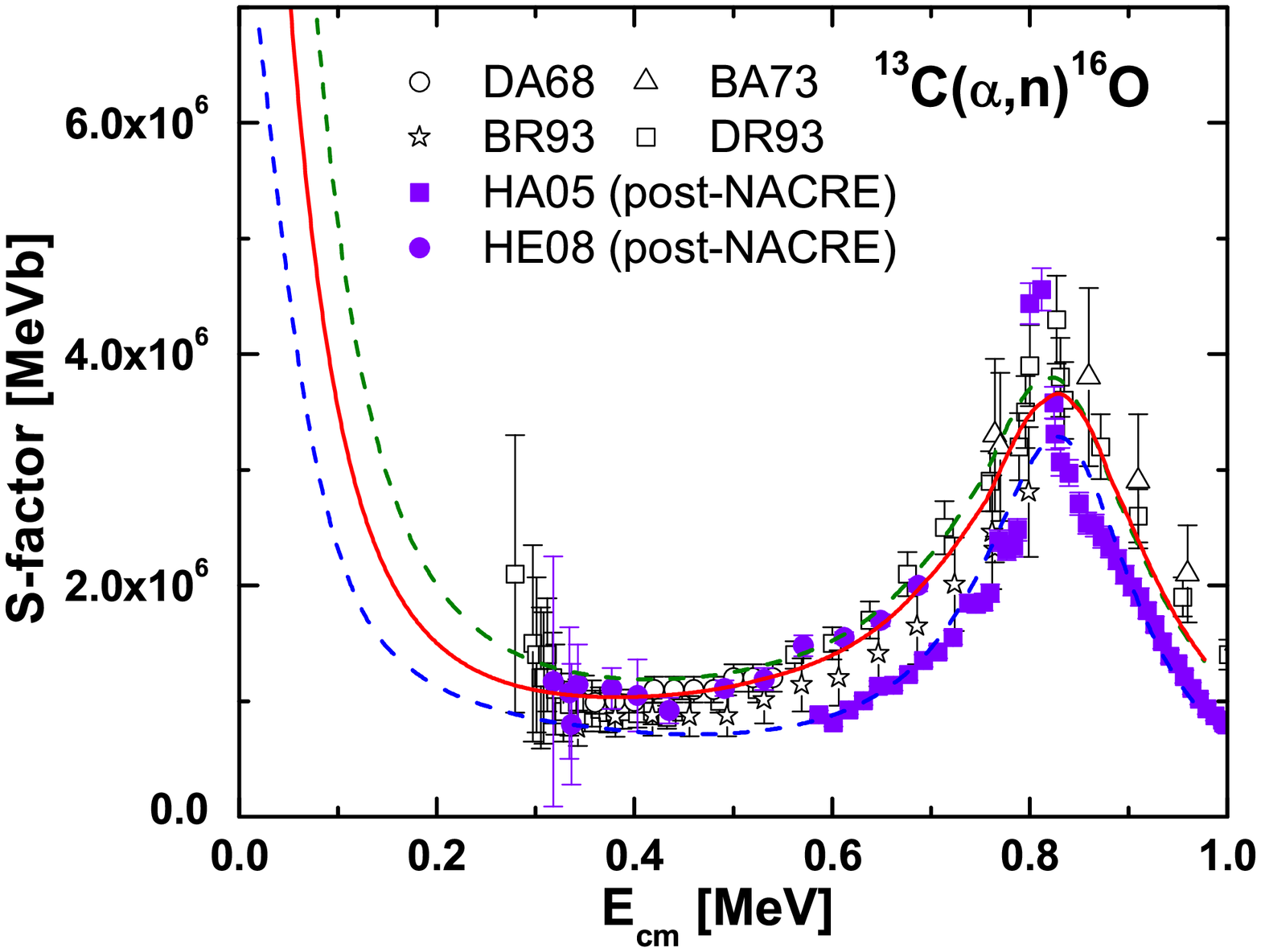}
\vspace{-0.5truecm}
\caption{The $S$-factor for \reac{13}{C}{\alpha}{n}{16}{O}.}
\label{c13anFig1a}
}
\end{figure}

\begin{figure}[t]
\centering{
\includegraphics[height=0.450\textheight,width=0.90\textwidth]{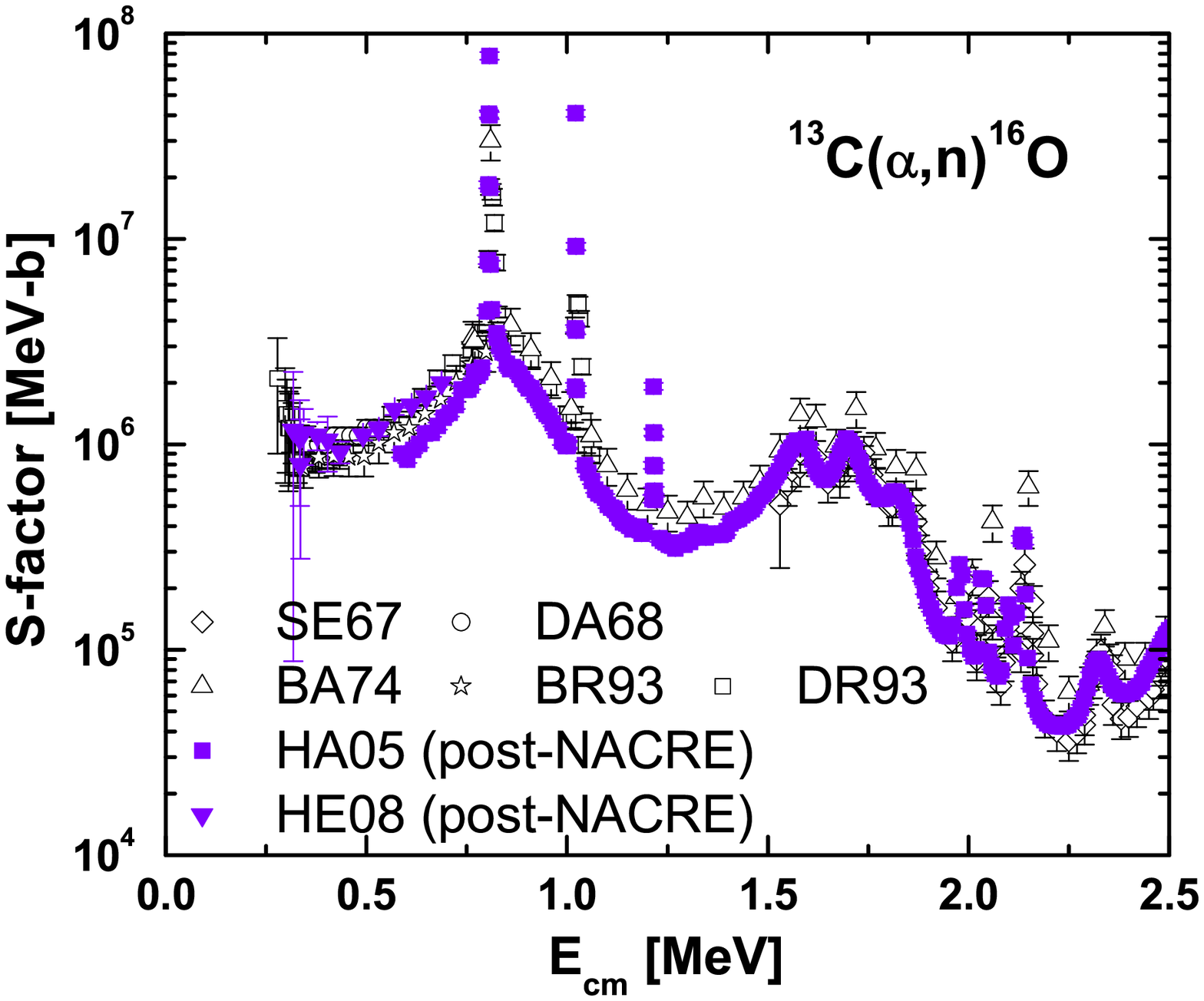}
\vspace{-0.5truecm}
\caption{Experimental $S$-factor for \reac{13}{C}{\alpha}{n}{16}{O} below $E_{\rm cm}$ = 3 MeV.}
\label{c13anFig1b}
}
\end{figure}

\begin{figure}[t]
\centering{
\includegraphics[height=0.450\textheight,width=0.90\textwidth]{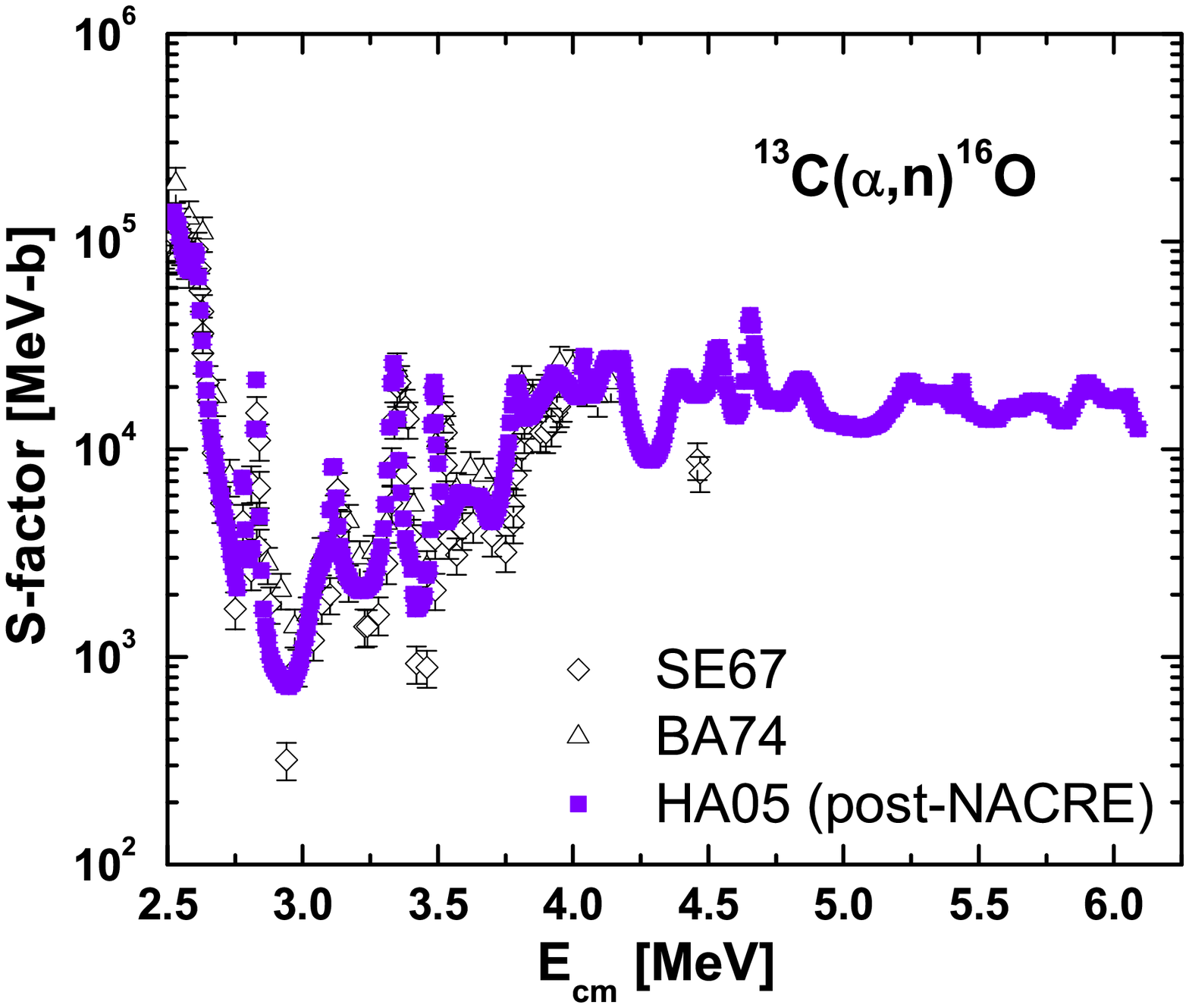}
\vspace{-0.5truecm}
\caption{Experimental $S$-factor for \reac{13}{C}{\alpha}{n}{16}{O} above $E_{\rm cm}$ = 3 MeV.}
\label{c13anFig1c}
}
\end{figure}

Figure \ref{c13anFig1a} compares the DWBA and experimental $S$-factors,  which are extended in Figs.\,\ref{c13anFig1b} and  \ref{c13anFig1c}.
The data for $E_{\rm cm} \lsimeq$ 0.9 MeV are used for the DWBA fit. They exhibit the $3/2^{+}$ resonance at $E_{\rm R} \simeq$ 0.84 MeV.
Possible contributions from the  sub-threshold states, such as the $1/2^{-}$ and $1/2^{+}$ states at $E_{\rm R} = - 0.419$ and $- 0.003$ MeV, are considered.
The adopted parameter values are given in Table \ref{c13anTab1}.
The present $S$(0.2 MeV) = $1.5^{+0.5}_{-0.4} \times 10^{6}$ MeV\,b.
In comparison, $S$(0.2 MeV) = 2.5 $\times 10^{6}$ MeV\,b [NACRE, parametrised sub-threshold contribution].

\begin{figure}[t]
\centering{
\includegraphics[height=0.33\textheight,width=0.90\textwidth]{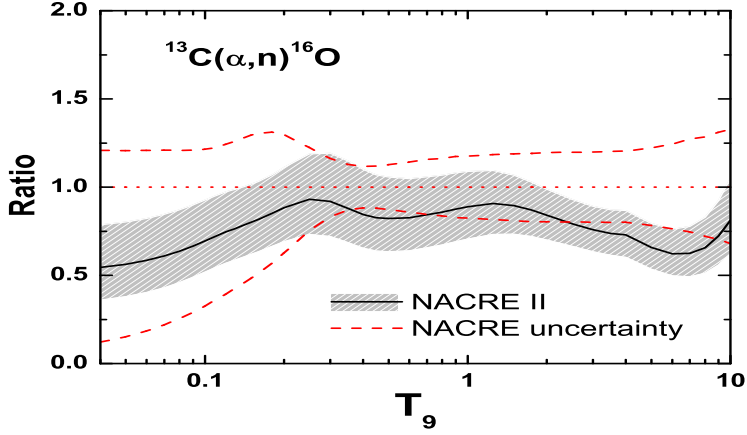}
\vspace{-0.4truecm}
\caption{\reac{13}{C}{\alpha}{n}{16}{O} rates in units of the NACRE (adopt) values.  The reduction at low temperatures owes mainly to the NACRE choice of a very steep $S$-factor assumed for the $1/2^{+}$ sub-threshold resonance contributions. For very high temperatures, we rely on the new experimental $S$ factors at high energies \cite{HA05}, whereas NACRE adopts a Hauser-Feshbach rate calculation. 
}
\label{c13anFig2}
}
\end{figure}

Table \ref{c13anTab2} gives the reaction rates at $0.04 \le T_{9} \le 10$, for which the DWBA-predicted and the experimental cross sections are used below and above $E_{\rm cm} \simeq$ 0.4 MeV, respectively.
Figure \ref{c13anFig2} compares the present and the NACRE rates.

{\footnotesize{See \cite{DU05} for a cluster model calculation; \cite{AD05} for a DWBA analysis. Also see 
 [346\,-\,351] 
 for indirect measurements by $\alpha$-transfer reactions in relation to the sub-threshold contributions.  
}}

\clearpage

\clearpage

\begin{table}[ht]
\caption{\reac{13}{C}{\alpha}{n}{16}{O} rates in units of $\rm{cm^{3}mol^{-1}s^{-1}}$.} \footnotesize\rm
\begin{tabular*}{\textwidth}{@{\extracolsep{\fill}}  l c c c | l c c c}
\hline
$T_{9}$ & adopted & low & high & $T_{9}$ & adopted & low & high \\
\hline
  0.04 & 2.91E$-$24 & 1.92E$-$24 & 4.20E$-$24 &     0.5 & 6.92E$-$02 & 5.34E$-$02 & 8.90E$-$02 \\
  0.05 & 1.82E$-$21 & 1.23E$-$21 & 2.60E$-$21 &     0.6 & 8.76E$-$01 & 6.78E$-$01 & 1.11E+00 \\
  0.06 & 2.45E$-$19 & 1.69E$-$19 & 3.44E$-$19 &     0.7 & 6.06E+00 & 4.72E+00 & 7.60E+00 \\
  0.07 & 1.22E$-$17 & 8.60E$-$18 & 1.69E$-$17 &     0.8 & 2.73E+01 & 2.14E+01 & 3.40E+01 \\
  0.08 & 3.06E$-$16 & 2.20E$-$16 & 4.19E$-$16 &     0.9 & 9.10E+01 & 7.18E+01 & 1.12E+02 \\
  0.09 & 4.65E$-$15 & 3.42E$-$15 & 6.29E$-$15 &     1. & 2.44E+02 & 1.94E+02 & 2.99E+02 \\
  0.1 & 4.86E$-$14 & 3.62E$-$14 & 6.48E$-$14 &     1.25 & 1.57E+03 & 1.27E+03 & 1.89E+03 \\
  0.11 & 3.78E$-$13 & 2.86E$-$13 & 4.98E$-$13 &     1.5 & 6.04E+03 & 4.94E+03 & 7.20E+03 \\
  0.12 & 2.33E$-$12 & 1.77E$-$12 & 3.04E$-$12 &     1.75 & 1.73E+04 & 1.43E+04 & 2.05E+04 \\
  0.13 & 1.18E$-$11 & 9.09E$-$12 & 1.53E$-$11 &     2. & 4.07E+04 & 3.39E+04 & 4.78E+04 \\
  0.14 & 5.15E$-$11 & 3.97E$-$11 & 6.58E$-$11 &     2.5 & 1.48E+05 & 1.24E+05 & 1.74E+05 \\
  0.15 & 1.96E$-$10 & 1.51E$-$10 & 2.48E$-$10 &     3. & 3.70E+05 & 3.09E+05 & 4.35E+05 \\
  0.16 & 6.63E$-$10 & 5.15E$-$10 & 8.38E$-$10 &     3.5 & 7.24E+05 & 6.03E+05 & 8.55E+05 \\
  0.18 & 5.77E$-$09 & 4.49E$-$09 & 7.25E$-$09 &     4. & 1.21E+06 & 9.99E+05 & 1.44E+06 \\
  0.2 & 3.72E$-$08 & 2.91E$-$08 & 4.68E$-$08 &     5. & 2.54E+06 & 2.06E+06 & 3.06E+06 \\
  0.25 & 1.58E$-$06 & 1.24E$-$06 & 2.02E$-$06 &     6. & 4.37E+06 & 3.49E+06 & 5.35E+06 \\
  0.3 & 2.89E$-$05 & 2.26E$-$05 & 3.74E$-$05 &     7. & 6.87E+06 & 5.42E+06 & 8.55E+06 \\
  0.35 & 3.14E$-$04 & 2.45E$-$04 & 4.10E$-$04 &     8. & 1.04E+07 & 8.09E+06 & 1.31E+07 \\
  0.4 & 2.43E$-$03 & 1.88E$-$03 & 3.17E$-$03 &     9. & 1.53E+07 & 1.18E+07 & 1.96E+07 \\
  0.45 & 1.45E$-$02 & 1.12E$-$02 & 1.88E$-$02 &    10. & 2.21E+07 & 1.69E+07 & 2.86E+07 \\
\hline
\end{tabular*}
\begin{tabular*}{\textwidth}{@{\extracolsep{\fill}} l c }
REV  = 
$ 5.79\,{\rm exp}(-25.712/T_{9}) $
 & \\
\end{tabular*}
\label{c13anTab2}
\end{table}
\clearpage
\subsection{\reac{13}{N}{p}{\gamma}{14}{O}}
\label{n13pgSect}
No experimental cross section data are found. The rates are dominated by the  contribution of the 1$^{-}$ resonance at $E_{\rm R} \simeq$  0.53 MeV.
NACRE refers to 
 [352\,-\,354] 
 for its total width, to \cite{FE89,SM93} for the $\gamma$-branching ratio, and to  \cite{DE93a} for the $\gamma$-width.  The Coulomb break-up measurements \cite{MO91,KI93} are rejected.

\begin{figure}[hb]
\centering{
\includegraphics[height=0.50\textheight,width=0.90\textwidth]{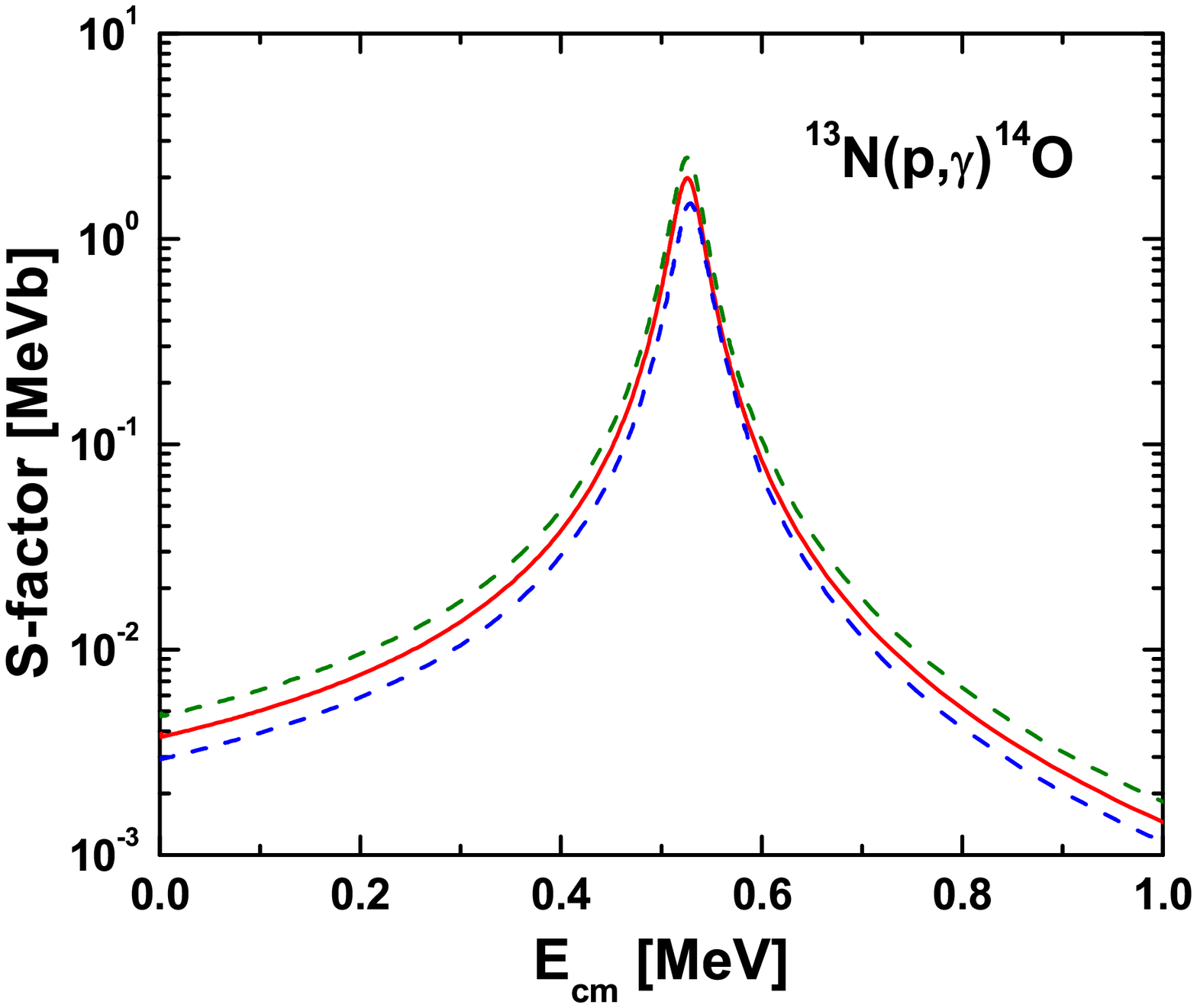}
\vspace{-0.5truecm}
\caption{The $S$-factor of \reac{13}{N}{p}{\gamma}{14}{O}.}
\label{n13pgFig1}
}
\end{figure}

Figure \ref{n13pgFig1} presents the PM prediction of the $S$-factors.
The resonance energy and the widths of that $1^{-}$ state, and their uncertainties are taken from NACRE in order to fix the height of the resonance to normalise the PM $S$-factors.
The non-resonant contributions are estimated from the same nuclear potentials as for \reac{13}{C}{p}{\gamma}{14}{N}, and turn out to be insignificant at low temperatures. Note that the s-wave ($0^{-}$) transition is forbidden to the $^{14}$O ground state, save the unlikely possibility of very strong (non-resonant) cascade transitions via the $1^{-}$ state from the higher energy range.  Similarly, the E1 cascade transition, possibly from the third excited ($0^{+}$) state, is not included here.  
The adopted parameter values are given in Table \ref{n13pgTab1}.
The present $S(0)$ = 3.8$_{-0.8}^{+1.0}$ keV\,b. 
In comparison, $S(0)$ =  5.77 keV\,b [RAD10].

\begin{figure}[t]
\centering{
\includegraphics[height=0.33\textheight,width=0.90\textwidth]{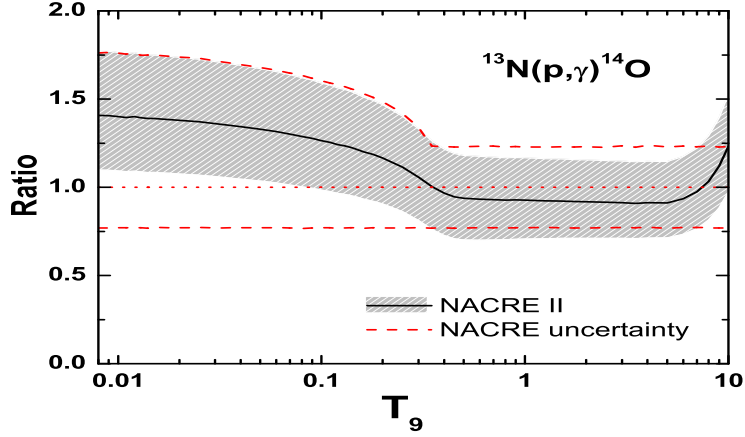}
\vspace{-0.4truecm}
\caption{\reac{13}{N}{p}{\gamma}{14}{O} rates in units of the NACRE (adopt) values. In the $1 \lsimeq T_{9} \lsimeq 5$ range, the rates are basically determined by the 1$^{-}$ resonance. The present rates in that range are slightly lower than NACRE, but agrees very well with those derived from the original experimental study \cite{DE93b}. 
The coincidence of the upper limits at low temperatures is accidental since NACRE (as well as \cite{DE93b}) introduces an unwarranted "interference" term (see \cite{Descbook}). At very high temperatures, some effects of the difference in the assumed non-resonant contributions are seen.
} 
\label{n13pgFig2}
}
\end{figure}

Table \ref{n13pgTab2} gives the reaction rates at 0.008 $\le T_{9} \le$ 10.
Figure \ref{n13pgFig2} compares the present and the NACRE rates.

{\footnotesize{See \cite{DE99} for a cluster model calculation; \cite{LA85} for a potential model prediction. 
}}

\clearpage

\begin{table}[hb]
\caption{\reac{13}{N}{p}{\gamma}{14}{O} rates in $\rm{cm^{3}mol^{-1}s^{-1}}$.}\footnotesize\rm
\begin{tabular*}{\textwidth}{@{\extracolsep{\fill}}  l c c c | l c c c}
\hline
$T_{9}$ & adopted & low & high & $T_{9}$ & adopted & low & high \\
\hline
   0.008&    1.48E$-$24&    1.15E$-$24&    1.86E$-$24&   0.2&    1.56E$-$03&    1.21E$-$03&    1.97E$-$03\\
   0.009&    2.56E$-$23&    2.00E$-$23&    3.22E$-$23&   0.25&    1.03E$-$02&    7.96E$-$03&    1.30E$-$02\\
   0.01&    2.98E$-$22&    2.33E$-$22&    3.76E$-$22&   0.3&    4.87E$-$02&    3.74E$-$02&    6.14E$-$02\\
   0.011&    2.55E$-$21&    1.99E$-$21&    3.22E$-$21&   0.35&    2.03E$-$01&    1.54E$-$01&    2.56E$-$01\\
   0.012&    1.71E$-$20&    1.33E$-$20&    2.15E$-$20&   0.4&    7.71E$-$01&    5.80E$-$01&    9.71E$-$01\\
   0.013&    9.34E$-$20&    7.28E$-$20&    1.18E$-$19&   0.45&    2.53E+00&    1.89E+00&    3.18E+00\\
   0.014&    4.32E$-$19&    3.37E$-$19&    5.44E$-$19&   0.5&    6.99E+00&    5.23E+00&    8.80E+00\\
   0.015&    1.74E$-$18&    1.36E$-$18&    2.19E$-$18&   0.6&    3.42E+01&    2.57E+01&    4.31E+01\\
   0.016&    6.20E$-$18&    4.84E$-$18&    7.81E$-$18&   0.7&    1.08E+02&    8.14E+01&    1.36E+02\\
   0.018&    5.89E$-$17&    4.60E$-$17&    7.42E$-$17&   0.8&    2.53E+02&    1.92E+02&    3.19E+02\\
   0.02&    4.10E$-$16&    3.19E$-$16&    5.16E$-$16&   0.9&    4.85E+02&    3.69E+02&    6.11E+02\\
   0.025&    1.99E$-$14&    1.55E$-$14&    2.50E$-$14&   1.&    8.07E+02&    6.16E+02&    1.02E+03\\
   0.03&    3.84E$-$13&    2.99E$-$13&    4.83E$-$13&   1.25&    1.94E+03&    1.49E+03&    2.44E+03\\
   0.04&    2.86E$-$11&    2.23E$-$11&    3.60E$-$11&   1.5&    3.33E+03&    2.56E+03&    4.19E+03\\
   0.05&    6.14E$-$10&    4.78E$-$10&    7.73E$-$10&   1.75&    4.74E+03&    3.67E+03&    5.97E+03\\
   0.06&    6.37E$-$09&    4.96E$-$09&    8.01E$-$09&   2.&    6.03E+03&    4.68E+03&    7.60E+03\\
   0.07&    4.13E$-$08&    3.21E$-$08&    5.19E$-$08&   2.5&    8.05E+03&    6.26E+03&    1.01E+04\\
   0.08&    1.93E$-$07&    1.50E$-$07&    2.43E$-$07&   3.&    9.31E+03&    7.25E+03&    1.17E+04\\
   0.09&    7.13E$-$07&    5.55E$-$07&    8.97E$-$07&   3.5&    1.00E+04&    7.80E+03&    1.26E+04\\
   0.1&    2.20E$-$06&    1.71E$-$06&    2.77E$-$06&   4.&    1.03E+04&    8.05E+03&    1.30E+04\\
   0.11&    5.90E$-$06&    4.59E$-$06&    7.42E$-$06&   5.&    1.03E+04&    8.07E+03&    1.30E+04\\
   0.12&    1.41E$-$05&    1.10E$-$05&    1.78E$-$05&   6.&    1.00E+04&    7.85E+03&    1.26E+04\\
   0.13&    3.10E$-$05&    2.41E$-$05&    3.90E$-$05&   7.&    9.74E+03&    7.66E+03&    1.22E+04\\
   0.14&    6.30E$-$05&    4.90E$-$05&    7.93E$-$05&   8.&    9.62E+03&    7.58E+03&    1.20E+04\\
   0.15&    1.20E$-$04&    9.34E$-$05&    1.51E$-$04&   9.&    9.67E+03&    7.62E+03&    1.20E+04\\
   0.16&    2.17E$-$04&    1.69E$-$04&    2.74E$-$04&  10.&    9.85E+03&    7.78E+03&    1.22E+04\\
   0.18 & 6.25E$-$04 & 4.85E$-$04 & 7.86E$-$04\\
\hline
\end{tabular*}
\begin{tabular*}{\textwidth}{@{\extracolsep{\fill}} l c }
REV  = 
 $ 3.57 \times 10^{10}T_{9}^{3/2}\,{\rm exp}(-53.697/T_{9})\,[1.0+\,{\rm exp}(-27.445/T_{9})] $
 & \\
\end{tabular*}
\label{n13pgTab2}
\end{table}
\clearpage
\subsection{\reac{14}{N}{p}{\gamma}{15}{O}}
\label{n14pgSect}
The experimental cross section data sets referred to in NACRE are LA57b \cite{LA57b}, PI57 \cite{PI57}, HE63 \cite{HE63} and SC87 \cite{SC87}$^\dag$, covering the 0.20 $\lsimeq E_{\rm cm} \lsimeq$ 3.3  MeV range$^\ddag$.  
Added are the post-NACRE data sets FO04 \cite{FO04}$^{\dag\dag}$ IM05 \cite{IM05}, RU05 \cite{RU05}, BE06b \cite{BE06b}, LE06b \cite{LE06b} and MA08 \cite{MA08}$^{\dag\dag}$, extending the range down to $E_{\rm cm} \simeq 0.07$ MeV.
[{\footnotesize{$^\dag$to be corrected for summing effects \cite{TR05};  $^\ddag$For $E_{\rm cm} \gsimeq 1$ MeV, NACRE used resonance strengths; $^{\dag\dag}$to g.s. only.}}]

\begin{figure}[hb]
\centering{
\includegraphics[height=0.50\textheight,width=0.9\textwidth]{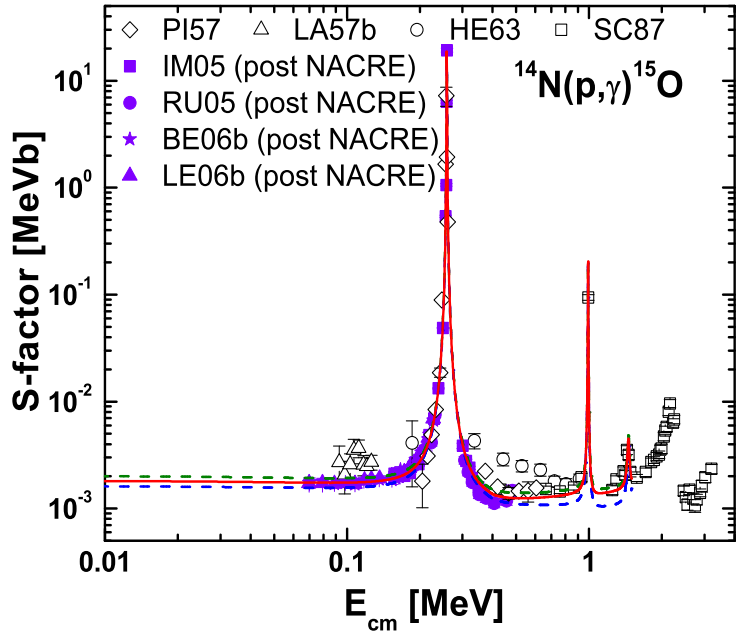}
\vspace{-0.5truecm}
\caption{The total $S$-factor for \reac{14}{N}{p}{\gamma}{15}{O}.  The data sets SC87, IM05 and RU05 are the sums of the partial $S$-factors shown in Figs.\,\ref{n14pgFig1b} and \ref{n14pgFig1c}. In the $E_{\rm cm} \simeq$ 0.3 MeV range, the PM fits are performed with the data shown here, whereas the sums of the partial $S$-factors are used at higher energies. See SC87 for data in the high-energy range.}
\label{n14pgFig1a}
}
\end{figure}

\begin{figure}[hb]
\centering{
\includegraphics[height=0.7\textheight,width=0.95\textwidth]{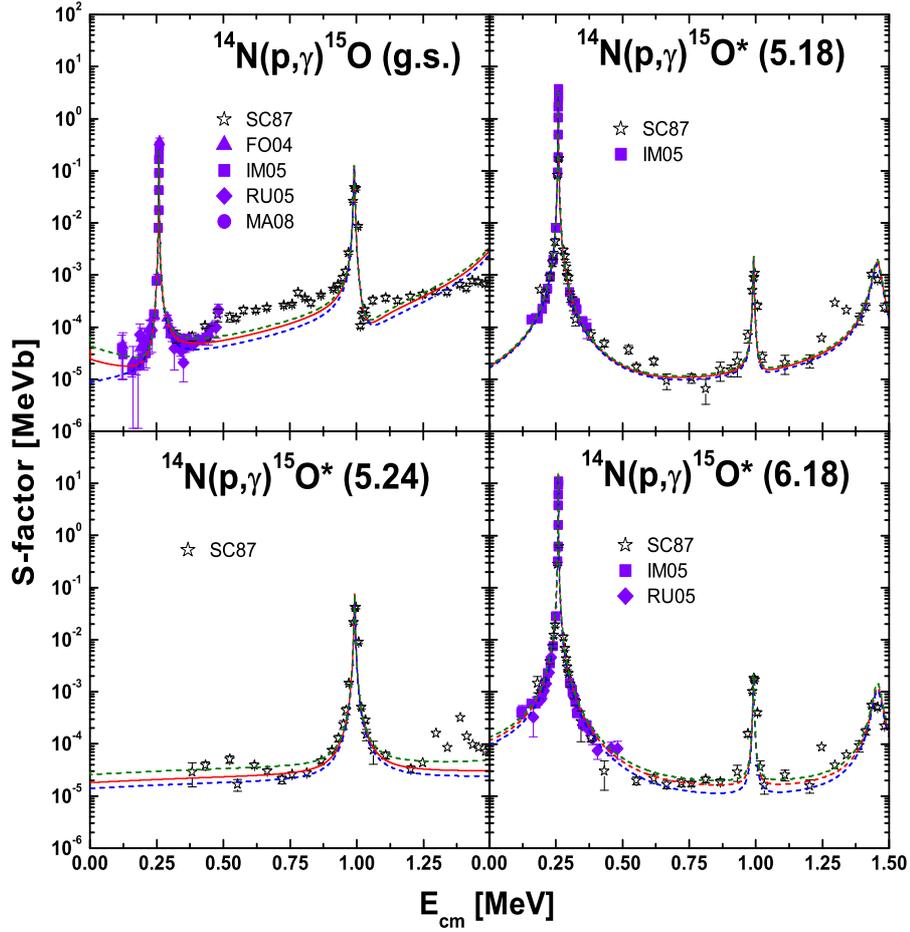}
\caption{The partial $S$-factors for the \reac{14}{N}{p}{\gamma}{15}{O} transitions to the $1/2^{-}$ ground state, and to the $1/2^{+}$, $5/2^{+}$
 and $3/2^{-}$ excited states at $E_{\rm x}$ = 5.183, 5.241 and 6.176 MeV, respectively.  
For the ground-state transitions, PM clearly fails to reproduce the general trend of SC87 toward high energies, unless a twist of the potential parameter values were made. [In R-matrix fits, this trouble is avoided by the addition of a background pole of choice, typically at 5 $\sim$ 6 MeV with a width of comparable magnitude.] 
 The present $S_{\rm g.s.}(0)$ = 0.04 $\pm$ 0.03 keV\,b is on the lower side of the R-matrix values such as $0.08_{-0.06}^{+0.13}$ keV\,b \cite{AN01} and  0.15 $\pm$ 0.07 keV\,b \cite{MU03} in contrast, e.g., to $0.20 \pm{0.05}$ keV\,b \cite{MA08}.
}
\label{n14pgFig1b}
}
\end{figure}

\begin{figure}[hb]
\centering{
\includegraphics[height=0.7\textheight,width=0.95\textwidth]{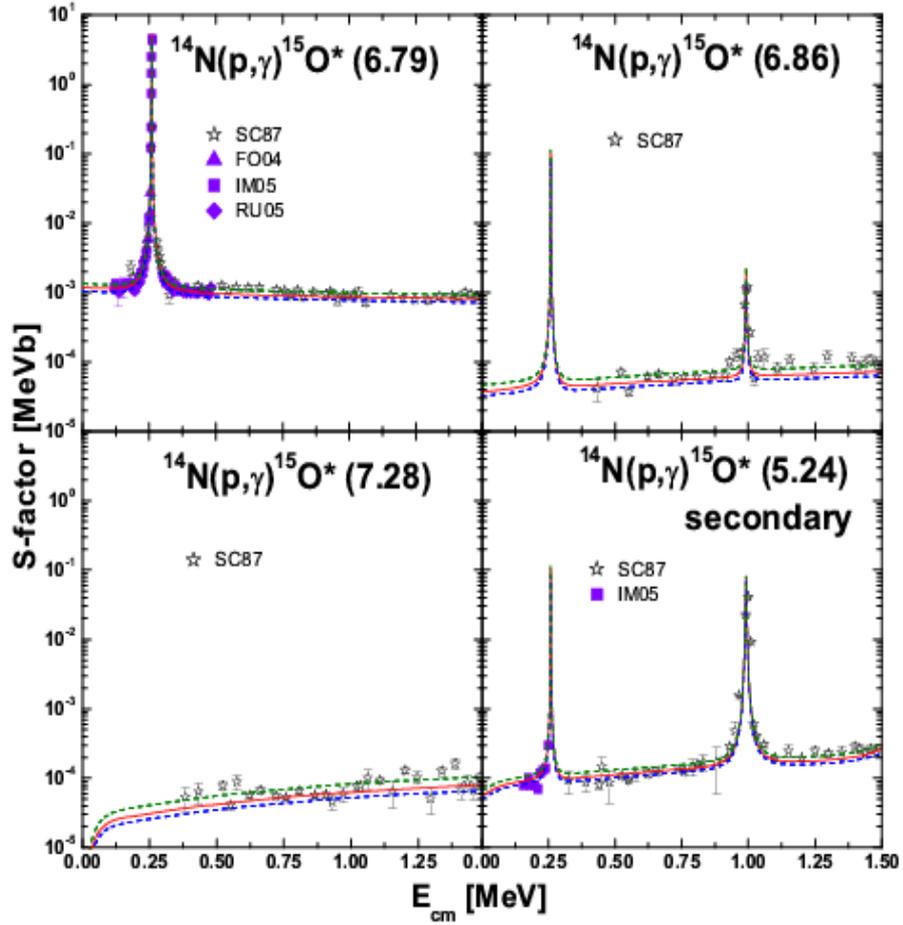}
\vspace{-0.5truecm}
\caption{The partial $S$-factors for the \reac{14}{N}{p}{\gamma}{15}{O} transitions to the  $3/2^{+}$, $5/2^{+}$ and $7/2^{+}$ excited states at $E_{\rm x}$ =  6.793, 6.859 and 7.276 MeV, respectively, with the last two contributing to the secondary transitions from the $5/2^{+}$ state at $E_{\rm x}$ = 5.241 MeV.
}
\label{n14pgFig1c}
}
\end{figure}

Figure \ref{n14pgFig1a} compares the PM and experimental total $S$-factors. 
 The partial $S$-factors in the  $E_{\rm cm} \lsimeq$ 1.5 MeV range for the transitions to the ground and five excited states are used for the PM fit (Figs.\,\ref{n14pgFig1b} and \ref{n14pgFig1c}). The data exhibit the $1/2^{+}$, $3/2^{+}$ and $1/2^{+}$ resonances at $E_{\rm cm} \simeq$ 0.26, 0.99 and 1.45 MeV, respectively. The possible contributions from the $3/2^{+}$ sub-threshold state at $E_{\rm R}\simeq  - 0.50$ MeV are additionally considered.  
The adopted parameter values are given in Table \ref{n14pgTab1}.  
The present  $S(0)$ = 1.8 $\pm$  0.2 keV\,b. 
In comparison, $S(0)$ = 3.2 $\pm$ 0.8 keV\,b  [NACRE], 1.47 keV\,b [RAD10], and 1.66 $\pm$ 0.12 keV\,b [SUN11, R-matrix].

\begin{figure}[t]
\centering{
\includegraphics[height=0.33\textheight,width=0.90\textwidth]{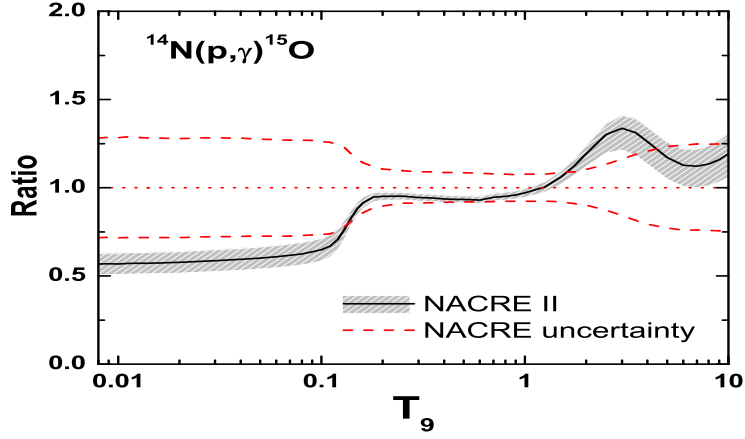}
\vspace{-0.4truecm}
\caption{\reac{14}{N}{p}{\gamma}{15}{O} rates in units of the NACRE (adopt) values. The use of the 'modern' data and the lower subthreshold contribution leads to the reductions of the ratios at the lowest energies. The increased ratios at $T_{9} > $ 1 reflect the seeming neglect by NACRE of the non-resonant contributions near the second resonance.}
\label{n14pgFig2}
}
\end{figure}

Table \ref{n14pgTab2} gives the reaction rates at $0.008 \le T_{9} \le 10$, for which the PM and the experimental cross sections are used below and above $E_{\rm cm} \lsimeq$ 0.3 MeV, respectively. 
Several very narrow resonances in the 2.18 $ \lsimeq E_{\rm cm} \lsimeq $ 4.55 MeV range with measured strengths 
 [374\,-\,376] 
 are also considered. Figure \ref{n14pgFig2} compares the present and the NACRE rates.

{\footnotesize{See \cite{GR08} for continuum shell model calculations. 
}}
\vspace{-1.0truecm}
\clearpage

\clearpage

\clearpage

\begin{table}[hb]

\caption{\reac{14}{N}{p}{\gamma}{15}{O} rates in $\rm{cm^{3}mol^{-1}s^{-1}}$}\footnotesize\rm
\begin{tabular*}{\textwidth}{@{\extracolsep{\fill}}  l c c c | l c c c}
\hline
$T_{9}$ & adopted & low & high & $T_{9}$ & adopted & low & high \\
\hline
   0.008 & 5.84E$-$25 & 5.21E$-$25 & 6.47E$-$25 &    0.2 & 7.85E$-$03 & 7.65E$-$03 & 8.05E$-$03  \\
   0.009 & 1.01E$-$23 & 9.03E$-$24 & 1.12E$-$23 &    0.25 & 1.09E$-$01 & 1.06E$-$01 & 1.11E$-$01  \\
   0.01 & 1.18E$-$22 & 1.05E$-$22 & 1.31E$-$22 &    0.3 & 6.05E$-$01 & 5.91E$-$01 & 6.19E$-$01  \\
   0.011 & 1.01E$-$21 & 9.01E$-$22 & 1.12E$-$21 &    0.35 & 2.00E+00 & 1.95E+00 & 2.04E+00  \\
   0.012 & 6.74E$-$21 & 6.02E$-$21 & 7.46E$-$21 &    0.4 & 4.77E+00 & 4.66E+00 & 4.88E+00  \\
   0.013 & 3.68E$-$20 & 3.28E$-$20 & 4.07E$-$20 &    0.45 & 9.20E+00 & 8.99E+00 & 9.41E+00  \\
   0.014 & 1.70E$-$19 & 1.52E$-$19 & 1.88E$-$19 &    0.5 & 1.53E+01 & 1.50E+01 & 1.57E+01  \\
   0.015 & 6.82E$-$19 & 6.09E$-$19 & 7.55E$-$19 &    0.6 & 3.18E+01 & 3.11E+01 & 3.25E+01  \\
   0.016 & 2.43E$-$18 & 2.17E$-$18 & 2.69E$-$18 &    0.7 & 5.18E+01 & 5.06E+01 & 5.31E+01  \\
   0.018 & 2.30E$-$17 & 2.05E$-$17 & 2.55E$-$17 &    0.8 & 7.31E+01 & 7.13E+01 & 7.48E+01  \\
   0.02 & 1.59E$-$16 & 1.42E$-$16 & 1.76E$-$16 &    0.9 & 9.39E+01 & 9.15E+01 & 9.62E+01  \\
   0.025 & 7.63E$-$15 & 6.82E$-$15 & 8.45E$-$15 &    1. & 1.14E+02 & 1.10E+02 & 1.17E+02  \\
   0.03 & 1.45E$-$13 & 1.30E$-$13 & 1.61E$-$13 &    1.25 & 1.58E+02 & 1.53E+02 & 1.62E+02  \\
   0.04 & 1.06E$-$11 & 9.45E$-$12 & 1.17E$-$11 &    1.5 & 2.00E+02 & 1.92E+02 & 2.07E+02  \\
   0.05 & 2.21E$-$10 & 1.98E$-$10 & 2.45E$-$10 &    1.75 & 2.50E+02 & 2.37E+02 & 2.59E+02  \\
   0.06 & 2.24E$-$09 & 2.01E$-$09 & 2.47E$-$09 &    2. & 3.11E+02 & 2.92E+02 & 3.24E+02  \\
   0.07 & 1.42E$-$08 & 1.27E$-$08 & 1.57E$-$08 &    2.5 & 4.81E+02 & 4.42E+02 & 5.05E+02  \\
   0.08 & 6.50E$-$08 & 5.83E$-$08 & 7.17E$-$08 &    3. & 7.21E+02 & 6.53E+02 & 7.61E+02  \\
   0.09 & 2.36E$-$07 & 2.12E$-$07 & 2.60E$-$07 &    3.5 & 1.03E+03 & 9.26E+02 & 1.09E+03  \\
   0.1 & 7.20E$-$07 & 6.48E$-$07 & 7.92E$-$07 &    4. & 1.40E+03 & 1.26E+03 & 1.49E+03  \\
   0.11 & 1.97E$-$06 & 1.78E$-$06 & 2.16E$-$06 &    5. & 2.31E+03 & 2.06E+03 & 2.48E+03  \\
   0.12 & 5.21E$-$06 & 4.75E$-$06 & 5.66E$-$06 &    6. & 3.37E+03 & 3.01E+03 & 3.65E+03  \\
   0.13 & 1.41E$-$05 & 1.31E$-$05 & 1.52E$-$05 &    7. & 4.52E+03 & 4.03E+03 & 4.92E+03  \\
   0.14 & 4.02E$-$05 & 3.79E$-$05 & 4.25E$-$05 &    8. & 5.70E+03 & 5.07E+03 & 6.22E+03  \\
   0.15 & 1.14E$-$04 & 1.09E$-$04 & 1.20E$-$04 &    9. & 6.84E+03 & 6.08E+03 & 7.50E+03  \\
   0.16 & 3.11E$-$04 & 3.00E$-$04 & 3.23E$-$04 &   10. & 7.93E+03 & 7.04E+03 & 8.71E+03  \\
   0.18& 1.83E$-$03 & 1.77E$-$03 & 1.88E$-$03 & \\
\hline
\end{tabular*}
\begin{tabular*}{\textwidth}{@{\extracolsep{\fill}} l c }
REV  = 
 $ 2.70 \times 10^{10}T_{9}^{3/2}\,{\rm exp}(-84.679/T_{9})  [1.0+0.333\,{\rm exp}(-26.840/T_{9})] $
 & \\
\end{tabular*}
\label{n14pgTab2}
\end{table}
\clearpage
\subsection{\reac{15}{N}{p}{\gamma}{16}{O}}
\label{n15pgSect}
The experimental data set referred to in NACRE is RO74b \cite{RO74b}, covering the 0.13 $\lsimeq E_{\rm cm} \lsimeq$ 2.3  MeV range.  
Added are the post-NACRE data sets BE09 \cite{BE09}, LE10 \cite{LE10} and CA11 \cite{CA11}, extending the range down to $E_{\rm cm} \simeq 0.07$ MeV.
HE60 \cite{HE60b}, rejected in NACRE in favour of RO74b, is resurrected here by following the argument in \cite{RMP}, and is partially used.

\begin{figure}[hb]
\centering{
\includegraphics[height=0.50\textheight,width=0.90\textwidth]{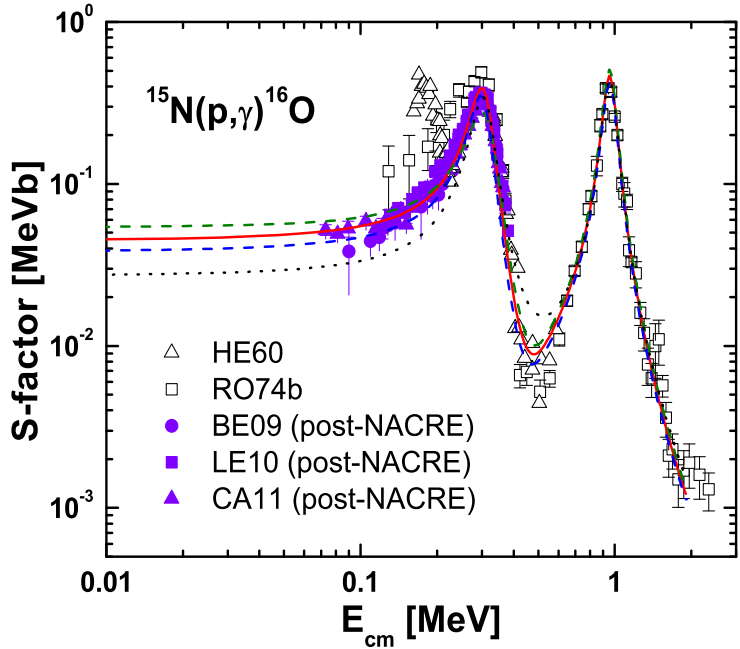}
\vspace{-0.5truecm}
\caption{The $S$-factor for \reac{15}{N}{p}{\gamma_0}{16}{O}. [See text for the contributions of the cascade transitions.] The dotted line corresponds to the solid ("adopt") curve without the interference between the two $1^{-}$  (s-wave) resonances. HE60 and RO74b data in the $E_{\rm cm} \lsimeq$ 0.3 MeV range are not considered in the fit because of the clear deviations from the 'modern' data.}
\label{n15pgFig1}
}
\end{figure}

Figure \ref{n15pgFig1} compares the PM and experimental $S$-factors for the transitions to the ground state.
The data for $E_{\rm cm} \lsimeq$ 1.3 MeV are used for the PM fit. They exhibit the predominant broad $1^{-}$ resonances at $E_{R} \simeq$  0.31 and 0.96 MeV. 
The adopted parameter values are given in Table \ref{n15pgTab1}.
The present  $S(0)$ = 45$_{-7}^{+9}$ keV\,b. 
In comparison, $S(0)$ = 64 $\pm$ 6 keV\,b  [NACRE, from RO74b], 22.1 keV\,b [RAD10], and 36 $\pm$ 6 keV\,b [SUN11, R-matrix  from \cite{MU08}].

\begin{figure}[t]
\centering{
\includegraphics[height=0.33\textheight,width=0.90\textwidth]{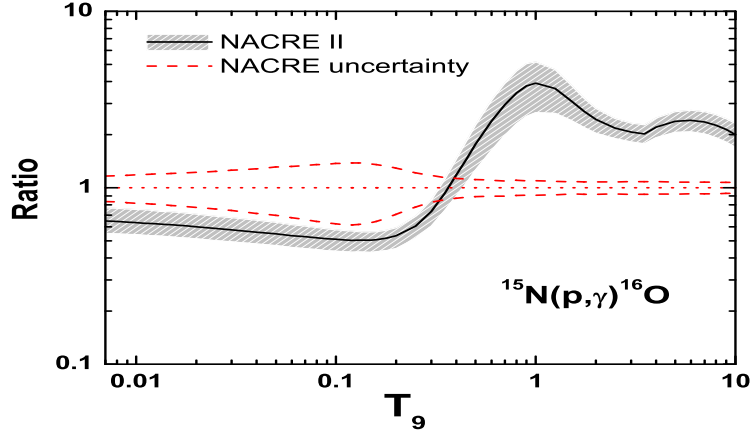}
\vspace{-0.4truecm}
\caption{\reac{15}{N}{p}{\gamma}{16}{O} rates in units of the NACRE (adopt) values. The use of the 'modern' data results in the reduced rates at the low temperatures. The enhancement of the ratio at $T_9 \gsimeq 1$ is owing to the contribution of the cascade transitions, which are not included in NACRE. 
The small hump around $T_9 \approx 6$ reflects a peculiar behaviour of unknown origin in the NACRE rates.
}  
\label{n15pgFig2}
}
\end{figure}

Table \ref{n15pgTab2} gives the reaction rates at $0.007 \le T_{9} \le 10$, for which the PM-predicted and the experimental cross sections are used below and above $E_{\rm cm} \simeq$ 0.2 MeV, respectively.
 The cascade transitions via two very narrow $2^{-}$ resonances as well as the $0^{-}$ and $3^{-}$ resonances in the 0.40 $\lsimeq E_{R} \lsimeq$ 1.14 MeV range 
 ([319, 384\,-\,386])
 are additionally considered for the rate calculations.
 Figure \ref{n15pgFig2} compares the present and the NACRE rates.

{\footnotesize{See \cite{IM12} for a very recent report on the partial $S$-factors of the cascade transitions. 
}}

\clearpage

\begin{table}[hb]
\caption{\reac{15}{N}{p}{\gamma}{16}{O} rates in $\rm{cm^{3}mol^{-1}s^{-1}}$}\scriptsize\rm
\footnotesize{
\begin{tabular*}{\textwidth}{@{\extracolsep{\fill}} l c c c |l c c c}
\hline
$T_{9}$ & adopted & low & high & $T_{9}$ & adopted & low & high \\
\hline
   0.007 & 4.54E$-$25 & 3.87E$-$25 & 5.39E$-$25 &    0.18 & 7.00E$-$03 & 5.95E$-$03 & 7.77E$-$03  \\
   0.008 & 1.34E$-$23 & 1.14E$-$23 & 1.59E$-$23 &    0.2 & 1.88E$-$02 & 1.59E$-$02 & 2.09E$-$02  \\
   0.009 & 2.34E$-$22 & 2.00E$-$22 & 2.78E$-$22 &    0.25 & 1.48E$-$01 & 1.26E$-$01 & 1.68E$-$01  \\
   0.01 & 2.75E$-$21 & 2.35E$-$21 & 3.26E$-$21 &    0.3 & 7.72E$-$01 & 6.37E$-$01 & 8.96E$-$01  \\
   0.011 & 2.36E$-$20 & 2.02E$-$20 & 2.81E$-$20 &    0.35 & 2.95E+00 & 2.35E+00 & 3.54E+00  \\
   0.012 & 1.59E$-$19 & 1.36E$-$19 & 1.89E$-$19 &    0.4 & 8.85E+00 & 6.73E+00 & 1.09E+01  \\
   0.013 & 8.73E$-$19 & 7.46E$-$19 & 1.03E$-$18 &    0.45 & 2.17E+01 & 1.59E+01 & 2.75E+01  \\
   0.014 & 4.05E$-$18 & 3.47E$-$18 & 4.80E$-$18 &    0.5 & 4.55E+01 & 3.24E+01 & 5.85E+01  \\
   0.015 & 1.64E$-$17 & 1.40E$-$17 & 1.94E$-$17 &    0.6 & 1.41E+02 & 9.63E+01 & 1.85E+02  \\
   0.016 & 5.86E$-$17 & 5.01E$-$17 & 6.93E$-$17 &    0.7 & 3.16E+02 & 2.12E+02 & 4.19E+02  \\
   0.018 & 5.60E$-$16 & 4.78E$-$16 & 6.60E$-$16 &    0.8 & 5.79E+02 & 3.87E+02 & 7.72E+02  \\
   0.02 & 3.90E$-$15 & 3.34E$-$15 & 4.60E$-$15 &    0.9 & 9.30E+02 & 6.24E+02 & 1.24E+03  \\
   0.025 & 1.91E$-$13 & 1.63E$-$13 & 2.24E$-$13 &    1. & 1.37E+03 & 9.33E+02 & 1.81E+03  \\
   0.03 & 3.70E$-$12 & 3.17E$-$12 & 4.34E$-$12 &    1.25 & 2.91E+03 & 2.10E+03 & 3.72E+03  \\
   0.04 & 2.78E$-$10 & 2.38E$-$10 & 3.24E$-$10 &    1.5 & 5.20E+03 & 4.02E+03 & 6.40E+03  \\
   0.05 & 5.97E$-$09 & 5.12E$-$09 & 6.94E$-$09 &    1.75 & 8.35E+03 & 6.78E+03 & 9.92E+03  \\
   0.06 & 6.21E$-$08 & 5.32E$-$08 & 7.17E$-$08 &    2. & 1.23E+04 & 1.03E+04 & 1.42E+04  \\
   0.07 & 4.03E$-$07 & 3.45E$-$07 & 4.63E$-$07 &    2.5 & 2.18E+04 & 1.90E+04 & 2.47E+04  \\
   0.08 & 1.89E$-$06 & 1.62E$-$06 & 2.16E$-$06 &    3. & 3.21E+04 & 2.82E+04 & 3.60E+04  \\
   0.09 & 6.98E$-$06 & 5.99E$-$06 & 7.95E$-$06 &    3.5 & 4.16E+04 & 3.66E+04 & 4.68E+04  \\
   0.1 & 2.16E$-$05 & 1.85E$-$05 & 2.44E$-$05 &    4. & 5.00E+04 & 4.37E+04 & 5.64E+04  \\
   0.11 & 5.82E$-$05 & 5.00E$-$05 & 6.57E$-$05 &    5. & 6.22E+04 & 5.38E+04 & 7.05E+04  \\
   0.12 & 1.40E$-$04 & 1.20E$-$04 & 1.58E$-$04 &    6. & 6.90E+04 & 5.94E+04 & 7.87E+04  \\
   0.13 & 3.11E$-$04 & 2.67E$-$04 & 3.47E$-$04 &    7. & 7.22E+04 & 6.17E+04 & 8.26E+04  \\
   0.14 & 6.41E$-$04 & 5.49E$-$04 & 7.14E$-$04 &    8. & 7.29E+04 & 6.20E+04 & 8.37E+04  \\
   0.15 & 1.24E$-$03 & 1.06E$-$03 & 1.38E$-$03 &    9. & 7.21E+04 & 6.11E+04 & 8.30E+04  \\
   0.16 & 2.30E$-$03 & 1.97E$-$03 & 2.55E$-$03 &   10. & 7.03E+04 & 5.94E+04 & 8.12E+04  \\
\hline
\end{tabular*}
\begin{tabular*}{\textwidth}{@{\extracolsep{\fill}} l c }
REV  = 
 $ 3.63 \times 10^{10}T_{9}^{3/2}\,{\rm exp}(-140.74/T_{9}) $
 & \\
\end{tabular*}
}
\label{n15pgTab2}
\end{table}
\clearpage
\subsection{\reac{15}{N}{p}{\alpha}{12}{C}}
\label{n15paSect}
The experimental data sets referred to in NACRE are SC52 \cite{SC52}, ZY79 \cite{ZY79} and RE82 \cite{RE82}, covering the 0.07 $\lsimeq E_{\rm cm} \lsimeq$ 1.5 MeV range.
Added is the post-NACRE data set LA07 \cite{LA07}$^\dag$, extending the range down to $E_{\rm cm} \simeq$ 0.02 MeV. 
[{\footnotesize{$^\dag$from $^{2}$H($^{15}$N,\,$\alpha$$^{12}$C)n (THM)}}]

\begin{figure}[hb]
\centering{
\includegraphics[height=0.50\textheight,width=0.90\textwidth]{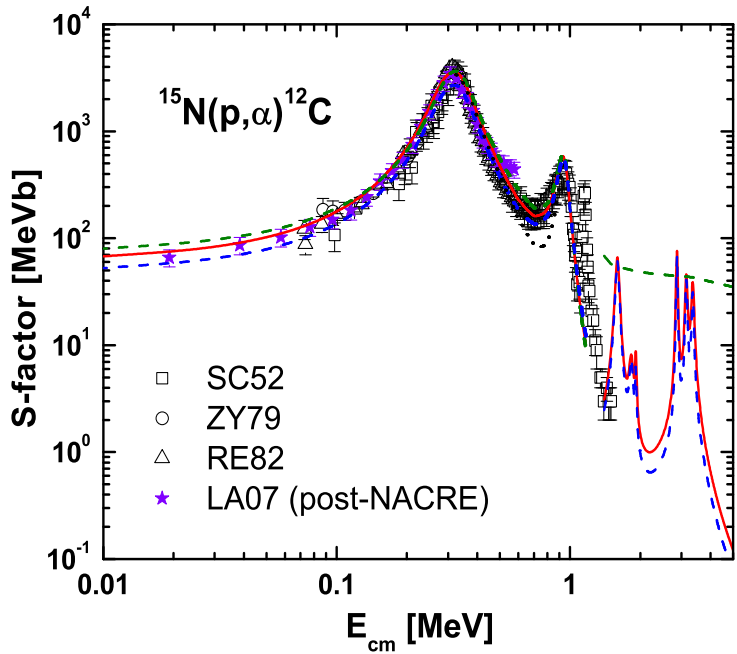}
\vspace{-0.5truecm}
\caption{The $S$-factor for \reac{15}{N}{p}{\alpha}{12}{C}. 
The dots showing a minimum at $E_{\rm cm} \approx$ 0.7 MeV correspond to the "adopt" case without the interference between the two $1^{-}$ resonances. In the low energy range, they closely follow the "low" curve.
The curves displayed in the $E_{\rm cm} \gsimeq$ 1.6 MeV range are from \cite{BA59} (structured curves), and the Hauser-Feshbach predictions \cite{Talys} (upper, monotonous dashed line) }
\label{n15paFig1}
}
\end{figure}

Figure \ref{n15paFig1} compares the DWBA and experimental $S$-factors.
The data for $E_{\rm cm} \lsimeq$ 1.1 MeV are used for the DWBA fit. They exhibit the $1^{-}$ resonances at $E_{\rm R} \simeq$  0.31 and 0.96 MeV.
The adopted parameter values are given in Table \ref{n15paTab1}.
The present $S(0)$ = 67 $\pm$ 14 MeV\,b.
In comparison, $S(0)$ = 64 $\pm$ 6 keV\,b [NACRE, from RE82], and 73 $\pm$ 5 MeV\,b [SUN11, R-matrix fit by \cite{LA09}].
At the high-energy end, the data set \cite{BA59} is added, and the $S$-factor from Hauser-Feshbach calculations \cite{Talys} is used as a measure for the upper limit. 
In this energy range the \reac{15}{N}{p}{\alpha_{1}\gamma}{12}{C^{*}}(4.44 MeV, $2^{+}$) becomes significant.
 
Table \ref{n15paTab2} gives the reaction rates at $0.005 \le T_{9} \le 10$, for which the DWBA-predicted and the experimental cross sections below and above $E_{\rm cm} \simeq$ 0.06 MeV are used, respectively. 
The additional contributions from the two very narrow $2^{-}$ resonances at $E_{\rm R} \lsimeq$ 0.40 and 0.84 MeV (\cite{isotopes}) are included. 

\begin{figure}[t]
\centering{
\includegraphics[height=0.33\textheight,width=0.90\textwidth]{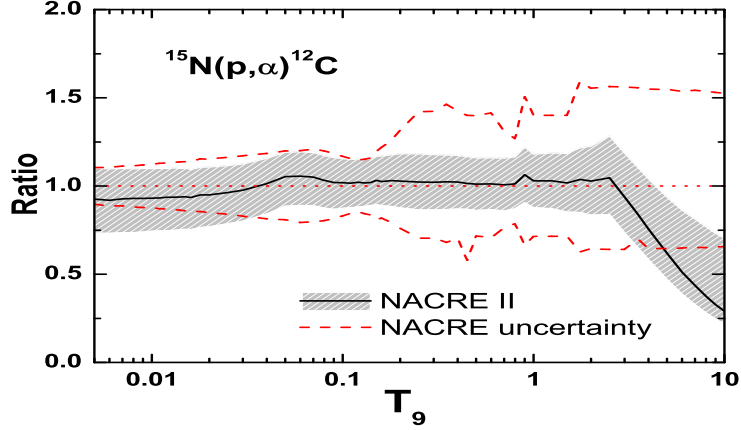}
\vspace{-0.4truecm}
\caption{\reac{15}{N}{p}{\alpha}{12}{C} rates in units of the NACRE (adopt) values. The large and systematic deviation at $T_9 \ge 2.5$ has its origin in the extrapolation made by NACRE with the help of the {\it rates} from a Hauser-Feshbach calculation. The zig-zagged fine structures in the NACRE rates are of unknown origin.}
\label{n15paFig2}
}
\end{figure}

Figure \ref{n15paFig2} compares the present and the NACRE rates.

{\footnotesize{See \cite{LA09,Barker08} for discussions on the subtleties of R-matrix fits.
}}

\clearpage

\begin{table}[hb]
\caption{\reac{15}{N}{p}{\alpha}{12}{C} rates in $\rm{cm^{3}mol^{-1}s^{-1}}$}\scriptsize\rm
\footnotesize{
\begin{tabular*}{\textwidth}{@{\extracolsep{\fill}} l c c c |l c c c}
\hline
$T_{9}$ & adopted & low & high & $T_{9}$ & adopted & low & high \\
\hline
   0.005 & 6.59E$-$26 & 5.22E$-$26 & 7.85E$-$26 &    0.16 & 1.00E+01 & 8.67E+00 & 1.14E+01  \\
   0.006 & 1.14E$-$23 & 9.07E$-$24 & 1.36E$-$23 &    0.18 & 3.38E+01 & 2.90E+01 & 3.86E+01  \\
   0.007 & 7.04E$-$22 & 5.59E$-$22 & 8.35E$-$22 &    0.2 & 9.91E+01 & 8.45E+01 & 1.14E+02  \\
   0.008 & 2.11E$-$20 & 1.67E$-$20 & 2.49E$-$20 &    0.25 & 9.00E+02 & 7.61E+02 & 1.04E+03  \\
   0.009 & 3.73E$-$19 & 2.97E$-$19 & 4.41E$-$19 &    0.3 & 4.76E+03 & 4.03E+03 & 5.49E+03  \\
   0.01 & 4.43E$-$18 & 3.54E$-$18 & 5.23E$-$18 &    0.35 & 1.70E+04 & 1.44E+04 & 1.95E+04  \\
   0.011 & 3.86E$-$17 & 3.09E$-$17 & 4.55E$-$17 &    0.4 & 4.58E+04 & 3.90E+04 & 5.26E+04  \\
   0.012 & 2.63E$-$16 & 2.10E$-$16 & 3.09E$-$16 &    0.45 & 1.00E+05 & 8.57E+04 & 1.15E+05  \\
   0.013 & 1.46E$-$15 & 1.17E$-$15 & 1.72E$-$15 &    0.5 & 1.90E+05 & 1.62E+05 & 2.18E+05  \\
   0.014 & 6.85E$-$15 & 5.50E$-$15 & 8.04E$-$15 &    0.6 & 4.97E+05 & 4.25E+05 & 5.69E+05  \\
   0.015 & 2.80E$-$14 & 2.25E$-$14 & 3.28E$-$14 &    0.7 & 9.92E+05 & 8.47E+05 & 1.14E+06  \\
   0.016 & 1.01E$-$13 & 8.14E$-$14 & 1.18E$-$13 &    0.8 & 1.66E+06 & 1.43E+06 & 1.91E+06  \\
   0.018 & 9.87E$-$13 & 7.96E$-$13 & 1.15E$-$12 &    0.9 & 2.50E+06 & 2.13E+06 & 2.87E+06  \\
   0.02 & 7.02E$-$12 & 5.68E$-$12 & 8.18E$-$12 &    1. & 3.47E+06 & 2.96E+06 & 3.98E+06  \\
   0.025 & 3.60E$-$10 & 2.94E$-$10 & 4.17E$-$10 &    1.25 & 6.38E+06 & 5.40E+06 & 7.36E+06  \\
   0.03 & 7.30E$-$09 & 5.99E$-$09 & 8.41E$-$09 &    1.5 & 9.88E+06 & 8.28E+06 & 1.15E+07  \\
   0.04 & 5.95E$-$07 & 4.94E$-$07 & 6.79E$-$07 &    1.75 & 1.39E+07 & 1.14E+07 & 1.63E+07  \\
   0.05 & 1.38E$-$05 & 1.16E$-$05 & 1.56E$-$05 &    2. & 1.82E+07 & 1.48E+07 & 2.17E+07  \\
   0.06 & 1.53E$-$04 & 1.29E$-$04 & 1.73E$-$04 &    2.5 & 2.71E+07 & 2.17E+07 & 3.33E+07  \\
   0.07 & 1.05E$-$03 & 8.89E$-$04 & 1.19E$-$03 &    3. & 3.57E+07 & 2.80E+07 & 4.52E+07  \\
   0.08 & 5.20E$-$03 & 4.41E$-$03 & 5.91E$-$03 &    3.5 & 4.32E+07 & 3.35E+07 & 5.68E+07  \\
   0.09 & 2.03E$-$02 & 1.73E$-$02 & 2.31E$-$02 &    4. & 4.93E+07 & 3.80E+07 & 6.79E+07  \\
   0.1 & 6.65E$-$02 & 5.71E$-$02 & 7.54E$-$02 &    5. & 5.84E+07 & 4.44E+07 & 8.84E+07  \\
   0.11 & 1.90E$-$01 & 1.64E$-$01 & 2.15E$-$01 &    6. & 6.40E+07 & 4.84E+07 & 1.08E+08  \\
   0.12 & 4.87E$-$01 & 4.21E$-$01 & 5.50E$-$01 &    7. & 6.76E+07 & 5.08E+07 & 1.26E+08  \\   0.13 & 1.14E+00 & 9.91E$-$01 & 1.29E+00 &    8. & 6.98E+07 & 5.23E+07 & 1.43E+08  \\
   0.14 & 2.49E+00 & 2.17E+00 & 2.82E+00 &    9. & 7.12E+07 & 5.31E+07 & 1.59E+08  \\
   0.15 & 5.13E+00 & 4.45E+00 & 5.81E+00 &   10. & 7.19E+07 & 5.36E+07 & 1.74E+08  \\
\hline
\end{tabular*}
\begin{tabular*}{\textwidth}{@{\extracolsep{\fill}} l c }
REV  = 
$ 0.706 \times {\rm exp}(-57.625/T_{9}) $
 & \\
\end{tabular*}
}
\label{n15paTab2}
\end{table}
\clearpage
\section{Summary}

The NACRE II update of NACRE reported in this work includes: (1) the collection of experimental data published, in the first instance, in the major journals of the field by early 2013; (2) the extrapolation of astrophysical $S$-factors to very low energies based on potential models, with a systematic evaluation of uncertainties; (3) the presentation in tabular form of adopted reaction rates along with their low and high limits  for temperatures in the $10^6 \lsimeq T \leq 10^{10}$ range.
The new reaction rates are also available electronically as part of the regularly improved and enlarged Brussels Library (BRUSLIB) of nuclear data. The NACRE II rates also supersede the NACRE ones in the Nuclear Network Generator (NETGEN) for astrophysics (http://www. astro.ulb.ac.be/databases.html).

\vspace{8truemm}
The authors are much obliged to anonymous referees for their careful reading of this long manuscript and for many valuable and constructive comments.  
This compilation was originally planned within the Konan-ULB convention "Construction of an Extended Nuclear Database for Astrophysics". The initiatives and the financial support by the Konan Board of Trustees and Professors H. Yoshizawa and  Y. Sugimura are warmly thanked. It has also been supported by the Interuniversity Attraction Pole IAP 5/07 of the Belgian Federal Science Policy, and by the Communaut\'e Fran\c caise de Belgique (Actions de Recherche Concert\'ees). The authors acknowledge with thanks the contributions to the early phase of this project by  M. Aikawa, K. Arai, and M. Katsuma, and the help of G. L. Chen for the preparation of the reference list.  Last but not least, their special thanks go to Alain Jorrisen for his continued interest in the present work.  
\vspace{8truemm}
\begin{center}
{\bf Appendix A}
\end{center}
\vspace{2truemm}

The adopted values of the potential parameters and of the overall renormalisation constants are collected here.  
The potential parameters are described in Sects.\,2.5.1 - 2.5.3, and are given in units of MeV or fm. 
The overall renormalisation constant for transfer reactions is $D_0^2S_{\rm F}$ in Eq.\,(\ref{eqdwbacross}). It is given in units of $10^{4}$ MeV$^2$\,fm$^3$, which is denoted as $\cal{D}_{\rm t,4}$. [$D_0^2$ is known to be $1.55 \times 10^{4}$ MeV$^2$\,fm$^3$for (d,p) reactions.] The overall renormalisation constant for capture reactions is $S_{\rm F}$ in Eq.\,(\ref{eqpotcross}). We denote this dimensionless quantity as $\cal{C}_{\rm c}$ in order to stress again the purely empirical nature of its determination.
The asterisk on $l_{i}$ in the non-resonant (N.R.) case means the exclusion of $J^{\pi}$ value(s) that is(are) exhausted by the resonance capture(s).
The sets of parameter values are in many cases not unique in that some different combinations may lead to fits that are equally good.  On the other hand, the limitation of the applicability of the model in certain cases may be self-evident from derived "unreasonable" sets of parameter values. 
We also note that the values given below are often rounded, so that some fine tuning may be necessary to reproduce, for example, the resonance energy to the desired accuracy. 
\clearpage

\noindent
{\bf Transfer Reactions}

%
\begin{table}[!htbp]
\caption{DWBA parameter values for \reac{2}{H}{d}{n}{3}{He}} \footnotesize\rm
\begin{tabular*}{\textwidth}{@{}l*{15}{@{\extracolsep{0pt plus12pt}}l}}

\hline
\multicolumn{1}{l}{}&\multicolumn{6}{l}{$i$: $d$+$^{2}$H}&\multicolumn{3}{l}{$f$: $n$+$^{3}$He}&\multicolumn{3}{l}{form f.: $p$+$^{2}$H}\\
\ N.R.&$V_{\rm R}$&$r_{\rm R}$&$a_{\rm R}$&$V_{\rm S}$&$r_{\rm S}$&$a_{\rm S}$&$V_{\rm R}$&$r_{\rm R}$&$a_{\rm R}$&$r_{\rm R}$&$a_{\rm R}$&$\cal{D}_{\rm t,4}$\\
\hline

ad & -78.25 & 1.17 & 0.81 & -0.010 & 1.33 & 0.52 & -35.11 & 1.02 & 0.68 & 0.98 & 0.68 & 3.96 \\

lo   & -79.97 &\ \ $r_{\rm C}$&  &  &  &  & -35.75 & & &\ \ $r_{\rm C}$&   & 4.47 \\

hi   & -75.53 &\ \ 1.70     &  &  &  &  & -34.03 & & &\ \ 5.73&         &   3.71 \\

\hline
\end{tabular*}
\label{ddnTab1}
\end{table}
%
\begin{table}[!htbp]
\caption{DWBA parameter values for \reac{2}{H}{d}{p}{3}{H}} \footnotesize\rm
\begin{tabular*}{\textwidth}{@{}l*{14}{@{\extracolsep{0pt plus12pt}}l}}

\hline
\multicolumn{1}{l}{}&\multicolumn{6}{l}{$i$: $d$+$^{2}$H}&\multicolumn{3}{l}{$f$: $p$+$^{3}$H}&\multicolumn{3}{l}{form f.: $n$+$^{2}$H}\\
\ N.R.&$V_{\rm R}$&$r_{\rm R}$&$a_{\rm R}$&$V_{\rm S}$&$r_{\rm S}$&$a_{\rm S}$&$V_{\rm R}$&$r_{\rm R}$&$a_{\rm R}$&$r_{\rm R}$&$a_{\rm R}$&$\cal{D}_{\rm t,4}$\\
\hline

ad & -78.95 & 1.17 & 0.81 & -0.010 & 1.33 & 0.52 & -34.45 & 1.02 & 0.68 & 0.98 & 0.68 & 4.18 \\

lo   & -81.20 &\ \ $r_{\rm C}$&  &  &  &  & -35.36 &\ \ $r_{\rm C}$&  &  &  & 4.72 \\

hi  & -77.14 &\ \ 1.70       &  &  &  &  & -33.75 &\ \ 3.62&         &  &  & 3.74 \\

\hline
\end{tabular*}
\label{ddpTab1}
\end{table}
%
\begin{table}[!htbp]
\caption{DWBA parameter values for \reac{3}{H}{d}{n}{4}{He}} \footnotesize\rm
\begin{tabular*}{\textwidth}{@{}l*{14}{@{\extracolsep{0pt plus12pt}}l}}

\hline
\multicolumn{1}{l}{}&\multicolumn{6}{l}{$i$: $d$+$^{3}$H}&\multicolumn{3}{l}{$f$: $n$+$^{4}$He}&\multicolumn{3}{l}{form f.: $p$+$^{3}$H}\\
&$V_{\rm R}$&$r_{\rm R}$&$a_{\rm R}$&$V_{\rm S}$&$r_{\rm S}$&$a_{\rm S}$&$V_{\rm R}$&$r_{\rm R}$&$a_{\rm R}$&$r_{\rm R}$&$a_{\rm R}$&$\cal{D}_{\rm t,4}$\\
\hline

\ R.R. &\ $l_{i}$ = s &\ [$J^{\pi}$,& $E_{\rm R}$] = &\ $[3/2^{+}$, & 0.05] & \\

ad & -67.92 & 1.17 & 0.81 & -2.88 & 1.33 & 0.53 & -12.75 & 1.05 & 0.687 & 1.02 & 0.68 & 0.95\\

lo & -67.92 &\ \ $r_{\rm C}$&  & -2.89 &  &  & -29.11 &  &  &\ \ $r_{\rm C}$&  & 0.39\\

hi & -67.92 &\ \ 1.70&  & -3.09 &  &  & -90.31 &  &  &\ \ 3.62&  & 0.41\\

\hline
\end{tabular*}
\label{tdnTab1}
\end{table}
%
\begin{table}[!htbp]
\caption{DWBA parameter values for \reac{3}{He}{d}{p}{4}{He}} \footnotesize\rm
\begin{tabular*}{\textwidth}{@{}l*{15}{@{\extracolsep{0pt plus12pt}}l}}

\hline
\multicolumn{1}{l}{}&\multicolumn{6}{l}{$i$: $d$+$^{3}$He}&\multicolumn{3}{l}{$f$: $p$+$^{4}$He}&\multicolumn{3}{l}{form f.: $n$+$^{3}$He}\\
&$V_{\rm R}$&$r_{\rm R}$&$a_{\rm R}$&$V_{\rm S}$&$r_{\rm S}$&$a_{\rm S}$&$V_{\rm R}$&$r_{\rm R}$&$a_{\rm R}$&$r_{\rm R}$&$a_{\rm R}$&$\cal{D}_{\rm t,4}$\\
\hline

\ R.R. &\ $l_{i}$ = s &\ [$J^{\pi}$,&\ $E_{\rm R}$] = &\ $[3/2^{+}$, &\ 0.21] & \\

ad & -68.71 & 1.17 & 0.81 & -3.37 & 1.33 & 0.53 & -39.16 & 1.05 & 0.68 & 1.02 & 0.68 & 0.51\\

lo   & -68.82 &\ \ $r_{\rm C}$&  & -3.111 &  &  & -73.06 &\ \ $r_{\rm C}$&  &  &  & 0.36\\

hi  & -68.72 &\ \ 1.70&  & -3.36 &  &  & -27.49 &\ \ 2.76&  &  &  & 0.70\\

\hline
\end{tabular*}
\label{he3dpTab1}
\end{table}
%
\begin{table}[!htbp]
\caption{DWBA parameter values for $^{3}${\rm He}($^{3}${\rm He},2{\rm p})$^{4}${\rm He}} \footnotesize\rm
\begin{tabular*}{\textwidth}{@{}l*{14}{@{\extracolsep{0pt plus12pt}}l}}

\hline
\multicolumn{1}{l}{}&\multicolumn{6}{l}{$i$: $\tau$+$^{3}$He}&\multicolumn{3}{l}{$f$: $2p$+$^{4}$He}&\multicolumn{3}{l}{form f.: $n$+$^{3}$He}\\
\ N.R.&$V_{\rm R}$&$r_{\rm R}$&$a_{\rm R}$&$V_{\rm S}$&$r_{\rm S}$&$a_{\rm S}$&$V_{\rm R}$&$r_{\rm R}$&$a_{\rm R}$&$r_{\rm R}$&$a_{\rm R}$&$\cal{D}_{\rm t,4}$\\
\hline

ad & -41.65 & 0.97 & 0.82 & -0.077 & 1.22 & 0.84 & -89.98 & 1.17 & 0.81 & 1.02 & 0.68 & 2.08\\

lo & -45.65 &\ \ $r_{\rm C}$&  & -0.23 &  &  & -93.24 &\ \ $r_{\rm C}$&  &  &  & 1.93\\

hi  & -42.98 &\ \ 1.32       &  & -23.06 &  &  & -90.22 &\ \ 1.70       &  &  &  & 3.85\\

\hline
\end{tabular*}
\label{he3he3Tab1}
\end{table}
%
\begin{table}[!htbp]
\caption{DWBA parameter values for \reac{6}{Li}{p}{\alpha}{3}{He}} \footnotesize\rm
\begin{tabular*}{\textwidth}{@{}l*{15}{@{\extracolsep{0pt plus12pt}}l}}

\hline
\multicolumn{1}{l}{}&\multicolumn{6}{l}{$i$: $p$+$^{6}$Li}&\multicolumn{3}{l}{$f$: $\alpha$+$^{3}$He}&\multicolumn{3}{l}{form f.: $t$+$^{3}$He}\\
\ N.R.&$V_{\rm R}$&$r_{\rm R}$&$a_{\rm R}$&$V_{\rm S}$&$r_{\rm S}$&$a_{\rm S}$&$V_{\rm R}$&$r_{\rm R}$&$a_{\rm R}$&$r_{\rm R}$&$a_{\rm R}$&$\cal{D}_{\rm t,4}$\\
\hline

ad & -83.17 & 1.08 & 0.68 & -51.25 & 1.31 & 0.52 & -74.43 & 0.94 & 0.77 & 0.97 & 0.82 & 4.58\\

lo & -82.18 &\ \ $r_{\rm C}$&  & -66.66 &  &  & -24.48 &\ \ $r_{\rm C}$&  &\ \ $r_{\rm C}$  &  & 22.2\\

hi & -88.29 &\ \ 2.07         &  & -68.26 &  &  & -32.40 &\ \ 1.48    &  &\ \ 1.32         &  & 15.1\\

\hline
\end{tabular*}
\label{li6paTab1}
\end{table}
%
\begin{table}[!htbp]
\caption{DWBA parameter values for \reac{7}{Li}{p}{\alpha}{4}{He}} \footnotesize\rm
\begin{tabular*}{\textwidth}{@{}l*{15}{@{\extracolsep{0pt plus12pt}}l}}

\hline
\multicolumn{1}{l}{}&\multicolumn{6}{l}{$i$: $p$+$^{7}$Li}&\multicolumn{3}{l}{$f$: $\alpha$+$^{4}$He}&\multicolumn{3}{l}{form f.: $t$+$^{4}$He}\\
\ N.R.&$V_{\rm R}$&$r_{\rm R}$&$a_{\rm R}$&$V_{\rm S}$&$r_{\rm S}$&$a_{\rm S}$&$V_{\rm R}$&$r_{\rm R}$&$a_{\rm R}$&$r_{\rm R}$&$a_{\rm R}$&$\cal{D}_{\rm t,4}$\\
\hline

ad & -59.81 & 1.09 & 0.68 & -0.083 & 1.31 & 0.52 & -75.36 & 0.97 & 0.77 & 1.00 & 0.82 & 2.32\\

lo   & -61.88 &\ \ $r_{\rm C}$&  & -1.87 &  &  & -73.35 &\ \ $r_{\rm C}$&  &\ \ $r_{\rm C}$& & 3.78\\

hi  & -58.64 &\ \ 1.90       &  & -0.082 &  &  & -79.08 &\ \ 1.48       &  &\ \ 1.32      &  & 1.64\\

\hline
\end{tabular*}
\label{li7paTab1}
\end{table}
%
%
\begin{table}[!htbp]
\caption{DWBA parameter values for \reac{9}{Be}{p}{d}{8}{Be}} \footnotesize\rm
\begin{tabular*}{\textwidth}{@{}l*{15}{@{\extracolsep{0pt plus12pt}}l}}

\hline

\multicolumn{1}{l}{}&\multicolumn{6}{l}{$i$: $p$+$^{9}$Be}&\multicolumn{3}{l}{$f$: $d$+$^{8}$Be}&\multicolumn{3}{l}{form. f: $n$+$^{8}$Be}\\
&$V_{\rm R}$&$r_{\rm R}$&$a_{\rm R}$&$V_{\rm S}$&$r_{\rm S}$&$a_{\rm S}$&$V_{\rm R}$&$r_{\rm R}$&$a_{\rm R}$&$r_{\rm R}$&$a_{\rm R}$&$\cal{D}_{\rm t,4}$\\
\hline

\ R.R. &\ $l_{i}$ = s &\ [$J^{\pi}$,&\ $E_{\rm R}$] = &\ $[1^{-}$, &\ 0.29] & \\

ad & -73.78  & 1.11  & 0.68  & -2.461  & 1.31  & 0.52  &-72.14 & 1.17 &0.81  & 1.10  & 0.61  &0.80\\

lo   &-73.96  &\ \ $r_{\rm C}$&  &-2.492  &1.309  &0.52  &-73.74 &\ \ $r_{\rm C}$&  &  &  &1.38\\

hi  &-73.67  &\ \ 1.69  &  &-1.88  &  &  &-70.03 &\ \ 1.48 &  &  &  &0.34\\

\hline
\end{tabular*}
\label{be9pdTab1}
\end{table}
%
%
\begin{table}[!htbp]
\caption{DWBA parameter values for \reac{9}{Be}{p}{\alpha}{6}{Li}} \footnotesize\rm
\begin{tabular*}{\textwidth}{@{}l*{15}{@{\extracolsep{0pt plus12pt}}l}}

\hline
\multicolumn{1}{l}{}&\multicolumn{6}{l}{$i$: $p$+$^{9}$Be}&\multicolumn{3}{l}{$f$: $\alpha$+$^{6}$Li}&\multicolumn{3}{l}{form f.: $t$+$^{6}$Li}\\
&$V_{\rm R}$&$r_{\rm R}$&$a_{\rm R}$&$V_{\rm S}$&$r_{\rm S}$&$a_{\rm S}$&$V_{\rm R}$&$r_{\rm R}$&$a_{\rm R}$&$r_{\rm R}$&$a_{\rm R}$&$\cal{D}_{\rm t,4}$\\
\hline

\ R.R. &\ $l_{i}$ = s &\ [$J^{\pi}$,&\ $E_{\rm R}$] = &\ $[1^{-}$, &\ 0.29] & \\

ad  &-73.99 &1.11 &0.68 &-3.07 &1.31 &0.52 &-22.07 &1.01 &0.78  &1.04 &0.82 &2.98\\

lo    &-74.71 &\ \ $r_{\rm C}$& &-3.64 & & &-20.18 &\ \ $r_{\rm C}$ &  &\ \ $r_{\rm C}$& &17.76\\

hi   &-73.30 &\ \ 1.69 & &-2.17 & & &-23.47 &\ \ 1.48 &  &\ \ 1.31 & &0.81\\

\hline
\end{tabular*}
\label{be9paTab1}
\end{table}
%
%
\begin{table}[!htbp]
\caption{DWBA parameter values for  \reac{9}{Be}{\alpha}{n}{12}{C}} \footnotesize\rm
\begin{tabular*}{\textwidth}{@{}l*{15}{@{\extracolsep{0pt plus12pt}}l}}

\hline
\multicolumn{1}{l}{}&\multicolumn{6}{l}{entrance channel}&\multicolumn{3}{l}{exit channel}&\multicolumn{3}{l}{form factor}\\
\multicolumn{1}{l}{}&\multicolumn{6}{l}{$i$: $\alpha$+$^{9}$Be}&\multicolumn{3}{l}{$f$: $n$+$^{12}$C}&\multicolumn{3}{l}{form f.: $\tau$+$^{9}$Be}\\
&$V_{\rm R}$&$r_{\rm R}$&$a_{\rm R}$&$V_{\rm S}$&$r_{\rm S}$&$a_{\rm S}$&$V_{\rm R}$&$r_{\rm R}$&$a_{\rm R}$&$r_{\rm R}$&$a_{\rm R}$&$\cal{D}_{\rm t,4}$\\
\hline

\ R.R. &\ $l_{i}$ = s &\ [$J^{\pi}$,&\ $E_{\rm R}$] = &\ $[5/2^{-}$, &\ 0.17] & \\
ad &-56.63 & 1.04 & 0.78 & -0.123 & 1.11 & 0.69 & -98.75 & 1.13 & 0.68 & 1.07 & 0.82 & 6.46\\

lo   &-56.62 &\ \ $r_{\rm C}$ & &-0.091 & & &-98.24 & & &\ \ $r_{\rm C}$ & &4.16\\

hi  &-56.63 &\ \ 1.48 & &-0.116 & & &-99.70 &  & &\ \ 1.30 & &9.02\\

\ R.R. &\ $l_{i}$ = p &\ [$J^{\pi}$,&\ $E_{\rm R}$] = &\ $[1/2^{+}$, &\ 0.35] & \\

ad &-37.00 & & &0.303 & & &-38.13 & & & & &0.48\\

lo   &-36.27 &       &      &0.302 &      &      &-38.21 &      &    & & &0.39\\

hi  &-37.11 &  &      &0.303 &      &      &-38.12 &      &    &   & &0.57\\

\ R.R.  &\ $l_{i}$ = d &\ [$J^{\pi}$,&\ $E_{\rm R}$] = &\ [$7/2^{-}$, & 0.105] & (see text) \\

\ SUB R.&\ $l_{i}$ =  s &\ [$J^{\pi}$,&\ $E_{\rm R}$] = &\ [$3/2^{-},$& $-0.751$] &  ad/lo/hi&  & $a$ = & 5.1 & $\theta_{\alpha}^{2}$ = & 0.1/0.0/0.3  \\

\hline
\end{tabular*}
\label{be9anTab1}
\end{table}
%
%
\begin{table}[!htbp]
\caption{DWBA parameter values for \reac{10}{B}{p}{\alpha}{7}{Be}} \footnotesize\rm
\begin{tabular*}{\textwidth}{@{}l*{15}{@{\extracolsep{0pt plus12pt}}l}}

\hline
\multicolumn{1}{l}{}&\multicolumn{6}{l}{$i$: $p$+$^{10}$B}&\multicolumn{3}{l}{$f$: $\alpha$+$^{7}$Be}&\multicolumn{3}{l}{form f.: $t$+$^{7}$Be}\\
&$V_{\rm R}$&$r_{\rm R}$&$a_{\rm R}$&$V_{\rm S}$&$r_{\rm S}$&$a_{\rm S}$&$V_{\rm R}$&$r_{\rm R}$&$a_{\rm R}$&$r_{\rm R}$&$a_{\rm R}$&$\cal{D}_{\rm t,4}$\\
\hline

\ R.R. &\ $l_{i}$ = s &\ [$J^{\pi}$,&\ $E_{\rm R}$] = &\ $[5/2^{+}$, &\ 0.01] & \\

ad & -73.55 & 1.12 &0.68 & -0.143  &1.31 &0.52 &-39.60 &1.02  & 0.78   &1.05 & 0.82 &3.01 \\

lo   &-73.54 & & &-0.226 & & &-61.99 & & & & & 0.083\\

hi  &-73.55 &\ \ $r_{\rm C}$ & &-0.140 & & &-39.60 &\ \ $r_{\rm C}$& &\ \ $r_{\rm C}$& &3.59 \\
      &        & \ \ 1.63      &       &       & & &        &\ \ 1.48       & &\ \ 1.30       & & \\

\hline
\end{tabular*}
\label{b10paTab1}
\end{table}
%
%
\begin{table}[!htbp]
\caption{DWBA parameter values for \reac{11}{B}{p}{\alpha}{8}{Be}} \footnotesize\rm
\begin{tabular*}{\textwidth}{@{}l*{17}{@{\extracolsep{0pt plus12pt}}l}}

\hline
\multicolumn{1}{l}{}&\multicolumn{6}{l}{$i$: $p$+$^{11}$B}&\multicolumn{6}{l}{$f$: $\alpha$+$^{8}$Be}&\multicolumn{3}{l}{form f.: $t$+$^{8}$Be}\\
&$V_{\rm R}$&$r_{\rm R}$&$a_{\rm R}$&$V_{\rm S}$&$r_{\rm S}$&$a_{\rm S}$&$V_{\rm R}$&$r_{\rm R}$&$a_{\rm R}$&$V_{\rm S}$&$r_{\rm S}$&$a_{\rm S}$&$r_{\rm R}$&$a_{\rm R}$&$\cal{D}_{\rm t,4}$\\
\hline

\ R.R. &\ $l_{i}$ = s &\ [$J^{\pi}$,&\ $E_{\rm R}$] = &\ $[2^{-}$, &\ 0.61] & \\

ad & -64.13 & 1.12  & 0.68  & -4.04 & 1.31 & 0.52 & -44.29 & 1.03 & 0.78 & -1.110 & 1.10 & 0.69 & 0.98 & 0.82 & 17.87 \\

lo   &-64.67  &\ \ $r_{\rm C}$       &       & -3.432 &      &      &-45.65  &\ \ $r_{\rm C}$      &      &-0.660  &      &      &\ \ $r_{\rm C}$      &      &20.99\\

hi  &-63.25  &\ 1.58       &       & -4.49 &      &      &-41.95  &\  1.48  &      &-0.120  &      &      &\  1.30      &      &13.21\\

\hline
\end{tabular*}
\label{b11paTab1}
\end{table}
%
%
\begin{table}[!htbp]
\caption{DWBA parameter values  for \reac{11}{B}{\alpha}{n}{14}{N}} \footnotesize\rm
\begin{tabular*}{\textwidth}{@{}l*{15}{@{\extracolsep{0pt plus12pt}}l}}
\hline

\multicolumn{1}{l}{}&\multicolumn{6}{l}{$i$: $\alpha$+$^{11}$B}&\multicolumn{3}{l}{$f$: $n$+$^{14}$N}&\multicolumn{3}{l}{form f.: $\tau$+$^{11}$B}\\
&$V_{\rm R}$&$r_{R}$&$a_{\rm R}$&$V_{\rm S}$&$r_{\rm S}$&$a_{\rm S}$&$V_{\rm R}$&$r_{\rm R}$&$a_{\rm R}$&$r_{\rm R}$&$a_{\rm R}$&$\cal{D}_{\rm t,4}$\\
\hline

\ R.R. &\ $l_{i}$ = p &\ [$J^{\pi}$,&\ $E_{\rm R}$] = &\ $[1/2^{+}$, &\ 0.44] & \\

ad &-87.83 &1.06 &0.79 &-0.349  &1.12 &0.70 &-67.02 &1.14 &0.68 &1.08 &0.82 &0.033\\

lo   &-87.81 &\ \ $r_{\rm C}$ & &-0.300  & & &-97.36 & & &\ \ $r_{\rm C}$ & &0.21\\

hi  &-87.86 &\ \ 1.48 & &-0.410  & & &-64.16 && &\ \ 1.29 & &0.042\\

\ R.R. &\ $l_{i}$ = p &\ [$J^{\pi}$,&\ $E_{\rm R}$] = &\ $[3/2^{+}$, &\ 0.77] & \\

all &-64.06 & & &-0.401 & & &-28.77 & & & &  &0.0077\\

\hline
\end{tabular*}
\label{b11anTab1}
\end{table}
%
%
\begin{table}[!htbp]
\caption{DWBA parameter values for the  \reac{13}{C}{\alpha}{n}{16}{O}} \footnotesize\rm
\begin{tabular*}{\textwidth}{@{}l*{15}{@{\extracolsep{0pt plus12pt}}l}}

\hline
\multicolumn{1}{l}{}&\multicolumn{6}{l}{i: $\alpha$+$^{13}$C}&\multicolumn{3}{l}{f: $n$+$^{16}$O}&\multicolumn{3}{l}{form f: $\tau$+$^{13}$C}\\
&$V_{\rm R}$&$r_{\rm R}$&$a_{\rm R}$&$V_{\rm S}$&$r_{\rm S}$&$a_{\rm S}$&$V_{\rm R}$&$r_{\rm R}$&$a_{\rm R}$&$r_{\rm R}$&$a_{\rm R}$&$\cal{D}_{\rm t,4}$\\
\hline

\ R.R. &\ $l_{i}$ = p &\ [$J^{\pi}$, &\ $E_{\rm R}$] = &\ $[3/2^{+}$, &\ 0.84] & \\

ad &-62.67 &1.07 &0.79 &-1.470 &1.13 &0.70 &-65.57 &1.14 &0.68 &1.1 &0.82 &18.6\\

lo    & &\ \ $r_{\rm C}$&&  & & && & &\ \ $r_{\rm C}$ & & 16.6 \\

hi       & &\ \ 1.48 & & & & & & & &\ \ 1.29 & &19.7\\

\ SUB R.& \ $l_{i}$ = p &\ [$J^{\pi}$, &\ $E_{\rm R}$] = & [$1/2^{+}$, & $-0.003$] &   ad/lo/hi& &   $a$ = & 7.5 &$\theta_{\alpha}^{2}$ = &  & 8./5./15. E-3  \\

       & \ $l_{i}$ = s  &\ [$J^{\pi}$, &\ $E_{\rm R}$] = &  [$1/2^{-}$, & $ -0.419$] &           &       &     &  &          &              & 0.65/0.50/0.85 \\

\hline
\end{tabular*}
\label{c13anTab1}
\end{table}
%
%
\begin{table}[!htbp]
\caption{DWBA parameter values for \reac{15}{N}{p}{\alpha}{12}{C}.} \footnotesize\rm
\begin{tabular*}{\textwidth}{@{}l*{15}{@{\extracolsep{0pt plus12pt}}l}}

\hline
\multicolumn{1}{l}{}&\multicolumn{6}{l}{$i$: $p$+$^{15}$N}&\multicolumn{3}{l}{$f$: $\alpha$+$^{12}$C}&\multicolumn{3}{l}{form f.: $t$+$^{12}$C}\\
&$V_{\rm R}$&$r_{\rm R}$&$a_{\rm R}$&$V_{\rm S}$&$r_{\rm S}$&$a_{\rm S}$&$V_{\rm R}$&$r_{\rm R}$&$a_{\rm R}$&$r_{\rm R}$&$a_{\rm R}$&$\cal{D}_{\rm t,4}$\\
\hline

\ R.R. &\ $l_{i}$ = p &\ [$J^{\pi}$,&\ $E_{\rm R}$] = &\ $[1^{-}$, &\ 0.31] & \\

ad &-59.16 &1.139 &0.68 &-0.93 &1.30 &0.53 &-43.70 &1.07 &0.79 &1.09 &0.82 &13.\\

lo   &-59.17 &\ \ $r_{\rm C}$ & &-0.93 & & &-43.69 &\ \ $r_{\rm C}$ & &\ \ $r_{\rm C}$ & &16.\\

hi  &-59.12 &\ \ 1.46 & &-0.88 & & &  -43.92 &\ \ 1.48 & &\ \ 1.29 & &12.\\

\ R.R. &\ $l_{i}$ = p &\ [$J^{\pi}$,&\ $E_{\rm R}$] = &\ $[1^{-}$, &\ 0.96] & \\

all &-54.87 & & &-2.026 & & &-44.10 & & & & &0.26\\

\hline
\end{tabular*}
\label{n15paTab1}
\end{table}
\clearpage

\noindent
{\bf Capture Reactions}

%
\begin{table}[!htbp]
\caption{PM parameter values for \reac{2}{H}{p}{\gamma}{3}{He}} \footnotesize\rm
\begin{tabular*}{\textwidth}{@{}l*{15}{@{\extracolsep{0pt plus12pt}}l}}

\hline
   & $l_i$  & & $ V_{i}$ & $r_{0}$ & $a_{0}$ & $\cal{C}_{\rm c}$ \\
\hline

\ N.R. &  sp  & ad/lo/hi & -50.0/-48.0/-52.0 & 0.84 & 0.35 & 1.00 \\

\hline
\end{tabular*}
\label{dpgTab1}
\end{table}

\begin{table}[!htbp]
\caption{PM parameter values for \reac{2}{H}{d}{\gamma}{4}{He}} \footnotesize\rm
\begin{tabular*}{\textwidth}{@{}l*{15}{@{\extracolsep{0pt plus12pt}}l}}
\hline
    & $l_i$ & & $V_{i}$ & $r_{0}$ & $a_{0}$ & $\cal{C}_{\rm c}$ \\
\hline

\ N.R. & d  & ad/lo/hi & -54.0/-52.2/-56.0 & 1.22/1.25/1.25 & 1.25 & 0.95/0.83/1.87 ($l_{f}=0$)\\

 & s & & & & & 0.035/0.023/0.081 ($l_{f}=2$)\\

\hline
\end{tabular*}
\label{ddgTab1}
\end{table}

\begin{table}[!htbp]
\caption{PM parameter values for \reac{2}{H}{\alpha}{\gamma}{6}{Li}} \footnotesize\rm
\begin{tabular*}{\textwidth}{@{}l*{16}{@{\extracolsep{0pt plus12pt}}l}}
\hline
   & $l_{i}$ & [$J^{\pi}, E_{\rm R}$]&  & $V_{i}$ & $r_{0}$ & $a_{0}$ & $\cal{C}_{\rm c}$ \\
\hline

\ R.C. & d  & [$3^{+}$, 0.711] & ad/lo/hi & -34.77  & 1.10 & 0.45 & 0.99/0.92/1.02   \\

     & d  & [$2{+}$, 3.892] &          & -77.67  & 0.70 & 0.16 & 2.6/1.9/2.4   \\

\ N.R. & p     & & ad/lo/hi      & -36.35  & 1.11 & 0.64 & 1.0/0.375/1.5 \\

     & d*    & &               &       &         &      & 1.5/1.4/1.5  \\

\hline
\end{tabular*}
\label{dagTab1}
\end{table}

\begin{table}[!htbp]
\caption{PM parameter values for \reac{3}{H}{\alpha}{\gamma}{7}{Li}} \footnotesize\rm
\begin{tabular*}{\textwidth}{@{}l*{15}{@{\extracolsep{0pt plus12pt}}l}}
\hline
   &  & $l_{i}$ & $V_{i}$ & $r_{0}$ & $a_{0}$ & $\cal{C}_{\rm c}$  \\
\hline

\ N.R. & s/d &  ad  & -38.6/30.6      & 1.24 & 0.425/0.43 & 0.73/1.0\\

     &     &  lo  & -39.6/36.3 &    & 0.42/0.52 & 0.64/0.51\\

     &     &  hi  & -36.3      &     & 0.52     & 0.71/0.67\\

\hline
\end{tabular*}
\label{tagTab1}
\end{table}

\begin{table}[!htbp]
\caption{PM parameter values for \reac{3}{He}{\alpha}{\gamma}{7}{Be}} \footnotesize\rm
\begin{tabular*}{\textwidth}{@{}l*{15}{@{\extracolsep{0pt plus12pt}}l}}
\hline
    & $l_{i}$ & & $V_{i}$ & $r_{0}$ & $a_{0}$ & $\cal{C}_{\rm c}$  \\
\hline

\ N.R. & s/d &  ad     & -38.6/36.3  & 1.24 & 0.425/0.52  & 0.82/0.71\\

     &     &  lo     & -39.6/36.3  &      & 0.42/0.52   & 0.71/0.67\\

     &     &  hi     & -36.3       &      & 0.52        & 0.76/0.75\\

\hline
\end{tabular*}
\label{he3agTab1}
\end{table}

\begin{table}[!htbp]
\caption{PM parameter values for \reac{6}{Li}{p}{\gamma}{7}{Be}} \footnotesize\rm
\begin{tabular*}{\textwidth}{@{}l*{15}{@{\extracolsep{0pt plus12pt}}l}}
\hline
   & $l_{i}$ & [$J^{\pi}, E_{\rm R}$]&  & $V_{i}$ & $r_{0}$ & $a_{0}$ & $\cal{C}_{\rm c}$ \\

\hline

\ R.C. & s    & [$3/2^{+}$, 0.2](see text) &   ad & -80.0 & 1.15 & 1.2 & 8.88E-4\\

     & p    & [$5/2^{-}$, 1.6](see text) &   hi & -74.5 & 0.65 & 0.2 & 0.03\\

\ N.R.  & s* & & ad    & -50.0  & 0.90 & 0.50 & 0.37\\
 
      & s  & & lo    &   &  &  & 0.165\\

      & s  & & hi/(ad/lo)$^{\dag}$    & -85.0  & 1.10 & 0.45 & 0.68/(0.52/0.60)$^{\dag}$\\

\hline
\end{tabular*}
[{\footnotesize{$^\dag$for Fig.\,25 without PA99.}}]
\label{li6pgTab1}
\end{table}
\clearpage
\begin{table}[!htbp]
\caption{PM parameter values for \reac{7}{Li}{p}{\gamma}{8}{Be}} \footnotesize\rm
\begin{tabular*}{\textwidth}{@{}l*{16}{@{\extracolsep{0pt plus12pt}}l}}
\hline
   & $l_{i}$ & [$J^{\pi}, E_{\rm R}$] & $\rightarrow$ $E_{\rm x}$($^{8}$B)    &  & $V_{i}$&$r_{0}$&$a_{0}$& $\cal{C}_{\rm c}$ \\
\hline

\ R.C. & p    & [$1^{+}$, 0.375] & $\rightarrow$ 0  &  ad/lo/hi & -120.19 & 0.4926 & 0.059 & 4.0/3.8/4.2 (E-3)\\

  &   &  & $\rightarrow$ 3.04 & all  &    &  & & 1.2/1.1/1.3 (E-3)\\

  &  p & [$1^{+}$, 0.896] & $\rightarrow$ 0 &    &  -67.9  & 0.658  & 0.17  & 6.2E-6\\

  &   &  & $\rightarrow$ 3.04 & ad/lo/hi  &   &   &    & 0.60/1.9/0.6 (E-4)\\


\ N.R. & s  & & $\rightarrow$ 0  & ad/lo/hi & -50.0 & 0.92 & 0.50 & 8.7/7.3/9.6 \\

  &   &  &$\rightarrow$ 3.04  &           &  &   &  & 2.1/1.8/2.5\\

 &\ \ or & (see Fig.31): \\

  & s  &  & $\rightarrow$ 0  & ad & -50.0 & 0.80 & 0.50 & 6.6 \\

  & p* & & \ \ \ (E2)            &    &  &  &  & 18.0  \\

\ SUB R.& p & [$2^{+}, -0.629$] & &   hi&  $a$ = & 40. & $\theta_{\alpha}^{2}$ = & 0.3  \\

\hline
\end{tabular*}
\label{li7pgTab1}
\end{table}

\begin{table}[!htbp]
\caption{PM parameter values for \reac{7}{Li}{\alpha}{\gamma}{11}{B}} \footnotesize\rm
\begin{tabular*}{\textwidth}{@{}l*{16}{@{\extracolsep{0pt plus12pt}}l}}
\hline
   & $l_{i}$ & [$J^{\pi}, E_{\rm R}$] & & $V_{i}$ & $r_{0}$ & $a_{0}$ & $\cal{C}_{\rm c}$  \\
\hline

\ R.C. & p  & [$5/2^{+}$, 0.609] & ad/lo/hi & -55.55 & 0.70 & 0.40 & 0.097/0.078/0.145\\

     & d  & [$3/2^{-}$, 1.595] &  & -54.07 & 0.82 & 0.60 & 2.7/2.17/3.3\\

     & d  & [$5/2^{-}$, 1.665] &  & -35.75 & 1.08 & 0.16 & 0.087/0.070/0.105\\

     & f  & [$7/2^{+}$, 1.932] &  & -85.44 & 1.04 & 0.48 & 0.078/0.0062/0.0094\\


\ N.R. & p* & & ad/lo/hi & -50.0   & 0.9 & 0.5 & 1.9/0.95/3.8\\

\ SUB R.& s & [$3/2^{-}, -0.105$] & ad/lo/hi&  $a$ = & 40. & $\theta_{\alpha}^{2}$ = & 2.6/2.0/3.0 (E-3) \\

\hline
\end{tabular*}
\label{li7agTab1}
\end{table}

%
%
\begin{table}[!htbp]
\caption{PM parameter values for \reac{7}{Be}{p}{\gamma}{8}{B}} \footnotesize\rm
\begin{tabular*}{\textwidth}{@{}l*{16}{@{\extracolsep{0pt plus12pt}}l}}
\hline
   & $l_{i}$ & [$J^{\pi}, E_{\rm R}$] &  &$V_{i}$&$r_{0}$&$a_{0}$& $\cal{C}_{\rm c}$ \\
\hline

\ R.C. & p & [$1^{+}$, 0.633] & ad/lo/hi & -61.62 & 0.70  & 0.12 & 1.23/1.16/1.25\\
     & p & [$3^{+}$, 2.183] &  & -96.0   & 0.656 & 1.73 & 10.3/10.1/10.4\\


\ N.R.& spd  &  & ad  & -50.00 & 0.92 & 0.60 & 0.45,  0.20, 0.25\\
    &      &  & lo  &        &      &      & 0.42,    0.18, 0.18\\
    &      &  & hi  &        &      &      & 0.49,    0.22, 0.28\\

\hline
\end{tabular*}
\label{be7pgTab1}
\end{table}

\begin{table}[!htbp]
\vskip-0.5cm
\caption{PM parameter values for \reac{7}{Be}{\alpha}{\gamma}{11}{C}} \footnotesize\rm
\begin{tabular*}{\textwidth}{@{}l*{15}{@{\extracolsep{0pt plus12pt}}l}}
\hline
   & $l_{i}$  & [$J^{\pi}, E_{\rm R}$]& & $V_{i}$&$r_{0}$&$a_{0}$&$\cal{C}_{\rm c}$ \\
\hline

\ R.C. & p & [$5/2^{+}$, 1.155] & ad/lo/hi & -56.94 & 0.70 & 0.40 & 0.13/0.10/0.15\\

    & d & [$3/2^{-}$, 2.106] &  & -55.45 & 0.82 & 0.60 & 3.6/2.8/4.3\\

    & d & [$5/2^{-}$, 2.236] &  & -36.24 & 1.08 & 0.16 & 0.29/0.23/0.34\\

    & d & [$7/2^{-}$, 2.426] &  hi & -94.33 & 0.93 & 0.10 & 1.92\\

    & f & [$7/2^{+}$, 2.539] & ad/lo/hi & -85.70 & 1.04 & 0.48 & 0.094/0.0075/0.0113\\


\ N.R. & p* & & ad/lo/hi & -50.0   & 0.9 & 0.5 & 1.7/0.9/3.4\\

\hline
\end{tabular*}
\label{be7agTab1}
\end{table}

\begin{table}[!htbp]
\vskip-0.5cm
\caption{PM parameter values for \reac{9}{Be}{p}{\gamma}{10}{B}} \footnotesize\rm
\begin{tabular*}{\textwidth}{@{}l*{16}{@{\extracolsep{0pt plus12pt}}l}}
\hline
   & $l_{i}$ & [$J^{\pi}, E_{\rm R}$]& & $V_{i}$&$r_{0}$&$a_{0}$& $\cal{C}_{\rm c}$ \\
\hline

\ R.C. & s  & [$1^{-}$, 0.290] & ad/lo/hi & -31.8 & 0.40 & 2.6 & 0.018/0.017/0.018\\

      & p  & [$2^{+}$, 0.892] &           & -62.68 & 0.65 & 0.3 & 1.0/0.84/1.18\\


\ N.R. & s*   &    & ad/lo/hi   & -53.9/-56.9/-53.9 & 1.0  & 0.70 & 0.30/1.9/0.36\\

\hline
\end{tabular*}
\label{be9pgTab1}
\end{table}
\clearpage

\begin{table}[!htbp]
\caption{PM parameter values for \reac{10}{B}{p}{\gamma}{11}{C}} \footnotesize\rm
\begin{tabular*}{\textwidth}{@{}l*{16}{@{\extracolsep{0pt plus12pt}}l}}
\hline
   & $l_{i}$ & [$J^{\pi}, E_{\rm R}$]&   & $V_{i}$ &$r_{0}$&$a_{0}$& $\cal{C}_{\rm c}$ \\
\hline

\ R.C. & s & [$5/2^{+}$, 0.01] & (see text) &  &  &  & \\

       & p & [$3/2^{-}$, 0.961] & ad/lo/hi & -88.72 & 1.35 & 0.76 & 0.67/0.8/0.8\\


 N.R. &  p*     &    & all & -76.09 & 0.76 & 0.68 & 9.6\\

\hline
\end{tabular*}
\label{b10pgTab1}
\end{table}

\begin{table}[!htbp]
\caption{PM parameter values for \reac{11}{B}{p}{\gamma}{12}{C}} \footnotesize\rm
\begin{tabular*}{\textwidth}{@{}l*{15}{@{\extracolsep{0pt plus12pt}}l}}
\hline
   & $l_{i}$ &[$J^{\pi}, E_{\rm R}$]&  & $V_{i}$&$r_{0}$&$a_{0}$& $\cal{C}_{\rm c}$ \\
\hline

\ R.C. & p & [$2^{+}$, 0.149] & ad/lo/high & -69.79 & 1.10 & 3.23 & 8.0/6.0/10.0\\

     & s & [$2^{-}$, 0.613] & all        & -38.82  & 1.10 & 0.60 & (see text) \\


\ N.R. & s*p*/d &      & all        & -62.0/75.0 & 1.10 & 0.60 & 1.00\\

\hline
\end{tabular*}
\label{b11pgTab1}
\end{table}

\begin{table}[!htbp]
\caption{PM parameter values for \reac{12}{C}{p}{\gamma}{13}{N}}\footnotesize\rm
\begin{tabular*}{\textwidth}{@{}l*{15}{@{\extracolsep{0pt plus12pt}}l}}
\hline
   & $l_{i}$ & [$J^{\pi}, E_{\rm R}$] &   & $V_{i}$&$r_{0}$&$a_{0}$& $\cal{C}_{\rm c}$ \\

\hline

\ R.C. & s  & [$1/2^{+}$, 0.421] & ad  & -36.00 & 1.15 & 0.65 & 0.33\\
     &    &         & lo  & -57.57 & 0.90 & 0.50 & 0.29\\
     &    &         & hi  & -60.72 & 1.45 & 0.40 & 0.33\\

\hline
\end{tabular*}
\label{c12pgTab1}
\end{table}

\begin{table}[!htbp]
\caption{PM parameter values for \reac{12}{C}{\alpha}{\gamma}{16}{O}} \footnotesize\rm
\begin{tabular*}{\textwidth}{@{}l*{8}{@{\extracolsep{0pt plus12pt}}l}}
\hline
   & $l_{i}$ & [$J^{\pi}, E_{\rm R}$] &$\rightarrow$ $E_{\rm x}$($^{16}$O)    &   & $V_{i}$&$r_{0}$&$a_{0}$& $\cal{C}_{\rm c}$ \\

\hline

\ R.C.& p  &[$1^{-}$, 2.423] &  $\rightarrow$ 0   & ad/lo/hi & -55.67 & 1.15 & 0.30 & 31.4/29.7/33.0 \\

    &    &       & $\rightarrow$ 6.05 &          &         &      &      & 66.0/57.8/74.3 \\

    &    &       & $\rightarrow$ 6.92 &          &         &      &      & 18./18./230. \\

    &    &       & $\rightarrow$ 7.12 &          &         &      &      & 0.34/0.25/0.49 \\

    & d  & [$2^{+}$, 2.683] & $\rightarrow$ 0   & ad/lo/hi & -99.86  & 0.337 & 0.021 & 6.0/3.0/3.0 \\

    &    &         & $\rightarrow$ 6.05 &        &         &       &     & 120./90./150. \\

    & g  & [$4^{+}$, 3.194] & $\rightarrow$ 6.92 & ad/lo/hi & -85.48 & 1.0 & 0.4 & 1.8/2.5/5.0 \\

    & d  & [$2^{+}$, 4.358] & $\rightarrow$ 0  & ad/lo/hi & -39.80  & 0.532 & 0.01 & 2.4/2.1/3.0 \\

    &    &         & $\rightarrow$ 6.05 &         &          &       &   & 3.0/2.4/3.9 \\

    &    &         & $\rightarrow$ 6.92 &         &          &       &   & 0.12/0.12/0.18 \\

   & f  & [$3^{-}$, 4.432] & $\rightarrow$ 6.13 & ad/lo/hi & -65.75 & 1.0 & 0.39 & 5.5/3.1/8.6 E-3 \\

   &    &         & $\rightarrow$ 6.92 &          &         &     &      &  90./130./300. \\

   &    &         & $\rightarrow$ 7.12 &          &         &     &      & 0.28/0.21/0.35 \\

\ SUB R.& p & [$1^{-}, -0.045$] &   & add/lo/hi &  $a$ = & 6.5 & $\theta_{\alpha}^{2}$ = & 0.008/0.006/0.010  \\

      & d & [$2^{+},-0.245$] &   &           &             & 7.5 & & 0.019/0.013/0.025 \\

\hline
\end{tabular*}
\label{c12agTab1}
\end{table}

\begin{table}[!htbp]
\caption{PM parameter values for \reac{13}{C}{p}{\gamma}{14}{N}} \footnotesize\rm
\begin{tabular*}{\textwidth}{@{}l*{16}{@{\extracolsep{0pt plus12pt}}l}}
\hline

   & $l_{i}$ & [$J^{\pi}, E_{\rm R}$]  & $\rightarrow$ $E_{\rm x}$($^{15}$O)    &  & $V_{i}$ & $r_{0}$ & $a_{0}$ & $\cal{C}_{\rm c}$ \\

\hline

\ R.C. & s &[$1^{-}$, 0.511] & $\rightarrow$ 0    & ad/lo/hi & -25.82 & 0.58 & 0.04  & 0.28/0.18/0.37 \\

     &   &        & $\rightarrow$ 2.31 & all      & -20.54 & 0.67 & 0.24  & 0.027 \\

     &   &        & $\rightarrow$ 3.95 &          & -25.82  & 0.58 & 0.04 & 0.28 \\

     &   &        & $\rightarrow$ 4.91 &          &         &      &      & 1.0 \\

     &   &        & $\rightarrow$ 5.11 &          &         &      &      & 224. \\

     &   &        & $\rightarrow$ 5.69 &          &         &       &     & 2.1  \\

N.R.  & s*pd &     & $\rightarrow$  0    & ad/lo/hi & -30.48 & 1.12 & 0.39 & 1.0/0.8/1.2 \\

      &  p   &     & $\rightarrow$ 2.31 & all      & -36.88 & 4.16 & 1.01 & 0.45 \\

      & s*p  &     & $\rightarrow$ 3.95 &          &        & 1.15 & 0.16 & 0.0015 \\

      & pd   &     & $\rightarrow$ 4.91 &          & -30.48 & 1.12 & 0.39 & 1.0 \\

      & s*p/d &    & $\rightarrow$ 5.11 &          & -34.48 &      &      & 0.05/1.0 \\

      &       &    & $\rightarrow$ 5.69 &          & -30.48 &      &      & 0.8 \\

      &       &    & $\rightarrow$ 5.83 &          &        &      &      & 0.5 \\

\hline
\end{tabular*}
\label{c13pgTab1}
\end{table}

\begin{table}[!htbp]
\vskip-0.5cm
\caption{PM parameter values for \reac{13}{N}{p}{\gamma}{14}{O}} \footnotesize\rm
\begin{tabular*}{\textwidth}{@{}l*{16}{@{\extracolsep{0pt plus12pt}}l}}

\hline

   & $l_{i}$ & [$J^{\pi}, E_{\rm R}$] & & $V_{i}$&$r_{0}$&$a_{0}$& $\cal{C}_{\rm c}$ \\
\hline

\ R.C. & s  & [$1^{-}$, 0.528] & ad/hi & -53.92 & 0.96 & 0.30 & 0.977/1.23\\
     &    &         & lo    & -53.91 &      &      & 0.775\\

\ N.R.  & pd &      & ad/lo/hi & -30.48 & 1.12 & 0.39 & 1.0/0.8/1.2 \\

\hline
\end{tabular*}
\label{n13pgTab1}
\end{table}

\begin{table}[!htbp]
\vskip-0.5cm
\caption{PM parameter values for \reac{14}{N}{p}{\gamma}{15}{O}} \footnotesize\rm
\begin{tabular*}{\textwidth}{@{}l*{15}{@{\extracolsep{0pt plus12pt}}l}}

\hline

   & $l_{i}$ & [$J^{\pi}, E_{\rm R}$]  & $\rightarrow$ $E_{\rm x}$($^{15}$O)    &  & $V_{i}$ & $r_{0}$ & $a_{0}$ & $\cal{C}_{\rm c}$ \\

\hline

\ R.C.  & s &[$1/2^{+}$, 0.260] & $\rightarrow$ 0  & all & -57.53 & 0.88 & 0.52 & 3.4 E-5 \\

& & & $\rightarrow$ 5.18  & ad/lo/hi  &  & & & 6.0/5.8/6.3 E-6 \\

& & & $\rightarrow$ 6.18  &           &  & & & 2.4/2.1/2.7 E-5 \\

& & & $\rightarrow$ 6.79  &           &  & & & 8.6/8.2/8.9 E-5 \\

& & & $\rightarrow$ 6.86  &           &  & & & 1.7/1.4/1.9 E-7 \\

& d  &[$3/2^{+}$, 0.987] & $\rightarrow$ 0   & ad/lo/hi & -57.78 & 0.88 & 0.36 & 3.4/3.2/3.6 E-3 \\

&    &   & $\rightarrow$ 5.18 &     & & & & 1.7/1.5/1.9 E-5 \\

&    &   & $\rightarrow$ 5.24 &     & & & & 6.1/5.5/6.5 E-4 \\

&    &   & $\rightarrow$ 6.18 &     & & & & 1.7/1.5/1.9 E-5 \\

&    &   & $\rightarrow$ 6.86  &          & & & & 1.7/1.2/1.9 E-5 \\

& d  &[$1/2^{+}$, 1.446] & $\rightarrow$ 5.18  & ad/lo/hi & -57.07 & 0.88 & 0.64 & 5.0/4.5/5.5 \\

&    &        & $\rightarrow$ 6.18  &     & & & & 5.0/4.5/6.0 E-3 \\
 

\ N.R. & s* & &$\rightarrow$ 0 & ad,hi & -35.0 & 1.00 & 0.55 & 0.015\\

     &    & &$\rightarrow$ 5.18  & all      & & & & 4.E-4 \\

     &    & &$\rightarrow$ 6.18  & ad/lo/hi & & & & 0.8/0.5/1.0 E-3 \\

     &    & &$\rightarrow$ 6.79  &          & & & & 7.8/7.5/8.0 E-2 \\

     & p  & &$\rightarrow$ 0     & ad/lo/hi & -40.0 & 1.00 & 0.55 & 11./11./12.\\

     &    & &$\rightarrow$ 5.24  &          & &  & & 0.09/0.07/0.12 \\

     &    & &$\rightarrow$ 6.79  &          & &  & & 0.50/0.45/0.60 \\

     &    & &$\rightarrow$ 6.86  &          & &  & & 0.40/0.35/0.50 \\

     &    & &$\rightarrow$ 7.28  &          & &  & & 0.95/0.76/1.25 \\

\ SUB R.& s & [$3/2^{+}, -0.504$] &   & add/lo/hi &  $a$ = & 5.8 & $\theta_{\rm p}^{2}$ = & 0.15/0.0/0.30 \\

\hline
\end{tabular*}
\label{n14pgTab1}
\end{table}

\begin{table}[!htbp]
\vskip-0.5cm
\caption{PM parameter values for \reac{15}{N}{p}{\gamma}{16}{O}.} \footnotesize\rm
\begin{tabular*}{\textwidth}{@{}l*{15}{@{\extracolsep{0pt plus12pt}}l}}
\hline
   & $l_{i}$ & [$J^{\pi}, E_{\rm R}$]&  &$V_{i}$&$r_{0}$&$a_{0}$& $\cal{C}_{\rm c}$ \\

\hline

\ R.C. & s & [$1^{-}$, 0.313] & ad       & -73.5  & 0.5  & 3.4  & 51.\\

      & &           & lo       & -68.0  & 0.6  & 3.4  & 61.\\   

      & &           & hi       & -72.0  & 0.4  & 3.6  & 6.7\\

      & s & [$1^{-}$, 0.963] & ad/lo/hi & -60.43 & 0.35 & 0.16 & 1.3/1.2/1.4\\

& &cascades&(see text)&  & & &\\

\ R.C. & s & [$2^{-}$, 0.669] & all      & -26.66 & 0.57 & 0.43 & 0.012\\

      & d & [$3^{-}$, 1.138] & all      & -42.71 & 1.0  & 0.86 & 0.040\\

\hline
\end{tabular*}
\label{n15pgTab1}
\end{table}

\clearpage

\begin{center}
{\bf Appendix B}
\end{center}
\vspace{3truemm}
Tabulated here are the reverse (endoergic) reaction rates that are defined by the product 
 REV$\times N_{\rm A}<\sigma\,v>$ of  REV [Eq.(\ref{eqrev})] and the forward reaction rates  $N_{\rm A}<\sigma\,v>$.
 Recall that they have different dimensions for transfer and photo-disintegration processes.
 In particular, the direct comparison of photo-disintegration rates with the forward capture rates is meaningless.
 In contrast, REV is dimensionless as far as the two-body transfer reactions considered in this compilation are concerned. 
The cases in which the forward reactions lead to three-body systems or $^{8}$Be are omitted.
Only the rates derived with the use of the "adopted" forward reaction rates are shown.
 
\vspace{3truemm}
\noindent
{\bf Transfer reactions}
\begin{table}[!htbo]
\caption{Backward two-body endoergic transfer reaction rates in $\rm{cm^{3}mol^{-1}s^{-1}}$}\footnotesize
\begin{tabular*}{\textwidth}{@{}l*{7}{@{\extracolsep{0pt plus12pt}}l }}
\hline
$T_{9}$&$^{3}$H(p,d)$^{2}$H & $^{3}$He(n.d)$^{2}$H & $^{3}$He($\alpha$,p)$^{6}$Li & $^{4}$He(n,d)$^{3}$H & $^{4}$He(p,d)$^{3}$He &  $^{4}$He ($\alpha$,p)$^{7}$Li & \\
\hline
   0.5  &          &  1.08E$-$26 &           &           &           &           &\\
   0.6  & 1.51E$-$27 &  4.36E$-$21 &  4.04E$-$28 &           &           &           &\\
   0.7  & 1.27E$-$22 &  4.48E$-$17 &  3.94E$-$23 &           &           &           &\\
   0.8  & 6.35E$-$19 &  4.62E$-$14 &  2.22E$-$19 &           &           &           &\\
   0.9  & 4.84E$-$16 &  1.04E$-$11 &  1.88E$-$16 &           &           &           &\\
   1.  & 9.91E$-$14 &  7.93E$-$10 &  4.16E$-$14 &           &           &           &\\
   1.25  & 1.47E$-$09 &  2.01E$-$06 &  7.17E$-$10 &           &           &           &\\
   1.5  & 9.08E$-$07 &  3.81E$-$04 &  4.94E$-$07 &           &           &           &\\
   1.75  & 9.05E$-$05 &  1.64E$-$02 &  5.39E$-$05 &           &           &           &\\
   2.  & 2.89E$-$03 &  2.79E$-$01 &  1.85E$-$03 &           &           &           &\\
   2.5  & 3.77E$-$01 &  1.51E+01 &  2.71E$-$01 &  6.77E$-$27 &  8.69E$-$29 &  7.37E$-$29 &\\
   3.  & 9.88E+00 &  2.19E+02 &  7.81E+00 &  4.99E$-$21 &  1.31E$-$22 &  7.19E$-$23 &\\
   3.5  & 1.03E+02 &  1.50E+03 &  8.85E+01 &  7.60E$-$17 &  3.32E$-$18 &  1.45E$-$18 &\\
   4.  & 6.04E+02 &  6.41E+03 &  5.56E+02 &  1.03E$-$13 &  6.60E$-$15 &  2.57E$-$15 &\\
   5.  & 7.32E+03 &  4.95E+04 &  7.54E+03 &  2.45E$-$09 &  2.68E$-$10 &  9.84E$-$11 &\\
   6.  & 3.93E+04 &  1.94E+05 &  4.41E+04 &  1.99E$-$06 &  3.10E$-$07 &  1.18E$-$07 &\\
   7.  & 1.32E+05 &  5.16E+05 &  1.58E+05 &  2.37E$-$04 &  4.67E$-$05 &  1.92E$-$05 &\\
   8.  & 3.28E+05 &  1.10E+06 &  4.17E+05 &  8.41E$-$03 &  1.99E$-$03 &  8.77E$-$04 &\\
   9.  & 6.68E+05 &  1.97E+06 &  8.96E+05 &  1.35E$-$01 &  3.62E$-$02 &  1.71E$-$02 &\\
  10.  & 1.18E+06 &  3.17E+06 &  1.66E+06 &  1.24E+00 &  3.64E$-$01 &  1.85E$-$01 &\\
\hline
  & & & & & & \\
\end{tabular*}
\end{table}
\vspace{-5truemm}
\begin{table}[!htbo]
\caption{Backward two-body endoergic transfer reaction rates in $\rm{cm^{3}mol^{-1}s^{-1}}$}\footnotesize
\begin{tabular*}{\textwidth}{@{}l*{7}{@{\extracolsep{0pt plus12pt}}l }}
\hline
$T_{9}$ & $^{6}$Li($\alpha$,p)$^{9}$Be & $^{7}$Be($\alpha$,p)$^{10}$B & $^{12}$C(n,$\alpha$)$^{9}$Be & $^{12}$C($\alpha$,p)$^{15}$N & $^{14}$N(n,$\alpha$)$^{11}$B & $^{16}$O(n,$\alpha$)$^{13}$C &  \\
\hline
   0.08 &    &           &    &    &  9.30E$-$24 &    &\\
   0.09 &    &           &    &    &  4.85E$-$21 &    &\\
   0.1 &    &           &    &    &  7.44E$-$19 &    &\\
   0.11 &    &           &    &    &  4.55E$-$17 &    &\\
   0.12 &    &           &    &    &  1.39E$-$15 &    &\\
   0.13 &    &           &    &    &  2.49E$-$14 &    &\\
   0.14 &    &           &    &    &  2.94E$-$13 &    &\\
   0.15 &    &           &    &    &  2.46E$-$12 &    &\\
   0.16 &    &           &    &    &  1.59E$-$11 &    &\\
   0.18 &    &           &    &    &  3.50E$-$10 &    &\\
   0.2 &    &  1.53E$-$26 &    &    &  4.18E$-$09 &    &\\
   0.25 &    &  2.54E$-$20 &    &    &  4.48E$-$07 &    &\\
&(to continue)&\\
\end{tabular*}
\end{table}
\clearpage
\begin{table}[!htbo]
\footnotesize
\begin{tabular*}{\textwidth}{@{}l*{7}{@{\extracolsep{0pt plus12pt}}l }}
$T_{9}$ & $^{6}$Li($\alpha$,p)$^{9}$Be & $^{7}$Be($\alpha$,p)$^{10}$B & $^{12}$C(n,$\alpha$)$^{9}$Be & $^{12}$C($\alpha$,p)$^{15}$N & $^{14}$N(n,$\alpha$)$^{11}$B & $^{16}$O(n,$\alpha$)$^{13}$C &  \\
\hline
   0.3  &          &  3.88E$-$16 &           &           &  1.62E$-$05 &           &\\
   0.35  & 8.97E$-$26 &  4.00E$-$13 &           &           &  2.83E$-$04 &           &\\
   0.4  & 1.13E$-$21 &  7.64E$-$11 &           &           &  2.65E$-$03 &           &\\
   0.45  & 1.79E$-$18 &  4.70E$-$09 &           &           &  1.56E$-$02 &  1.29E$-$26 &\\
   0.5  & 6.61E$-$16 &  1.30E$-$07 &           &           &  6.61E$-$02 &  1.86E$-$23 &\\
   0.6  & 4.82E$-$12 &  2.06E$-$05 &           &           &  6.33E$-$01 &  1.24E$-$18 &\\
   0.7  & 2.81E$-$09 &  8.28E$-$04 &           &           &  3.84E+00 &  3.92E$-$15 &\\
   0.8  & 3.35E$-$07 &  1.43E$-$02 &           &  6.11E$-$26 &  1.83E+01 &  1.74E$-$12 &\\
   0.9  & 1.39E$-$05 &  1.38E$-$01 &           &  2.75E$-$22 &  7.17E+01 &  2.06E$-$10 &\\
   1.  & 2.74E$-$04 &  8.92E$-$01 &  7.77E$-$25 &  2.31E$-$19 &  2.35E+02 &  9.63E$-$09 &\\
   1.25  & 5.84E$-$02 &  2.90E+01 &  1.87E$-$18 &  4.29E$-$14 &  2.36E+03 &  1.06E$-$05 &\\
   1.5  & 2.09E+00 &  3.30E+02 &  4.68E$-$14 &  1.44E$-$10 &  1.21E+04 &  1.26E$-$03 &\\
   1.75  & 2.70E+01 &  1.97E+03 &  7.33E$-$11 &  4.91E$-$08 &  4.05E+04 &  4.17E$-$02 &\\
   2.  & 1.84E+02 &  7.78E+03 &  1.89E$-$08 &  3.94E$-$06 &  1.03E+05 &  6.15E$-$01 &\\
   2.5  & 2.72E+03 &  5.54E+04 &  4.63E$-$05 &  1.87E$-$03 &  3.98E+05 &  2.93E+01 &\\
   3.  & 1.64E+04 &  2.11E+05 &  8.47E$-$03 &  1.15E$-$01 &  1.03E+06 &  4.06E+02 &\\
   3.5  & 5.93E+04 &  5.55E+05 &  3.52E$-$01 &  2.16E+00 &  2.11E+06 &  2.70E+03 &\\
   4.  & 1.56E+05 &  1.15E+06 &  5.81E+00 &  1.93E+01 &  3.76E+06 &  1.13E+04 &\\
   5.  & 6.02E+05 &  3.24E+06 &  3.03E+02 &  4.07E+02 &  9.04E+06 &  8.59E+04 &\\
   6.  & 1.47E+06 &  6.47E+06 &  4.39E+03 &  3.05E+03 &  1.72E+07 &  3.48E+05 &\\
   7.  & 2.80E+06 &  1.05E+07 &  3.05E+04 &  1.27E+04 &  2.81E+07 &  1.01E+06 &\\
   8.  & 4.49E+06 &  1.54E+07 &  1.34E+05 &  3.67E+04 &  4.10E+07 &  2.42E+06 &\\
   9.  & 6.49E+06 &  2.03E+07 &  4.27E+05 &  8.33E+04 &  5.50E+07 &  5.09E+06 &\\
  10.  & 8.69E+06 &  2.55E+07 &  1.10E+06 &  1.60E+05 &  6.97E+07 &  9.78E+06 &\\
\hline
\end{tabular*}
\end{table}
\noindent
{\bf Photo-disintegrations}
\begin{table}[!htbo]
\caption{Backward photodisintegration reaction rates in $\rm{s^{-1}}$}\footnotesize
\begin{tabular*}{\textwidth}{@{}l*{7}{@{\extracolsep{0pt plus12pt}}l }}
\hline
$T_{9}$&$^{3}$He($\gamma$,p)$^{2}$H & $^{4}$He($\gamma$,d)$^{2}$H & $^{6}$Li($\gamma$,$\alpha$)$^{2}$H & $^{7}$Li($\gamma$,$\alpha$)$^{3}$H & $^{7}$Be($\gamma$,p)$^{6}$Li & $^{7}$Be ($\gamma$,$\alpha$)$^{3}$He & \\
\hline
   0.35 &           &           &  5.85E$-$15 &           &           &  4.25E$-$15 &\\
   0.4 &           &           &  5.09E$-$12 &           &           &  7.38E$-$12 &\\
   0.45 &           &           &  1.04E$-$09 &  2.43E$-$17 &           &  2.62E$-$09 &\\
   0.5 &           &           &  7.68E$-$08 &  2.18E$-$14 &           &  3.00E$-$07 &\\
   0.6 &           &           &  5.45E$-$05 &  6.29E$-$10 &           &  4.09E$-$04 &\\
   0.7 &           &           &  6.66E$-$03 &  1.03E$-$06 &           &  7.81E$-$02 &\\
   0.8 &           &           &  2.71E$-$01 &  2.80E$-$04 &           &  4.28E+00 &\\
   0.9 &           &           &  5.29E+00 &  2.26E$-$02 &           &  1.01E+02 &\\
   1. &  1.11E$-$15 &   &  6.05E+01 &  7.79E$-$01 &  1.50E$-$16 &  1.32E+03 &\\
   1.25 &  7.37E$-$10 &   &  5.76E+03 &  4.98E+02 &  1.41E$-$10 &  1.51E+05 &\\
   1.5 &  6.16E$-$06 &   &  1.34E+05 &  4.02E+04 &  1.46E$-$06 &  3.93E+06 &\\
   1.75 &  4.16E$-$03 &   &  1.34E+06 &  9.75E+05 &  1.12E$-$03 &  4.32E+07 &\\
   2. &  5.77E$-$01 &   &  7.79E+06 &  1.11E+07 &  1.70E$-$01 &  2.74E+08 &\\
   2.5 &  6.34E+02 &   &  9.63E+07 &  3.61E+08 &  2.05E+02 &  4.01E+09 &\\
   3. &  7.30E+04 &   &  5.39E+08 &  3.95E+09 &  2.46E+04 &  2.64E+10 &\\
   3.5 &  2.31E+06 &   &  1.94E+09 &  2.31E+10 &  7.90E+05 &  1.08E+11 &\\
   4. &  3.23E+07 &  2.77E$-$18 &  5.25E+09 &  8.99E+10 &  1.10E+07 &  3.28E+11 &\\
   5. &  1.41E+09 &  5.72E$-$12 &  2.33E+10 &  6.47E+11 &  4.69E+08 &  1.72E+12 &\\
   6. &  1.90E+10 &  1.03E$-$07 &  6.91E+10 &  2.62E+12 &  6.13E+09 &  5.70E+12 &\\
   7. &  1.29E+11 &  1.20E$-$04 &  1.62E+11 &  7.05E+12 &  4.05E+10 &  1.42E+13 &\\
   8. &  5.68E+11 &  2.51E$-$02 &  3.22E+11 &  1.54E+13 &  1.74E+11 &  2.92E+13 &\\
   9. &  1.85E+12 &  1.65E+00 &  5.69E+11 &  2.85E+13 &  5.59E+11 &  5.24E+13 &\\
  10. &  4.91E+12 &  4.78E+01 &  9.23E+11 &  4.70E+13 &  1.46E+12 &  8.50E+13 &\\
\hline
\end{tabular*}
\end{table}
%
\begin{table}[!htbo]
\caption{Backward photodisintegration reaction rates in $\rm{s^{-1}}$}\footnotesize
\begin{tabular*}{\textwidth}{@{}l*{7}{@{\extracolsep{0pt plus12pt}}l }}
\hline
$T_{9}$& $^{8}$B($\gamma$,p)$^{7}$Be &$T_{9}$& $^{8}$B($\gamma$,p)$^{7}$Be &$T_{9}$& $^{8}$B($\gamma$,p)$^{7}$Be & \\
\hline
   0.04&      1.01E$-$16&    0.12&      8.64E$-$01  &    0.3&      1.27E+06              &              \\
   0.05&      3.05E$-$12&    0.13&      4.43E+00    &    0.35&      6.65E+06           &                \\
   0.06&      3.72E$-$09&    0.14&      1.86E+01    &    0.4&      2.48E+07           &                 \\
   0.07&      6.99E$-$07&    0.15&      6.56E+01    &    0.45&      7.32E+07            &                         \\
   0.08&      3.98E$-$05&    0.16&      2.03E+02    &    0.5&      1.82E+08            &                         \\
   0.09&      1.00E$-$03&    0.18&      1.40E+03    &    0.6&      7.90E+08           &                         \\
   0.1&      1.42E$-$02&     0.2&      6.92E+03     &    0.7&      2.49E+09           &                        \\
   0.11&      1.30E$-$01&    0.25&      1.44E+05    &    0.8&      6.40E+09             &                         \\
\hline
\end{tabular*}
\end{table}
\begin{table}[!htbo]
\caption{Backward photodisintegration reaction rates in $\rm{s^{-1}}$}\footnotesize
\begin{tabular*}{\textwidth}{@{}l*{7}{@{\extracolsep{0pt plus12pt}}l }}
\hline
$T_{9}$& $^{8}$B($\gamma$,p)$^{7}$Be & $^{10}$B($\gamma$,p)$^{9}$Be & $^{11}$B($\gamma$,$\alpha$)$^{7}$Li & $^{11}$C($\gamma$,p)$^{10}$B &  $^{11}$C($\gamma$,$\alpha$)$^{7}$Be & $^{12}$C($\gamma$,p)$^{11}$B & \\
\hline
   0.9 &  1.42E+10 &  &  &  &  &   &\\
   1. &  2.80E+10 &  &  &  &  &   &\\
   1.25 &  1.08E+11 &  6.49E$-$14 &           &           &           &           &\\
   1.5 &  2.95E+11 &  3.89E$-$09 &  1.93E$-$16 &  4.01E$-$17 &  5.95E$-$13 &           &\\
   1.75 &  6.42E+11 &  1.12E$-$05 &  5.47E$-$12 &  1.18E$-$12 &  5.73E$-$09 &           &\\
   2. &  1.20E+12 &  4.67E$-$03 &  1.22E$-$08 &  2.82E$-$09 &  5.72E$-$06 &           &\\
   2.5 &  3.13E+12 &  2.32E+01 &  6.26E$-$04 &  1.67E$-$04 &  9.52E$-$02 &  1.67E$-$17 &\\
   3. &  6.42E+12 &  7.02E+03 &  9.24E$-$01 &  2.73E$-$01 &  6.48E+01 &  6.42E$-$12 &\\
   3.5 &  1.12E+13 &  4.22E+05 &  1.81E+02 &  5.66E+01 &  7.01E+03 &  6.53E$-$08 &\\
   4. &  1.79E+13 &  9.15E+06 &  9.82E+03 &  3.22E+03 &  2.42E+05 &  6.79E$-$05 &\\
   5. &  3.77E+13 &  6.88E+08 &  2.86E+06 &  9.89E+05 &  3.59E+07 &  1.20E+00 &\\
   6. &  6.77E+13 &  1.23E+10 &  1.33E+08 &  4.85E+07 &  1.05E+09 &  8.58E+02 &\\
   7. &  1.10E+14 &  9.82E+10 &  2.13E+09 &  8.33E+08 &  1.21E+10 &  9.76E+04 &\\
   8. &  1.65E+14 &  4.69E+11 &  1.74E+10 &  7.35E+09 &  7.71E+10 &  3.55E+06 &\\
   9. &  2.37E+14 &  1.60E+12 &  8.98E+10 &  4.13E+10 &  3.29E+11 &  6.06E+07 &\\
  10. &  3.24E+14 &  4.28E+12 &  3.37E+11 &  1.70E+11 &  1.06E+12 &  6.07E+08 &\\
\hline
\end{tabular*}
\end{table}
\begin{table}[!htbo]
\caption{Backward photodisintegration reaction rates in $\rm{s^{-1}}$}\footnotesize
\begin{tabular*}{\textwidth}{@{}l*{7}{@{\extracolsep{0pt plus12pt}}l }}
\hline
$T_{9}$ & $^{13}$N($\gamma$,p)$^{12}$C & $^{14}$N($\gamma$,p)$^{13}$C & $^{14}$O($\gamma$,p)$^{13}$N & $^{15}$O($\gamma$,p)$^{14}$N & $^{16}$O($\gamma$,p)$^{15}$N &$^{16}$O($\gamma$,$\alpha$)$^{12}$C &  \\
\hline
   0.4 &  2.32E$-$15 &    &    &    &    &    &\\
   0.45 &  3.94E$-$12 &    &    &    &    &    &\\
   0.5 &  1.58E$-$09 &    &    &    &    &    &\\
   0.6 &  1.32E$-$05 &    &   &    &    &    &\\
   0.7 &  8.64E$-$03 &    &   &    &    &    &\\
   0.8 &  1.13E+00 &           &  4.57E$-$17 &    &    &   &\\
   0.9 &  5.04E+01 &           &  1.81E$-$13 &   &    &   &\\
   1. &  1.05E+03 &           &  1.38E$-$10 &   &    &   &\\
   1.25 &  2.54E+05 &  5.11E$-$17 &  2.14E$-$05 &  2.26E$-$17 &    &   &\\
   1.5 &  9.88E+06 &  1.34E$-$11 &  6.20E$-$02 &  3.02E$-$12 &           &  2.00E$-$17 &\\
   1.75 &  1.36E+08 &  1.01E$-$07 &  1.85E+01 &  1.51E$-$08 &           &  2.57E$-$13 &\\
   2. &  9.70E+08 &  8.15E$-$05 &  1.33E+03 &  9.72E$-$06 &  3.47E$-$16 &  3.66E$-$10 &\\
   2.5 &  1.54E+10 &  9.77E$-$01 &  5.34E+05 &  1.00E$-$01 &  1.11E$-$09 &  1.30E$-$05 &\\
   3. &  9.84E+10 &  5.19E+02 &  2.91E+07 &  5.58E+01 &  2.56E$-$05 &  1.78E$-$02 &\\
   3.5 &  3.76E+11 &  4.66E+04 &  5.08E+08 &  5.66E+03 &  3.40E$-$02 &  3.65E+00 &\\
   4. &  1.04E+12 &  1.38E+06 &  4.35E+09 &  1.94E+05 &  7.61E+00 &  2.25E+02 &\\
   5. &  4.46E+12 &  1.66E+08 &  8.95E+10 &  3.08E+07 &  1.51E+04 &  9.54E+04 &\\
   6. &  1.22E+13 &  4.24E+09 &  6.88E+11 &  9.97E+08 &  2.39E+06 &  7.01E+06 &\\
   7. &  2.54E+13 &  4.45E+10 &  3.06E+12 &  1.27E+10 &  9.00E+07 &  1.73E+08 &\\
   8. &  4.49E+13 &  2.66E+11 &  9.76E+12 &  8.91E+10 &  1.37E+09 &  2.02E+09 &\\
   9. &  7.04E+13 &  1.09E+12 &  2.50E+13 &  4.16E+11 &  1.14E+10 &  1.42E+10 &\\
  10. &  1.01E+14 &  3.41E+12 &  5.51E+13 &  1.45E+12 &  6.23E+10 &  6.83E+10 &\\
\hline
\end{tabular*}
\end{table}
 
\vskip1.0cm

{\it Note added in-proof}:  The details of Ref.\,[175] are now available, for which see
 J.J. He, S.Z. Chen, C.E. Rolfs, S.W. Xu, J. Hu, X.W. Ma, M. Wiescher, R.J. deBoer, T. Kajino, M. Kusakabe, L.Y. Zhang, S.Q. Hou, X.Q. Yu, N.T. Zhang, G. Lian, Y.H. Zhang, X.H. Zhou, H.S. Xu, G.Q. Xiao, W.L. Zhan, Phys. Lett. B 725 (2013) 287-291.

\end{document}